\documentclass[final,a4paper]{unswthesis}
\pdfoutput=1
\usepackage{comment}
\usepackage{mythesis}
\usepackage[round]{natbib}
\usepackage{amsmath}
\usepackage{amssymb}
\usepackage[pdftex]{graphics}
\usepackage{booktabs}
\usepackage{rotating}

\allowdisplaybreaks[1]

\begin{document}
\frontmatter

\maketitle

\abstract{Magnetic fields play an important role in star formation by
regulating the removal of angular momentum from collapsing molecular cloud
cores. Hall diffusion is known to be important to the magnetic field behaviour
at many of the intermediate densities and field strengths encountered 
during the gravitational collapse of molecular cloud cores into protostars,
and yet its role in the star formation process is not well-studied. This
thesis describes a semianalytic self-similar model of the collapse of rotating
isothermal molecular cloud cores with both Hall and ambipolar diffusion,
presenting similarity solutions that demonstrate that the Hall effect has a
profound influence on the dynamics of collapse. 

Two asymptotic power law similarity solutions to the collapse equations on the
inner boundary are derived. The first of these represents a Keplerian disc in
which accretion is regulated by the magnetic diffusion; with an appropriate
value of the Hall diffusion parameter a stable rotationally-supported disc
forms, but when the Hall parameter has the opposite sign disc formation is
suppressed by the strong diffusion. The second solution describes the infall
when the magnetic braking is so efficient at removing angular momentum from
the core that no disc forms and the matter free falls onto the protostar. 

The full similarity solutions show that the size and sign of the Hall
parameter can change the size of the protostellar disc by up to an order of
magnitude and the accretion rate onto the protostar by $1.5 \times 10^{-6}$
M$_\odot$ yr$^{-1}$ when the ratio of the Hall to ambipolar diffusion
parameters moves between the extremes of $-0.5 \le \tilde{\eta}_H /
\tilde{\eta}_A \le 0.2$. These variations (and their dependence upon the
orientation of the magnetic field with respect to the axis of rotation)
create a preferred handedness to the solutions that could be observed in
protostellar cores using next-generation instruments such as ALMA. 

Hall diffusion also determines the strength of the magnetic diffusion and
centrifugal shocks that bound the pseudo and rotationally-supported discs, and
can introduce subshocks that further slow accretion onto the protostar. In
cores that are not initially rotating Hall diffusion can even induce rotation,
which could give rise to disc formation and resolve the magnetic braking
catastrophe. The Hall effect clearly influences the dynamics of gravitational
collapse and its role in controlling the magnetic braking and radial diffusion
of the field would be worth exploring in future numerical simulations of star
formation.} 

\setcounter{page}{1}
\newpage
\statement{\vspace{5mm} Catherine Braiding (30615399) \hspace{20mm} 
	15 / 7 / 2011 }
\newpage

\acknowledgements{\noindent Writing this thesis has been the hardest thing I
have ever done; over the course of the last few years my self-esteem and sense
of self-worth have plummeted, and I could not have finished without the
support of a great number of people. To all of them I give my heartfelt
thanks, but in particular I would like to acknowledge: 

\begin{itemize}
    \item my supervisor, Mark Wardle, who is brilliant and funny and has
always supported me, even when he was overburdened with too many PhD students
to supervise and theses to read. You inspire me, and if I continue on this
path it will be thanks to you. 
    \item Gemma and James, for all the hugs, tears and unconditional love, as
well as the gifts of backup drives. You make me a better person; when times
were rocky you were there for me and I love you both. 
    \item Korinne, Sarah and Anna, the best PhD buddies and officemates anyone
could ask for. Thanks for the songs, the cupcakes, the hugs, and making it
through to the end with me --- I'm proud of all of us.  
    \item Korinne again, because she proofread this beast twice. Thanks, crazy
lady. 
    \item my family, for everything. I promise I'll get a real job soon.
    \item the department of physics and astronomy and others, particularly
Alan, Judith and Carol, for the moral and occasional monetary support.
    \item Raquel, Pandey, and the participants of the \textit{DDP} and
\textit{CC2YSO} workshops in 2009 and 2010, for the engaging and stimulating
discussions. 
    \item finally, my closest friends: Brendan, Heidi, Tony, Nilanka, Roberto
and Drew; the West Ryde knitters; and all the other friends, online and off,
who have listened to my rants and offered sympathy as needed. May all of you
continue to be awesome. 
\end{itemize}

\noindent Thank you all,

Catherine}
\newpage
\thispagestyle{empty}
\cleardoublepage

\tableofcontents
\cleardoublepage

\listoffigures
\listoftables
\newpage
\thispagestyle{empty}
\cleardoublepage

\quotepage{\begin{quote}
``It seems to me, that if the matter of our sun and planets, and all the
matter of the universe, were evenly scattered throughout all the heavens, and
every particle had an innate gravity towards all the rest, and the whole space
throughout which this matter was scattered, was finite, the matter on the
outside of this space would by its gravity tend towards the matter on the
inside, and by consequence fall down into the middle of the whole space, and
there compose one great spherical mass. But if the matter were evenly disposed
throughout an infinite space, it could never convene into one mass; but some
of it would convene into one mass and some into another, so as to make an
infinite number of great masses, scattered great distances from one to another
throughout all that infinite space. And thus might the sun and fixed stars be
formed, supposing the matter were of a lucid nature."
\end{quote}
- Newton to Bentley (December 10, 1692), quoted by \citet{j1928}}
\newpage
\thispagestyle{empty}
\cleardoublepage

\mainmatter

\chapter{Star Formation}\label{ch:litrev}

In the seventeenth century Sir Isaac Newton pondered gravity's influence on
the formation of stars and planets and the consequences of such star formation
on an unbounded interstellar medium \citep[Newton to Bentley, December 10,
1692; as quoted by][]{j1928}. His assertion that the gradual growth of
inhomogeneities in the material that forms stars could lead to runaway
gravitational collapse remains at the heart of star formation theory today,
and it is from this basic description that our modern understanding of star
formation by the gravitational condensation of diffuse matter in space under
the influence of rotational and magnetic effects has evolved. 

Understanding star formation is critical to our understanding of the universe,
as stars are the fundamental objects of astronomy. Star formation determines
the structure, evolution and luminosity of galaxies \citep[e.g.][]{fb2002};
planet formation and evolution occurs in protoplanetary discs as a result of
the star formation process \citep{s1967} and most of the elements that are not
hydrogen are made in stars \citep{h1946} --- so it is no surprise that the
intricacies of star formation are the focus of many studies in astronomy. The
questions surrounding how molecular clouds are formed and supported against
gravity \citep[e.g.][]{tt2009}; what triggers their collapse
\citep[e.g.][]{cf1953, f1978, ns1980}; the detailed dynamics of the collapse
throughout the different stages of the star formation process
\citep[e.g.][]{l1969, p1969, s1977, tsc1984, sal1987, sp2004, mo2007} and the
importance of the magnetic field and the angular momentum of the initial cloud
in determining the final properties of the protostar and its protostellar disc
\citep[e.g.][]{ms1956, s1966, b1998, db2010} are of vital importance to our
understanding of the universe. 

In the last fifty years the increasing availability of computers and numerical
techniques enabled astronomers to simulate gravitational collapse and star
formation with increasing resolution and complexity \citep{kkh2009, m2010},
while observations at infrared and radio wavelengths have started to unveil
the physical processes at work in the molecular cloud cores from which stars
form and determine the characteristics that distinguish the youngest stars
from their host clouds \citep[e.g.][]{fetal2007, wetal2007}. Both theoretical
and observational studies of star formation have gradually converged on the
consensus theories that low-mass stars form as a result of the gravitational
collapse of dense molecular cloud cores, while higher mass stars form under
the influence of more complicated processes, such as the fragmentation of
molecular cloud clumps, the merger of smaller stars, and turbulent motions
within the molecular clouds \citep[e.g.][]{bc2004, mk2004, bkmv2007}. 

In the ``standard model'' of star formation proposed by \citet{sal1987} it is
assumed that magnetic fields and thermal pressure initially provide pressure
support against gravity in these molecular cloud cores and carry away excess
angular momentum; however, as the gas is weakly ionised it is not in a
strictly-steady state and gradually contracts. The neutral molecules are
pulled inwards by gravity and drift inward through the magnetic field (which
is supported by ions) in the process of ambipolar diffusion \citep{s1978}.
Once the density is high enough that the magnetic field is no longer able to
support the core, it dynamically collapses to form a protostar, usually
(although not always) surrounded by a protostellar disc from which it accretes
further mass. The protostar quickly comes to gravitationally dominate a
progressively larger region of the molecular cloud in which it has formed
\citep{sal1987}. 

This introductory chapter outlines the previous research and current theories
of primarily low-mass star formation, to motivate the research into the
influence of Hall diffusion on the dynamics of gravitational collapse and 
accretion disc physics that is described within the rest of this thesis. The
properties of molecular clouds and the star forming cores within them are
described first, to provide an overview of the initial conditions of collapse,
followed by a brief description of Hall magnetohydrodynamics, emphasising the
importance of magnetic field diffusion on the dynamics of weakly ionised gases
such as those found in molecular clouds. 

The current state of gravitational collapse studies is then explored across
several stages of the star formation process, with particular focus on the
effects of rotation and the magnetic field on the collapsing core. The current
major problem of star formation simulations --- the ``magnetic braking
catastrophe'', in which all of the angular momentum is removed from the
collapsing core by magnetic braking --- is outlined, along with recent
approaches towards solving this problem and predictions for how Hall
magnetohydrodynamics shall affect the magnetic braking behaviour in the flow.
Finally, an overview of the thesis, its aims and primary results shall be
presented with reference to existing theories of star formation. Should
readers wish for more detailed information on the current state of star
formation research than is presented in this chapter, they are directed to the
thorough reviews by \citet{mo2007} and \citet{m2010}, and the references cited
therein. 

\section{Molecular Clouds}\label{lr:mc}

Stars form in molecular clouds, generally dark regions of the interstellar
medium in which the density and temperature of the gas permit the formation of
molecules. The gas is primarily composed of molecular hydrogen, which is
difficult to detect observationally, and so the gas is traced and the mass
inferred from its luminosity in the $J = 1 - 0$ line of $^{12}$CO or
$^{13}$CO, which are the dominant carbon-bearing species \citep{ldbbtvw2000}.

Giant molecular clouds make up most of the mass of the interstellar medium;
they have masses $10^4$--$10^6$ M$_{\odot}$, diameters $\sim30$--$50$ pc and
volume-averaged number densities of $n_\text{H} \sim 10^2$--$10^4$ cm$^{-3}$
\citep[][and the references within these]{awb2000, wbm2000, fetal2007}. They
are in general not gravitationally bound \citep{dbp2011}, and are surrounded
by a layer of less dense atomic gas which shields the molecules from the
interstellar UV radiation field that is capable of dissociating molecules and
leads to a low rate of heating by external radiation \citep{e1993}. Giant
molecular clouds may contain several sites of star formation, and while
smaller molecular clouds with masses {$\lesssim 10^2$ M$_\odot$} may also form
stars, their contribution to the total star formation rate in the Galaxy is
negligible \citep{mcbb1995}. 

Molecular clouds have a hierarchical structure and the terminology used to
describe the differently-scaled features varies across papers; the conventions
outlined in \citet{mo2007} are followed here. Overdense coherent regions in
$l$-$b$-$v$ (galactic longitude, galactic latitude and radial velocity) space
identified from spectral line maps of molecular emission within molecular
clouds are referred to as \textit{clumps}. \textit{Cluster-forming clumps} are
the massive clumps out of which stellar clusters form; they are
gravitationally bound or bound by the pressure of the interclump medium even
though most clusters are unbound \citep{wbs1995}. Molecular cloud
\textit{cores} are the particularly dense self-gravitating regions out of
which single stars (or small multiple systems such as binaries) form. Not all
of the material that goes into forming a star must come from the core --- some
may be accreted from the surrounding clump or cloud as the protostar moves
through it \citep{bbcp1997}. The physical properties of molecular clouds,
clumps and cores are summarised in Table \ref{tab-mcprop}
\citep[from][]{kkh2009}. 
\begin{table}
\begin{center}
\vspace{-1mm}
\begin{tabular}{llll}
\toprule
  & molecular clouds & cluster-forming & protostellar\\
  & 		    & clumps 	      &  cores\\
 \midrule
 Size (pc) 		& 2--50 	& 0.1--2	& $\lesssim 0.1$\\
 Density ($n_\text{H}$ cm$^{-3}$)	& $10^2$--$10^4$ & $10^3$--$10^5$ & $> 10^5$\\
 Mass (M$_\odot$)	& $10^2$--$10^6$ & 10--$10^3$ 	& 0.1--10\\
 Temperature (K)	& 10--30 	& 10--20	& 7--12\\ 
 Line width (km s$^{-1}$) & 1--10	& 0.3--3	& 0.2--0.5\\
 Column density (g cm$^{-2}$) & 0.03	& 0.03--1.0	& 0.3--3\\
 Column density (M$_\odot$ pc$^{-2}$) & $\sim$144 & 144--5000	& 1500--15000\\
 Crossing time (Myr)	& 2--10		& $\lesssim$ 1	& 0.1--0.5\\
 Free fall time (Myr)	& 0.3--3	& 0.1--1	& $\lesssim$ 0.1\\
 Examples		& Taurus,	& L1641, L1709	& B68, L1544\\
 &  Ophiuchus & & \\
 \bottomrule
 \end{tabular}
\end{center}
\vspace{-5mm}
\caption[Physical properties of clouds, clumps and cores
\citep{kkh2009}]{Physical properties of molecular clouds, cluster-forming
clumps and isolated protostellar cores from \citet{kkh2009}.} 
\label{tab-mcprop}
\vspace{-2mm}
\end{table}

Molecular cloud cores that form single or binary low-mass stars (as opposed to
clumps that form clusters and very massive stars) such as those in the Taurus,
Orion and Ophiuchus molecular clouds have typical diameters $\lesssim0.1$ pc,
number densities $n_\text{H} \gtrsim3\times10^{4}$ cm$^{-3}$ and masses
ranging from a fraction of a solar mass to $10$ M$_{\odot}$, although the
masses can be uncertain by a factor of $\sim3$ due to uncertainties in the
measured distance, dust opacity, molecular abundances and calibration
\citep{fetal2007}. The core morphologies depend on the intensity level chosen
to mark the boundary between the core and the host cloud, as well as the
wavelength range and angular resolution of the observations. In general, cores
that are round (in projection against the sky) are interpreted as being
spherical in shape, while cores that are elongated (again, in projection) are
typically interpreted as being either oblate or prolate spheroids; statistical
analyses have shown that most isolated cores are oblate rather than prolate
\citep[in contrast with the results of Ryden (1996) who assumed axisymmetry in
their analysis]{jbd2001,jb2002}; and that starless cores have more extreme
axial ratios than protostellar cores \citep[cores in which a protostar has
formed;][]{gww2002}. 

Mappings of molecular cloud cores show that they often display nonzero initial
infall velocities before they contain infrared sources (protostars).
\citet{lmt2001} mapped 53 targets and identified 19 candidate collapsing cores
that demonstrated infall based upon the shape of the CS lines and analysis of
both the thick CS and the thin N$_2$H$^+$ peak velocities. They derived
one-dimensional infall speeds for these starless cores to obtain typical
values of 0.05--0.09 km s$^{-1}$, corresponding to mass infall rates of about
10$^{-6}$--10$^{-5}$ M$_\odot$ yr$^{-1}$. These observations were consistent
with both of the dominant models for the removal of support against gravity
causing the contraction of molecular cloud cores: ambipolar diffusion and
dissipation of turbulence theories \citep[][these theories shall be explored
in detail in Section \ref{lr:col}]{fam2004}. 

Approximately 1\% of the mass of a molecular cloud is in the form of dust
particles or grains, on which hydrogen molecules form by condensation
\citep{s1978}. The rate at which this process occurs increases with density,
so that the abundance of molecules within the gas depends also upon density
\citep{l2003}. The opacity of the dust shields the gas from the UV radiation
that would normally dissociate the molecules; the continued existence of
molecules requires that the host cloud possess a column density of at least
$20$ M$_\odot$ pc$^{-2}$ \citep{e1993}. Typical molecular clouds are quite
opaque and have column densities in excess of $100$ M$_\odot$ pc$^{-2}$ (see
Table \ref{tab-mcprop}). 

About half of the known interstellar molecular species have been detected in
the single core TMC-1 \citep[Taurus Molecular Cloud-1;][]{ok1998}. Surveys of
isolated molecular cloud cores show that they are chemically differentiated
with NH$_3$ and N$_2$H$^+$ dominant in the denser inner regions, and C$_2$S,
CS and other carbon-rich molecules more abundant in the diffuse outer regions
\citep{klv1996, lbah2003}; the ratio of C$_2$S to N$_2$H$^+$ abundances has
been proposed as an indicator of the time since the gas was atomic carbon-rich
\citep{bl1997, ldbbtvw2000}. In regions of massive star and cluster formation
hot cores are produced (in which the temperature is $> 50$ K); these may
contain more complex organic molecules such as CH$_3$OH, CH$_3$CH$_2$OH and
HCOOCH$_3$ and typically have a large deuterium fraction \citep{klv1996,
i1999}. The chemical differences observed between regions of isolated and
clustered star formation, such as the presence of complex organic molecules,
may (if retained in comets) provide a way of tracing the formation conditions
of the solar system \citep{i1999}. 

Molecular clouds are generally cold, with constant temperatures of around $T
\simeq 10$ K across a wide range of densities \citep[e.g.][]{hn1965, l1985,
mi2000}. In regions of cluster formation compressional heating and radiative
heating from new stars can cause the temperature of individual cores to rise
to 100 K or higher, and when the density is particularly high the collapsing
gas becomes thermally-coupled to the dust; despite this the temperature of
cores typically remains constant so long as the density remains below $\sim
10^{10}$ cm$^{-3}$, above which radiative trapping occurs and the temperature
increases \citep[e.g.][]{wn1980, wk1993, ck1998}. This corresponds to the
formation of the ``first'' (opaque) core, which is followed later by dynamic
collapse to the ``second'' core (the protostar) once molecular hydrogen
dissociates \citep[e.g.][]{l1969}. The rate at which molecular clouds are
cooled by collisionally-excited atomic and molecular emission processes is
high --- this emission primarily takes the form of far-infrared radiation from
molecules such as CO, which is the dominant coolant in molecular clouds
\citep{mswg1982, g1984}. 

The near-constant low temperature across molecular clouds is an important
feature of the star formation process because of its influence on the Jeans
mass, and it is what makes possible the collapse of prestellar cloud cores
with masses as small as one solar mass \citep{l2003}. The Jeans mass is the
critical mass at which a cloud becomes unstable and starts to collapse, as it
possesses insufficient pressure support to balance the force of gravity. It is
given by the equation 
\begin{equation}
    M_J = \left(\frac{\pi k T}{\mu m_\text{H}G}\right)^{3/2}\rho^{-1/2}
	= 18 \text{M}_\odot \left(\frac{T}{10 \text{ K}}\right)^{3/2}
	  \!\biggl(\frac{n_\text{H}}{50 \text{ cm}^{-3}}\biggr)^{-1/2}
\label{jeans}
\end{equation}
where $\rho$ is the density and $\mu$ is the mean atomic mass per particle,
which is usually taken to be $2.29$ in a fully molecular cloud that is 25\% 
Helium by mass and 10\% by number \citep{j1928}. In the absence of pressure or
other support, gravitational collapse of such a cloud will occur in a free
fall time, 
\begin{equation} 
    t_{f\!f} = \left(\frac{3\pi}{32G\rho}\right)^{3/2} 
	= 3.4\times 10^7
	\biggl(\frac{n_\text{H}}{50 \text{ cm}^{-3}}\biggr)^{-1/2}
	\text{ years}
\label{fftime}
\end{equation}
\citep{s1978}. For a cloud with typical temperature $T = 10$ K and density
$n_\text{H} \ge 50$ cm$^{-3}$, the Jeans mass is $M_J \le 80$ M$_\odot$ and
the free fall time is calculated to be $t_{f\!f} \le 5 \times 10^6$ years.
This suggests that molecular clouds are highly unstable; if thermal pressure
support was the only mechanism holding up molecular clouds then free fall
collapse would lead to a galactic star formation rate of $\dot{M}_\star \ge
200$ M$_\odot$ yr$^{-1}$, which is far in excess of the observed galactic
average of $\sim 3$ M$_\odot$ yr$^{-1}$. 

Molecular clouds appear to have lifetimes of about $\sim 10^6$--$10^7$ years;
longer lifetimes would imply that they cannot all be collapsing at free fall
speeds (as this would give a much higher observed star formation rate) --- in
general, they must be supported by another mechanism \citep{zp1974, e1999,
awb2000, wetal2007}. Various alternatives to thermal pressure support of
molecular clouds have been theorised: magnetic fields \citep[e.g.][]{cf1953,
bm1994, as2007}, rotation \citep[e.g.][]{f1978, e1999} and turbulence
\citep[e.g.][]{ns1980, mk2004, bkmv2007}. All of these likely contribute in
some measure, with magnetic fields the most important --- without magnetic
fields it would be difficult to explain the generation of observed levels of
rotation and turbulence in molecular clouds. 

The interstellar medium is strongly magnetised, and magnetic fields are
important to the dynamics and evolution of molecular clouds \citep{mzgh1993}.
The most useful measure of the magnetic field is the Zeeman effect, which
measures the line-of-sight component of the field, as opposed to the
Chandrasekhar-Fermi method which measures the component of the field in the
plane of the sky. The morphology of the field is measured from dust
polarisation and the linear polarisation of spectral lines. The largest
compilation of Zeeman measurements of the magnetic field strengths in
molecular clouds was performed by \citet{c1999}, who found that the median value
of the Alfv\'en Mach number across 27 clouds of varying mass was
$\mathcal{M}_A \simeq 1$ (although the observations also found a median
temperature of 40 K, which is much higher than the average temperature typically
measured in molecular clouds). From this the median value of the magnetic
field was calculated to be  
\begin{equation}
B_\text{med} \simeq 30 \left(\frac{n_\text{H}}{10^3 \text{ cm}^{-3}}\right)^{1/2}
\left(\frac{\sigma_{\text{nt}}}{1\text{ km s}^{-1}}\right) \mu \text{G}
\end{equation}
where $n_{\text{H}} \gtrsim 2 \times 10^3$ cm$^{-3}$ and $\sigma_{\text{nt}}$
is the non-thermal velocity dispersion of the cloud. $\sigma_{\text{nt}}$ is
used as a measure of the amount of turbulence in a volume, and is typically
observed to be $\propto r^q$ on large scales, where $q \simeq 1/2$
\citep{gbwh1998}. 

\citet{c1999} also found that the observed molecular cloud clumps and cores
were in approximate virial equilibrium, that is, that the kinetic and magnetic
energies were in approximate equipartition; this implies that the gas motions
must provide a significant component of the support of molecular clouds
against gravity. The gas velocities, calculated by examining the
Doppler-broadening of spectral linewidths, were found to be supersonic (by
about a factor of five) and approximately equal to the Alfv\'en velocity,
which suggests that the internal motions in the gas were likely the result of
magnetohydrodynamic (MHD) waves. 

The average ratio of thermal to magnetic pressures across the 15 cores for
which such measurements were possible was found to be $\beta_\text{mag}
\approx 0.04$, where $\beta_\text{mag} < 1$ implies that magnetic effects
dominate thermal effects, showing that magnetic fields are therefore very
important to the physics of molecular clouds \citep{c1999}. These observations
\citep[as well as the later ones by][]{ct2007, cht2009} fit the relation $B
\propto n_\text{H}^{0.47}$, which is consistent with simulations of molecular
clouds supported by the magnetic field, in particular ambipolar diffusion of
the field (where the flux is frozen into the ions, which scale as $n_i \propto
n_\text{H}^{0.5}$; this process is described in more detail in Section 
\ref{lr:mhd}). However, this scaling is also expected from simulations of
turbulent motions that are constrained so that they are comparable to the
Alfv\'en velocity \citep{bm1992, fm1993, e1999}. 

MHD star formation theory holds that a self-gravitating cloud can be supported
by a purely static field with no associated waves \citep[see e.g.][]{sp2004}.
Flux freezing within molecular clouds signifies both that the magnetic field
is tied to the motion of the fluid and that the gas itself is constrained by
the field configuration; magnetic field lines in cores have been indirectly
observed to possess bent or hour-glass shapes \citep{cc2006}. Many giant
molecular cloud complexes possess a stratified appearance that could well
indicate alignment along the large-scale ambient field, as could the prolate
shapes of dense cores \citep{gww2002, mo2007}. 

Rotation is a stabilising influence that raises the Jeans mass for a fixed
temperature and background radiation rate, and it flattens numerical models of
molecular clouds and cores if the rotational kinetic energy is an appreciable
fraction of the gravitational potential energy. Molecular line observations
such as those of \citet{getal1993}, \citet{kc1997} and \cite{pzcjm2003} have
shown that a majority of dense molecular cloud cores present evidence of
rotation. \citet{lbah2003} in particular found evidence for differential
rotation in the B68 core, observing velocity gradients of 3.4 km s$^{-1}$
pc$^{-1}$ from C$^{18}$O and 4.8 km s$^{-1}$ pc$^{-1}$ from N$_2$H$^+$
emission, which, as mentioned previously, trace the outer and inner regions of
the core respectively. 

The typical angular momentum of cores is small and characterised by the ratio
of the rotational kinetic energy to the gravitational binding energy, which is
given by 
\begin{equation}
    \beta_\text{rot} = \frac{1}{3}\left(\frac{V_\phi^2}{GM/R}\right)
\label{beta_rot}
\end{equation}
for a uniformly rotating sphere of constant density \citep{getal1993}. If
$\beta_\text{rot}$ is large the core is stable against gravitational
instability and collapse, however, if $\beta_\text{rot}$ is very small the
core will never have enough rotational energy to support it against collapse
and it cannot develop any instabilities that are driven by rotation, such as 
fragmentation. \citet{getal1993} found that the cores they studied all had
$\beta_\text{rot} \le 0.18$, with a typical value of $\beta_\text{rot} \sim
0.02$ on scales of $0.1$ pc, confirming that rotation was not rapid enough to
support the cores on its own. This low value of $\beta_\text{rot}$ could
inhibit binary formation mechanisms such as that caused by the
rotation-induced fragmentation of molecular cloud cores, however as all
rotating cores are expected to form centrifugal discs and discs may fragment
to form a binary system this value is not regarded as problematic.  

The role of turbulence in supporting molecular cloud cores against collapse is
an increasing source of contention within the star formation community, as the
supersonic motions observed in cores by \citet{gbwh1998}, \citet{c1999} and
others could be formed by magnetohydrodynamical turbulence that is in rough
equipartition with self-gravity in the core. MHD turbulence may be more
important than ambipolar diffusion in triggering the formation and collapse of
molecular cloud cores, as turbulent support decays quickly and speeds star
formation, while stronger turbulence could cause cores to fragment and form
multiple star systems. This scenario would suggest that star-forming clouds
are transient and that star formation is a rapid process, in direct contrast
with the quasistatic slow process that is star formation driven by ambipolar
diffusion \citep[][and the references within these reviews; see also Section
\ref{lr:col}]{mk2004, e2007, wetal2007, bkmv2007}. Recent observations by
\citet{cht2009} were unable to prove which of these mechanisms of core support
and collapse is truly dominant in molecular clouds; studies of the magnetic
field strength and orientations in cores with better resolution and higher
signal-to-noise ratios are required to determine between them \citep{mt2009}.

Depending on the local environment of a molecular cloud, different model
scenarios for the support and structure of clumps and cores may be relevant in
different regions of the cloud. The effect of local density, pressure,
temperature and magnetic field variations, as well as the presence or absence
of other nearby stars and protostars, likely all contribute to determining
which forces dominate the formation and evolution of dense molecular cloud
cores \citep{wetal2007}. 

\section{Magnetic Diffusion}\label{lr:mhd}

Simulations of star formation have typically approximated the magnetic field
behaviour by ideal magnetohydrodynamics (IMHD), where the mass-to-flux ratio
is held constant and the magnetic field is regarded as being frozen into the
neutral medium \citep[e.g.][]{glsa2006, ml2008, mmi2008}. In this situation
the magnetic field and the particles move together in the collapsing flow,
however, this simplification only truly applies in the outermost regions of
gravitational collapse where the density is low. If IMHD were to hold true
throughout the collapse then the magnetic flux in the star would be
$10^3$--$10^5$ times larger than that observed in young stars \citep[this is
the ``magnetic flux problem'', described in more detail in][and Section
\ref{lr:mfp}]{cf1953}. A mechanism for allowing the field to move 
against the inward flow of the neutral particles is required to reduce the
field in the protostar to observed values; at densities higher than those
encountered at the edge of molecular cloud cores flux freezing breaks down,
so that the magnetic field diffusion depends upon the coupling of the field to
the charged particles and the drift of these against the neutral gas. 

The diffusion of a magnetic field through a molecular cloud core is determined
by the drift of charged particles through the dominant neutral component in
response to the electric field in the neutral rest frame. When the gas is
weakly ionised, the charged particle species develop a drift velocity with
respect to the neutral fluid velocity. The Lorentz force (which only acts on
the charged particles) is transmitted to the neutral gas through the drag
forces caused by collisions between the neutral and charged particles
\citep[e.g.][]{ks2010}. These collisions determine the efficiency of the
angular momentum transport by the field in weakly ionised gas, and the
outwards diffusion of the magnetic field in a molecular cloud core, which
erodes the magnetic support until the core becomes gravitationally unstable
and undergoes collapse \citep{ms1956}. 

The degree of coupling between charged species and the neutral gas is measured
by the Hall parameter, $\beta_j$, the ratio of the gyrofrequency to the
frequency of collisions between charged species $j$ and the neutrals (where
$j$ are typically ions or electrons, denoted by $i$ and $e$ respectively). The
Hall parameter measures the relative importance of the Lorentz and drag forces
in balancing the electric force, and is defined by 
\begin{equation}
    \beta_j = \frac{|Z_j|eB}{m_jc}\frac{1}{\gamma_j\rho}
\label{Hall-param}
\end{equation}
for a particle of mass $m_j$, charge $Z_je$ and collision frequency
$\gamma_j\rho$, where
\begin{equation}
    \gamma_j \equiv \frac{<\!\sigma\nu\!>_j}{(m + m_j)};
\label{gamma}
\end{equation}
$<\!\!\sigma\nu\!\!>_j$ is the rate coefficient for collisional momentum
transfer between the charged particles and neutrals of mass $m$. 

The relative drifts of different charged species with respect to the neutral
particles delineate three different magnetic diffusivity regimes (aside from
IMHD): 
\begin{itemize}
{\item 
\begin{figure}[!b]
  \centering
  \includegraphics[width=4.5in]{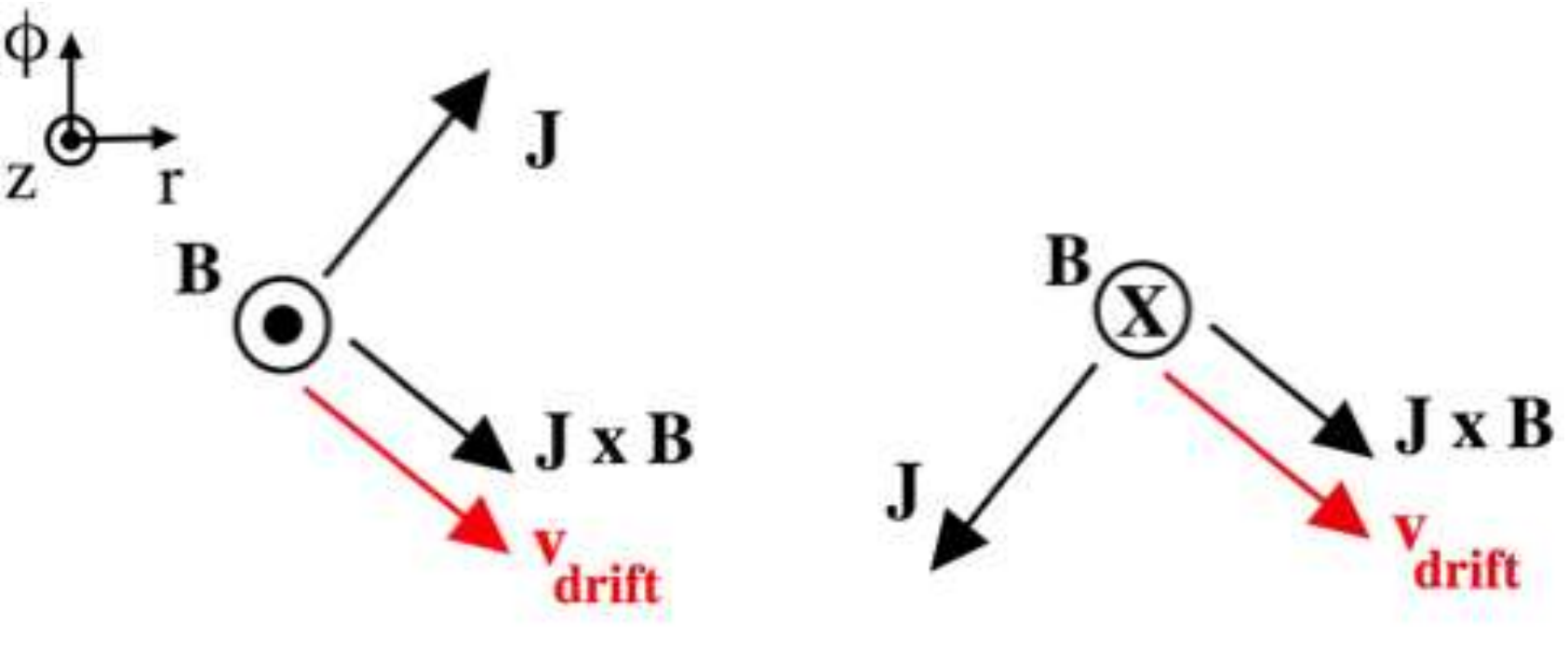}
  \vspace{-5mm}
  \caption[Vector diagram of ambipolar and Ohmic drift \citep{w2009talk}]
{Vector diagram for ambipolar and Ohmic diffusion. When the orientation of the
magnetic field is reversed, as in the second panel, the current $\mathbf{J}$
changes direction; however, $\mathbf{J} \times \mathbf{B}$, and the drift of
the ions, electrons and the magnetic field through the neutral particles
retain the same direction \citep[from][]{w2009talk}.} 
\label{ambi-vector}
\end{figure}
the \textit{Ohmic (resistive) diffusion} limit, which dominates in high
density regions where the ionisation fraction is low. The ions and the
electrons frequently collide with the neutrals over the electron gyration
period, and the magnetic field is decoupled from all charged particles. In
this limit $\beta_i \ll \beta_e \ll 1$ and the ion and electron drifts are not
affected by the presence of the magnetic field. Ohmic diffusion is thought to
be important in the innermost regions of the protostellar disc where the
density and collisional rates are high \citep[e.g.][]{sglc2006, mim2008}.} 
{\item the \textit{ambipolar diffusion} limit, which dominates in regions of
relatively low density where the fractional ionisation is high. In this limit
$1 \ll \beta_i \ll \beta_e$ and the magnetic field is tied to the charged
particles by electromagnetic stresses. The ionised component drifts with the
field through the neutrals, redistributing the matter in the flux tubes.
Ambipolar diffusion is dominant in molecular clouds \citep{w2007}, in
protostellar discs at radial distances beyond $\sim 10$ AU and close to the
surface of these discs nearer to the protostar \citep{s2009}. It is the type 
of diffusion most commonly included in simulations of star formation that go
beyond ideal MHD \citep[e.g.][see also Section \ref{lr:col}]{ck1998, kk2002,
as2007, ml2009}. In both the Ohmic and ambipolar diffusion limits, when the
magnetic field is globally reversed the magnetic response of the disc is
unchanged, as demonstrated in Figure \ref{ambi-vector}. }
{\item 
\begin{figure}[!b]
  \centering
  \includegraphics[width=4.5in]{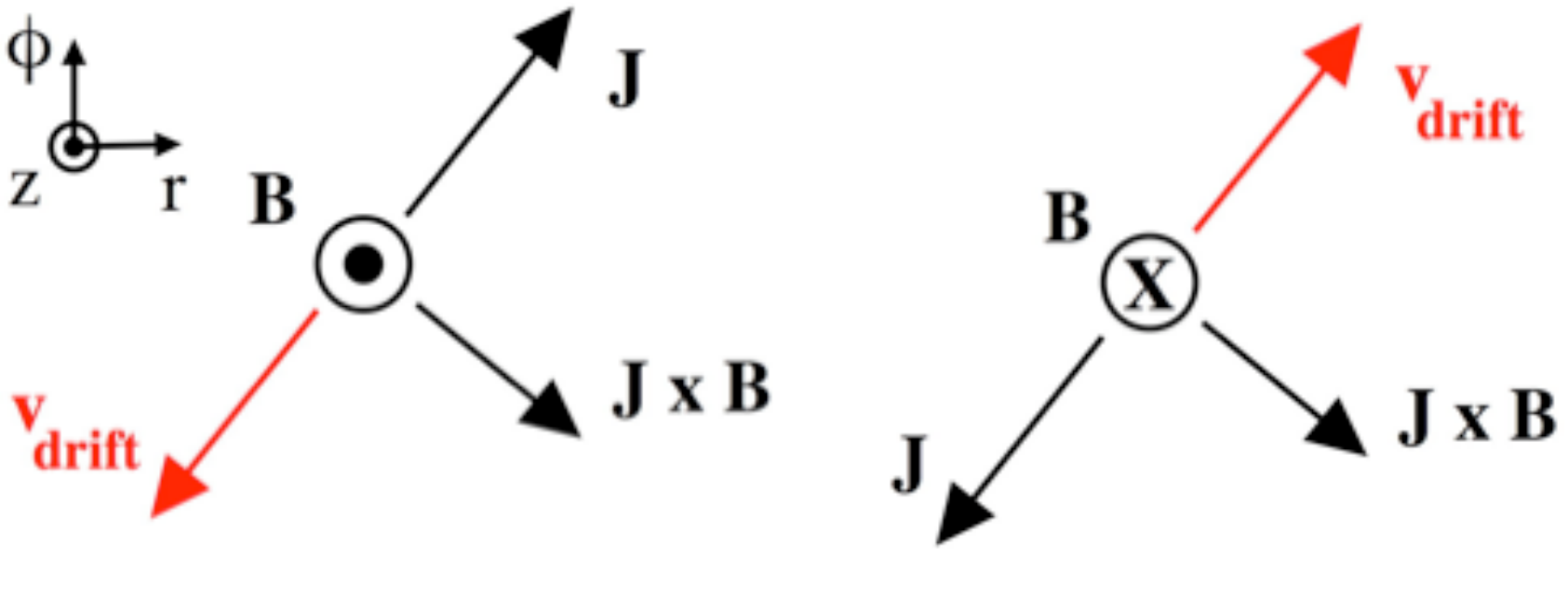}
  \vspace{-5mm}
  \caption[Vector diagram of Hall drift \citep{w2009talk}.]
{Vector diagram showing drift caused by Hall diffusion. Unlike in the
ambipolar and Ohmic diffusion cases shown in Figure \ref{ambi-vector}, when the
field is reversed in the second panel, the direction of the drift of electrons
and the magnetic field through the neutral particles and ions is also
reversed. From \citet{w2009talk}.} 
\label{hall-vector}
\end{figure}
the \textit{Hall diffusion} limit, which dominates in the intermediate
regimes between ambipolar and Ohmic diffusion. In this limit $\beta_i \ll 1
\ll \beta_e$ and the charged species have a varied degree of coupling to the
magnetic field, typically with the electrons tied to the magnetic field. The
more massive particles such as ions and charged dust grains are decoupled from
the magnetic field and are instead collisionally-coupled to the neutral gas.
In the Hall regime, the magnetic response of the disc is no longer invariant
under a global reversal of the magnetic field, as shown in Figure
\ref{hall-vector}, and plane-polarised damped Alfv\'{e}n waves do not exist
\citep{wn1999}. Hall diffusion is expected to dominate in many regions of
molecular clouds as they undergo gravitational collapse \citep{w2004}, and in
protostellar discs \citep{ss2002a, ss2002b}.} 
\end{itemize}

The degree of coupling between the magnetic field and the charged particles
depends upon the fractional ionisation of the gas and the momentum transfer
cross sections for collisions between the charged particles and the neutrals.
In a weakly ionised gas, such as that encountered in molecular clouds, the
abundances of charged species are so low that their inertia and thermal
pressure are negligible. Typically in molecular clouds, molecular ionisation
by cosmic rays is balanced by the rapid dissociative recombination of
molecular ions with the metal ions that are the dominant positive charge
carriers \citep{un1990, wn1999}. 

For grains of radius $a = 0.1$ $\mu$m at a temperature of $10$ K in a cloud
with cosmic ray ionisation rate $\xi = 10^{-17}$ s$^{-1}$, the ion density is
usually taken to scale as $\rho_i \propto \rho_n^{1/2}$ when $10^4 \lesssim
n_\text{H} \lesssim 10^7$ cm$^{-3}$ \citep{e1979, kn2000}. This behaviour is
an oversimplification, as \citet{cm1998} showed that for typical cloud and
grain parameters the proportionality of the ion density cannot be
parameterised by a single power law exponent $k$, as $k > 1/2$ for densities 
$n_\text{H} \lesssim 10^5$ cm$^{-3}$ and $k \ll 1/2$ for densities $n_\text{H}
\gg 10^5$ cm$^{-3}$, but it is still a reasonable and widely-adopted
approximation to the ion density in collapsing cores on scales $\gtrsim 10^3$
AU \citep[see e.g.][]{sal1987, gs1993a, ck1998, cck1998, kk2002}. This shall
be discussed further in Chapter \ref{ch:derivs}. 

Outside of the central $\sim0.1$ AU of a protostellar system the ionisation of
the gas is driven by stellar X-ray and UV radiation, as well as interstellar
cosmic rays \citep{h1981,gks2005}. At the higher densities found here and more
generally in the inner regions of the collapsing core, grains are typically
the dominant carriers of both positive and negative charge and their densities
scale as $n_\text{H}^{1/2}$ \citep{nnu1991, t2005}. In these innermost regions
of protostellar systems, the fractional ionisation is low, as the high density
leads to a very rapid recombination rate and the disc column density shields
the gas from cosmic rays and X-rays. 

The degree of the coupling between the material and the magnetic field depends
also upon the abundance and size distribution of grains in the gas. As dust
grains have large cross sections they typically become decoupled from the
field at lower densities than other ions, reducing the diffusivity of the gas.
If grains are important and decoupled from the field then Ohmic diffusion
dominates, however if the grains have settled or aggregated then Hall
diffusion is important. In molecular clouds that are not overdense, the ions
and electrons are tied to the field while large grains are not, so that
ambipolar diffusion dominates the magnetic field behaviour \citep{w2007}.

\begin{figure}[t]
  \centering
  \includegraphics[width=4in]{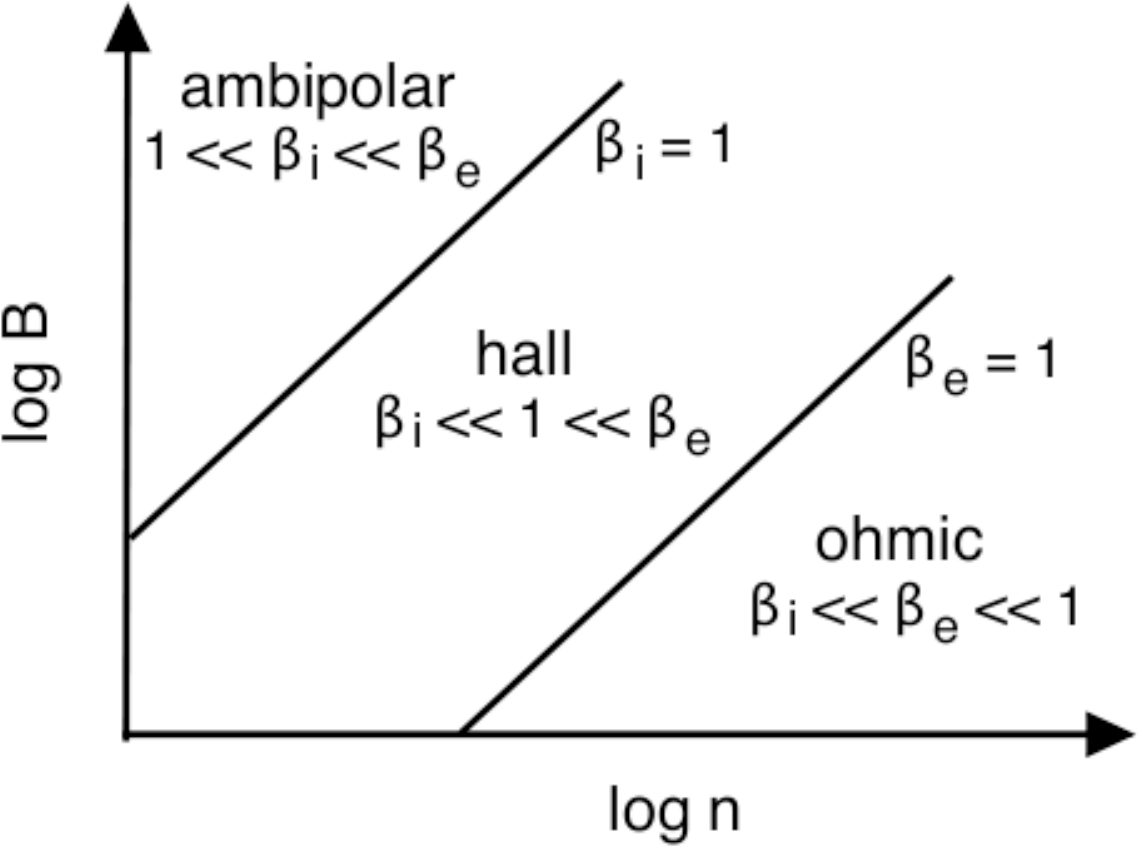}
  \vspace{-5mm}
  \caption[Magnetic diffusion regimes \citep{w2007}]
{The magnetic diffusion regimes of a weakly ionised three-component plasma are
determined by the ion and electron Hall parameters $\beta_i$ and $\beta_e$,
which are proportional to $B/n_\text{H}$, with $\beta_e/\beta_i$ typically
$\sim 1000$ \citep[from][]{w2007}.} 
\label{eta-regimes}
\end{figure}

The regions of parameter space in which each of the three types of coupling
are dominant in a weakly ionised three-component plasma are shown on a
simplified $\log B$-$\log n_\text{H}$ plane in Figure \ref{eta-regimes}
\citep[from][]{w2007}. It is clear from the figure that Ohmic diffusion
dominates at high densities with weak fields, while the opposite is true for
ambipolar diffusion. The intermediate region of parameter space between these
two limits is dominated by Hall diffusion. In particular, the Hall term is
significant for molecular gas densities in the range $\sim 10^8$--$10^{11}$
cm$^{-3}$ (when $B$ scales as $B \propto n_\text{H}^{1/4}$), although the
presence and particular distribution of grains complicates the calculation of
the diffusivities \citep{wn1999}. 

The nature of the coupling (ambipolar, Hall, Ohmic) determines the magnetic
field direction as the field lines emerge from the surface of the protostellar
disc. This in turn controls the amount of material that is able to slide along
the field lines and be flung outwards from the surface instead of being
accreted in disc-driven wind models \citep{wk1993, w2004-2}. 

It was noted by \citet{nh1985} that the Hall component of the conductivity
tensor could be important in cloud regions where the neutral density and the
temperature are weakly variable. They dismissed the effects of the Hall term
in their discussion of the resistivity of molecular clouds, arguing that the
Hall current leads to a charge separation that generates an electrostatic
field which modifies the direction of the total field until the Hall current
vanishes. 

However, this logic does not hold in situations where the boundaries are not
idealised, and \citet{wn1999} countered this argument using the specific
example of the quasistatic collapse of an axisymmetric cloud core through a
poloidal magnetic field. In their example the current is toroidal and the
electric field in the neutral fluid frame is poloidal so that if the Hall and
ambipolar diffusion components of the conductivity are of the same order then
the toroidal components of the velocity and magnetic field will both
contribute to the gravitational support of the core. For this simple model, 
the boundaries on which \citet{nh1985} presumed that the charge would build up
would be surfaces of constant azimuth, which cannot exist under axisymmetry.

More generally, it is not possible in MHD to build up charged surfaces such as
those proposed by \citet{nh1985}, as typically $E \sim \frac{V}{c}{B}$. As the
gas velocities in molecular clouds are very low compared to the speed of light
\citep[typically around $0.1$--$10$ km s$^{-1}$; see Section \ref{lr:mc}
and][]{c1999, lmt2001} then $E \ll B$ and the effects of charge separation are
negligible compared to the magnetic field effects. As the current $\mathbf{J}
\propto \nabla \times \mathbf{B}$ in MHD then $\nabla \cdot \mathbf{J} = 0$,
and the Hall current cannot cause a build up of charge that would generate an
electrostatic field. The argument that electrostatic fields would negate the
Hall current cannot then be used to dismiss the Hall effect in collapse
calculations, and it will be shown to be important to the infall dynamics. 

The magnitude and type of coupling that occurs between the fluid and magnetic
fields in molecular clouds and protostellar discs is uncertain due to the
difficulty in obtaining detailed observations in these regions, particularly
of the magnetic field. Calculations of the ionisation equilibrium and
resistivity by \citet{w2004} suggested that the Hall diffusion term is
important and may dominate the magnetic field diffusion at many of the
densities and field strengths encountered in molecular clouds and protostellar
discs, and presumably also during the collapse stages between these two
evolutionary phases. Further calculations of the resistivity in protoplanetary
discs by \citet{w2007} showed that Hall diffusion dominates in the innermost
regions of protoplanetary discs. The Hall effect also imparts an implicit
handedness to the fluid dynamics (as illustrated in Figure \ref{hall-vector})
that is sensitive to a global reversal of the magnetic field direction and
could be important in calculations of gravitational collapse and protostellar
disc formation \citep{w2007}.  

As will be discussed in the following sections, in those simulations of star
formation that take a more sophisticated approach to the magnetic field
diffusion than adopting the IMHD approximation, the breakdown of flux freezing
in molecular clouds and protostellar discs is usually approximated by either
ambipolar diffusion in simulations of the early stages of collapse where the
density is low or Ohmic diffusion in the higher density late stages of
protostellar disc evolution. In most numerical or semianalytic simulations of
star formation the Hall diffusion term is dismissed as insignificant, using
the same arguments as \citet{nh1985}; however, the consequences of including
Hall diffusion in calculations of star formation by quasistatic gravitational
collapse and of the subsequent evolution of protostellar discs are likely to
be profound, in a similar manner to that found for the magnetorotational
instability \citep[the MRI;][]{w1999, ss2002a, ss2002b}. 

The coupling between the field and the charged particles determines the
dynamics of the collapse and can help resolve the angular momentum and
magnetic flux problems of star formation, in which young stars have rotation
rates and field strengths much smaller than their equivalent mass in a
molecular cloud \citep{cf1953, ms1956, s1978}. These problems are discussed in
more detail in Sections \ref{lr:amp} and \ref{lr:mfp}; while the relevance of
Hall diffusion in simulations of gravitational collapse, and its ability to
enhance or mitigate the magnetic braking catastrophe (which prevents disc
formation), are explored further in Section \ref{lr:hall}.

\section{Gravitational Collapse}\label{lr:col}

\begin{figure}[tp]
  \centering
  \vspace{-5mm}
  \includegraphics[width=5in]{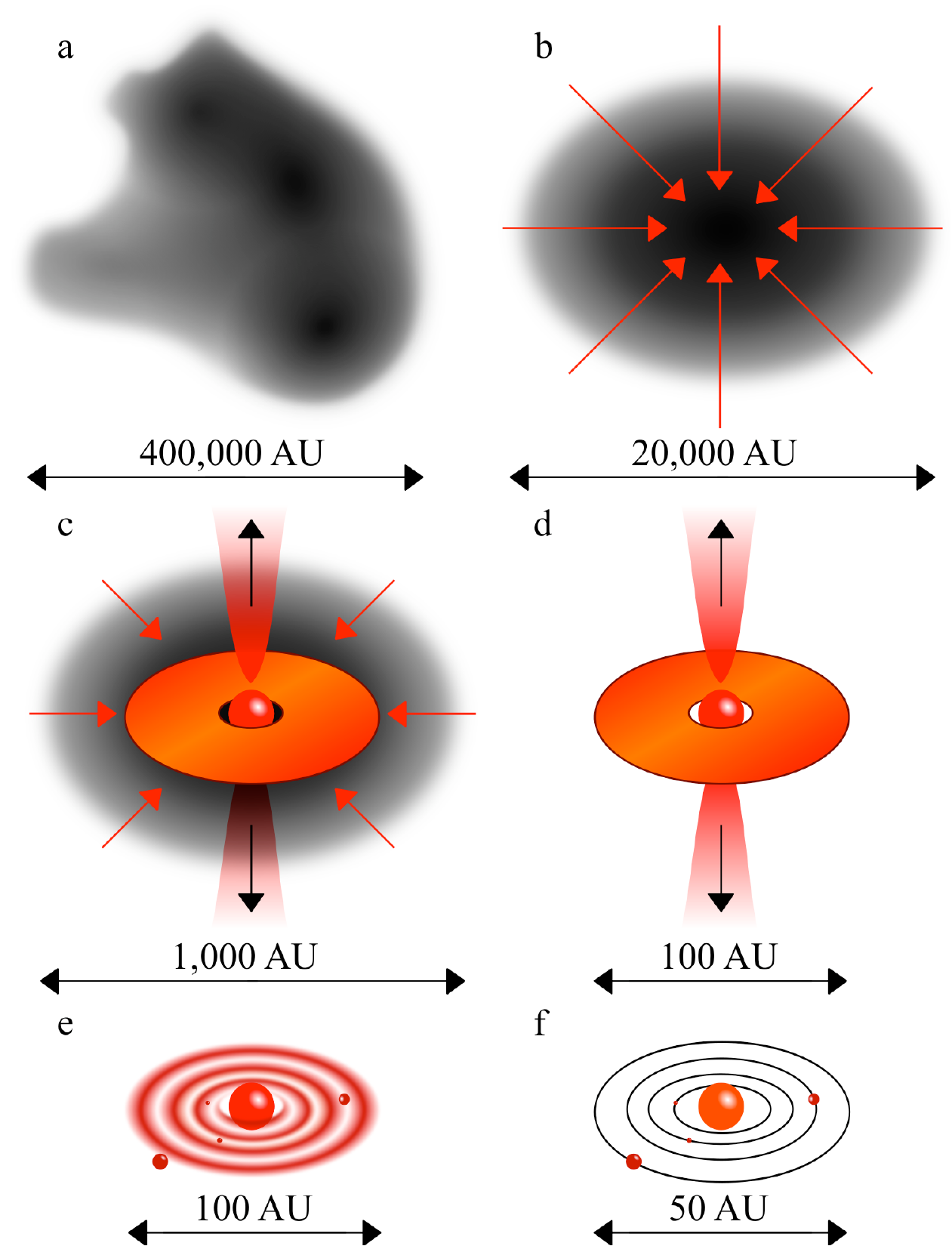}
  \caption[Star formation by gravitational collapse]
{Developmental stages and scales of low-mass star formation \citep[adapted
from figure 7 of Shu et al., 1987 and figure 2 of][]{g2001}.\ (a) Star 
formation begins when cores form as magnetic and turbulent support in
molecular clouds is lost through ambipolar diffusion.\ (b) A core becomes
unstable and collapses dynamically into a pseudodisc.\ (c) As centrifugal
forces balance gravitational forces, a protostar and disc form from the
collapsing envelope, and a disc wind or jet may start.\ (d) The envelope
dissipates or is accreted and the protostar becomes visible at optical
wavelengths as a T Tauri star, with associated outflow and protoplanetary
disc.\ (e) The protoplanetary disc starts to form planets and may be visible as
a debris disc, while the pre-main sequence star continues to contract.\ (f)
Nuclear fusion begins and the star joins the main sequence, and stellar winds
blow away the remainders of the disc.} 
\label{gcillus}
\end{figure}

Low-mass star formation by the gravitational collapse of molecular clouds
takes place in stages over many orders of magnitude in size and density as
illustrated in Figure \ref{gcillus} \citep[adapted from figure 7 of Shu et
al., 1987 and figure 2 of][]{g2001}. Firstly, cores form within the cloud as a
result of turbulent fluctuations, and they gradually contract as ambipolar
diffusion erodes the magnetic support \citep[Figure \ref{gcillus} a, b;
][]{tt2009, fbck2010}. Turbulence may also support the core against collapse,
and its decay can aid the magnetic diffusion in triggering star formation
\citep{mk2004, bkmv2007}.  

The core becomes gravitationally unstable and collapses dynamically into what
is termed a ``pseudodisc'' \citep{gs1993a, gs1993b}, which has a flattened
shape due to the material falling in preferentially along the magnetic field
lines that support it against collapse in the radial direction. A protostar
arises at the centre of the pseudodisc, which may be surrounded by a
centrifugally-supported protostellar disc \citep{tsc1984, sal1987}. Material 
collapsing from the envelope onto the protostar must pass through the
pseudodisc and protostellar disc, building up material and flux at the
boundaries of these that take the form of shock fronts as the
dynamically-infalling gas collides with the slower-moving disc material. A
disc wind or jet may form, launched from the inner regions of the collapse
\citep[Figure \ref{gcillus} c;][]{wk1993, t2002, asl2003b}. 

When the density becomes larger than $\simeq 10^{10}$ cm$^{-3}$ the gas
becomes optically thick and the thermal structure of the collapsing core is
nearly adiabatic, forming a thermally supported inner core \citep[the
``first'' or "opaque" core; e.g.][]{l1969, mim2007} which collapses
dynamically into the protostar \citep[the ``second core'';][]{l1969, mim2006}.
Eventually the remainders of the collapsing envelope and pseudodisc are
accreted or dissipated by the wind, and the protostar becomes visible at
optical wavelengths as a T Tauri star with associated protoplanetary disc and
outflow \citep[Figure \ref{gcillus} d;][]{ns1980, ketal2002, aw2007, dm2010}.
The system continues to contract, and planets may accrete from or fragment the
disc around the pre-main sequence star. Gaps begin to appear in the
protoplanetary disc as the planets sweep up the gas, and the system is visible
as a protostar with debris disc \citep[Figure \ref{gcillus} e;][]{bw2008,
w2008}. Finally, the star joins the main sequence and the stellar wind
dissipates the remainder of the protoplanetary disc, leaving behind a young
stellar system \citep[Figure \ref{gcillus} f;][]{tsc1984, wt2002}. 

It is impossible to cover all of the work that has been done both
observationally and in theoretical simulations of star formation in the space
available here. Instead, this section shall focus primarily on the isothermal
collapse of a molecular cloud core to an adiabatic protostar, potentially
surrounded by a centrifugally-supported disc (Figure \ref{gcillus} b--c) for
low-mass molecular cloud cores. For further information on the other stages of
star formation, the evolution and dynamics of protostellar discs, or the
processes involved in the formation of higher-mass stars such as fragmentation
and turbulence, the reader is directed to the review of \citet{sp2004} for a
basic overview and the more detailed reviews by \citet{l2003}, \citet{mo2007}
and \citet{m2010} for a more complete description.

\subsection{Core formation and support}\label{lr:mccf}

Molecular clouds are transient objects, with irregular structures and
internal motions that suggest they are dynamic and rapidly changing
\citep{fetal2007}. They are likely formed by compressive motions of
gravitational or turbulent origin (or some combination of the two). Giant
molecular clouds are believed to be formed by gravitational instability
\citep{tt2009}, or the agglomeration of smaller clouds \citep{d2008}; and they
are concentrated towards the spiral arms of galaxies \citep{sl2006}. Smaller
molecular clouds can be formed by ram pressure from supersonic flows such as
those driven by supernova blast waves and the MRI in galaxies \citep[][and the
references therein]{ns1980,bkmv2007}. 

Molecular cloud cores are the gravitationally-bound regions of higher density
within molecular clouds that are the progenitors of protostars. There are
currently two competing models for the formation of molecular cloud cores
which are most likely extremes on the continuum of collapse models. The first
is a slow, quasistatic core formation in which a dense region of the molecular
cloud becomes centrally-condensed by an ambipolar diffusion-driven phase of
collapse \citep[e.g.][]{sal1987,bc2004} or by the gradual dissipation of
low-level turbulent fields \citep[e.g.][]{m1999}. At the other extreme is a
more dynamic formation process in which highly turbulent flows create large
scale fluctuations and inhomogeneities in the gas of the molecular cloud, some
of which become gravitationally unstable and collapse to form stars, with MHD
waves carrying away any excess turbulent energy. This behaviour has yet to be
seen in simulations of nonmagnetic molecular clouds, however the models of
\citet{tt2009} and \citet{fbck2010} both demonstrated gravitational
instabilities in nonmagnetic clouds, and gravitational instabilities have been
observed in simulations of large isolated cores and small clouds with a few
magnetic cores \citep[e.g.][]{bc2004, pb2007}.

Observations of isolated cores show that mild turbulence and magnetic fields
play an approximately equal role in the evolution of molecular cloud cores
\citep{c1999}, while observations of protoclusters and molecular cloud clumps
show supersonic infall velocities that require strong external compression and
are inconsistent with self-initiated forms of collapse \citep[][and the
references therein]{fmwom2001,wetal2007}. These suggest that each of the core
formation models may be valid under differing circumstances, with the
turbulence-driven model applying to cluster-forming clumps and the ambipolar
diffusion model to isolated low-mass star forming cores. 

Assuming that the ionisation rate in a core scales with density as 
\begin{equation}
    \rho_i = C_i\rho_n^{1/2},
\label{ionisation}
\end{equation}
where $C_i$ is the numerical coefficient of the ionisation with typical value
$C_i = 3 \times 10^{-16}$ cm$^{-3/2}$ g$^{1/2}$ \citep{e1979}, the ratio of
the ambipolar diffusion time scale to the time scale for dynamical collapse is 
approximately given by
\begin{equation}
    \frac{t_{AD}}{t_{dyn}} \sim \frac{\gamma_iC_i}{2\sqrt{2\pi G}}. 
\label{lrtime}
\end{equation}
For a typical value of the ion-neutral drag coefficient $\gamma_i = 3.5 \times
10^{13}$ cm$^3$ g$^{-1}$ s$^{-1}$ \citep{drd1983}, this ratio is evaluated as
$t_{AD}/t_{dyn} \approx 8$, demonstrating that ambipolar diffusion takes place
at a relatively slow rate until the density is sufficiently high that the
ionisation fraction begins to depart from the relationship described in
Equation \ref{ionisation} \citep{cm1998}. As ambipolar diffusion is a slow
process, once the magnetic support of the core has been eroded, the core will
undergo rapid dynamic collapse and form a protostar before the more gradual
contraction under ambipolar diffusion could cause a protostar to appear at the
origin. 

Star formation is an inefficient process \citep{e1999} and the ratio of the
ambipolar diffusion to dynamical time scales in Equation \ref{lrtime} provides
an explanation for this, as a slow rate of contraction caused by ambipolar
diffusion could see many cores dissipating before they have the chance to form
stars \citep{sal1987}. On the other hand, numerical simulations by
\citet{cb2004} showed that clouds with large turbulent motions also result in
low star formation efficiencies as the bulk of the cloud escapes due to the
initial supersonic motion. Both of the dominant theoretical mechanisms for
star formation (ambipolar diffusion and turbulence) could then be responsible
for the observed low rate of star formation. 

Molecular cloud cores are supported against gravitational collapse by the
magnetic field as long as the core is thermally and magnetically
``subcritical'', that is, the mass-to-flux ratio is less than the critical value,
\begin{equation}
    \left(\frac{M}{\Phi}\right)_{crit} = \frac{C_\Phi}{G^{1/2}}
\label{mfratio}
\end{equation}
where $\Phi$ is the magnetic flux threading the cloud and $C_\Phi$ is a
dimensionless (in cgs units) numerical coefficient that depends upon the
internal distribution of the magnetic field and density \citep{mzgh1993}.
Discs with only thermal support along the field lines have $C_\Phi \approx
0.126$ \citep{ms1976}; an infinite cold sheet has $C_\Phi = 1/2\pi \simeq
0.16$ \citep{nn1978}, and simulations of cold clouds with different
distributions of a poloidal field require $C_\Phi \simeq 0.17$--$0.18$ for the
central flux tube to be critical \citep{tin1988b}. 

Taking a value of $C_\Phi = 0.12$, \citet{c1999} found that the mass-to-flux
ratio in their observed molecular cloud cores was typically twice the critical
value, suggesting that static magnetic fields are insufficient to support
molecular cloud cores against gravity on their own. They also found that cores
were in near-virial balance, which was used as evidence to suggest that the
cores were supported by turbulence and that it is the decay of such turbulence
that triggers collapse \citep{m1999}. Simulations have shown that for cores
that are close to the critical value the dynamical and ambipolar diffusion
time scales are comparable \citep{ln2004, kkh2009}. It is likely to be the
case that both the decay of turbulence and the loss of magnetic support in
cores are important in initiating dynamic collapse, although only magnetic
support is considered in this work. 

Magnetic braking, by which the torque exerted on the rotating gas by the
twisting of magnetic field lines transports angular momentum from the
collapsing core to the molecular cloud, is very effective during this phase of
molecular cloud core contraction. This process reduces the angular momentum of
cores to the low values observed \citep{cck1998}, helping to resolve the
traditional angular momentum problem of star formation in which young stars
have less angular momentum than the initial core from which they formed during
this early, isothermal phase of collapse (see Sections \ref{lr:amp} and
\ref{lr:mfp} and the references therein).  

\subsection{Dynamic collapse}\label{lr:dcol}

Once ambipolar diffusion has caused enough of the neutral material to move
inwards with respect to the magnetic field, the mass-to-flux ratio exceeds the
critical value defined in Equation \ref{mfratio}. The core is then said to be
``supercritical'' and the central density of the molecular cloud core formally
tries to achieve infinite values. It contracts dynamically to form a protostar
surrounded by a slowly-rotating pseudodisc, while the envelope remains
magnetically subcritical. The ambipolar diffusion time scale remains relevant
to the evolution of the core envelope so long as it is subcritical. 

During this contraction, the field is effectively frozen into the cloud as
IMHD holds true, and the infalling material is deflected by the field lines
towards the equatorial plane, creating a pseudodisc that is not
rotationally-supported. The pseudodisc contracts dynamically in the radial
direction, dragging the field lines into a split monopole configuration
\citep{gs1993a, gs1993b}. The resulting build up in magnetic pressure acts as
an impediment to further collapse, however, the magnetic tension in the
envelope never suffices to suspend the envelope against the gravity of the
growing protostar \citep{asl2003a, asl2003b}. 

Disregarding the effects of rotation and the magnetic field on the core for
the moment, a slowly contracting molecular cloud core that is not artificially
separated from its surroundings will tend to acquire the density distribution
of a singular isothermal sphere, 
\begin{equation}
    \rho = \frac{c_s^2}{2\pi Gr^2}\,,
\label{sis}
\end{equation}
where $c_s$ is the isothermal sound speed \citep{l1969, sal1987}. There are
two limiting cases to the gravitational collapse of a singular isothermal 
sphere. The first is that presented in the solutions of \citet{l1969} and
\citet{p1969}, in which the collapse begins near the outer radius of a
marginally unstable core and the $r^{-2}$ density gradient is created as the
wave of collapse propagates inward, leaving every scale marginally unstable as
the collapse accelerates. At the time when the protostar first forms the
collapse is highly dynamic with an infall speed of $3.3c_s$, and the accretion
rate onto the star is high. 

The second case is that of \citet{s1977}, in which the evolution of the core
is much slower and the infall velocity at the moment of protostar formation is 
negligible. The collapse is initiated in the centre of the core, and the
``expansion wave'', the point at which gas begins to fall inward, propagates
outward at the sound speed. This process is referred to as ``inside-out
collapse'' and inside of this rarefaction wave the gas accelerates until it is
nearly free falling onto the protostar. 

These solutions are referred to as examples of ``self-similar'' or 
scale-invariant collapse, as the properties of the collapsing core are similar
(in the mathematical sense) to those properties at an earlier period of time
at smaller radii (self-similarity is a typical property of fractals). The
self-similar collapse models represent a semianalytic method of exploring
collapse at higher resolution than is possible in numerical calculations,
which are limited by either the number of particles or the scaling of the box
in which the collapse takes place. Although they require many assumptions and
simplifications to the gas dynamics (see Chapter \ref{ch:derivs} for more
information on the formulation of a self-similar model), the calculation of
similarity solutions is a useful analytic technique for studying star
formation. Self-similar behaviour is often observed in numerical simulations
of collapse \citep[e.g.][and others]{mi2000, t2002, ml2009, db2010}; yet
similarity solutions can follow the collapse to regions of high density close
to the protostar, including inside the boundary of the sink cells that are
often used to represent the inner point mass in numerical calculations. 

The evolution of star-forming molecular clouds has been studied through
numerical simulations, primarily of the ambipolar diffusion-initiated 
formation of supercritical cores and the early stages of dynamical collapse
\citep[e.g.][]{mim2008, km2009, km2010}. These numerical simulations of
collapse tend towards the similarity solutions during the dynamical collapse
stage despite being started with different initial conditions in the core
\citep[e.g.][]{t2002, ml2008, ml2009, fbck2010}, even though collapsing flows
may never actually approach the similarity solutions in reality. 

The Larson--Penston and Shu similarity solutions represent extreme cases of
the continuum of nonmagnetic, nonrotating solutions (``fast'' and ``slow''
collapse), and their applicability is determined by the initial conditions
chosen to describe the core. The more general family of nonmagnetic,
nonrotating self-similar gravitational collapse solutions were explored by
\citet{ws1985}; the outer density profile in this type of solution is $\propto
r^{-2}$ whereas in the innermost regions near the central singularity the
density is $\propto r^{-3/2}$. 

The similarity solutions have been duplicated by numerical simulations: for
example, \citet{fbck2010} were able to reproduce the density gradients and
mass accretion rates of the isothermal sphere modelled by \citet{s1977} in
their calculations of cluster formation. This was achieved using sink
particles in their hydrodynamic simulations to represent the central point
masses which are notoriously difficult to model otherwise due to their high
concentration of mass in a small region of the calculation grid. The
simulations of \citet{t2002} and \citet{ml2008, ml2009} also demonstrated 
self-similarity in their collapsing flows for single cores; this behaviour is
regarded as an important test of the models. 

These early similarity solutions did not include the effects of rotation and
magnetic fields on the core, nor did they include turbulence, which becomes
more important as the core mass increases. It is the addition of a magnetic
field that leads to the formation of a pseudodisc \citep{gs1993a, gs1993b},
while rotation may lead to the formation and growth of a
rotationally-supported disc at the centre of the pseudodisc \citep{tsc1984}.

In rotating collapse simulations the conservation of angular momentum during
the near-free fall dynamical collapse results in a progressive increase in 
the centrifugal force that eventually becomes important and creates a
centrifugal barrier to collapse, forming a rotationally-supported disc around
the central protostar. If the centrifugal force becomes strong enough it may
trigger the fragmentation of the contracting core and cause the formation of a
binary or multiple star system \citep[e.g. the solutions of][]{wt2002, cb2006,
kboy2007, pb2007, mmi2008, kkh2009, abi2009}. 

Early similarity solutions that included rotation showed that it causes a
centrally-condensed disc to form around the protostar, disrupting the
near-spherical dynamic collapse. For the singular isothermal sphere, most of
the collapsing envelope settles into a centrifugally-supported disc around the
protostar \citep{tsc1984, sal1987}. Numerical simulations by \citet{mhn1997}
showed that the behaviour of collapsing cores with rotation approached that of
analytic self-similar models of a singular isothermal disc
\citep[e.g.][]{hnm1982, t1982}, or a rotationally-flattened (due to the
conservation of angular momentum) fast isothermal collapse \citep{sh1998}.
This same self-similar behaviour is seen in numerical simulations with
rotation and magnetic fields \citep[e.g.][and the references in Section
\ref{lr:mfp}]{mim2008, ml2009}. 

When a magnetic field (aligned with the axis of rotation) is present during
the dynamic collapse phase, the protostar develops in a nearly self-similar
way. The collapse propagates as a fast MHD wave, travelling faster in the
direction perpendicular to the field, creating a prolate shape immediately
inside the head of the expansion wave. Further inside the collapsing core, the
tangential component of the Lorentz force produced by the curvature of the
magnetic field lines deflects the gas towards the equatorial plane, forming an
oblate pseudodisc of flattened infalling material \citep{gs1993a, gs1993b}.
The magnetic field of the core takes on a characteristic hourglass shape in
which the field lines flare above and below the pseudodisc (also called a
split monopole) that is consistent with observations \citep{cc2006, grm2006, 
ggg2008, aetal2009}. 

Calculations of magnetic collapse that continued through to the stage of
accretion by a central point mass, such as those by \citet{cm1993, cm1994,
cm1995}, \citet{ck1998}, and \citet{ml2009}, showed that ambipolar diffusion,
which is unimportant during the dynamic collapse phase, is revitalised once a
protostar starts to grow in the centre of the core. This leads to a decoupling
of the magnetic flux from the inflowing gas, which takes place within an
outwardly-propagating magnetohydrodynamic shock, the existence of which was
first proposed by \citet{lm1996}. This MHD shock (referred to as the magnetic
diffusion shock in the results presented in this work) takes the form of a
continuous transition in the calculations of \citet{dm2001} and \citet{kk2002}
because the shock structure depends upon ambipolar diffusion, which is
explicitly determined in their equations. Inwards of the shock the magnetic
braking reduces the angular momentum, and ambipolar diffusion reduces the
magnetic flux as the neutral material continues to fall inward at a near-free
fall speed. The centrifugal force starts to become important, gradually
triggering the formation of a hydrodynamic shock that strongly decelerates the
infalling matter and allows a Keplerian disc to form. 

Accretion through the pseudodisc to the free fall region and inner disc may
occur in bursts, as mass piled up at the magnetic barriers opposes rapid
accretion. As the split monopole field inside of the pseudodisc grows, mass
builds at the boundary of the disc until it becomes too heavy to be supported.
The excess mass rushes inside towards the origin, dragging in more flux to be
assimilated by the split monopole; the magnetic barrier grows and mass must
then build up over a longer time interval before overcoming the barrier and
becoming part of the protostar. These oscillations first appeared in the
numerical simulations of \citet{asl2003b, asl2003a} as a result of
instabilities that were caused by the simplified physics in their simulations,
however they were later reproduced in the ``magnetic wall'' that inhibits
collapse in the calculations of \citet{tm2005a, tm2005b}, and in the
magnetogyrosphere of \citet{ml2008, ml2009}. \citet{asl2003a} suggested that
this magnetic behaviour may be responsible for observed FU Orionis outbursts
that are not well-explained at present.

\subsection{Late stages of collapse}\label{lr:rest}

Once the central density reaches $n_\text{H} \gtrsim 10^{10}$ cm$^{-3}$ the
assumption of isothermality breaks down as the gas becomes optically thick
against the thermal radiation from dust grains. The gas becomes adiabatic, and
once it is hot enough to dissociate molecular hydrogen, the protostar
undergoes a second dynamic collapse \citep{t2002, mim2007, mim2008}. At 
$n_\text{H} \gtrsim 10^{12}$ cm$^{-3}$ the magnetic field is effectively
decoupled from the gas and Ohmic diffusion becomes dominant \citep{nnu2002}.
This expected behaviour was not seen in the numerical simulations of
\citet{dm2001} where the gas decoupled from the field before the breakdown of
isothermality. During the decoupling stage in their simulations the electrons
were very well attached to the magnetic field instead of the expected ions;
the ions had detached from the field at lower densities. \citet{dm2001} found
that Ohmic dissipation was unimportant during the decoupling stage because of
the high electron mobility and the small $e-$H$_{2}$ cross section, although
it does become important in the innermost regions of the adiabatic core and
protostellar disc where the density is particularly high
\citep[e.g.][]{db2010, m2010}. 

Accretion onto the adiabatic core and protostar may be stopped by
hydrodynamic, magnetohydrodynamic and photoevaporative processes. Because of
the ubiquity of bipolar outflows in star forming regions, \citet{tsc1984}
implicated young stellar object (YSO) winds and jets as the dominant practical
mechanisms by which forming stars define their own masses, particularly in
isolated regions of low-mass star formation. It is believed that YSO outflows
arise because of a fundamental interaction between the rapid rotation in a
Keplerian disc and a strongly magnetised object such as a protostar
\citep[e.g.][]{gl1979a, bp1982, wk1993}. 

The collapsing pseudodisc, protostellar disc and adiabatic core can lose
angular momentum and collapse further when the magnetic field threading the
collapsing material is able to generate an outflow. Many numerical simulations
have now demonstrated that both fast and slow jets and winds can be generated
from the thin disc both before and after the formation of the adiabatic core
\citep[for example, the simulations of][]{t2002, asl2003b, mim2007, mim2008,
mmi2008, ml2008, ml2009, b2010, ch2010}. The weakly ionised gas in the
rotating collapsing flow induces a toroidal field that is able to accelerate
the gas and form an outflow, the intensity of which depends upon the rotation
rate and field strength. \citet{kk2002} were able to demonstrate how a disc
wind could be included into their semianalytic solutions by including a mass
loss term in their asymptotic inner disc solutions (which satisfy the
conditions necessary for wind launching), however at present this work has yet
to be performed \citep[see][and Section \ref{future} for further
details]{wk1993}. 

Studies of three-dimensional radiation hydrodynamical calculations have shown
that the formation of the stellar core drives a shock wave through the disc,
and dramatically decreases the accretion rate onto the stellar core
\citep{b2010} --- this behaviour was also present in earlier one-dimensional
simulations of radiative, hydrodynamical collapse of molecular cloud cores
\citep[e.g.][]{mi2000}. Radiation produced in the first stars in a molecular
cloud reduces the ability of cores to fragment, which can suppress formation
of low mass objects by increasing the temperature in the high density material
\citep{km2008, pb2009, k2010, km2009, km2010}. Numerical simulations of
molecular clouds with strong (though supercritical) magnetic fields and
radiative feedback demonstrate an inefficient star formation process with a
star formation rate that approaches the observed rate in molecular clouds
\citep{pb2009}. 

\section{Rotation and the Angular Momentum Problem}\label{lr:amp}

\begin{table}[t]
\begin{center}
\begin{tabular}{lc}
\toprule
 Object & $J/M$ (cm$^2$ s$^{-1}$)\\
 \midrule
 Molecular cloud (scale $1$ pc) 	& $10^{23}$\\
 Molecular cloud core (scale $0.1$ pc)	& $10^{21}$\\
 Binary ($10^4$ yr period)		& $4\times10^{20}$--$10^{21}$\\
 Binary ($10$ yr period) 		& $4\times10^{19}$--$10^{20}$\\ 
 Binary (3 day period)			& $4\times10^{18}$--$10^{19}$\\
 100 AU disc (1 M$_\odot$ central star) & $4.5 \times 10^{20}$\\
 Jupiter (orbit)			& $10^{20}$\\
 T Tauri star (spin)			& $5\times10^{17}$\\
 Present Sun				& $10^{15}$\\
 \bottomrule
 \end{tabular}
\end{center}
\vspace{-5mm}
\caption[Characteristic values of specific angular momentum
\citep{b1995}]{Characteristic values of the specific angular momentum for 
molecular clouds, cores, binaries, discs, and stars, indicating the many
orders of magnitude difference between the initial cloud and the present day 
Sun \citep{b1995}.}
\label{tab-angmom}
\end{table}
One of the classic problems of star formation is the requirement that nearly
all of the initial angular momentum of the molecular cloud core must be
removed or redistributed during the formation process. Stars typically have
far less angular momentum than the equivalent mass in the interstellar medium;
even molecular cloud cores rotate much more slowly than would be expected if
they had condensed from the diffuse interstellar medium with no loss of
angular momentum \citep[e.g.][]{getal1993}. Table \ref{tab-angmom} lists
characteristic values of the angular momentum for many of the stages of star
formation and evolution to illustrate this point. 

The basic angular momentum problem is well illustrated by the example
presented in \citet{s1978}: as an extreme case, consider a filamentary
interstellar cloud in the form of a cylinder with length $10$ pc and a radius
$0.2$ pc, rotating around its long axis at the galactic value of angular
velocity $\Omega = 10^{-15}$ s$^{-1}$. This cloud will have a mass of about
$1$ M$_\odot$ when the number density is $n_\text{H} = 20$ cm$^{-3}$. The
rotational effects will not impede collapse parallel to the long axis, but in
order to form a star of solar density the radius must decrease by a factor of
about 10$^{-7}$, while the conservation of angular momentum requires that
$\Omega$ must increase by $10^{14}$. The resulting rotation period would then
be roughly a minute, with the rotational velocity of the star becoming 20\% of
the speed of light, and the centrifugal force exceeds gravity at the equator
by four orders of magnitude. 

While this is a simplified example, the argument demonstrates the problem
well. In Table \ref{tab-angmom} the discrepancy between the specific angular
momentum of a T Tauri star ($10^{17}$ cm$^2$ s$^{-1}$) and that of a molecular
cloud core on the scale of 0.1 pc ($10^{21}$ cm$^2$ s$^{-1}$) is shown to be
around four orders of magnitude. The angular momentum problem is also
highlighted by observations that show many young stars are rotating slowly in
comparison to molecular clouds \citep[e.g.][]{getal1993, fmwom2001, lbah2003,
fetal2007}. It is possible that young stars could be braked after
formation by such mechanisms as stellar winds or interactions between the
magnetic field of the star and the protoplanetary disc \citep[e.g.][]{gl1979a,
k1991, mp2008}, but some braking must occur earlier to allow the protostar and
disc to form. 

Magnetic forces, gravitational forces and pressure forces may all play a role
in transporting angular momentum in star forming clouds. As outlined in
Section \ref{lr:mc}, the magnetic forces can exceed thermal forces in
molecular clouds and cores, and magnetic braking can remove much of the
initial angular momentum from a large scale cloud and determine the amount of
angular momentum remaining in the dense core \citep{bm1994, bm1995a, bm1995b}.
The basic process of magnetic braking is that the torque exerted on the
rotating fluid by the twisting of magnetic field lines causes the vertical
propagation of Alfv\'en waves. These waves carry angular momentum from the
cloud to material external to the cloud, in a manner first described by
\citet{ms1956} and numerically calculated by \citet{mp1979, mp1980}. This
braking of the rotational motions by the magnetic field is believed to be the
dominant mechanism for reducing the angular momentum in collapsing flows to
the low values observed in YSOs \citep{mzgh1993, mo2007}, and shall be
explained more fully in Chapter \ref{ch:derivs}. 

There are three major episodes of angular momentum transfer during the
formation and evolution of a young star. The first is the formation of
molecular cloud cores from the host cloud, where the usual solution to the
angular momentum problem is magnetic braking and the transfer of angular
momentum by Alfv\'en waves from the centre to the outer regions \citep{m1991}. 
This tends to produce centrally-condensed uniformly-rotating cloud cores that
are stable against fragmentation \citep{bm1994}. This solution has been
considered somewhat problematic as it is difficult to form binaries from such
stable cores and young binary stars are very common, however, stable cores
form centrifugal discs that may fragment and form binary systems
\citep{b1995}. 

More recent research has shown that the formation of cores may also be
triggered by fragmentation that is regulated by magnetic fields and ambipolar
diffusion, and that the rate of core growth is dependant upon the magnetic
field strength in the core \citep{kboy2007}. \citet{b2009} showed further that
the shape of molecular cloud cores is important to the angular momentum
transport and fragmentation. Their radiative hydrodynamical code, which
included prescriptions for the magnetic braking and ambipolar diffusion,
showed that oblate cores collapse to form rings that are susceptible to
fragmentation; and that the fragmentation of prolate cores depends upon the
density profile in the core --- cores that possess shallow density profiles
being more likely to fragment than those with steep profiles.

As is expected, angular momentum is approximately conserved during the dynamic
near-free fall collapse of a core into a flattened pseudodisc and so this
phase of star formation does not add to the angular momentum problem
\citep{gs1993a, gs1993b}. However inside the pseudodisc at the centre of the
dynamic collapse the magnetic field has built up to the point where it is
possible for magnetic braking to be the dominant force reducing the angular
momentum in the collapsing fluid. 

Fragmentation into a wide binary is likely during the phase of isothermal
pseudodisc evolution if the initial angular momentum is high; the spin angular
momentum of the cloud is converted into the orbital angular momentum of the
binary \citep{ms1956, l1985, b1995}. The occurrence of fragmentation during
collapse depends on the initial ratio of thermal to gravitational energy
\citep{ti1999a, ti1999b} and in a wide range of cases the final outcome of
gravitational collapse is the formation of a binary or multiple system via the
formation and fragmentation of a ring or bar structure in the fluid instead
of, or as well as, the expected disc configuration \citep{mh2003}. 

If the initial angular momentum of the core is low then the flattened
pseudodisc will be unable to fragment and will instead form a central
protostellar core surrounded by an optically thick disc where the angular 
momentum must be removed by magnetic fields, turbulent viscosity or the
formation of a massive planet that may interact with the disc to allow
material to accrete onto the star. 

The ideal MHD simulations of \citet{ml2008} showed that their strongly-braked
disc was surrounded by a vertically-extended structure, referred to as a
magnetogyrosphere, where the angular momentum was parked; this was supported
by a combination of the (toroidal) magnetic field and rotation. The collapsing
fluid was channelled by the magnetic wall surrounding the magnetogyrosphere
into the equatorial region, and the infalling material was braked as it
crossed a series of centrifugal barriers. It is possible that such a
magnetogyrosphere has been observed in the Class 0 source IRAM 04191
\citep[see the references and interpretation within][]{ml2008}; however higher
resolution observations are needed to confirm this. 

The magnetorotational instability (MRI) has been shown to generate turbulence
that can remove angular momentum from the infalling gas in the pseudodisc and
later Keplerian disc \citep{bh1998}. The MRI acts by converting the free
energy of differential rotation into turbulent motions that transfer angular 
momentum radially outward via the Maxwell stress of small scale, disordered
magnetic fields. Its properties have been studied in both the linear and
nonlinear stages and it has been shown to be very efficient at removing
angular momentum from discs \citep[e.g.][]{ss2002a, ss2002b, sw2003, sw2004,
sw2005}. \citet{skw2007a,skw2007b} showed that radial angular momentum
transport by the MRI operates where $2\eta\alpha^2 < 1$, where $\eta$ is the
ratio of the Keplerian rotation time to the neutral-ion momentum exchange time
and $\alpha$ is the midplane ratio of the Alfv\'en speed to the sound speed.
They also found that radial and vertical angular momentum transport operate in
different regions of the disc, with the MRI dominant when $\alpha \ll 1$. 

The semianalytic self-similar collapse solutions of \citet{kk2002} produced
rotationally-supported discs that were magnetorotationally stable, however,
their calculations were limited to one dimension and contained only ambipolar
diffusion in their approximation to the magnetic field diffusion, so they may
have oversimplified the calculations to the point where the disc could not be
magnetorotationally unstable. The MRI has been shown to be important in
protostellar and protoplanetary discs \citep{ss2002a, ss2002b} and may be
important at other points in the star formation process. 

Another possible way to remove angular momentum from lower-mass flattened
pseudodiscs is to form a massive planet or companion star in the disc. The
torques exerted on the disc by a planet can transport angular momentum from
the inner regions to the outer parts of the disc, and can also drive inflow
towards the central protostar. This behaviour has been observed in numerical
simulations of gravitational collapse such as those by \citet{bb2005},
\citet{mmi2008} and others. 

The final evolutionary stage in which the angular momentum problem is evident
is the necessary removal of angular momentum from the fluid that is accreting
from the innermost rotationally-supported disc, as it typically carries enough
angular momentum that it will spin up the star in a relatively short period of
time \citep{b1995}. In this regime the problem is likely solved by the
formation of a disc wind or stellar winds, as the protostellar wind blows away
the majority of the remaining non-accreted envelope of the core independent of
the magnetic braking \citep{sal1987}. It has been shown that a rotating disc
wind can remove angular momentum as well as mass from the innermost
partially-ionised region of the disc very efficiently \citep{mim2007,
mim2008}; the simulations of \citet{t2002} showed that 99\% of the total
angular momentum of their protostellar systems was transferred from the system
by outflows during the adiabatic second phase of collapse. 

Stellar dynamo activity may also contribute to the formation of a
magnetospheric region around the protostar that can generate a bipolar jet;
this could remove further angular momentum in order to resolve the angular
momentum problem \citep{l2003}. Alternatively, the interaction of the stellar
dipole magnetic field with the accretion disc could also remove angular
momentum from the protostar (or spin it up, if the initial rotation rate is
low). Such accretion would take place in a steady state along the stellar
magnetic field lines, where the spin-up torque on the star is balanced by a 
spin-down torque transmitted by field lines that thread the disc beyond the
corotation radius \citep{gl1979a, gl1979b, k1991}.

During the later stages of pre-main sequence contraction, when the interaction
between the star and disc is no longer important, the YSO evolves with
near-conservation of angular momentum, and the angular velocity of the
pre-main sequence protostar decouples from that of the envelope \citep{b1995}.

The angular momentum problem has in recent years been so well solved by
simulations that include the transport of angular momentum by magnetic braking
that it has given rise to the magnetic braking catastrophe in which all of the
angular momentum has been removed from the collapsing material so that it free
falls with no rotation onto the protostar. This behaviour has been observed in
many simulations with different coupling between the field and the matter;
generally it occurs in simulations that adopt IMHD \citep[e.g.][]{asl2003a,
asl2003b, ml2008} or non-ideal MHD in the form of ambipolar diffusion
\citep[e.g.][]{kk2002, ml2009}. This catastrophe has also been seen in
simulations in which the magnetic field and the rotational axis of the core
are not initially aligned \citep{pb2007, hc2009, ch2010}. 

\section{Magnetic Fields and the Magnetic Flux Problem}\label{lr:mfp}

In a similar puzzle to the angular momentum problem, there is a several orders
of magnitude discrepancy between the observed values and upper limits on the
magnetic flux of T Tauri stars and the flux associated with the corresponding
mass in the precollapse molecular cloud \citep{cf1953, ms1956}. The
mass-to-flux ratio in stars is very large, with $\mu \sim 10^4$--$10^5$ in
magnetic stars and $\mu \sim 10^8$ in the Sun, whereas the mass-to-flux ratio
in molecular clouds is typically $\mu \sim 1$ \citep{n1983,c1999}. The
magnetic flux problem of protostellar collapse is a complex puzzle that has
not yet been completely solved. 

The problem is illustrated using Spitzer's example of a cylinder-shaped cloud
containing 1 M$_\odot$ of constant density gas from Section \ref{lr:amp} once
more. Assuming conservation of flux, the magnetic field in the collapsed
protostar exceeds that in the original cloud by $10^{14}$, the same factor as
$\Omega$, so that if $B$ were 3 $\mu$G in the original cloud, the formed star
of solar type would possess a magnetic field that is equal to $3 \times 10^8$
G. The magnetic energy in the star exceeds the gravitational energy by several
times, but the difference between the field in the model star and that in a
real young stellar object ($\sim 10^3$--$10^5$ times) is not as large a
discrepancy as that which defines the angular momentum problem \citep[see also
the previous section]{s1978}. 

In order to reduce the magnetic flux in the collapsing cloud, there are
several processes that can be invoked: the first is local reconnection of 
field lines. This process occurs when two lines of force intersect at a point
of zero magnetic field strength or where two oppositely-directed lines of 
magnetic force are pushed together by gravitational or hydromagnetic forces
\citep{s1978}. This alters the topology of the field, and reconnection can
displace the region in which the flux crosses the forming or accreting disc.
\citet{gs1993b} observed reconnection as a numerical artefact in their
calculations, arguing that the pinched configuration of the magnetic field in
the equatorial plane of their pseudodisc could be subject to several
instabilities that would cause reconnection. Numerical reconnection has been 
observed in the collapse calculations of \citet{glsa2006, gcls2009} and
\citet{kls2010}; this is not true reconnection and can lead to anomalous
effects in their solutions \citep[such as the large resistivity required to
form centrifugal discs in][]{kls2010}.

The second and more commonly discussed mechanism for solving the magnetic flux
problem is ambipolar diffusion \citep[see e.g.][and many others]{ms1956,
m1991, fm1993, bm1994, sl1997}. Simulations have shown that ambipolar
diffusion is important during the phase of collapse pre-point mass formation,
when the mass-to-flux ratio is still subcritical and the contraction is slow
(see Section \ref{lr:col}); however, the flux contained in a 1 M$_\odot$
region of a molecular cloud core at point mass formation is still
$10^3$--$10^5$ times larger than that of a typical 1 M$_\odot$ protostar
\citep{cm1994, cm1995}.  

Further dissipation must then occur after the cloud becomes supercritical and
enters into the dynamic, near-free fall phase of collapse, prior to or during
the formation of a circumstellar disc. Ambipolar diffusion allows the magnetic
field to decouple from the rapidly collapsing neutral particles at small
radii, reducing the amount of magnetic flux dragged into the origin
\citep{lm1996, l1998}; in the numerical simulations of \citet{dm2001} this was
demonstrated to such a degree that they concluded that the magnetic flux
problem could be solved by ambipolar diffusion alone. The earlier numerical
simulation of \citet{ck1998} had shown that the rate of ambipolar diffusion is
strongly increased during dynamical collapse and causes a decrease in the
magnetic field of over two orders of magnitude (relative to the flux
associated with the same mass of pre-collapse gas) in their central object.
The semianalytic solutions of \citet{l1998} demonstrated that it was possible
to find similarity solutions for which there was no central flux when certain
constraints upon the coupling parameters, the degree of magnetisation and the
initial conditions of the core held true. 
 
By considering the timescales for ambipolar and Ohmic diffusion at high
densities \citet{nu1986a, nu1986b}, building on the calculations of
\citet{pm1965}, suggested that significant flux loss could only occur during
the dynamical phase of core collapse by Ohmic dissipation, which is important
once densities of $\sim 10^{11}$ cm$^{-3}$ are achieved in the core. The ratio
of the ambipolar diffusion to Ohmic diffusion timescales depends upon the
inverse square of the magnetic field, so that as the field is weakened, Ohmic
dissipation will come to dominate over ambipolar diffusion at the later stages
when the density is high and the magnetic field has decoupled from the gas
(see Figure \ref{eta-regimes}). 

The Ohmic dissipation process has been shown to be more important than
vertical collapse (in which the gas infalls along the magnetic field lines) in
resolving the magnetic flux problem both analytically \citep{nnu2002} and
numerically \citep{mim2007}. \citet{sglc2006} showed that for their
semianalytic solution with a large spatially-uniform Ohmic diffusion
coefficient and numerical reconnection the magnetic flux problem was resolved,
and \citet{tm2007a, tm2007b, tm2007c} found that their numerical simulations
of collapse with ambipolar and Ohmic diffusion produced a magnetic field in
the central region ($r \lesssim 10$ AU) of about $0.1$ G when the central
protostar has a mass $\sim 0.01$ M$_\odot$, which is approaching that observed
in strongly-magnetic stars; their simulations were, however, unable to follow
the growth of the protostar to actual stellar masses. 

\citet{slgc2009} performed numerical simulations comparing the relative
importance of ambipolar diffusion and ``reconnection diffusion'': the removal
of flux from gravitating clouds by turbulent reconnection \citep[as outlined
by][]{l2005}. They showed that while cores with low turbulence will be
dominated by ambipolar diffusion, cores that are more active will be subject
to reconnection diffusion at many densities, and that this will speed the
quasistatic contraction of the cloud core before dynamic collapse occurs. They
also showed that turbulent diffusivity behaved in a similar manner to enhanced
Ohmic diffusivity in non-turbulent simulations, aiding the removal of magnetic
flux from the collapsing fluid. 

It seems likely that star formation requires all of these processes to a
greater or lesser degree. While individual solutions may demonstrate a solved
magnetic flux problem for particular controlled parameters, there has been no
complete solution that includes both ambipolar and Ohmic diffusion, which are
crucial to the magnetic field behaviour. Similarly, these studies have
disregarded the potential of Hall diffusion changing the dynamics of collapse
and aiding the resolution of the magnetic flux problem at those intermediate
densities between the ambipolar and Ohmic diffusion regimes where Hall
diffusion dominates \citep{wn1999}.  

It will be shown in Chapter \ref{ch:asymptotic} that no rotationally-supported
disc can form in certain Hall similarity solutions when the magnetic braking
is particularly strong. This ``magnetic braking catastrophe'' in which all of
the angular momentum of the collapsing material is removed so that there is no
rotational support and no inner Keplerian disc forms has been observed in many
numerical simulations \citep[e.g.][]{kk2002, ml2009} and shall be discussed in
more detail in Section \ref{catastrophe}. It may be resolved by the inclusion
of Ohmic dissipation, which reduces the amount of magnetic braking in the
inner high density regions of collapse, allowing rotationally-supported discs
to form \citep{db2010, mim2010}. It will also be demonstrated in Chapter
\ref{ch:discuss} that it is possible to form larger Keplerian discs in the
Hall similarity solutions when the magnetic field is reversed (with respect to
the axis of rotation); in this situation the magnetic field diffusion can
increase the angular momentum of the collapsing gas, spinning it up and
changing the dynamics of the angular momentum problem in molecular cloud cores
that have low initial rotation rates. 

\section{The Hall Effect in Star Formation}\label{lr:hall}

Figure \ref{eta-regimes} showed the magnetic diffusion regimes of a weakly
ionised plasma, and indicated that Hall diffusion is important at intermediate
densities and field strengths such as those encountered in molecular cloud
cores. \citet{w2007} showed that in molecular clouds ions and electrons were
coupled to the field while the largest grains are not. As the density
increases the smaller grains and the ions also decouple from the field,
suggesting that Hall diffusion would become important after molecular cloud
cores start to collapse. 

Within a fluid in which the Hall diffusion is dominant the ions (and grains)
are strongly tied to the neutral particles by collisions, which also are
responsible for transmitting the electromagnetic stresses to the neutral
particles. The current is dominated by the electrons, which drift
perpendicular to the magnetic and electric fields so that the net Lorentz
force is zero. Collisions with the neutrals and oppositely-charged grains have
negligible effect on the electron motion \citep{wn1999}.

The drift velocity of the magnetic field in a weakly ionised medium is given
by 
\begin{equation}
    \mathbf{V}_B = \frac{c}{4\pi B}\left[(\eta_\parallel + \eta_A)
	\mathbf{J}_\perp \times \mathbf{B} - \eta_H\mathbf{J}_\perp\right]
\label{V_b1}
\end{equation}
where $\mathbf{J}$ is the current density, defined by 
\begin{equation}
    \mathbf{J} = \frac{4\pi}{c} (\nabla \times \mathbf{B}),
\label{Jdef}
\end{equation}
and $\mathbf{J}_\perp$ is the component of $\mathbf{J}$ that is perpendicular
to the magnetic field. The diffusivities $\eta_{\parallel,A,H}$ are those for
Ohmic, ambipolar and Hall diffusion respectively --- these determine the
coupling between the gas and the field, and shall be discussed in more detail
in Chapter \ref{ch:derivs}. Clearly, this can be divided into two velocities,
that caused by Hall diffusion of the field, 
\begin{equation}
    \mathbf{V}_{H} \approx \eta_H \mathbf{J},
\label{V_bH}
\end{equation}
and that of ambipolar and Ohmic diffusion,
\begin{equation}
    \mathbf{V}_{AO} \approx (\eta_A + \eta_\parallel) \left(\mathbf{J} \times
	\mathbf{B}\right). 
\label{V_bAO}
\end{equation}

\begin{figure}[t]
  \centering
  \includegraphics[width=3in]{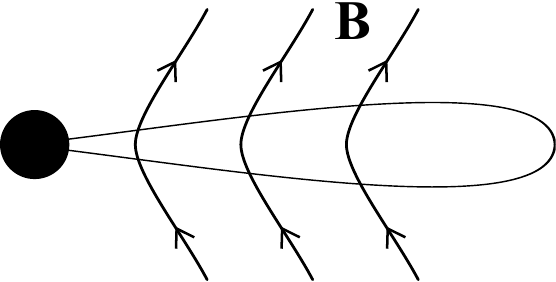}
  \vspace{-2mm}
  \caption[Cartoon of a thin disc with pinched magnetic field]
{Cartoon of a thin disc with a protostar at the centre. The disc has a
pinched magnetic field $\mathbf{B}$ with an azimuthal component that is
out of the page. The magnetic field is strictly vertical at the midplane of
the disc, and the current at the equator is toroidal.} 
\label{disc-cartoon}
\end{figure}
For a thin disc such as that illustrated in Figure \ref{disc-cartoon}, the
magnetic field has been pulled into a pinched configuration by the initial
collapse under IMHD; this field shape is expected from simulations of
gravitational collapse under a variety of conditions \citep[e.g.][]{gs1993a,
gs1993b} and observations \citep[e.g.][]{cc2006, ggg2008}. It is only once the
density has built up that ambipolar and Hall diffusion start to become
important \citep[e.g.][]{dm2001, w2004}. This pinching produces a strong 
toroidal current as indicated by Equation \ref{Jdef}, which causes Hall
diffusion in the azimuthal direction (Equation \ref{V_bH}). 

Hall diffusion twists up the field and in turn changes the angular momentum of
the neutral fluid, which in a rotationally-supported disc causes the gas to
fall inwards if it loses angular momentum, or outwards if it gains it. The
direction in which the field diffuses is determined by the initial orientation
of the field with respect to the axis of rotation, and the field drifts in the
opposite direction if the initial direction of the field is reversed.
Ambipolar and Ohmic diffusion, however, always cause the field to move in the
radial direction against the flow of the neutrals --- reversing the direction
of the field does not affect the direction of the field diffusion.

Under Hall diffusion any radial field drift causes an increase in the
azimuthal field and correspondingly the angular momentum, tying together the
angular momentum and magnetic flux problems outlined in previous sections.
It provides a mechanism for resolving both of these problems in simulations of
star formation, yet because of the numerical difficulties it incurs and the
previous assumption that it would not be important, Hall diffusion is only 
just starting to be included in simulations of collapsing molecular cloud
cores \citep[e.g.][]{kls2010-2}. The inclusion of Hall diffusion in the
semianalytic collapse model will provide a better understanding of the
collapse and disc dynamics in observed cores, and shall show that the Hall
effect can help resolve the magnetic braking catastrophe, as will be discussed
further in Section \ref{catastrophe}. 

\section{Project Outline}\label{lr:po}

This project aims to find similarity solutions for the  gravitational collapse
of rotating, isothermal, magnetic molecular cloud cores that are similar to
those of \citet{kk2002} but with Hall diffusion included in the conductivity
tensor. \citet{kk2002} adopted a conductivity that contained only ambipolar
diffusion, but it has been shown that the Hall diffusion term is important at
many of the intermediate densities found during gravitational collapse
\citep{w2004}. Hall diffusion also significantly alters the vector evolution
of the magnetic field (see Figure \ref{hall-vector}), and it is Hall diffusion
that is of particular interest here, especially in contrast with the ambipolar
and Ohmic diffusion terms.

In Chapter \ref{ch:derivs} the problem of gravitational collapse is properly
set out and the self-similar MHD equations are derived under the assumptions
of isothermality, axisymmetry, that the forming pseudodisc is thin and others
describing the magnetic field behaviour. These assumptions are discussed in
detail and the outer boundary conditions (corresponding also to the initial
conditions of the core, due to the self-similar nature of the problem) are
derived with values chosen to be compatible with both observations and
three-dimensional numerical collapse simulations. 

Chapter \ref{ch:asymptotic} derives the inner asymptotic solutions, which
function as the second set of boundary conditions for the collapse problem (at
the inner edge and infinite time). These inner similarity solutions are power
law descriptions of a Keplerian accretion disc which is supported by the
influence of both Hall and ambipolar diffusion, and a rapidly infalling
solution in which the material falls directly onto the central star without
forming a disc. These similarity solutions demonstrate the restrictions that
must be placed upon the Hall and ambipolar diffusion parameters in order to
find the true solutions, as the handedness of the magnetic response (shown in
Figure \ref{hall-vector}) can lead to unphysical solutions if the adopted
nondimensional Hall diffusion parameter is too large with respect to the
ambipolar diffusion parameter and the orientation of the magnetic field. 

Chapter \ref{ch:nohall} describes the construction of the self-similar model
in stages of increasing complexity, duplicating the results of \citet{kk2002}
and comparing these with numerical simulations. The iterative code that
numerically integrates the self-similar equations and solves the boundary
condition problem is outlined, as well as the approximations used to estimate
the positions of the magnetic diffusion and centrifugal shocks. The similarity
solutions to collapse models that have no magnetic field, ideal
magnetohydrodynamics or only ambipolar diffusion are then presented and
analysed; these are used to demonstrate that the size of the
rotationally-supported disc and the accretion rate onto the protostar depend
upon the initial rotation rate and the amount of magnetic braking in the core.

The main results of the project are presented in Chapter \ref{ch:hall}, where
the similarity solutions for the expanded model with the full conductivity
tensor are explored. Further details of the modelling process are described,
including simplifications adopted to calculate initial values of the variables
at a matching point (to increase the likelihood of convergence onto the true
solutions) and modifications to the equation set that are necessary near the
inner boundary. The importance of Hall diffusion to the collapse process is
compared with the ambipolar term that was previously expected to dominate the
magnetic field behaviour in the earlier stages of collapse. In particular, Hall
diffusion is shown to introduce additional shock fronts interior to the
magnetic diffusion and centrifugal shocks that alter the dynamics of the
collapse. 

The primary results of this thesis are summarised and discussed in more detail
in Chapter \ref{ch:discuss}, where the limitations of the model and the
assumptions supporting it are considered in the context of the results. The
difference in the behaviour and size of the protostar and Keplerian disc are
examined with respect to the boundary conditions of the core and the diffusion
coefficients. Depending upon the orientation of the magnetic field with
respect to the axis of rotation, Hall diffusion is shown to change the size of
the Keplerian disc by an order of magnitude, and can increase the accretion
rate onto the protostar. Hall diffusion is also shown to have a clear impact
on the magnetic braking catastrophe --- enhancing it by further decreasing the
angular momentum of the collapsing flow so that no disc forms, or changing the
magnetic braking behaviour to the point where Hall diffusion can
\textit{induce} rotation in initially nonrotating flows, again depending upon
the orientation of the field. Observational tests of this model are proposed
and suggestions for future work expanding the model by better calculating the
vertical angular momentum transport and adopting a different scaling in
self-similar space for the nondimensional Hall diffusion parameter are
presented. Finally, the conclusions of this work and its implications will be
summarised, emphasising that the Hall effect is indeed important in studies of
star formation. 

\cleardoublepage

\chapter{Self-Similar Gravitational Collapse}\label{ch:derivs}

Simulations of gravitational collapse are some of the most important tools
available for studying star formation. Numerical simulations, while
increasingly able to model complex physics across a wide range of densities,
are limited in resolution. Due to the computational costs of these, only the
most typical regions of parameter space are able to be explored; novel or
otherwise interesting solutions may be overlooked as typical values of the
model parameters are adopted. 

Semianalytic simulations are instead able to model solutions with very high
resolution rapidly, so that it is possible to explore larger regions of
parameter space. This is achieved by reducing the dimensionality of the 
problem and sacrificing some of the featured physics. A semianalytic model
then is the optimal choice for studying a simplified collapse problem with
Hall diffusion, so that the parameter space can be quickly explored and the
prospects for future research in numerical simulations identified. 

The goal of this work is to construct a semianalytic model of gravitational
collapse similar to that of \citet{kk2002} but including terms for Hall
diffusion in the equations for the magnetic field diffusion and braking. This
will allow the calculation of similarity solutions that show the importance of
Hall diffusion in molecular cloud cores and collapsing flows, as well as
comparisons between the influence of the Hall and ambipolar diffusion terms.
These solutions will then be used to motivate the inclusion of Hall diffusion
in numerical models of star formation and protostellar discs.

This chapter describes the derivation of the self-similar equations, and the
set of initial (outer) boundary conditions that describe a molecular cloud
core which is magnetically supercritical and in the process of contracting 
slowly in the radial direction. The MHD equations are presented in cylindrical
coordinates and the assumptions that must be adopted to ensure that
self-similarity holds are outlined. These assumptions include isothermality,
which guarantees that the sound speed in the core is constant, and
axisymmetry, which is adopted to reduce the dimensionality of the problem, 
The core is then assumed to be a thin disc, so that the equations may be
averaged over the scale height of the collapsing flow --- this further reduces
the dimensionality of the problem so that the MHD variables then depend only
on the radius and time. The prescriptions for the radial and azimuthal field
components are derived, and the azimuthal term is shown to control the
magnetic braking of angular momentum, which slows the rotation of the inner
regions of the core by transporting angular momentum to the external envelope.

Finally, the equations are self-similarised by choosing a similarity variable
that is a function of $r$ and $t$ and adopting nondimensional scalings of the
fluid variables so that they become functions of the similarity variable. The
MHD equations can then be written in terms of the nondimensional variables,
and once the outer boundary conditions are modified to reflect this
self-similarity, it is then possible to integrate the one-dimension equations
and find similarity solutions that shall provide new insight into the star
formation process. 

\section{Basic Equations}\label{eqeq}

The methods of \citet{kk2002} are used to simplify and rearrange the MHD
equations into the self-similar form in which the equations depend only upon
the nondimensional similarity variable, $x = r/c_st$, and are one-dimensional
ordinary differential equations. The principal difference between this work
and theirs is the inclusion of the full conductivity tensor in the induction
equation, rather than just the ambipolar diffusion term. The vertical angular
momentum transport is also modified by the inclusion of a Hall diffusion term. 

The magnetohydrodynamics equations are: 
\begin{align}
    \frac{\partial\rho}{\partial{t}} + \nabla &\cdot (\rho\mathbf{V}) = 0,\\
    \rho\frac{\partial{\mathbf{V}}}{\partial{t}} + \rho(\mathbf{V}
	\cdot\nabla)\mathbf{V} &+ \nabla{P} = \rho\mathbf{g} 
         + \frac{\mathbf{J} \times \mathbf{B}}{c}\,,\\
    \nabla^2\Phi &= 4\pi{G\rho},\\
    \nabla\cdot\mathbf{B} &= 0
\label{vectors1}
\end{align}
\begin{align}
    \text{and }\frac{\partial\mathbf{B}}{\partial{t}} 
    &= \nabla\times(\mathbf{V}\times\mathbf{B})
    - \nabla\times\left[\eta_\parallel(\nabla\times\mathbf{B})
     + \eta_H(\nabla\times\mathbf{B})\times\hat{\mathbf{B}}
     + \eta_A(\nabla\times\mathbf{B})_\perp\right]
\label{vectors2}
\end{align}
where $\rho$ is the gas density, $\mathbf{V}$ the velocity field, $P$ the
pressure, $\mathbf{g}$ the gravitational field, $\Phi$ the gravitational
potential, $c$ the speed of light, $\mathbf{J}$ the current, $\mathbf{B}$ the
magnetic field, $\mathbf{\hat{B}}$ is the unit vector in the direction of
$\mathbf{B}$, and $\eta_{\parallel, H, A}$ are the respective diffusion
coefficients for the Ohmic, Hall and ambipolar terms in the induction
equation. 

To simplify the calculations of the collapsing cloud core, the axis of
rotation is aligned with the background magnetic field. The external cloud
medium is characterised by constant low density, thermal pressure and angular
velocity. 

Using cylindrical coordinates, under the assumptions of isothermality (that $P
= \rho c_s^2$) and axisymmetry (that the derivatives with respect to $\phi$ are
equal to zero), the mass, radial momentum, and angular momentum conservation
relations, as well as the hydrostatic equilibrium equation and the induction
equation are given by: 
\begin{equation}
    \frac{\partial\rho}{\partial{t}} + 
    \frac{1}{r}\frac{\partial}{\partial{r}}(r\rho{V_{r}}) = 
    -\frac{\partial}{\partial{z}}(\rho{V_{z}}),
    \label{m1}
\end{equation}
\begin{align}
    \rho\frac{\partial{V_{r}}}{\partial{t}} + 
    \rho{V_{r}}\frac{\partial{V_{r}}}{\partial{r}} 
    &= \rho{g_{r}} - c_s^{2}\frac{\partial{\rho}}{\partial{r}} 
    + \rho\frac{V^{2}_{\phi}}{r}
    + \frac{B_{z}}{4\pi}\frac{\partial{B_{r}}}{\partial{z}} \nonumber \\
    &- \frac{\partial}{\partial{r}}\left(\frac{B_{z}^{2}}{8\pi}\right) - 
    \frac{1}{8\pi{r^{2}}}\frac{\partial}{\partial{r}}(rB_{\phi})^{2} 
    - \rho{V_{z}}\frac{\partial{V_{r}}}{\partial{z}},
    \label{rm1}
\end{align}
\begin{equation}
    \frac{\rho}{r}\frac{\partial}{\partial{t}}(rV_{\phi}) + 
    \frac{\rho{V_{r}}}{r}\frac{\partial}{\partial{r}}(rV_{\phi}) = 
    \frac{B_{z}}{4\pi}\frac{\partial{B_{\phi}}}{\partial{z}} + 
    \frac{B_{r}}{4\pi{r}}\frac{\partial}{\partial{r}}(rB_{\phi}) - 
    \rho{V_{z}}\frac{\partial}{\partial{z}}(r{V_{\phi}}),
    \label{am1}
\end{equation}
\begin{equation}
    \rho\frac{\partial V_z}{\partial t} 
    + \rho V_r\frac{\partial V_z}{\partial r}
    + \rho V_z\frac{\partial V_z}{\partial z}
    + c_s^{2}\frac{\partial{\rho}}{\partial{z}} = \rho{g_{z}} - 
    \frac{\partial}{\partial{z}}\left(\frac{B^{2}_{\phi}}{8\pi} + 
    \frac{B^{2}_{r}}{8\pi}\right) + 
    \frac{B_{r}}{4\pi}\frac{\partial{B_{z}}}{\partial{r}}
    \label{he1}
\end{equation} 
and
\begin{equation}
    \frac{\partial{\mathbf{B}}}{\partial{t}} = \nabla \times (\mathbf{V} 
    \times \mathbf{B}) -  \nabla \times 
    \left[\eta_\parallel\left(\nabla \times \mathbf{B}\right) + 
    \eta_{H}\left(\nabla \times \mathbf{B}\right)\times \mathbf{\hat{B}} 
    + \eta_{A}\left(\nabla \times \mathbf{B}\right)_{\perp}\right],
    \label{in1}
\end{equation}
where $g_{r}$ and $g_{z}$ are the radial and vertical components of the
gravitational field, and $c_s$ is the isothermal sound speed given by  $c_s =
(k_BT/m_n)^{1/2} \approx 0.19$ km s$^{-1}$ (with $k_B$ the Boltzmann constant,
$T$ the gas temperature, typically taken to be $10$ K, and $m_n$ the mean mass
of a gas particle). For the sake of comparison, the sound speed is denoted $C$
in many papers \citep[including Krasnopolsky and K\"onigl, 2002, and others
such as][etc]{cm1993, km2010}, and $a$ in others \citep[e.g.][etc]{sal1987,
glsa2006, ml2009}. The notation adopted here matches that of \citet{nh1985},
\citet{wn1999} and \citet{pw2008}, which was originally chosen in order to
differentiate the sound speed from the speed of light ($c$, no subscript).  

\section{Assumptions}\label{eqassume}

A number of assumptions must be made to simplify the equations to the point
that they are ordinary differential equations that can be solved
semianalytically. These assumptions are justified below. 

The cloud, pseudodisc and inner disc are all rotating slowly, and because of
this it is possible to assume that axisymmetry (that the partial derivatives
with respect to the azimuthal angle are equal to zero) holds true; this is not
expected to introduce significant errors to the simulations. This assumption
makes it impossible to include direct calculations of nonaxisymmetric effects
such as fragmentation or turbulence, which are important in studies of
high-mass star formation \citep[e.g.][]{mk2004, bkmv2007} but are less
relevant to the collapse of smaller isolated cores into low-mass stars such as
those studied in this work. 

The nonaxisymmetric core collapse calculations of \citet{bc2004} found that
the cores formed were near-oblate or triaxial rather than prolate, however,
the timescales, infall speeds and mass-to-flux ratios in their cores were
compatible with those from axisymmetric calculations. The three-dimensional
non-ideal MHD calculations of \citet{mim2007} showed that in their
rapidly-rotating collapses the thin disc is occasionally transformed into a
bar by nonaxisymmetric effects, and that this bar may fragment at later stages
of the collapse, however, they too were able to form discs with behaviour that
did not vary with azimuth. The assumption of axisymmetry in this model is
required in order to reduce the dimensionality of the problem to a 
one-dimensional self-similar equation set, and as near-axisymmetric collapse
occurs in three-dimensional models this assumption is considered generally to
be appropriate everywhere, except potentially in the inner disc regions.  

The magnetic field and the axis of rotation are assumed to be aligned, which
is not necessarily supported by observations of molecular cloud cores,
although it is a common and accepted simplification of the problem in
theoretical studies. Theoretical simulations have shown that the combination
of aligned rotation axis and magnetic field leads to the formation of
molecular cloud cores that have oblate shapes that match observations
\citep{cm1993, gww2002}. Observations of the magnetic field orientation and
the rotation of cores \citep[e.g.][]{vdhou2005} have yet to provide a decisive
answer on this matter, but the observations of the binary protostellar system
NGC 1333 IRAS 4A by \citet{grm2006} showed that the axis normal to the
envelope surrounding the binary lay between the outflow and the magnetic field
axes, which suggests that the spin and magnetic field axes were not in
alignment when collapse was initiated. These analyses were supported by
further observations by \citet{aetal2009} and modelling by \citet{ggg2008},
which showed that the field in this core possesses the classic hourglass
shape, even though it is not aligned with the outflow from the binary star. 

The simulations of \citet{pb2007}, \citet{hc2009} and \citet{ch2010} have
shown that the angle between the rotation axis and the magnetic field
influences the dynamics of collapse, and could prevent the formation of a
rotationally-supported disc. Disc formation in calculations with high initial
magnetic flux occurred most often when the magnetic field and axis of rotation
were orthogonal due to the reduced magnetic braking caused by the field. Given
these results, it would appear that a better choice for the initial shape of
the magnetic field would see it lie at an angle to the rotational axis,
however by definition this would violate axisymmetry, which, as stated
previously, is necessary for one-dimensional self-similarity. 

The assumption that the pseudodisc is thin is based upon the results of
\citet{fm1992, fm1993} and the many other simulations of Mouschovias and
coworkers, which have shown that an initially uniform, self-gravitating,
magnetised molecular cloud core that is spherically or cylindrically-symmetric
rapidly flattens along magnetic field lines. This assumption allows for a
further reduction in the dimensionality of the problem, although again it
implies that effects that depend on variation in the density or magnetic
field with height within the disc cannot be included in the collapse
calculation. Such processes may include turbulence and interactions between
active and dead zones in the disc that are caused by the differential (with
respect to $r$ and $z$) ionisation of the disc by external sources
\citep{pofb2007}. They are not expected to have a large effect on the overall
dynamics of early collapse, but they may again become important in the
innermost regions of protostellar discs in the later stages of collapse once
the adiabatic core and protostar have formed.

Isothermality is required for self-similarity, as the similarity variable $x$
is defined in terms of the isothermal sound speed, $c_s$. This assumption
breaks down due to radiative trapping when the central density reaches values
$\sim10^{10}$ cm$^{-3}$ \citep{g1963}. For a typical simulation of the
collapse of a nonrotating molecular cloud core this occurs on scales $r
\lesssim 5$ AU, which is several orders of magnitude smaller than the size of
the average core, and so nonmagnetic or nonrotating simulations
\citep[e.g.][]{ck1998} and numerical simulations \citep[e.g.][]{tm2005a,
tm2005b} often treat the inner region as point-like. 

By treating the forming protostar and the region around it as a central sink
cell into which the infalling matter disappears, it is possible to avoid
having to deal with complications such as the breakdown of isothermality and
the difficulty of tracking regions of high density in grid-based calculations.
The breakdown of isothermality is usually treated by adopting an adiabatic
equation of state once a critical density is exceeded in the inner region of
the collapse, as a way of avoiding the computational expense of performing
radiative transfer calculations \citep[e.g.][]{mim2007, mim2008, tm2007a,
tm2007b, tm2007c}. It is now possible to perform some full radiative transfer
simulations, but usually at the expense of magnetic field behaviour or grid
resolution \citep[e.g.][]{km2008, km2009, km2010, pb2009}. 

Isothermality breaks down in the inner disc region of the self-similar
collapse calculations of \citet{kk2002}, as the surface density in their
quasistatic Keplerian disc is much higher than that in the preceding
near-free fall stage. They estimated that radiative trapping occurs on scales
$\lesssim 10^2$ AU, although irradiation by the central protostar would
mitigate this to maintain vertical isothermality in the outer regions of the
disc. However, this same irradiation establishes a $T \propto r^{-1/2}$
variation in temperature in the innermost regions of the disc
\citep{dccl1998}. While this obviously means that self-similarity could break
down in the innermost regions of this disc, thermal stresses do not play a
significant role in the larger-scale core collapse dynamics and the assumption
of isothermality is not expected to introduce significant inaccuracies into
the overall results. 

The pressure support that could be provided by internal turbulence 
is neglected, as it cannot be directly calculated due to the dimensional
simplifications adopted and any parameterisation possible within the
self-similar framework would be so simplistic as to render it useless as a
measure of the influence of turbulence on collapse simulations. As with all
the other assumptions on the collapsing material, this is unlikely to
introduce large errors into the calculations.

\section{Vertical Averaging}\label{eqzaverage}

In order to simplify the calculations and reduce the problem to one that
depends only on $r$ and $t$, the pseudodisc is assumed to be geometrically
thin with a half-thickness $H(r) \ll r$, and Equations \ref{m1}--\ref{in1} are
vertically averaged by integrating over $z$. 

The density, radial velocity, azimuthal velocity and radial gravity are taken
to be constant with height, and the surface density of the pseudodisc is
defined as 
\begin{equation}
    \Sigma = \int^{\infty}_{-\infty}\rho{dz} = 2H\rho,
\label{Sigma}
\end{equation}
while the specific angular momentum is given by the equation
\begin{equation}
    J = rV_\phi.
\label{J}
\end{equation}

It is assumed that the thin disc is threaded by an open magnetic field 
configuration possessing an even symmetry: 
\begin{align}
    B_r(r,-z) &= -B_r(r,z),\\
    B_\phi(r,-z) &= -B_\phi(r,z),\\
    \text{and }B_z(r,-z) &= B_z(r,z);
\label{evenfield}
\end{align}
clearly $B_r = B_\phi = 0$ at the midplane of the disc. Where they are not
held constant with height the other physical variables are assumed to be
reflection-symmetric about the midplane. 

The solenoidal condition on the magnetic field (derived from Gauss' law for
magnetic fields, Equation 2.4, under the assumption of axisymmetry) is given
by
\begin{equation}
    \frac{\partial{B_z}}{\partial{z}} = 
    -\frac{1}{r}\frac{\partial}{\partial{r}}(rB_r).
    \label{solenoid}
\end{equation}
This implies that 
\begin{equation}
    \frac{\Delta{B_r}}{r} \approx \frac{\Delta{B_z}}{H};
\label{soldel}
\end{equation}
and the variation of $B_z$ from $z = 0$ to $z = H$ is then 
\begin{equation}
    \Delta{B_z} \approx \frac{H}{r}\, B_{r,s}
\label{delbz}
\end{equation}
where $B_{r,s}$ is the value of $B_r$ at the surface of the disc. Assuming
that $B_z$ is the dominant field component (that $B_{r,s} \lesssim B_z$, which
is not always true although the terms are always of the same order of
magnitude), and that $H \ll r$, then 
\begin{equation}
    \Delta B_z \ll B_z,
\label{varbz}
\end{equation}
so the variation of $B_z$ within the disc is small \citep{lrn1994}. It is then
possible to treat $B_z$ in the thin disc as being constant with height during
the vertical averaging; the solenoidal condition is used to average any terms
in which $\partial{B_z}/\partial z$ appears. 

The other field components are assumed to scale as 
\begin{equation}
    B_r(r,z) = B_{r,s}(r)\frac{z}{H(r)} 
    \label{b_r}
\end{equation}
and
\begin{equation}
    B_\phi(r,z) = B_{\phi,s}(r)\frac{z}{H(r)}
    \label{b_phi}
\end{equation}
where $B_{\phi,s}$ is the surface value of the azimuthal field component,
$B_\phi$; these scalings are motivated by the field configuration of a
rotationally-supported thin disc in which the field is comparatively
well-coupled to the gas \citep{wk1993}. While this holds true in the outer
regions of the collapse, this approximation is no longer adequate to describe
the field behaviour in the inner regions of the disc where the field is only
weakly-coupled to the fluid \citep{l1996, w1997}. However, as none of the
dominant terms in the equation set depend upon the particulars of the vertical
variation of the field within the disc, it is reasonable to adopt these
scalings across the domain of the self-similar collapse \citep{kk2002}. A
better method of handing the field variation with height in the disc should be
explored in future research. 

The radial component of gravity, $g_r$, is taken from the monopole expression
\begin{equation}
g_r = -\frac{GM(r)}{r^2},
\label{g_r}
\end{equation}
where the enclosed mass $M(r) \approx M_c$ when the central mass dominates.
This was found by \citet{cck1998} to be near enough to the value given for the
radial gravitational force by an iterative calculation method that it could be
used without correction terms in their self-similar model. Following their 
work, the monopole expression for gravity is similarly adopted here \citep[as
in][]{kk2002}, with little expectation that it will introduce significant
errors into the calculation. 

In the interest of keeping this text readable, each of the MHD equations is
treated separately below. 

\subsection{Conservation of mass}\label{cmsubsection}

Using Equation \ref{Sigma}, the equation of continuity (Equation \ref{m1})
integrates to 
\begin{equation}
    \frac{\partial\Sigma}{\partial{t}} + 
    \frac{1}{r}\frac{\partial}{\partial{r}}(r\Sigma{V_{r}}) = 
    -\frac{1}{2\pi{r}}\frac{\partial\dot{M_{w}}}{\partial{r}},
    \label{m2}
\end{equation}
where the term on the right hand side represents the mass flux in a disc wind
with total outflow rate $\dot{M_{w}}(r)$ within a radius $r$, defined by 
the relation
\begin{equation}
   \frac{\partial\dot{M}_w}{\partial r} = 2\pi r\Sigma V_z.
\label{m_w}
\end{equation}
The calculations presented here do not include a disc wind, so this term is
now set to zero. A discussion of how a nonzero outflow mass flux could be
incorporated into the work in future is included in Chapter \ref{ch:discuss}
\citep[see also appendix C of][]{kk2002}. 

\subsection{Conservation of radial momentum}\label{crmsubsection}

With the approximation that $V_r$ is constant with height, the final term in 
the radial momentum conservation equation (Equation \ref{rm1}), $\rho{V_z}
\frac{\partial V_r}{\partial z}$, vanishes. Applying the approximations
outlined above and integrating the rest of the radial momentum conservation 
equation over $z$ gives 
\begin{align}
    \Sigma\frac{\partial{V_{r}}}{\partial{t}} + 
    \Sigma{V_{r}}\frac{\partial{V_{r}}}{\partial{r}} &= \Sigma{g_{r}} 
    - c_s^{2}\frac{\partial\Sigma}{\partial{r}} + 
    \Sigma\frac{V_{\phi}^{2}}{r} \nonumber\\
    &+ \frac{1}{8\pi}\int^{\infty}_{\infty}\left[2B_{z}
    \frac{\partial{B_{r}}}{\partial{r}} 
    - \frac{\partial{(B^{2}_{z}})}{\partial{r}} - 
    \frac{1}{r^{2}}\frac{\partial}{\partial{r}}(rB_{\phi})^{2}\right]dz,
    \label{rm2}
\end{align}
where the final integral cannot be integrated as easily as the terms on the
first line of the equation. 

The first term in the integral is integrated by rewriting the inner derivative
as 
\begin{equation} 
    \int^\infty_{-\infty} \frac{B_z}{4\pi}\frac{\partial{B_r}}{\partial{z}} dz
    = \frac{1}{4\pi}\int^\infty_{-\infty} 
    \left[\frac{\partial}{\partial{z}}(B_{z}B_{r}) 
    	- B_{r}\frac{\partial{B_z}}{\partial{z}}\right]dz;
\label{rmint1-1}
\end{equation}
the scaling of $B_r$ given in Equation \ref{b_r} and the solenoidal condition
(Equation \ref{solenoid}) are then applied so that
\begin{equation}
    \int^\infty_{-\infty} \frac{B_z}{4\pi}\frac{\partial{B_r}}{\partial{z}} dz
    = \frac{1}{4\pi}\int^\infty_{-\infty} 
    \left[\frac{\partial}{\partial{z}}\left(\frac{B_{z}B_{r,s}z}{H}\right) 
    + \frac{B_{r}}{r}\frac{\partial}{\partial{r}}(rB_r)\right] dz,
\label{rmint1-2}
\end{equation}
and finally the first term in this equation can be integrated over the disc
scale height and the second term simplified to give
\begin{equation}
    \int^\infty_{-\infty}\frac{B_z}{4\pi}\frac{\partial{B_r}}{\partial{z}}dz
    = \frac{B_{z}B_{r,s}}{2\pi} + \frac{1}{8\pi{r^2}}\int^\infty_{-\infty}
    \frac{\partial}{\partial{r}}(rB_r)^{2} dz.
\label{rmint1-3}
\end{equation}
The second term in the integral in \ref{rm2} is then integrated using the
assumption that $B_z$ is constant with height:
\begin{equation}
    \int^\infty_{-\infty}\frac{\partial}{\partial{r}}
    \left(\frac{B_z^2}{8\pi}\right) dz 
    = \frac{H}{4\pi}\frac{\partial}{\partial{r}}(B_z^2);
\label{rmint2-1}
\end{equation}
and this and Equation \ref{rmint1-3} are then substituted into Equation
\ref{rm2}: 
\begin{align}
    \Sigma\frac{\partial{V_{r}}}{\partial{t}} + 
    \Sigma{V_{r}}\frac{\partial{V_{r}}}{\partial{r}} 
    &= \Sigma{g_{r}} - c_s^{2}\frac{\partial\Sigma}{\partial{r}} 
    + \Sigma\frac{V_{\phi}^{2}}{r} + \frac{B_{z}B_{r,s}}{2\pi} 
    - \frac{H}{4\pi}\frac{\partial{B_z^2}}{\partial{r}} \nonumber \\
    &+ \frac{1}{8\pi{r^2}}\int^{\infty}_{-\infty}
    \frac{\partial}{\partial{r}}[r^2(B_r^2 - B_{\phi}^{2})]dz.
\label{rm3}
\end{align}

The last term of Equation \ref{rm3} is evaluated over the finite interval of
the disc height, $[{-H(r)},{+H(r)}]$, as the mass is distributed only between
the boundaries of the disc. This integral is then solved by parts: 
\begin{equation} 
    \int^{\infty}_{-\infty} 
    \frac{\partial}{\partial{r}}[r^2(B_r^2 - B_{\phi}^{2})] dz
    = \frac{\partial}{\partial{r}} 
     \left[r^{2} \int^{H}_{-H} (B_r^2 - B_\phi^2) dz \right] 
     - 2r^2 (B_{r,s}^2 - B_{\phi,s}^2)\left(\frac{dH}{dr}\right)
\label{rmint4-1}
\end{equation}
\citep{lrn1994}. The vertical scalings for the field components
(\ref{b_r}--\ref{b_phi}) are substituted into this, and the integral on the
right hand side of Equation \ref{rmint4-1} then becomes 
\begin{equation} 
    \int^{H}_{-H} \frac{\partial}{\partial{r}}[r^2(B_r^2 - B_{\phi}^{2})] dz
    = \frac{2}{3}\frac{\partial}{\partial{r}}[r^2H(B_{r,s}^2 - B_{\phi,s}^2)]
     - 2r^2 (B_{r,s}^2 - B_{\phi,s}^2)\left(\frac{dH}{dr}\right);
\label{rmint4-2}
\end{equation}
the derivative on the left of this equation is then expanded out so that the
integral becomes
\begin{equation} 
    \int^{H}_{-H} \frac{\partial}{\partial{r}}[r^2(B_r^2 - B_{\phi}^{2})] dz
    = \frac{2H}{3}\frac{\partial}{\partial{r}}[r^2(B_{r,s}^2 - B_{\phi,s}^2)]
     - \frac{4r^2}{3} (B_{r,s}^2 - B_{\phi,s}^2) \left(\frac{dH}{dr}\right).
\label{rmint4-3}
\end{equation}

All of the above terms are collected together so that the radial momentum
conservation equation is thus 
\begin{align}
    \Sigma\frac{\partial{V_{r}}}{\partial{t}} + 
    \Sigma{V_{r}}\frac{\partial{V_{r}}}{\partial{r}} 
    &= \Sigma{g_{r}} - c_s^{2}\frac{\partial\Sigma}{\partial{r}} 
    + \Sigma\frac{V_{\phi}^{2}}{r} + \frac{B_{z}B_{r,s}}{2\pi} 
    - \frac{H}{4\pi}\frac{\partial{B_z^2}}{\partial{r}} \qquad \nonumber \\
    &+ \frac{H}{12\pi{r^2}}
       \frac{\partial}{\partial{r}}[r^2(B_{r,s}^2 - B_{\phi,s}^2)]
     - \frac{1}{6\pi} (B_{r,s}^2 - B_{\phi,s}^2) \left(\frac{dH}{dr}\right).
    \qquad
\label{rm5}
\end{align}
Rearranging this and substituting in the monopole approximation for $g_r$
(given in Equation \ref{g_r}) gives the full vertically-averaged radial
momentum conservation equation: 
\begin{align}
    \frac{\partial{V_{r}}}{\partial{t}} + 
    {V_{r}}\frac{\partial{V_{r}}}{\partial{r}} 
    &= - \frac{GM}{r^2} 
    - \frac{c_s^{2}}{\Sigma}\frac{\partial\Sigma}{\partial{r}} 
    + \frac{J^2}{r^3} + \frac{B_{z}B_{r,s}}{2\pi\Sigma} 
    - \frac{HB_z}{2\pi\Sigma}\frac{\partial{B_z}}{\partial{r}} \qquad \nonumber\\
    &+ \frac{H}{12\pi\Sigma{r^2}}\frac{\partial}{\partial{r}}
    [r^2(B_{r,s}^2 - B_{\phi,s}^2)] 
    - \frac{1}{6\pi\Sigma}\left(\frac{dH}{dr}\right)
    (B_{r,s}^2 - B_{\phi,s}^2). \qquad 
\label{rm6}
\end{align}

\subsection{Conservation of angular momentum}\label{camsubsection}

As in the case of the radial momentum, the final term in the angular momentum
conservation equation (Equation \ref{am1}), $\rho{V_z}
\frac{\partial}{\partial{z}} (r{V_{\phi}})$, vanishes as $V_\phi$ is constant
with respect to $z$. By rearranging the equation and applying the solenoidal
condition to the partial derivatives of $B_z$ with respect to $z$, the angular
momentum equation becomes: 
\begin{equation} 
    \frac{\rho}{r}\frac{\partial}{\partial{t}}(rV_{\phi}) + 
    \frac{\rho{V_{r}}}{r}\frac{\partial}{\partial{r}}(rV_{\phi}) 
    = \frac{1}{4\pi}\frac{\partial}{\partial{z}}(B_zB_\phi)
    + \frac{B_r}{4\pi{r}}\frac{\partial}{\partial{r}}(rB_\phi)
    + \frac{B_\phi}{4\pi{r}}\frac{\partial}{\partial{r}}(rB_r);
    \label{am2}
\end{equation}
this equation is integrated over $z$ to give 
\begin{equation}
    \frac{\Sigma}{r}\frac{\partial}{\partial{t}}(rV_\phi) 
    + \frac{\Sigma{V_r}}{r}\frac{\partial}{\partial{r}}(rV_\phi)
    = \frac{B_zB_{\phi,s}}{4\pi}
    +  \int^\infty_{-\infty} 
     \left[\frac{B_r}{4\pi{r}}\frac{\partial}{\partial{r}}(rB_\phi)
    + \frac{B_\phi}{4\pi{r}}\frac{\partial}{\partial{r}}(rB_r)\right] dz.
\label{am3}
\end{equation}

Using the same methods as for the radial momentum conservation equation, the
integral on the right hand side of Equation \ref{am3} becomes
\begin{equation} 
    \int^\infty_{-\infty} 
     \left[\frac{B_r}{4\pi{r}}\frac{\partial}{\partial{r}}(rB_\phi)
    + \frac{B_\phi}{4\pi{r}}\frac{\partial}{\partial{r}}(rB_r)\right] dz
    = \frac{1}{6\pi{r^2}}\frac{\partial}{\partial{r}}(r^{2}B_{r,s}B_{\phi,s})
    - \frac{B_{r,s}B_{\phi,s}}{3\pi}\frac{dH}{dr}.
\label{am4-1}
\end{equation}
Substituting this and Equation \ref{J} for the specific angular momentum into
Equation \ref{am3}, and rearranging the terms once more gives the final
version of the vertically-averaged angular momentum conservation equation:
\begin{equation}
\frac{\partial{J}}{\partial{t}} + V_{r}\frac{\partial{J}}{\partial{r}}
    = \frac{rB_{z}B_{\phi,s}}{2\pi\Sigma}
    + \frac{H}{6\pi{r\Sigma}}\frac{\partial}{\partial{r}}
    (r^{2}B_{r,s}B_{\phi,s})
    - \frac{rB_{r,s}B_{\phi,s}}{3\pi\Sigma}\left(\frac{dH}{dr}\right).
    \label{am5}
\end{equation}

\subsection{Vertical hydrostatic balance}\label{hbsubsection}

The gas pressure is assumed to vanish at the disc surface. The pressure at the
midplane of the disc is approximated by
\begin{equation}
    p_c \approx \frac{\Sigma c_s^2}{2H}.
\label{p_c}
\end{equation}
Because the disc is thin, it is assumed that all accretion onto it in the
vertical direction has already taken place, so that $V_z = 0$ and all of the
terms in Equation \ref{he1} that feature it vanish. The vertical hydrostatic 
balance equation, when integrated, becomes 
\begin{equation}
    p_c = \Sigma{g_z} - \frac{B_{r,s}^{2} + B_{\phi,s}^2}{8\pi} 
    - \frac{HB_{r,s}}{8\pi}\frac{\partial{B_z}}{\partial{r}}.
\label{he2}
\end{equation}

The vertical component of gravity, $g_z$, is approximated by 
\begin{equation}
    g_z = \frac{\pi}{2}G\Sigma^2 + \frac{GM_{c}\Sigma{H}}{4r^3},
\label{g_z}
\end{equation}
where the first term is the local disc self-gravity, and the other is the
tidal squeezing by the central point mass. This is then substituted into
Equation \ref{he2}, which is rearranged to give
\begin{equation}
    \left(\frac{GM_{c}\Sigma}{4r^3} 
    - \frac{B_{r,s}}{8\pi}\frac{\partial{B_z}}{\partial{r}}\right)H^2
    + \left(\frac{B_{r,s}^{2} + B_{\phi,s}^2}{8\pi} 
	+ \frac{\pi}{2}G\Sigma^2\right)H -\frac{\Sigma{c_s^2}}{2} = 0,
\label{he3}
\end{equation}
the final form of the vertical hydrostatic equilibrium equation. 

\subsection{$z$-component of the induction equation}\label{bzsubsection}

The preceding four derivations produce the same equations as their
counterparts in \citet{kk2002}. As the focus of this work is to adopt a
more complete induction equation than has been modelled previously, which
contains terms for the ambipolar, Hall and Ohmic diffusion of the magnetic
field, the simplifications that affect the induction equation must be closely
examined. This will ensure that the similarity solutions obtained properly
explore how Hall diffusion of the field affects star formation. 

The $z$-component of the induction equation (Equation \ref{in1}) is 
\begin{equation} 
    \frac{\partial{B_{z}}}{\partial{t}} 
    = \Bigl[\nabla \times (\mathbf{V} \times \mathbf{B})\Bigr]_{z}\!\!
    -\left[\nabla \times \!
      \left(\!\eta_\parallel\!\left(\nabla \times \mathbf{B}\right) 
      + \eta_{H}\!\left(\nabla \times \mathbf{B}\right)\times \mathbf{\hat{B}} 
      + \eta_{A}\!\left(\nabla \times \mathbf{B}\right)_{\perp}\!\right)\right]_{z}
\label{in2}
\end{equation}
which is expanded into the form
\begin{align}
    \frac{\partial{B_{z}}}{\partial{t}}
    &= \frac{1}{r}\frac{\partial}{\partial{r}}
      \Bigl[r(V_{z}B_{r} - V_{r}B_{z})\Bigr] \nonumber\\
    & \quad - \left[\nabla \times \left( 
    \eta_\parallel(\nabla \times \mathbf{B})
    + \eta_{H}(\nabla \times \mathbf{B})\times \mathbf{\hat{B}}
    - \eta_{A}\left((\nabla \times \mathbf{B})\times \mathbf{\hat{B}}\right)
    \times \mathbf{\hat{B}}\right)\right]_{z}
    \label{in2e}
\end{align}
where $\mathbf{\hat{B}} = \mathbf{B}/B$ and $B = \sqrt{B^2_r + B^2_\phi +
B^2_z}$. As before, it is assumed that the solutions are axisymmetric, so that
all partial derivatives with respect to $\phi$ are zero. The vertical mass
flux was defined to be equal to zero, so the vertical component of the gas
velocity is $V_z = 0$. Applying these assumptions and expanding out the right
hand side further gives 
\begin{equation} 
    \frac{\partial{B_{z}}}{\partial{t}} 
    = -\frac{1}{r}\frac{\partial}{\partial{r}} \Biggl[r 
     \Bigl(\!V_{r}B_{z} + \!\Big[\eta_\parallel (\nabla \times \mathbf{B})
    + \frac{\eta_{H}}{B}(\nabla \times \mathbf{B}) \times \mathbf{B}
    - \frac{\eta_{A}}{B^{2}}\left((\nabla \times \mathbf{B}) \times 
    \mathbf{B}\right) \times \mathbf{B}\Big]_{\phi}\Bigr)\!\Biggr]
\label{in3}
\end{equation}
which becomes
\begin{align}
    \frac{\partial{B_{z}}}{\partial{t}} 
    = &- \frac{1}{r}\frac{\partial}{\partial{r}} 
      \Biggl[ r\biggl(V_{r}B_{z} + \eta_\parallel\!
      \left(\frac{\partial{B_{r}}}{\partial{z}} 
            - \frac{\partial{B_{z}}}{\partial{r}}\!\right)
    + \frac{\eta_{H}}{B}\!\left(\!B_{z}\frac{\partial{B_{\phi}}}{\partial{z}} 
      + \frac{B_r}{r}\frac{\partial}{\partial{r}}(rB_{\phi})\!\right) \nonumber\\
    & \quad- \frac{\eta_{A}}{B^{2}}\!\left(\!(B_{z}^{2} + B_{r}^{2})
      \left(\!\frac{\partial{B_{z}}}{\partial{r}} 
      - \frac{\partial{B_{r}}}{\partial{z}}\!\right) 
    - B_{r}B_{\phi}\frac{\partial{B_{\phi}}}{\partial{z}}
    + \frac{B_{z}B_{\phi}}{r}\frac{\partial}{\partial{r}}(rB_{\phi})\!\right) 
    \!\biggr)\!\Biggr]. 
    \label{in3e}
\end{align}

The magnetic flux enclosed within a radius $r$ is given by
\begin{equation}
    \Psi(r) = \Psi_c + 2\pi\int^{r}_{0}B_z(r')r'dr',
\label{intpsi}
\end{equation}
where $\Psi_c$ is the flux within the central point mass. This equation is
then rewritten in differential form as
\begin{equation}
    B_z = \frac{1}{2\pi{r}}\frac{\partial{\Psi}}{\partial{r}},
\label{difpsi}
\end{equation}
and its derivative with respect to time is
\begin{equation}
    \frac{\partial{B_{z}}}{\partial{t}} 
    = \frac{1}{2\pi{}r}\frac{\partial}{\partial{r}}
    \left(\frac{\partial{\Psi}}{\partial{t}}\right).
    \label{psi}
\end{equation}
This is substituted into Equation \ref{in3e} and the partial derivative with
respect to $r$ and the factor of $r^{-1}$ are cancelled to obtain 
\begin{align}
    \frac{1}{2\pi{}}\frac{\partial{\Psi}}{\partial{t}}
    = &- r\biggl[V_{r}B_{z} + \eta_\parallel\!
      \left(\frac{\partial{B_{r}}}{\partial{z}} 
            - \frac{\partial{B_{z}}}{\partial{r}}\!\right)
    + \frac{\eta_{H}}{B}\!\left(\!B_{z}\frac{\partial{B_{\phi}}}{\partial{z}} 
      + \frac{B_r}{r}\frac{\partial}{\partial{r}}(rB_{\phi})\!\right)
      \qquad\qquad\nonumber\\
    & \,- \frac{\eta_{A}}{B^{2}}\!\left(\!(B_{z}^{2} + B_{r}^{2})
      \left(\!\frac{\partial{B_{z}}}{\partial{r}} 
      - \frac{\partial{B_{r}}}{\partial{z}}\!\right) 
    - B_{r}B_{\phi}\frac{\partial{B_{\phi}}}{\partial{z}}
    + \frac{B_{z}B_{\phi}}{r}\frac{\partial}{\partial{r}}(rB_{\phi})\!\right) 
    \!\biggr]. 
\label{in4}
\end{align}

In dealing with the partial derivatives with respect to $z$, such as the $B_z
\frac{\partial B_\phi}{\partial z}$ term above, care must be taken to ensure
that the vertical scaling of all of the variables is properly accounted for. 
As such, the integration is presented with more intermediate steps than in the
previous calculations. 

The $\eta_{\parallel, H, A}$ terms depend on $B^{0,1,2}$ respectively, so the
leading fractions of the diffusive terms may be ignored as the integration
over $z$ is performed.  

The flux, magnetic force and Ohmic diffusion terms are integrated over $z$ to
give: 
\begin{equation}
    \int_{-\infty}^{+\infty}{\frac{1}{2\pi}}\frac{\partial\Psi}{\partial{t}}dz
    =\left[\frac{1}{2\pi}\frac{\partial\Psi}{\partial{t}}z\right]^{+H}_{-H}
     =\frac{2H}{2\pi}\frac{\partial\Psi}{\partial{t}};
\label{in5-1}
\end{equation}
\begin{equation}
    \int_{-\infty}^{+\infty}{V_rB_z}dz
    = \Biggl[V_rB_zz\Biggr]^{+H}_{-H} = 2HV_rB_z;
\label{in5-2}
\end{equation}
and
\begin{align}
    \int_{-\infty}^{+\infty}\left(\frac{\partial{B_r}}{\partial{z}}
        - \frac{\partial{B_z}}{\partial{r}}\right)dz
    &= \int_{-\infty}^{+\infty}
        \frac{\partial}{\partial{z}}\left(\frac{B_{r,s}z}{H}\right)
        - \frac{\partial{B_z}}{\partial{r}}dz\nonumber\\
    &= \left[\frac{B_{r,s}z}{H} 
        - \frac{\partial{B_z}}{\partial{r}}z\right]^{+H}_{-H}\nonumber\\
    &= 2\left(B_{r,s} - H\frac{\partial{B_z}}{\partial{r}}\right).
\label{in5-3}
\end{align} 
The Hall diffusion terms are rearranged into the form 
\begin{equation}
    B_z\frac{\partial{B_\phi}}{\partial{z}}
        + \frac{B_r}{r}\frac{\partial}{\partial{r}}(rB_\phi{})
    = \frac{\partial}{\partial{z}}(B_zB_\phi) 
            - B_\phi\frac{\partial{B_z}}{\partial{z}}
            + \frac{B_r}{r}\frac{\partial}{\partial{r}}(rB_\phi);
\label{in5-4-1}
\end{equation}
and the vertical scaling of the azimuthal field component is substituted into
the first term and the solenoidal condition (Equation \ref{solenoid}) is
applied to the second:
\begin{equation}
    B_z\frac{\partial{B_\phi}}{\partial{z}}
        + \frac{B_r}{r}\frac{\partial}{\partial{r}}(rB_\phi{})
    = \frac{\partial}{\partial{z}}\left(\frac{B_zB_{\phi,s}z}{H}\right)
            + \frac{B_\phi}{r}\frac{\partial}{\partial{r}}(rB_r) 
            + \frac{B_r}{r}\frac{\partial}{\partial{r}}(rB_\phi).
\label{in5-4-2}
\end{equation}
The integral of the Hall terms may then by written as
\begin{equation} 
    \int_{-\infty}^{+\infty}\!\left[B_z\frac{\partial{B_\phi}}{\partial{z}}
        + \frac{B_r}{r}\frac{\partial}{\partial{r}}(rB_\phi{})\right]\!dz
    = \int_{-\infty}^{+\infty}
        \!\left[\frac{\partial}{\partial{z}}\left(\frac{B_zB_{\phi,s}z}{H}\right)
          + \frac{1}{r^2}\frac{\partial}{\partial{r}}(r^2B_rB_\phi)\right]\!dz,
\label{in5-4-3}
\end{equation}
which, after the vertical scalings of $B_r$ and $B_\phi$ are substituted into
it, is evaluated to give: 
\begin{align}
    \int_{-\infty}^{+\infty}\!\left[B_z\frac{\partial{B_\phi}}{\partial{z}}
      + \frac{B_r}{r}\frac{\partial}{\partial{r}}(rB_\phi{})\right]\!dz
    &= \left[\frac{B_zB_{\phi,s}z}{H} 
      + \frac{z^3}{3r^2}\frac{\partial}{\partial{r}}
      \left(\frac{r^2B_{r,s}B_{\phi,s}}{H^2}\right)\right]^{+H}_{-H}\nonumber\\
    &= 2B_zB_{\phi,s} + \frac{2H^3}{3r^2}\frac{\partial}{\partial{r}}
      \left(\frac{r^2B_{r,s}B_{\phi,s}}{H^2}\right)\!.
\label{in5-4-4}
\end{align}

Finally, the ambipolar diffusion terms are expanded out and integrated. The
first of these terms is straightforward, as $B_z$ is regarded as constant with
height unless specifically differentiated with respect to $z$ and may be taken
outside of the integral, which is solved to obtain
\begin{align}
    \int_{-\infty}^{+\infty} (B_r^2+B_z^2) \frac{\partial{B_z}}{\partial{r}}dz
    &= \frac{\partial{B_z}}{\partial{r}}\int_{-\infty}^{+\infty}
        \left(\frac{B_{r,s}^2z^2}{H^2} + B_z^2\right)dz \nonumber\\
    &= \frac{\partial{B_z}}{\partial{r}}
        \left[\frac{B_{r,s}^2z^2}{3H^2} + zB_z^2\right]^{+H}_{-H} \nonumber\\
    &= 2H\frac{\partial{B_z}}{\partial{r}}
        \left(\frac{B_{r,s}^2}{3} + B_z^2\right).
\label{in5-5}
\end{align}
The second of the ambipolar diffusion terms is rearranged into the form
\begin{equation} 
    (B_r^2+B_z^2) \frac{\partial{B_r}}{\partial{z}}
    = B_r^2\frac{\partial{B_r}}{\partial{z}} 
      + B_z\frac{\partial}{\partial{z}}(B_rB_z) 
      - B_rB_z\frac{\partial{B_z}}{\partial{z}}
\label{in5-6-1}
\end{equation}
to which the solenoidal condition (Equation \ref{solenoid}) and the scalings
for the other field components are applied. The integral of this term is then
\begin{align}
   \int_{-\infty}^{+\infty}\!&(B_r^2+B_z^2) \frac{\partial{B_r}}{\partial{z}} dz
   \nonumber\\
   &= \!\int_{-\infty}^{+\infty}\! \frac{B_{r,s}^2z^2}{H^2}
    \frac{\partial}{\partial{z}}\!\left(\!\frac{B_{r,s}z}{H}\!\right)\!
    + B_z\frac{\partial}{\partial{z}}\! \left(\!\frac{B_{r,s}B_{z}z}{H}\!\right)
    + \frac{B_{r,s}B_zz^2}{Hr}\frac{\partial}{\partial{r}}\!
     \left(\!\frac{rB_{r,s}}{H}\!\right)\! dz;
\label{in5-6-2}
\end{align}
this is evaluated over the height of the disc to give
\begin{align}
    \int_{-\infty}^{+\infty}(B_r^2&+B_z^2)
        \frac{\partial{B_r}}{\partial{z}}dz
    = \left[\frac{B_{r,s}^3z^3}{3H^3} + \frac{B_z^2B_{r,s}z}{H} 
        + \frac{B_{r,s}B_zz^3}{3rH}\frac{\partial}{\partial{r}}
        \left(\frac{rB_{r,s}}{H}\right)\right]^{+H}_{-H}\nonumber\\
    &= \frac{2}{3}B_{r,s}^3  + 2B_z^2B_{r,s}
        + \frac{2}{3}H^2B_{r,s}^2B_z
        \left[\frac{d}{dr}[\ln(rB_{r,s})] - \frac{d}{dr}[\ln{H}]\right].
    \label{in5-6-3}
\end{align}
The third of the ambipolar diffusion terms is again straightforward; it is
vertically-averaged by applying the vertical scalings to the radial and
azimuthal components to the field and then performing the integral over $z$ to
find 
\begin{align}
    \int_{-\infty}^{+\infty}B_rB_\phi\frac{\partial{B_\phi}}{\partial{z}}dz
    &= \int_{-\infty}^{+\infty}\frac{B_{r,s}B_{\phi,s}z^2}{H^2}
      \frac{\partial}{\partial{z}}\left(\frac{B_{\phi,s}z}{H}\right)dz\nonumber\\
    &= \frac{B_{r,s}B_{\phi,s}^2}{H^3} \int_{-H}^{+H}z^2 dz \nonumber\\
    &= \frac{2}{3}B_{r,s}B_{\phi,s}^2.
\label{in5-7}
\end{align}
Finally, the last of the ambipolar diffusion terms in Equation \ref{in4} is
averaged by substituting in the vertical scalings of the field components
and then performing the integral:
\begin{align}
    \int_{-\infty}^{+\infty}\frac{B_\phi{}B_z}{r}
        \frac{\partial}{\partial{r}}(rB_\phi)dz
        &= \int_{-\infty}^{+\infty}\frac{B_{\phi,s}B_zz}{rH}
        \frac{\partial}{\partial{r}}\left(\frac{rB_{\phi,s}z}{H}\right)dz\nonumber\\
    &= \frac{B_{\phi,s}B_z}{rH}
        \frac{\partial}{\partial{r}}\left(\frac{rB_{\phi,s}}{H}\right)
        \int_{-\infty}^{+\infty}z^2dz\nonumber\\
    &= \frac{2}{3}B_zB_{\phi,s}^2H
        \left[\frac{d}{dr}[\ln(rB_{\phi,s})] - \frac{d}{dr}[\ln{H}]\right].
\label{in5-8}
\end{align}

Collecting all of these integrated terms into the same order as in Equation
\ref{in4} then gives the full vertically-averaged induction equation: 
\begin{align}
    &\frac{H}{2\pi}\frac{\partial\Psi}{\partial{t}}
    = -r\Biggl[HV_rB_z 
     + \eta_\parallel\!\left(\!B_{r,s} - H\frac{\partial{B_z}}{\partial{r}}\!\right)\!
     + \frac{\eta_H}{B} \!\left(\!B_zB_{\phi,s} + \frac{H^3}{3r^2}
       \frac{\partial}{\partial{r}}
       \!\left(\!\frac{r^2B_{r,s}B_{\phi,s}}{H^2}\!\right)\!\right)\nonumber\\
    &\quad - \frac{\eta_A}{B^2}\!\biggl[\!
       -\!\left(\!B_{r,s} - H\frac{\partial{B_z}}{\partial{r}}\!\right)\!
       \!\left(B_z^2 + \frac{1}{3}B_{r,s}^2\right)\!
       - \frac{1}{3}B_{\phi,s}^2B_{r,s}    \label{in6}\\
    &\quad + \frac{1}{3}HB_zB_{\phi,s}^2
       \!\left(\!\frac{d}{dr}[\ln(rB_{\phi,s})] 
       - \frac{d}{dr}[\ln{H}]\!\right)\! - \frac{1}{3}HB_zB_{r,s}^2
	\left(\frac{d}{dr}[\ln(rB_{r,s})] - \frac{d}{dr}[\ln{H}]\!\right)\!
        \!\biggr]\!\Biggr] .\nonumber
\end{align}
It is clear from this equation that the azimuthal field is pivotal in causing
Hall drift in the radial direction; $B_{\phi,s}$ should not be neglected, even
in axisymmetric models.

\subsection[Radial field component, $B_{r,s}$]{Radial field component,
$\mathbf{B_{r,s}}$}\label{b_rsubsection}

\citet{cck1998} found that, as with the radial component of gravity, one could
calculate $B_{r,s}$ iteratively for more accurate similarity solutions,
however the behaviour of the disc at $t\ge0$ was sufficiently well described
by the monopole expression that they did not need to adopt a more complete
method of calculating the radial field component. Following their example the
monopole approximation to $B_{r,s}$ is also adopted here to simplify the
calculation of the field: 
\begin{align}
B_{r,s} &= \frac{\Psi(r,t)}{2\pi{r^2}}.
\label{b_rs}
\end{align}
This simplification, which was also used by \citet{kk2002}, is not expected to
introduce any significant errors to the calculations.

\section{Vertical Angular Momentum Transport}\label{eqb_phis}

The vertical angular momentum transport above and within the pseudodisc is
achieved by magnetic braking, especially during the dynamic collapse phase
inwards of the magnetic diffusion shock. It is assumed that magnetic braking
remains the dominant angular momentum transport mechanism during the
subsequent evolution of the core, although it is likely that a
centrifugally-driven disc wind may dominate in the innermost
rotationally-supported disc. The approach to modelling the magnetic braking
adopted here is adapted from that of \citet{bm1994} for the pre-point mass
formation collapse phase. This formulation is not well-defined in the
innermost rotationally-supported regions of the disc, where the calculated
magnetic braking becomes stronger than is expected and the angular momentum
transport is expected to be dominated by a disc wind (this is discussed in
more detail in Chapter \ref{ch:discuss}). A cap is then placed upon the
azimuthal magnetic field component in order to ensure that it does not greatly
exceed the vertical component; because of this the magnetic braking
prescription is not expected to introduce significant errors into the inner
regions of the calculations. 

\begin{figure}[ht]
  \centering
  \includegraphics[width=4.7in]{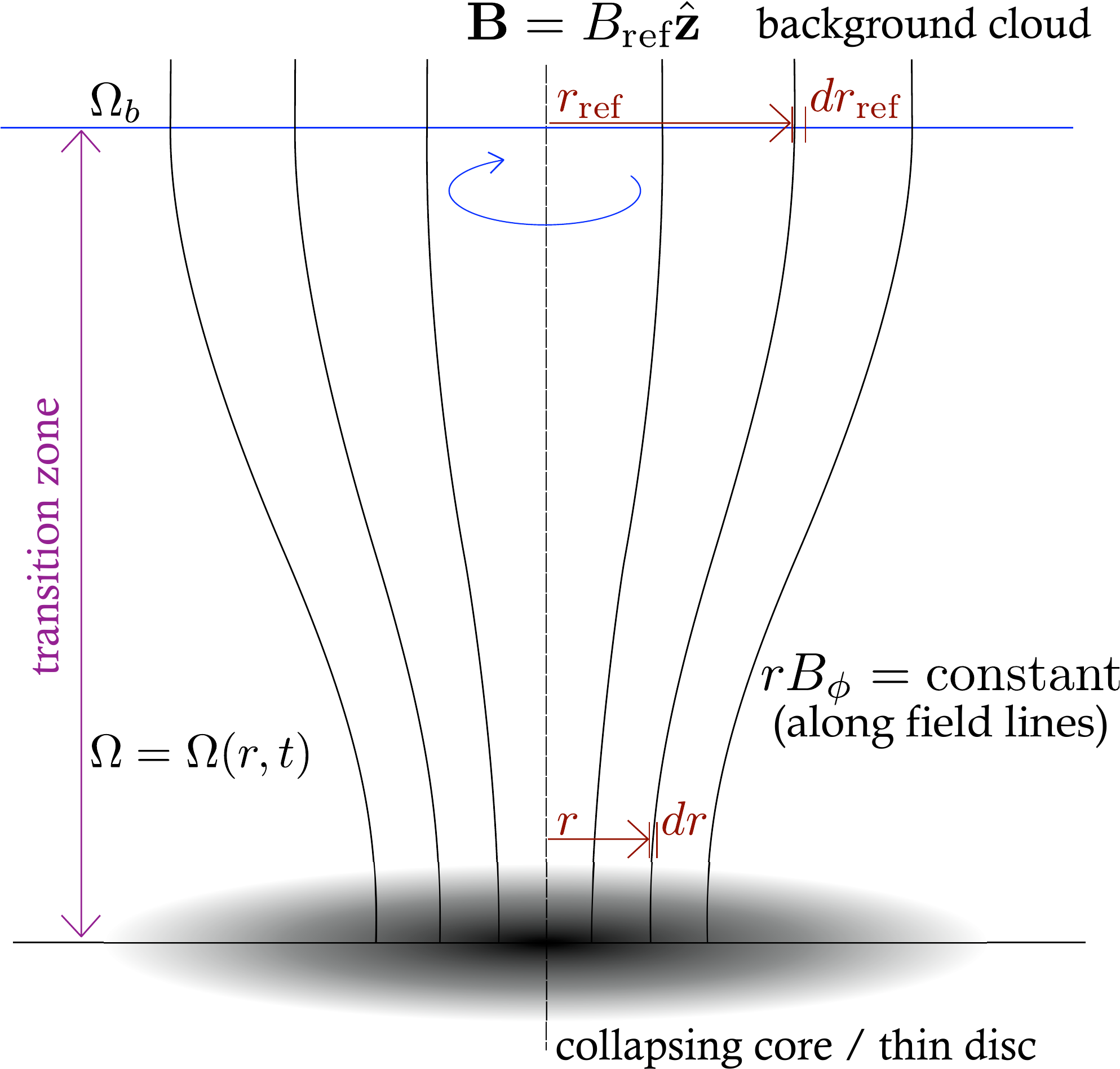}
  \caption[Vertical angular momentum transport]{The geometry of the thin
(pseudo) disc, background cloud and magnetic field lines, illustrating the
terms used to derive the angular momentum transport in the disc. The field is
assumed to be vertical at the disc midplane and in the background cloud.}
\label{angmom}
\end{figure}

The geometry of the magnetic field is illustrated in Figure \ref{angmom}; the
magnetic field is frozen into the low-density, constant-pressure external
medium, which has density $\rho_{ext}$ and angular velocity $\Omega_b$. Within
the external medium the magnetic field assumes the value $\mathbf{B} = B_{ref}
\hat{z}$, and the exterior flux tubes corotate with the core. Because the
transition region has a low moment of inertia relative to the core, and the
crossing time for Alfv\'en waves is always much smaller than the evolutionary
time of the core, the transition region can relax to a steady state during all
stages of contraction \citep{bm1994}. 

The induction equation under IMHD implies that 
\begin{equation}
    (B_p \cdot \nabla) \Omega = 0,
\label{Bp}
\end{equation}
where $B_p$ is the poloidal field, so that the angular velocity $\Omega$ is
constant on a magnetic surface. The force equation is similarly reduced to 
\begin{equation}
    (B_p \cdot \nabla) rB_\phi = 0,
\label{Bp2}
\end{equation}
which further implies that $rB_\phi$ does not change along the field lines.
The neutral particles carry the torque and angular momentum is carried upwards
by torsional Alfv\'en waves generated by the rotation of the disc. 

Over a period of time $dt$ an amount of material equal to $2\pi\rho\,r_{ref}
dr_{ref}$ moves from the undisturbed position $r_{ref}dr_{ref}$ in the
external medium along a flux tube with angular velocity $\Omega$ to a radius
$rdr$ at the disc surface. The angular momentum of the gas goes as 
\begin{equation}
    dJ = - [2\pi\rho_{ext}{r_{ref}}dr_{ref}] (V_{A,ext}dt) r_{ref}^2
        (\Omega - \Omega_b),
\label{dJ1}
\end{equation}
where $V_{A,ext}$, the external Alfv\'{e}n speed, is given by
\begin{equation}
    V_{A,ext} = \frac{B_{ref}}{\sqrt{4\pi{\rho_{ext}}}}.
\label{valf}
\end{equation}
For purely azimuthal motions in the external medium, the total angular
momentum in each flux tube is conserved. This angular momentum must be
removed from the disc at a rate equal to  
\begin{equation}
    \frac{dJ}{dt} = -2\pi{r_{ref}^2}v_{A,ext}\rho_{ext}(\Omega - 
       \Omega_b)r_{ref}dr_{ref},
\label{dJdt}
\end{equation}
which gives a torque on the disc equal to
\begin{equation}
    N = -\frac{2\pi{r_{ref}^2}(V_{A,ext}\rho_{ext})(\Omega - \Omega_b)
	r_{ref}dr_{ref}}{\pi{rdr}}.
\label{n1}
\end{equation}
The amount of flux remains constant along flux tubes, so that the flux through
the disc inside of a radius $r$ is equal to the amount of flux through the
cylindrical external cloud inside of the radius $r_{ref}$:
\begin{equation}
    \Psi = \int^{r}_{0}2\pi{r'}B_{z,eq}(r')dr' 
         = \pi{r_{ref}^2}B_{ref},
\label{psi2}
\end{equation}
where $B_{z,eq}$ is the value of $B_z$ at the midplane of the disc.
Thus 
\begin{equation}
    d\Psi = 2\pi{r}B_{z,eq}dr = 2\pi{r_{ref}}B_{ref}dr_{ref}
\label{dpsi}
\end{equation}
and
\begin{equation}
    \frac{r_{ref}dr_{ref}}{rdr} = \frac{B_{z,eq}}{B_{ref}},
\label{bzeq}
\end{equation}
so that the torque in Equation \ref{n1} becomes
\begin{align}
    N &=-\frac{2r^2_{ref}(\Omega - \Omega_b)B_{z,eq}}{B_{ref}}
         \left(\frac{B_{ref}\rho_{ext}}{\sqrt{4\pi\rho_{ext}}}\right),\nonumber\\
      &=-\frac{(\Omega - \Omega_b)B_{z,eq}(\Psi/2\pi)}{\pi{}V_{A,ext}}.
\label{n1-2}
\end{align}

The torque per unit area on the disc is given by
\begin{equation}
    N = \frac{rB_{z,eq}B_{\phi,s}}{2\pi},
\label{n2}
\end{equation}
and combining Equations \ref{n1-2} and \ref{n2} gives the steady state 
azimuthal component of the magnetic field at the surface of the disc:
\begin{equation}
    B_{\phi,s} = -\frac{\Psi}{\pi{r^2}}\frac{(r\Omega -
r\Omega_b)}{V_{A,ext}}.
\label{b_phis1}
\end{equation}
It is clear that the properties of the external medium determine the
conditions at the disc surface. This steady state approximation requires that
the ratio of the Alfv\'{e}n travel time in the external medium to the initial
radius of the cloud be less than the evolutionary timescale, which scales with
$r$ as $\sim r/|V_r|$. For the rotationally-supported discs presented in this
work $|V_r| \lesssim c_s$ (and $|V_r| \to 0$ as $r \to 0$), which is much
smaller than $V_{A,ext} \approx 5c_s$, the Alfv\'en speed for the adopted
temperature in the gas of 10K; this implies that the assumption of rapid
braking of the core should not introduce large errors into the solutions. 

The angular velocity $\Omega$ is given by the equation
\begin{equation}
    \Omega = \frac{1}{r}(V_\phi + V_{B\phi}),
\label{omega}
\end{equation}
where, using $\eta_P = \eta_A + \eta_\parallel$,
\begin{equation} 
    V_{B\phi} = -\frac{1}{B}\left[\eta_H(\nabla\times\mathbf{B})_\perp 
      -\eta_P(\nabla\times\mathbf{B})_\perp\times\mathbf{\hat{B}}\right]_\phi.
\label{vbphi1}
\end{equation}
This equation is then expanded out to become
\begin{align}
    V_{B\phi} &= - \frac{1}{B^2}\Biggl[\frac{\eta_H}{B}
      \left((B_z^2+B_r^2)\left(\frac{\partial{B_z}}{\partial{r}}
        -\frac{\partial{B_z}}{\partial{z}}\right) 
        + B_{\phi}B_z\frac{1}{r}\frac{\partial}{\partial{r}}(rB_\phi)
        - B_{\phi}B_r\frac{\partial{B_\phi}}{\partial{z}}\right)\nonumber\\
    &\qquad \qquad - \frac{\eta_P}{B^2}(B_r^2 + B_\phi^2 + B_z^2)
        \left(\frac{B_r}{r}\frac{\partial}{\partial{r}}(rB_\phi)
        + B_z\frac{\partial{B_\phi}}{\partial{z}}\right) \Biggr],
\label{vbphi2}
\end{align}
which is then vertically-integrated over the thin disc as in subsection
\ref{bzsubsection} to obtain 
\begin{align}
    H&V_{B\phi} \biggl(\!\frac{1}{3}B_{r,s}^2 + \frac{1}{3}B_{\phi,s}^2 
      + B_z^2\!\biggr) = \frac{\eta_P}{B^2} \Biggl[\!B_zB_{\phi,s}
       \left(\!\frac{1}{3}B_{r,s}^2 + \frac{1}{3}B_{\phi,s}^2 + B_z^2\!\right)
    \nonumber\\ &\qquad + HB_{r,s}B_{\phi,s}
    \biggl(\!\frac{B_{r,s}^2}{5} + \frac{B_{\phi,s}^2}{5} 
     + \frac{B_z^2}{3}\!\biggr)\!\! \biggl(\frac{d}{dr}[\ln(rB_{\phi,s})] 
     + \frac{d}{dr}[\ln(rB_{r,s})] - 2\frac{d}{dr}[\ln{H}]\!\biggr) 
    \!\Biggr]\nonumber\\ &- \frac{\eta_H}{B} \Biggl[\! 
    \left(\!\frac{B_{r,s}^2}{3} + B_z^2\!\right)\!\! \left(\!B_{r,s} 
     - H\frac{\partial{B_z}}{\partial{r}}\!\right) + \frac{1}{3}HB_zB_{r,s}^2
     \!\left(\!\frac{d}{dr}[\ln(rB_{r,s})] - \frac{d}{dr}[\ln{H}]\!\right)
    \nonumber\\ &\qquad \quad + \frac{1}{3}B_{r,s}B_{\phi,s}^2 
     - \frac{1}{3}HB_zB_{\phi,s}^2 \!\left(\!\frac{d}{dr}[\ln(rB_{\phi,s})] 
     - \frac{d}{dr}[\ln{H}]\!\right)\!\Biggr]\!.
\label{vb2}
\end{align}
Most of the terms in Equation \ref{vbphi2} have direct analogies in Equation
\ref{in4}, and the individual steps of the integration are not reproduced
here. 

This equation is simplified by omitting any terms of order ${\cal O}(H/r)$
save for the $[B_{r,s} - H(\partial{B_z}/\partial{r})]$ term, which is
important in refining the structure of the magnetic diffusion shock. This
simplification is explained in more detail and justified below in Section
\ref{eqsimple}; and the final form of $V_{B\phi}$ is then:
\begin{equation}
    V_{B\phi} = -\frac{1}{H}\left[\frac{\eta_H}{B}\left(B_{r,s} 
	- H\frac{\partial{B_z}}{\partial{r}}\right) 
	- \frac{\eta_P}{B^2}B_zB_{\phi,s}\right];
\label{vb3}
\end{equation}
this equation is equivalent to the ion-neutral drift velocity adopted by
\citet[][equation 9]{kk2002}, with the inclusion of terms describing the
effect of Hall diffusion. 

The $\Omega_b$ term is then dropped from Equation \ref{b_phis1} as the
molecular cloud rotation rate is slow compared with that of the collapsing
material. Rotation is dynamically important in the inner regions of the
solutions presented in this thesis, while it is not important in most
molecular clouds (see Section \ref{lr:mc} and the references therein), so it
is reasonable to declare that $\Omega \gg \Omega_b$ and dismiss $\Omega_b$ as
small. 

The external Alfv\'en speed, $V_{A,ext}$, is treated as a constant with
respect to the isothermal sound speed in these calculations, and is
parameterised by the formula 
\begin{equation}
    \alpha = \frac{c_s}{V_{A,ext}},
\label{alpha}
\end{equation}
where $\alpha$ is a constant, typically of order $0.1$ in the solutions. This
scaling of $V_{A,ext}$ is also a reasonable assumption, as the observations by
\citet{c1999} indicated that $V_{A} \approx 1$ km s$^{-1}$ over at least four
orders of magnitude in density ($\sim 10^3$--$10^7$ cm$^{-3}$) in their
observed molecular clouds. 

Equations \ref{vb3} and \ref{alpha} are substituted into \ref{b_phis1} to
find that
\begin{equation}
    B_{\phi,s} = -\frac{\Psi\alpha}{\pi{r^2}c_s}
     \left[\frac{J}{r} - \frac{\eta_H}{B}
       \left(B_{r,s} - H\frac{\partial{B_z}}{\partial{r}}\right)\right]
     \left[1 + \frac{\Psi\alpha}{\pi{r^2}c_s}\frac{\eta_P}{B^2}
       \frac{B_z}{H}\right]^{-1}.
\label{b_phis}
\end{equation}
Note that $B = \sqrt{B_{r,s}^2 + B_{\phi,s}^2 + B_z^2}$, so that the terms
featuring $\eta_{H,P}/B^{1,2}$ have an implied $B_{\phi,s}$ dependence. This
is typically solved for numerically when calculating the azimuthal field
component. 

For the inner solutions, $\Omega$ increases with decreasing $r$ (proportional
to $r^{-3/2}$ in the Keplerian disc solution and $r^{-1}$ in the free fall
solution); this would make $B_{\phi,s}$ the dominant field component at the
surface near to the central point mass. Such behaviour is not expected in a
real disc, where internal kinks of the field and magnetohydrodynamical
instabilities (for example, the magnetorotational instability) should reduce
the value of $B_{\phi}$ at the surface. An artificial limit on $B_{\phi,s}$ is
imposed so that 
\begin{equation}
|B_{\phi,s}| \leq \delta{B_z}, 
\label{b_phislim}
\end{equation}
where $\delta$ is a parameter of the model usually chosen to be $\delta = 1$
in order to ensure that the azimuthal field component does not exceed the
vertical component. \citet{kk2002} point out that this value quite
conveniently corresponds to that expected for a rotationally-supported 
ambipolar diffusive disc where the vertical angular momentum transport is
dominated by a centrifugally-driven wind. 

Applying this cap to Equation \ref{b_phis} then gives the final equation for
$B_{\phi,s}$:
\begin{equation}
    B_{\phi,s} = -\min\left[\!\frac{\Psi\alpha}{\pi{r^2}c_s}
     \left[\frac{J}{r} - \frac{\eta_H}{B}\!
       \left(\!B_{r,s} - H\frac{\partial{B_z}}{\partial{r}}\!\right)\!\right]
     \!\!\left[1 + \frac{\Psi\alpha}{\pi{r^2}c_s}\frac{\eta_P}{B^2}
       \frac{B_z}{H}\right]^{-1};\delta{B_z}\right].
\label{b_phisfinal}
\end{equation}

\section{Further Simplifications}\label{eqsimple}

The disc equations are further simplified by recognising that the thin disc
approximation adopted earlier states that $H \ll r$, this in turn means that 
terms of order ${\cal O}(H/r)$ are small in comparison with other terms and
can then be dropped from the equations. 

The only term of order ${\cal O}(H/r)$ that was found to influence the
solutions of \citet{kk2002} was the $[B_{r,s} - H(\partial{B_z}/\partial{r})]$
term in the radial momentum equation, which was necessary to refine the
structure of the magnetic diffusion shock. A similar result is obtained in
these solutions, and the term is retained in all of the equations in which it
appears, including Equation \ref{vb3} where it is the dominant component of
the Hall diffusion term in $V_{B\phi}$. 

It is expected that $B_z$ will generally be the dominant field component
within the disc, although due to the cap placed upon $B_{\phi,s}$ and the
monopole approximation adopted for $B_{r,s}$ it may not be the dominant term
at the disc surface. In order to reproduce the results and equations of
\citet{kk2002}, the $B_z^2$ terms are kept preferentially over those of the
other field components of the same order in the induction and azimuthal field
drift equations. 

Taking all of these into account gives the simplified set of equations:
\begin{equation}
    \frac{\partial{\Sigma}}{\partial{t}} 
    + \frac{1}{r}\frac{\partial}{\partial{r}}(rV_r\Sigma) = 0,
\label{cmfinal}
\end{equation}
\begin{equation}
    \frac{\partial{V_r}}{\partial{r}} + V_r\frac{\partial{V_r}}{\partial{t}}
    = g_r - \frac{c_s^2}{\Sigma}\frac{\partial{\Sigma}}{\partial{r}}
    + \frac{B_z}{2\pi{\Sigma}} 
    \left(B_{r,s} - H\frac{\partial{B_z}}{\partial{r}}\right) + \frac{J^2}{r^3},
\label{crmfinal}
\end{equation}
\begin{equation}
    \frac{\partial{J}}{\partial{t}} + V_r\frac{\partial{J}}{\partial{r}}
    = \frac{rB_zB_{\phi,s}}{2\pi\Sigma},
\label{camfinal}
\end{equation}
\begin{equation}
    \frac{\Sigma{c_s^2}}{2H} = \frac{\pi}{2}G\Sigma^2 
     + \frac{GM_c\rho{H}^2}{2r^3} + \frac{1}{8\pi}
     \left(B_{r,s}^2 + B_{\phi,s}^2  - B_{r,s}H
     \frac{\partial{B_z}}{\partial{r}}\right),
\label{hefinal}
\end{equation}
and
\begin{equation}
    \frac{H}{2\pi}\frac{\partial\Psi}{\partial{t}} = -rHV_rB_z
    - r\eta_\parallel \!\!\left[B_{r,s} 
    - H\frac{\partial{B_z}}{\partial{r}}\right]\! 
    - \frac{r\eta_H}{B}B_zB_{\phi,s} - \frac{r\eta_A}{B^2}
    \!\!\left[\!B_{r,s} - H\frac{\partial{B_z}}{\partial{r}}\!\right]\!\!B_z^2.
\label{infinal}
\end{equation}

\section{Self-Similar Form of the Equations}\label{eqssform}

With the assumption that the disc is axisymmetric and thin, the collapse
behaviour resembles the shape of the inside-out collapse models described in
Chapter \ref{ch:litrev}, where at any instant in time the solution looks like
a stretched version of itself at previous times; this fractal-like behaviour
is referred to as self-similarity. The pseudodisc forms as a collapse wave
(referred to as the magnetic diffusion shock) propagates outwards at the speed
of sound, and similarly the outer boundary of the inner rotationally-supported
disc or free fall collapse region moves outwards at the sound speed.

Gravitational collapse occurs over many orders of magnitude in radius and
density, so that the point mass has negligible dimensions in comparison with
the accretion flow. The self-similarity of the waves of infall arises because
of the lack of characteristic time and length scales in the flow. The only
dimensional quantities that effect the flow are the magnetic field
$\mathbf{B}$, the diffusion coefficients $\eta_{\parallel, H, A}$, the
gravitational constant $G$, the isothermal sound speed $c_s$, the local radius
$r$ and the instantaneous time $t$; this means that, except for scaling
factors, all of the flow variables may be written as functions of a similarity
variable $x$, which is defined as 
\begin{equation}
    x = \frac{r}{c_st}.
  \label{x}
\end{equation}
At a temperature of $T = 10$ K in a molecular cloud core of typical
composition, $c_s = 0.19$ km s$^{-1}$; \citet{kk2002} noted that for this
value of the sound speed $x = 1$ corresponds to a distance of $r \approx
6\times10^{15}$ cm (400 AU) when $t=10^4$ yr (which is the characteristic age
of a Class 0 YSO) and to a distance of $r\approx6\times10^{16}$ cm (4,000 AU)
when $t=10^5$ yr (the age of a Class 1 YSO). The Class 0 YSO IRAM 04191 has a
dense inner disc-like structure that resembles a tilted ring with an average
radius of $r_0 \sim 1400$ AU \citep{lhw2005} --- this is of the same order of
magnitude as the centrifugal shock radius in the disc-forming solutions at the
same age. 

The physical quantities are then expressed as the product of a nondimensional
flow variable that depends only upon $x$ and a dimensional part constructed
from $c_s$, $G$ and $t$. These are:  
\begin{align}
    \Sigma(r,t) &= \left(\frac{c_s}{2\pi{Gt}}\right)\sigma(x),&
       g_r(r,t) &= \left(\frac{c_s}{t}\right)g(x),\label{sssigmag}\\
    V_r(r,t) &= c_su(x), &
      H(r,t) &= c_sth(x), \label{ssuh}\\
    V_\phi(r,t) &= c_sv(x), &
         J(r,t) &= c_s^2tj(x), \label{ssvj}\\
    M(r,t) &= \left(\frac{c_s^3t}{G}\right)m(x), &
    \dot{M}(r,t)&= \left(\frac{c_s^3}{G}\right)\dot{m}(x),\label{ssmmdot}\\ 
    \mathbf{B}(r,t) &= \left(\frac{c_s}{G^{1/2}t}\right)\mathbf{b}(x),&
         \Psi(r,t) &= \left(\frac{2\pi{c_s^3t}}{G^{1/2}}\right)\psi(x),\label{ssbpsi}\\
    \text{and }\eta_{\parallel,H,A} &= c_s^2t\eta'_{\parallel,H,A}. \label{sseta}
\end{align}
These equations take the same form and notation as those in \citet{kk2002},
with the addition of extra diffusion coefficients to model the magnetic field
diffusion more completely.

The Ohmic and ambipolar diffusion terms scale together, to a zeroth-order
approximation, as they possess a similar dependence upon $B$ and appear in the
induction equation and the azimuthal field component equation multiplied with
the same magnetic field terms. Because the field within the thin disc is
effectively vertical, both ambipolar and Ohmic diffusion influence the field
drift in the same manner (see Figure \ref{ambi-vector}). While one type of
diffusion may dominate over the other at any individual point in the disc (in
general, ambipolar diffusion dominates in the outer regions where the density
is low and Ohmic diffusion dominates in the innermost regions where the
density is high), only one term is needed in order to study the change in the
disc behaviour introduced by the Hall diffusion term that is of most interest
in this work. The Ohmic and ambipolar diffusion terms are then combined into a
single term parameterised by the dimensionless constant $\tilde{\eta}_A$, that
for simplifying the discussion is referred to as the ambipolar diffusion
parameter. 

The ambipolar diffusion coefficient in a molecular cloud core without grains
is given (in cgs units) by the equation 
\begin{equation}
    \eta_A = \frac{B^2}{4\pi\gamma\rho_i\rho},
\label{etaa}
\end{equation}
where $1/\gamma\rho_i = \tau_{ni}$ is the neutral-ion momentum exchange
timescale, parameterised as
\begin{equation}
    \tau_{ni} = \frac{\tilde{\eta}_A}{\sqrt{4{\pi}G\rho}};
\label{tauni}
\end{equation}
the nondimensional ambipolar diffusion parameter $\tilde{\eta}_A$ is a
constant of the model \citep[simply denoted $\eta$ in the solutions of
][]{kk2002}. $\eta_A/B^2$ is then self-similarised using the scalings above to
give 
\begin{equation}
    \frac{\eta_A'}{b^2} = \tilde{\eta}_A\frac{h^{3/2}}{\sigma^{3/2}};
\label{ssetaa}
\end{equation}
it is important to note that the self-similarity of the solution depends upon
the relationship $\rho_i \propto \rho^{1/2}$ as discussed in Section
\ref{lr:mhd} \citep[and][]{e1979}. 

As a matter of pragmatism, a similar scaling with respect to the density and
scale height is adopted for the Hall diffusion parameter, $\eta_H$. By stating
that the self-similar Hall diffusion coefficient scales as 
\begin{equation}
    \frac{\eta'_H}{b} = \tilde{\eta}_Hb\frac{h^{3/2}}{\sigma^{3/2}},
\label{ssetah}
\end{equation}
where $\tilde{\eta}_H$ is the constant nondimensional Hall diffusion parameter
used to characterise the solutions, the ratio of the nondimensional ambipolar
and Hall diffusion parameters becomes the most important factor in determining
the magnetic behaviour of the similarity solutions. In truth, the Hall
diffusion coefficient could be scaled with respect to the density and field
strength by multiplying the nondimensional Hall parameter by any function of
the similarity variable $x$ and the fluid variables. This topic is discussed
in more detail in Section \ref{future}, where an alternate scaling is proposed
for future work on the self-similar collapse model. The scaling of $\eta'_H$
given in Equation \ref{ssetah} is appropriate for a molecular cloud core with
grains acting as the dominant positive charge carriers. 

For convenience the variable $w\equiv{}x-u$ is used to simplify the equation
set. The similarity variables are then used to rewrite Equations \ref{difpsi}
and \ref{cmfinal}--\ref{infinal} in self-similar form:
\begin{equation}
   \frac{d\psi}{dx} = xb_z,
\label{sspsi}
\end{equation}
\begin{equation}
   \frac{dm}{dx} = x\sigma,
\label{sscm}
\end{equation}
\begin{equation}
   (1-w^2)\frac{1}{\sigma}\frac{d\sigma}{dx}
   = g + \frac{b_z}{\sigma}\left(b_{r,s} - h\frac{db_z}{dz}\right)
   + \frac{j^2}{x^3} + \frac{w^2}{x},
\label{sscrm}
\end{equation}
\begin{equation}
   \frac{dj}{dx} = \frac{1}{w}\left(j 
   - \frac{xb_zb_{\phi,s}}{\sigma}\right),
\label{ssam}
\end{equation}
\begin{equation}
   \left(\frac{\sigma{m_c}}{x^3} - b_{r,s}\frac{db_z}{dx}\right)h^2
   + \left(b_{r,s}^2 + b_{\phi,s}^2 + \sigma^{2}\right)h - 2\sigma = 0,
\label{ssvhe}
\end{equation}
and
\begin{equation}
   \psi - xwb_z + \tilde{\eta}_Hxb_{\phi,s}b_zbh^{1/2}\sigma^{-3/2}
   + \tilde{\eta}_Axb_z^2h^{1/2}\sigma^{-3/2}\left(b_{r,s} 
                   - h\frac{db_z}{dx}\right) = 0.
\label{ssin}
\end{equation}
These equations are augmented by the self-similar definitions 
\begin{align}
   m &= xw\sigma,\label{ssmxw}\\
   \dot{m} &= -xu\sigma \\
   \text{and }g &= -\frac{m}{x^2},  
\label{ssextra}
\end{align}
while the other field components are given by 
\begin{equation}
   b_{r,s} = \frac{\psi}{x^2}
\label{ssb_r}
\end{equation}
and
\begin{equation}
   b_{\phi,s} = -\min\!\left[\! \frac{2\alpha\psi}{x^2}\!\left[\frac{j}{x} 
    -\frac{\tilde{\eta}_Hh^{1/2}b}{\sigma^{3/2}}\!\left(\!b_{r,s} 
    -h\frac{db_z}{dx}\!\right)\!\right]\!\!
    \left[1 + \frac{2\alpha\tilde{\eta}_Ah^{1/2}\psi{b_z}}{x^2\sigma^{3/2}}
    \right]^{-1}\!;\delta{b_z}\!\right]\!. 
\label{ssb_phis}
\end{equation}
These equations completely describe the collapse of the core into a pseudodisc
and the accretion onto the central point mass (potentially through a
rotationally-supported disc). The equations are the same as equations 20--32 
of \citet{kk2002}, save for the induction and azimuthal field equations which
contain new terms describing the effect of Hall diffusion. In the ambipolar
diffusion limit, $\tilde{\eta}_H = 0$, Equations \ref{sspsi}--\ref{ssb_phis}
reduce to those of \citet{kk2002}, allowing direct comparisons to be made
between the similarity solutions from both models. 

In order to solve the collapse equations a good understanding of the
asymptotic boundary conditions of the collapse (which take the form of power
law relations in $x$) is necessary to ensure that the solutions match onto
observations of collapsing cores and protostellar discs. While the innermost
boundary conditions are derived explicitly in the following chapter, the outer
boundary conditions describing a supercritical molecular cloud core at the
moment of point mass formation are discussed below.

\section{The Outer Solution}\label{eqouter}

The outer boundary of the self-similar collapse is chosen to match the density
and magnetic profile of a molecular cloud core contracting quasistatically
under ambipolar diffusion until it has just become supercritical and a point
mass forms at the centre \citep[e.g.][]{s1977, sal1987}. Such a description is
chosen because while the similarity variable $x = r / c_s t$ is undefined at
$t = 0$, $x \to \infty$ also corresponds to $r \to \infty$ and so the initial
conditions of the collapse may also be regarded as the outer boundary
conditions of a core that is just starting to collapse. Such cores are
described by the equations for singular isothermal spheres, which have a
density profile $\rho \propto r^{-2}$ \citep{l1969, p1969, s1977, ws1985,
mo2007}. The surface density is then given by 
\begin{equation}
     \Sigma (t = 0) = \frac{Ac_s^2}{2\pi Gr}
\label{outSigma}
\end{equation}
where $A$ is a constant parameter of the core. This density profile matches
observations of cores \citep[e.g.][and the references within]{awb2000,
wetal2007} and simulations of cores produced by turbulent compression
\citep{bkmv2007} or contracting under ambipolar diffusion \citep{bm1994}. This
is then self-similarised using Equation \ref{sssigmag} into the outer boundary
condition 
\begin{equation}
     \sigma (x \to \infty) = A\,x^{-1}; 
\label{outsigma}
\end{equation}
the value of $A$ is determined from the initial accretion rate of the core,
which is discussed further below. 

Magnetically supercritical cores that are contracting slowly have typical
observed radial infall speeds of around {$0.05$--$0.10$ km s$^{-1}$}
\citep{lmt2001}. The initial state of the model core is taken to have a
spatially-constant radial inflow velocity 
\begin{equation}
    V_r (t = 0) = u_0\,c_s,
\label{outVr}
\end{equation}
where $u_0$ is the constant nondimensional velocity, which has the
self-similar form:
\begin{equation}
    u (x \to \infty) = u_0
\label{outu}
\end{equation} 
(using Equation \ref{ssuh}). For an isothermal molecular cloud core at $T = 10$
K, the sound speed is given by $c_s = 0.19$ km s$^{-1}$, so $u_0$ must be of
order unity to match observed values of core inflow. The similarity solutions
presented in this work use values of $u_0 \in -[0.1,1.5]$ as the outer
boundary condition, which are within the range of values observed in molecular
cloud cores by \citet{lmt2001}. 

Because the infall velocity is constant, so too is the accretion rate, which
is determined by the equation
\begin{equation}
    \dot{M}_0 = -\frac{Au_0c_s^3}{G}\,.
\label{outMdot}
\end{equation}
A constant accretion rate may not be strictly realistic, as accretion onto
stars and cores are thought to be time-variable, however it is required to
preserve the self-similarity of the collapse solutions. Equation \ref{outMdot}
may be expressed in self-similar form using Equation \ref{ssmmdot} in the
limit of large $x$ to give the dimensionless enclosed mass 
\begin{equation}
    m (x \to \infty) = Ax.
\label{outm}
\end{equation}
Equation \ref{sscm} shows that this is equivalent to the boundary condition in
Equation \ref{outsigma}; because they are not unique only one of these
relations may be used to find the similarity solutions. In the numerical 
solutions of \citet{ck1998} for the collapse of a nonrotating magnetic core
with ambipolar diffusion the mass accretion rate at point mass formation was
found to be {$\dot{M} \simeq 5$ M$_{\odot}$ Myr$^{-1}$}. This corresponds to a
nondimensional parameter $A \simeq 3$, which is the value used in the
self-similar calculations of \citet{cck1998} and \citet{kk2002}. A typical
accretion rate for such cores is $A = 3$, as observations show that {$\dot{M}
\in [1,10]$ M$_\odot$ Myr$^{-1}$} across many cores \citep[these rates
correspond to values of {$A \in [0.6,6.1]$};][]{lmt2001}; and this value of
this parameter is held constant throughout this work in order to match the
solutions of \citet{kk2002}. 

In the outermost regions of the core where the density is low the magnetic
field behaviour is described best by ideal MHD; this means that the
mass-to-flux ratio is a constant, given by 
\begin{equation}
    \frac{M}{\Psi} = \frac{\mu_0}{2\pi\sqrt{G}},
\label{mfr}
\end{equation}
where $\mu_0$ is a constant parameter of the core. In particular, $\mu_0$ is
the dimensionless mass-to-flux ratio in the outer core, and is defined by 
\begin{equation} 
    \mu_0 = \frac{M/\Psi}{(M/\Psi)_{crit}};
\label{mu0}
\end{equation}
that is, $\mu_0$ is the ratio of the mass-to-flux ratio in the outermost
regions of the core to the critical value for support against gravity derived
in Equation \ref{mfratio} \citep{nn1978}. Again, to match \citet{kk2002}, the
value of $\mu_0 = 2.9$ is adopted for all of the similarity solutions in this
work; this particular value was taken from the numerical simulations of
\citet{ck1998}. It also matches the observations of \citet{c1999}, which
showed that most cores possess mass-to-flux ratios that are more than twice
critical. From Equations \ref{outsigma} and \ref{outm} it is then possible to
derive the magnetic boundary conditions for a supercritical core under ideal
MHD: 
\begin{align}
    b_z (x \to \infty) &= \frac{\sigma}{\mu_0} \label{outbz}\\
    \text{and } \psi (x \to \infty) &= \frac{m}{\mu_0} \label{outpsi};
\end{align}
again, these are equivalent and only one may be used to calculate the similarity
solutions.

The initial value of the rotational velocity in the outer regions of the
molecular cloud core is also spatially-uniform and given by 
\begin{equation}
    V_\phi = v_0\,c_s,
\label{outVphi}
\end{equation}
which is written in nondimensional form as
\begin{equation}
    v (x \to \infty) = v_0.
\label{outv}
\end{equation}
The constant $v_0$ is the initial dimensionless rotational velocity,
determined by the expression 
\begin{equation}
    v_0 \approx \frac{A\Omega_bc_s}{\sqrt{G}B_{ref}}
\label{v_0}
\end{equation}
\citep[which was derived using the $r^{-1}$ dependance of the core surface 
density and magnetic field in][]{b1997}. As outlined in Chapter
\ref{ch:litrev}, a typical value of the uniform background angular velocity of
a molecular cloud core is {$\Omega_b = 2 \times 10^{-14}$ rad s$^{-1}$}
\citep[and the other references in Section \ref{lr:mc}]{getal1993,kc1997} and
the background magnetic field may be taken to be $B_{ref} = 30$ $\mu$G
\citep{c1999}. Substituting these and the values of $A$ and $c_s$ given above
into Equation \ref{v_0} gives the value of the initial rotational velocity to
be $v_0 = 0.15$. This is a factor of 10 larger than that found in
\citet{b1997}, however, the range of observed core velocities is $v_0 \in
[0.01, 1.0]$, and the dependence of the similarity solutions on the initial
core rotational velocity is explored in Chapter \ref{ch:nohall}, where it is
shown that the rotational velocity directly influence the size of the inner
Keplerian disc. The Hall similarity solutions presented in Chapter
\ref{ch:hall} possess a constant value of $v_0 = 0.73$ for the rotational
velocity in order to facilitate comparison with the fiducial ambipolar
diffusion solution of \citet{kk2002}.

To summarise, the outer boundary conditions of the collapse, which also serve
as initial conditions at point mass formation, are described by the set of
unique self-similar asymptotic equations: 
\begin{align}
    \sigma &\to \frac{A}{x}, \label{outersig}\\
    b_z &\to \frac{\sigma}{\mu_0}, \label{outerbz}\\
    u &\to u_0 \label{outeru}\\
    \text{and } v &\to v_0, \label{outerv}
\end{align}
where $A$, $\mu_0$, $u_0$ and $v_0$ are the parameters that describe the outer
collapse. In the similarity solutions presented in this work, these constants
take on the values listed in Table \ref{tab-bc1} unless explicitly stated.
These conditions match those of \citet{kk2002}, as neither Hall nor ambipolar
diffusion are expected to be important in the low density outer regions where
ideal MHD is dominant. 
\begin{table}[htb]
\begin{center}
\begin{tabular}{cl}
\toprule
 boundary condition & value \\
\midrule
 $A$		& 3 \\
 $\mu_0$	& 2.9 \\
 $u_0$		& $-1$ \\
 $v_0$		& 0.15 \\
 \bottomrule
 \end{tabular}
\end{center}
\vspace{-5mm}
 \caption[Outer boundary conditions]{Outer boundary conditions describing a
molecular cloud core that is supercritical at point mass formation. The
fiducial value of $v_0$ listed in this table (derived above using Equation
\ref{v_0}) is lower than that used as a boundary condition for calculating the
Hall similarity solutions ($v_0 = 0.73$), but both are within the large range
of observed core rotational velocities, $v_0 \in [0.01, 1.0]$
\citep{getal1993}.} 
\label{tab-bc1}
\end{table}

The boundary conditions that describe the collapse behaviour in the innermost
regions of self-similar space (corresponding to $r \to 0$ or $t \to \infty$)
are less straightforward than those on the outer boundary. These are
influenced by both ambipolar and Hall diffusion, which determine the surface
density and the accretion rate onto the central mass, and by the strength of
the magnetic braking within the core, which slows the rotation of the gas and
may prevent the formation of a rotationally-supported disc around the
protostar. Two sets of inner boundary conditions satisfy the collapse
equations as $x \to 0$: these are similarity solutions both with and without
an inner Keplerian disc. The derivation of these asymptotic power law
solutions, and their individual applicability and dependence upon the magnetic
parameters of the core are the focus of the following chapter. 

\cleardoublepage

\chapter{The Inner Asymptotic Solutions}\label{ch:asymptotic}

Whether or not a rotationally-supported disc forms as a result of the initial
collapse of a molecular cloud core is a point of contention in current
simulations of star formation. The previously-accepted model for innermost
regions of the collapsing flow says that the conservation of angular momentum
during collapse results in the progressive increase of the centrifugal force,
which eventually halts the infalling gas and leads to the development of a
central mass surrounded by a flattened rotationally-supported disc. This has
recently been called into question by both numerical simulations and
observations of young stars. 

In particular, \citet{ml2009} have shown that no discs could form around the 
protostars in their numerical MHD simulations unless the calculations were
started with very dense cores or unreasonably low ionisation rates.
\citet{hc2009} and \citet{ch2010} examined the influence of the orientation of
the field with respect to the axis of rotation in the core, and found that
disc formation was suppressed for cores with an initial nondimensional
mass-to-flux ratio of $\mu_0 \simeq 3$ when the magnetic field and rotational
axes are perpendicular and for $\mu_0 \lesssim 5-10$ when the field was
aligned with the axis of rotation. However, the observations of \citet{c1999}
showed that molecular cloud cores typically possess mass-to-flux ratios that
are $\mu_0 < 5$, which would preclude disc formation based upon the results of
\citet{hc2009}. \citet{pb2007} compared hydrodynamical and
magnetohydrodynamical simulations and found that protostar formation was
delayed in calculations with a magnetic field; they also found that magnetic
pressure support causes the suppression of fragmentation within a forming
disc. These results present a complicated picture in which Keplerian disc
formation is not a certain consequence of the star formation process.

Observations have also cast doubt on the star-and-disc formation model.
\citet[following up on the observations of \citet{smmv1999}]{smvmh2001} found
in their observations a population of fifteen slowly-rotating ($P_{rot} > 4$
days) pre-main-sequence stars that show no evidence of circumstellar discs.
This result was unexpected, as it had previously been believed that such
slowing of the angular momentum of young stars could only occur so early in
the collapse process through interactions between the protostar and a
protostellar disc \citep[e.g.][]{k1991, gl1979a, gl1979b}.

The semianalytic solutions of \citet[as well as the similarity solutions
presented in Chapter \ref{ch:hall}]{kk2002} demonstrated disc formation for 
cores with $\mu_0 < 5$, however they were also able to show that no disc would
form if the magnetic braking is particularly strong or the initial cloud
rotation rate too slow. \citet{sglc2006} describe the problem of forming discs
as a further manifestation of the magnetic flux problem outlined in Chapter
\ref{ch:litrev} (while others and this work refer to it as the magnetic
braking catastrophe); they also showed that the addition of non-ideal MHD
processes make it possible to dissipate the flux and form a thin disc (in
their particular calculations with numerical reconnection they chose to
include Ohmic resistivity). 

For the self-similar collapse equations developed in the preceding chapter 
there are two distinct power law similarity solutions for the innermost
region: one in which a rotationally-supported disc forms and the fluid
accretes slowly onto the central object, and a second in which the material
supersonically falls onto the central protostar in a magnetically-diluted free
fall. This chapter presents these similarity solutions, the dominant physics
in each and the initial conditions in the molecular cloud core that determine
whether or not a disc may form during collapse. The implications of these
solutions on the dichotomy of results that form the magnetic braking
catastrophe are also discussed. 

\section{Derivation}\label{indev}

In the innermost regions of the gravitational collapse of a molecular cloud
core, the self-similar collapse equations derived in Chapter \ref{ch:derivs}
are satisfied by similarity solutions in which the variables take the form of
power laws with respect to the similarity variable, $x = r/c_st$. These
solutions are found by taking the limit of the equations as $x \to 0$; this
limit can also be thought of as either of the dimensional limits $r \to 0$ or
$t \to \infty$, which are the innermost region and the late stages of collapse
respectively. 

The similarity solutions to the full collapse process that are presented in
Chapters \ref{ch:nohall} and \ref{ch:hall} tend asymptotically towards these
power law solutions which are then imposed as boundary conditions on the inner
boundary of the collapse calculations. While there may be other mathematically
valid similarity solutions to the disc equations in the central regions of the
collapse, these are unphysical (typically with negative values of the scale
height or surface density) and therefore are not within the scope of this
work. 

The two similarity solutions presented here are found by assuming that the
collapse variables take the form of a set of power law relations defined by 
\begin{align}
    \sigma &= \sigma_1\,x^{-p}, \label{asym-sigma}\\
    b_z &= b_{z1}\,x^{-q}, \label{asym-bz}\\
    \text{and }j &= j_1\,x^{-r}, \label{asym-j}
\end{align}
where the exponents $p$, $q$ and $r$ are real numbers, and the coefficients
$\sigma_1$, $b_{z1}$ and $j_1$ are constants that are found by solving the
collapse equations once the exponents have been derived. The other variables
are calculated by substituting these equations into the MHD equations and
taking the limit as $x \to 0$ to find and equate the dominant terms. 

As the differential equation for $m$ depends only on $\sigma$ and $x$,
Equation \ref{asym-sigma} can be substituted into Equation \ref{sscm} and 
integrated to give 
\begin{equation}
    m = m_c + \frac{\sigma_1}{2-p}\,x^{2-p}.
    \label{asym-m}
\end{equation}
This equation cannot be simplified further until the value of $p$ is known, or
until limits are placed upon $p$ such that one term or the other is dominant.
Clearly, the term with the lower exponent of $x$ will be larger when $x \ll
1$, so that for the case when $p < 2$, $x^0 \gg x^{2-p}$, so that the enclosed
mass $m \to m_c$; and conversely when $p > 2$, $m \to \sigma_1 x^{2-p} /(2-p)$
as $x \to 0$. 

Similarly, the differential equation for the enclosed flux depends only on
$b_z$ and $x$, so that Equation \ref{sspsi}, when integrated, becomes
\begin{equation}
    \psi = \psi_c + \frac{b_{z1}}{2-q}\,x^{2-q}.
    \label{asym-psi}
\end{equation}
The vertical magnetic field is expected to scale with $x$ in the same manner
as the magnetic dominant field component, if it is not itself the dominant
component. When $q < 2$ then $\psi = \psi_c$ and $b_{r,s}$ scales as $x^{-2} >
x^{-q}$, so that it becomes the dominant scaling term in the magnetic field.
In order to maintain the dominance of $b_z$, the central flux is set to
$\psi_c = 0$. This is supported in the literature, as \citet{kk2002} also drop
this term in their description of the inner disc. 

Having accepted this rationale for the behaviour of the flux, the radial field
component is then defined by the monopole approximation (Equation
\ref{ssextra}) as
\begin{equation}
    b_{r,s} = \frac{b_{z1}}{(2-q)}\,x^{-q},
    \label{asym-b_rs}
\end{equation}
which clearly scales with the vertical field component. 

The exact scaling of the azimuthal field component cannot be so easily
determined, due to the artificial cap that keeps $|b_{\phi,s}| \le
\delta{b_z}$. This cap implies that $b_{\phi,s}$ can be regarded as scaling
with $b_z \sim x^{-q}$ as a first approximation, especially as $b_{\phi,s}$
typically appears in the disc equations summed with another component of the
magnetic field, with the precise value of $b_{\phi,s}$ to be determined once
the other scalings are known. 

\begin{figure}[ht]
  \centering
  \vspace{-3mm}
  \includegraphics[width=4.5in]{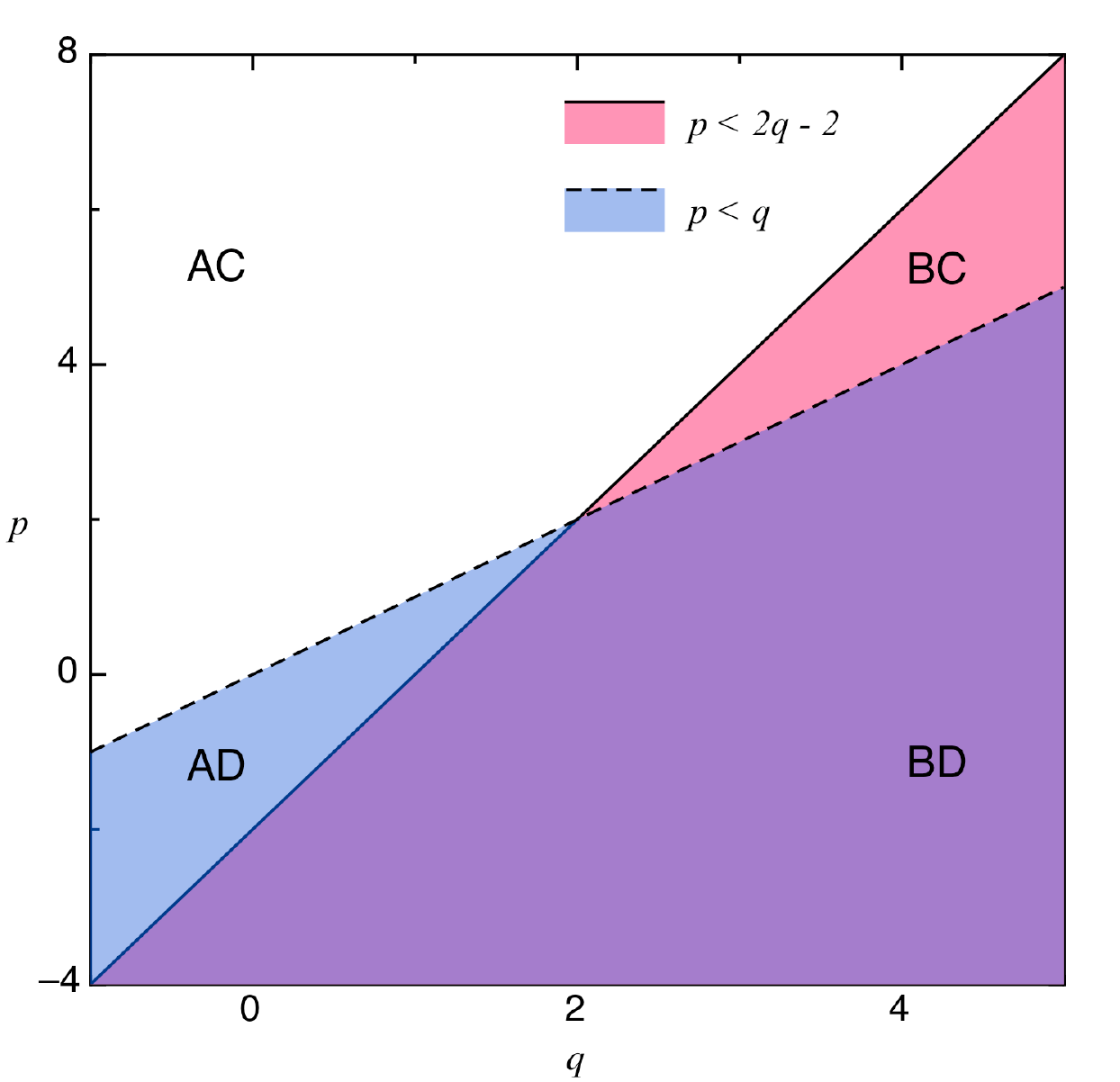}
  \vspace{-4mm}
  \caption[The $pq$-plane]{The $pq$-plane used to describe the different
regions of $pq$-space that dictate how the components of $h$ behave. The
different regions have been colour coded, for example, the white area in the
upper left of the plot represents the region AC where $p>2q-2$ and $p>q$.} 
\label{pq-plane}
\end{figure}

The scale height is written as the solution to the quadratic equation for the
vertical hydrostatic equilibrium (\ref{ssvhe}): 
\begin{equation}
    h = \frac{\hat{\sigma}x^3}{2\hat{m}_c}\left[-1
          + \sqrt{1+\frac{8\hat{m}_c}{\hat{\sigma}^2x^3}}\,\right],
\label{h-quad}
\end{equation}
where
\begin{align}
    \hat{m}_c &= m_c - x^3b_{r,s}\frac{db_z}{dx} \label{hat-m_c} 
\end{align}
and
\begin{align}
    \hat{\sigma} &= \sigma + \frac{b_{r,s}^2 + b_{\phi,s}^2}{\sigma}\,.
\label{hat-sigma}
\end{align}
Applying Equations \ref{asym-sigma}--\ref{asym-b_rs} to $\hat{\sigma}$ and
$\hat{m}_c$, these terms become
\begin{align} 
    \hat{m}_c &= m_c - \frac{b_{z1}x^{3-q}}{(2-q)\sigma_1x^{-p}}\,
		(-qb_{z1}x^{-q-1})\nonumber\\
 	    &= m_c + \frac{qb_{z1}^2}{(2-q)\sigma_1}\,x^{2-2q+p}
    \label{asym-mhat}
\end{align}
and
\begin{align}
    \hat{\sigma} &= \sigma_1x^{-p} + \frac{x^p}{\sigma_1}
      \left[\frac{b_{z1}^2}{(2-q)^2}x^{-2q} 
      + b_{\phi1}^2(\sim{x^{-2q}})\right];
    \label{aysm-sighat}
\end{align}
so that for any combination of $p$ and $q$ the behaviour of $h$ can be
determined. There are two terms that can be dominant for each of
$\hat{\sigma}$ and $\hat{m}_c$, and each applies when the following conditions
on $p$ and $q$ are satisfied:
\begin{align}
    &\text{A.} &\hat{m}_c &= m_c &(p &> 2q -2)\label{abcd-a}\\
    &\text{B.} &\hat{m}_c &= -\frac{x^3b_{r,s}}{\sigma}\left(\frac{db_z}{dx}\right)
      &(p &< 2q-2)\label{abcd-b}\\
    &\text{C.} &\hat{\sigma} &= \sigma &(p &> q)\label{abcd-c}\\
    &\text{D.} &\hat{\sigma} &= \frac{(b_{r,s}^2 + b_{\phi,s}^2)}{\sigma}
      &(p &< q)
    \label{abcd-d}
\end{align}
The different regions of $pq$-space that are delineated by these
inequalities are shown in Figure \ref{pq-plane}, where the combinations of
A/B and C/D describe particular regions of behaviour for $h$. 

The scaling of $h$, with respect to $\hat{m}_c$ and $\hat{\sigma}$, can have
one of two different values, depending upon the scalings of the terms within
the square root in Equation \ref{h-quad}; clearly the larger of 
$8\hat{m}_c/\hat{\sigma}^2x^3$ and $1$ will dominate and determine the
behaviour of $h$. Again, there are two possible cases: 
\begin{align}
    &\text{a.} &\frac{8\hat{m}_c}{\hat{\sigma}^2x^3} &\ll 1
      &h&\sim\frac{\hat{\sigma}x^3}{2\hat{m}_c}\left[-1 + \sqrt{1^+}\right] \to
0^+,\\
    \text{and }&\text{b.} &\frac{8\hat{m}_c}{\hat{\sigma}^2x^3} &\gg 1
      &h&\sim\sqrt{\frac{2x^3}{\hat{m}_c}}\,.
    \label{abcdh}
\end{align}
Case a, regardless of the region of $pq$-space in which it is calculated,
represents the trivial solution of $h$ and will not be examined further in
this work, as any solution that satisfies this condition, while mathematically
valid, would represent a state that is intrinsically unphysical. Case b, and
the boundary case where $8\hat{m}_c/\hat{\sigma}^2x^3$ scales as $x^0$, must
be examined in each of the four regions of the $pq$-plane in order to
determine all of the possible similarity solutions of interest for the
innermost regions of the gravitational collapse. 

This exploration of $pq$-space, and the derivation of the two physical
similarity solutions to the fluid equations, is presented in Appendix
\ref{ch:pq} for the interested reader who wishes to study the solutions in
more depth. The first of these is a slowly-accreting Keplerian disc, which is
discussed in the following section, while the second is a magnetically-diluted
free fall flow onto the central protostar, which is discussed in detail in
Section \ref{ff}. 

\section{Keplerian Disc Solution}\label{kepdisc}

The first solution (derived in Section \ref{sub:AD} and belonging to the
region AC of the $pq$-plane) is a Keplerian disc, where the material is
supported against gravity by its angular momentum, and only slowly accretes
onto the central mass. The scale height of the disc and the surface density
are both dependent upon the Hall and ambipolar diffusion coefficients,
represented here by the nondimensional constants $\tilde{\eta}_H$ and
$\tilde{\eta}_A$. 

The similarity solution in nondimensional form is given by the set of
relations:
\begin{align}
    m &= m_c, \label{kep-ssimm}\\  
    \dot{m}&= m_c, \label{kep-ssimmdot}\\
	\sigma &= \sigma_1\,x^{-3/2}, \label{kep-ssimsigma}\\
    h &= h_1\,x^{3/2}, \label{kep-ssimh}\\
    u &= -\frac{m_c}{\sigma_1}\,x^{1/2}, \label{kep-ssimu}\\
	v &= \sqrt{\frac{m_c}{x}}, \label{kep-ssimv}\\
	j &= \sqrt{m_cx}, \label{kep-ssimj}\\
    \psi &= \frac{4}{3}\,b_zx^2, \label{kep-ssimpsi}\\
	b_z &= \frac{m_c^{3/4}}{\sqrt{2\delta}}\,x^{-5/4}, \label{kep-ssimbz}\\
    b_{r,s} &= \frac{4}{3}\,b_z, \label{kep-ssimbphis}\\
	\text{and } b_{\phi,s} &= -\delta{b_z};
    \label{kep-ssim}
\end{align}
the constant $m_c$ is the nondimensional infall rate onto the central star and
$\delta$ is the constant parameterising the artificial cap placed upon
$b_{\phi,s}$ to prevent it from becoming the dominant field component in the
innermost regions of the disc. The coefficients $h_1$ and $\sigma_1$ are
constants, determined by the expressions 
\begin{equation}
    h_1 = \left(\frac{2}{m_c[1+(f/2\delta)^2]}\right)^{\!1/2}
\label{kep-h1}
\end{equation}
and
\begin{equation}
    \sigma_1 = \frac{\sqrt{2m_c}f}{2\delta\sqrt{(2\delta/f)^2 + 1}}\,,
\label{kep-sigma1}
\end{equation}
where $f$ is a function of the magnetic diffusion parameters, given by the
equation 
\begin{equation}
    f = \frac{4}{3}\tilde{\eta}_A 
     - \delta\tilde{\eta}_H\sqrt{\frac{25}{9} + \delta^2}.
\label{kep-f}
\end{equation}

Rather than continuing to study this solution in the nondimensional form, it
is more illuminating to convert these power law relations to dimensional form
using Equations \ref{sssigmag}--\ref{sseta}, so that the Keplerian disc
variables are 
\begin{align}
    M&=\frac{c_s^3m_c}{G}\,t, \label{kepfakeM}\\
	\dot{M} &= \frac{c_s^3m_c}{G}, \label{kepfakeMdot}\\ 
    \Sigma &= \frac{\sigma_1c_s^{5/2}}{2\pi{G}}\,\frac{t^{1/2}}{r^{3/2}},
\label{kepfakeSigma}\\
    H &= h_1\,\frac{r^{3/2}}{\sqrt{c_st}}, \label{kepfakeH}\\
    V_r &= -\frac{m_c}{\sigma_1}\sqrt{\frac{rc_s}{t}}, \label{kepfakeVr}\\
    V_\phi &= \sqrt{\frac{m_cc_s^3t}{r}}, \label{kepfakeVphi}\\
    J &= \sqrt{m_cc_s^3rt}, \label{kepfakeJ}\\
    \Psi &= \frac{8\pi{c_s}^{9/4}m_c^{3/4}}{3\sqrt{2{\delta}G}}\,r^{3/4}t^{1/4},
\label{kepfakePsi}\\
    B_z &= \frac{m_c^{3/4}c_s^{9/4}}{\sqrt{2\delta{G}}}\,\frac{t^{1/4}}{r^{5/4}}, 
\label{kepfakeBz}\\
    B_{r,s} &= \frac{4}{3}B_z, \label{kepfakeBrs}\\
    \text{and } B_{\phi,s} &= -\delta{B_z};
\label{kepfake}
\end{align}
the coefficients $\sigma_1$ and $h_1$ remain those in Equations
\ref{kep-h1}--\ref{kep-f}, $c_s$ is the isothermal sound speed and $G$ is the
gravitational constant. The disc matter is in Keplerian orbit, with $V_\phi$
given by the canonical value \citep[$V_\phi = \sqrt{GM/r}$ ;][]{n1687}. The
radial scaling of the surface density ($\Sigma \propto r^{-3/2}$) is that
expected from the minimum mass solar nebula \citep{w1977} and other
simulations of protostellar discs \citep[e.g.][]{cp1973}; and the magnetic
field scaling also matches that from theory, particularly of discs that
support disc winds \citep[$B_z \propto r^{-5/4}$;][]{bp1982}. It is
interesting to note that the coefficient of the vertical magnetic field,
$B_z$, does not depend upon the magnetic diffusion parameters, but rather the
nondimensional mass infall rate and the cap placed upon $B_{\phi,s}$. 

The azimuthal magnetic field component blows up with respect to the other
field components in this small $x$ limit because the azimuthal magnetic field
drift speed is slow compared with the Keplerian speed. The model adopted for
the vertical angular momentum transport is unable to properly account for the
effects of magnetic braking in the small $x$ limit, so $B_{\phi,s}$ takes on
the value of the cap placed upon it. If a different scaling for $\eta_H$ were
adopted then Hall diffusion could act to limit $B_{\phi,s}$ in such a way that
the cap becomes unnecessary; however, within the context of this disc solution
other scalings of $\eta_H$ lead either to a breakdown of self-similarity, or
to a solution in which the diffusion is so strong that the fluid falls rapidly
onto the protostar and no rotationally-supported disc may form, in a similar
manner to the second similarity solution presented in Section \ref{ff}.  

These power laws are the solutions to the following simplified equations: 
\begin{align}
    g_r + \frac{J^2}{r^3} &= 0,\\
    V_r\frac{\partial{J}}{\partial{r}} &= \frac{rB_zB_{\phi,s}}{2\pi\Sigma},\\
    HV_r + \frac{\eta_H}{B}B_{\phi,s} &+ \frac{\eta_A}{B^2}B_{r,s}B_z = 0,
       \label{kepeqnsid}\\
    \frac{GM}{2r^3}H^2 &+ {\pi}G{\Sigma}H - c_s^2 = 0,\\
    \dot{M} &= \text{constant},\\
    J &= rV_\phi,\\
    B_{r,s} &= \frac{4}{3}B_z,\label{kepeqnsbrs}\\
    B_{\phi,s} &= -\delta{B_z},\\
    \text{and }\Psi &= \frac{8}{3}\pi{r^2}B_z.
    \label{kepeqns}
\end{align}
The induction equation (\ref{kepeqnsid}) can be written quite simply as $V_r +
V_{Br} = 0$, that is, the inward radial velocity of the fluid is balanced by
the drift of the magnetic field with respect to the gas. Any accretion through
the centrifugally-supported disc is regulated by the outward diffusion of the
magnetic field against the flow. 

\begin{figure}
  \centering
  \includegraphics{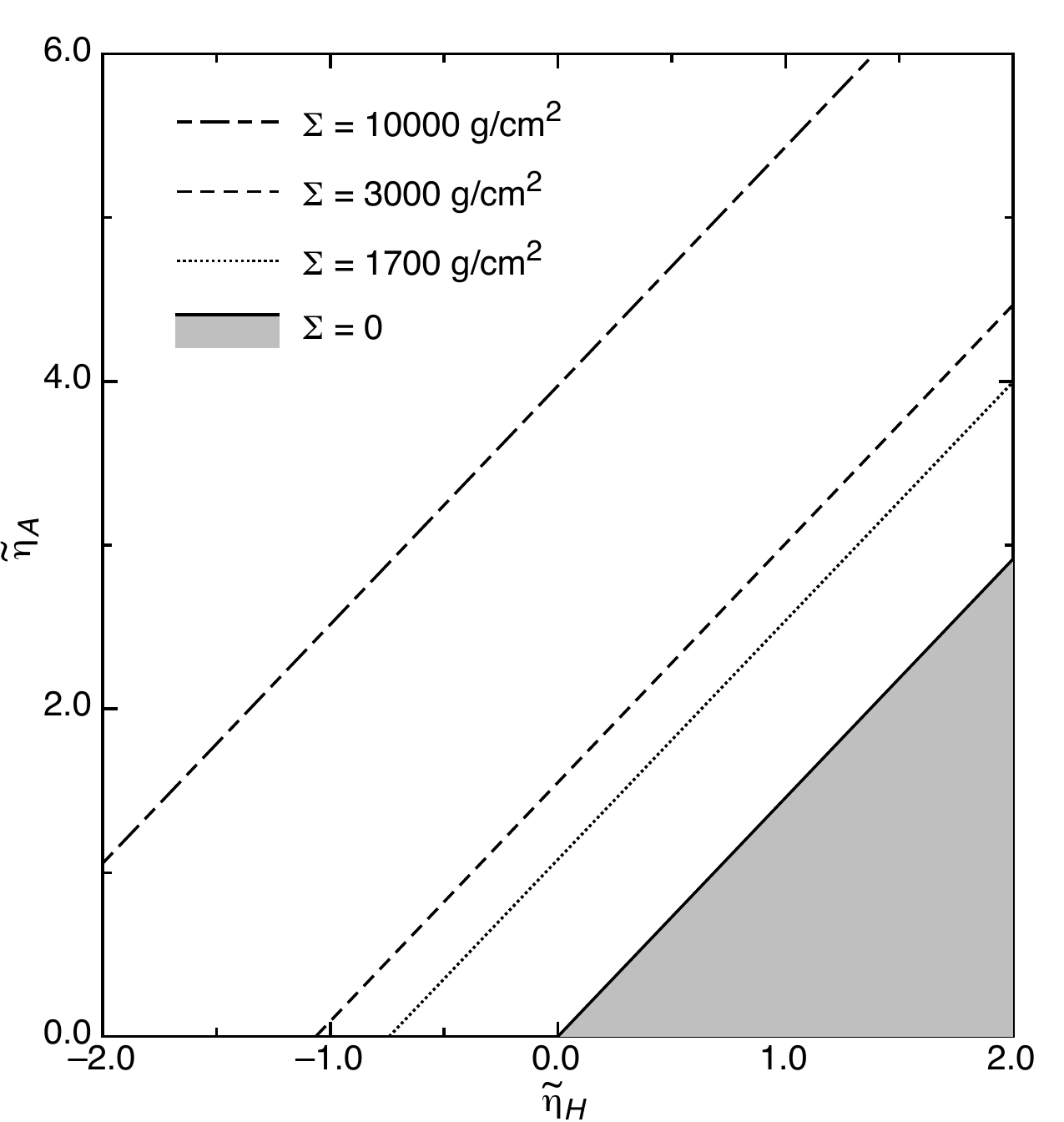}
  \caption[$\tilde{\eta}_A$ vs $\tilde{\eta}_H$ for the Keplerian disc
solution where $\dot{M} = 10^{-5}$ M$_{\odot}$ yr$^{-1}$]{Exploration of the
relationship between $\tilde{\eta}_H$ and $\tilde{\eta}_A$ in the Keplerian
disc solution, for various different values of $\Sigma$ at $r = 1$ AU and
$t = 10,000$ yr, when $\delta = 1$, $c_s = 0.19$ km s$^{-1}$ and $\dot{M} =
10^{-5}$ M$_{\odot}$ yr$^{-1}$ (which corresponds to $B_z = 1.15$ G). The
solid black line corresponds to $\Sigma = 0$, and the shaded region beneath it
is unphysical; no centrifugally-supported disc may form in this region of
parameter space. The dashed lines above this correspond to $\Sigma = 1700$,
$3000$ and $10000$ g cm$^{-2}$ respectively.} 
\label{etaKvsetaB}
\end{figure}
The size of the magnetic diffusion parameters determine the build up of fluid
relative to the magnetic field in the disc, and the sign of the Hall diffusion
term places limits on its size. Should the Hall parameter become too large,
and possess the wrong sign, relative to the other diffusion term, then $f$
becomes negative and the surface density with it, which is clearly unphysical.
In order to ensure that the induction equation is satisfied and a
rotationally-supported disc forms, the diffusion parameters must satisfy the
inequality $f \ge 0$, which becomes 
\begin{equation}
    \tilde{\eta}_A \ge \tilde{\eta}_H\sqrt{\frac{17}{8}}
\label{kep-fineq}
\end{equation}
for the typical value of $\delta = 1$. The forbidden region of parameter space
in which no disc may form is the shaded area in Figure \ref{etaKvsetaB}, with
the solid boundary line corresponding to $f = \Sigma = 0$. 

As is clear from Figure \ref{etaKvsetaB}, there can be no solution for the
core with purely Hall diffusion where the Hall diffusion parameter is
positive, as the positive Hall parameter acts to restrain the effects of
ambipolar and Ohmic diffusion. However, when the Hall diffusion parameter is
negative (corresponding to a reversal of the magnetic field) it acts in the
same radial direction as the ambipolar diffusion term, enhancing the diffusion 
of the magnetic field through the disc. In the case of pure ambipolar
diffusion the field moves inward slower than the neutral particles and the
solution reduces to that depicted in the inner asymptotic region of the 
fiducial solution of \citet[][equations 51--57]{kk2002}.

The plot in Figure \ref{etaKvsetaB} shows the two-dimensional area of magnetic
diffusion parameter space that gives feasible values of the surface density
$\Sigma$ at $r = 1$ AU when $t = 10,000$ years, $c_s = 0.19$ km s$^{-1}$ and
$\delta = 1$ for solutions with an accretion rate of $\dot{M}_c = 10^{-5}$
M$_{\odot}$ yr$^{-1}$ (which corresponds to $B_z = 1.15$ G). The dotted line
represents a surface density at 1 AU of $\Sigma = 1700$ g/cm$^{2}$, which is
the value from the minimum mass solar nebula model \citep{w1977} in which the
surface density of the solar nebula is estimated by adding sufficient hydrogen
and helium to the solid bodies in the solar system to recover standard
interstellar abundances, and spreading this material smoothly into a disc. The
dashed lines in Figure \ref{etaKvsetaB} correspond to higher surface densities
that are more like those realistically expected to occur in protostellar
discs. 

As $\tilde{\eta}_H$ decreases in comparison to a constant $\tilde{\eta}_A$ the
surface density becomes large due to the changed amount of field line drift as
the radial field diffusion slows the rate of infall. The ambipolar
diffusion-only fiducial solution of \citet{kk2002}, which shall be discussed
in detail in Section \ref{ad}, has an accretion rate of $\dot{M}_c = 7.66
\times 10^{-6}$ M$_\odot$ yr$^{-1}$, and the surface density at 1 AU in their
solution is $1310$ g/cm$^2$. Adding a positive value of $\tilde{\eta}_H$ to
this similarity solution causes the surface density to decrease, which is
problematic if one expects to form large planets from their disc, however, if
the Hall parameter is negative then its presence raises the surface density to
something more realistic. 

\begin{figure}
  \centering
  \includegraphics{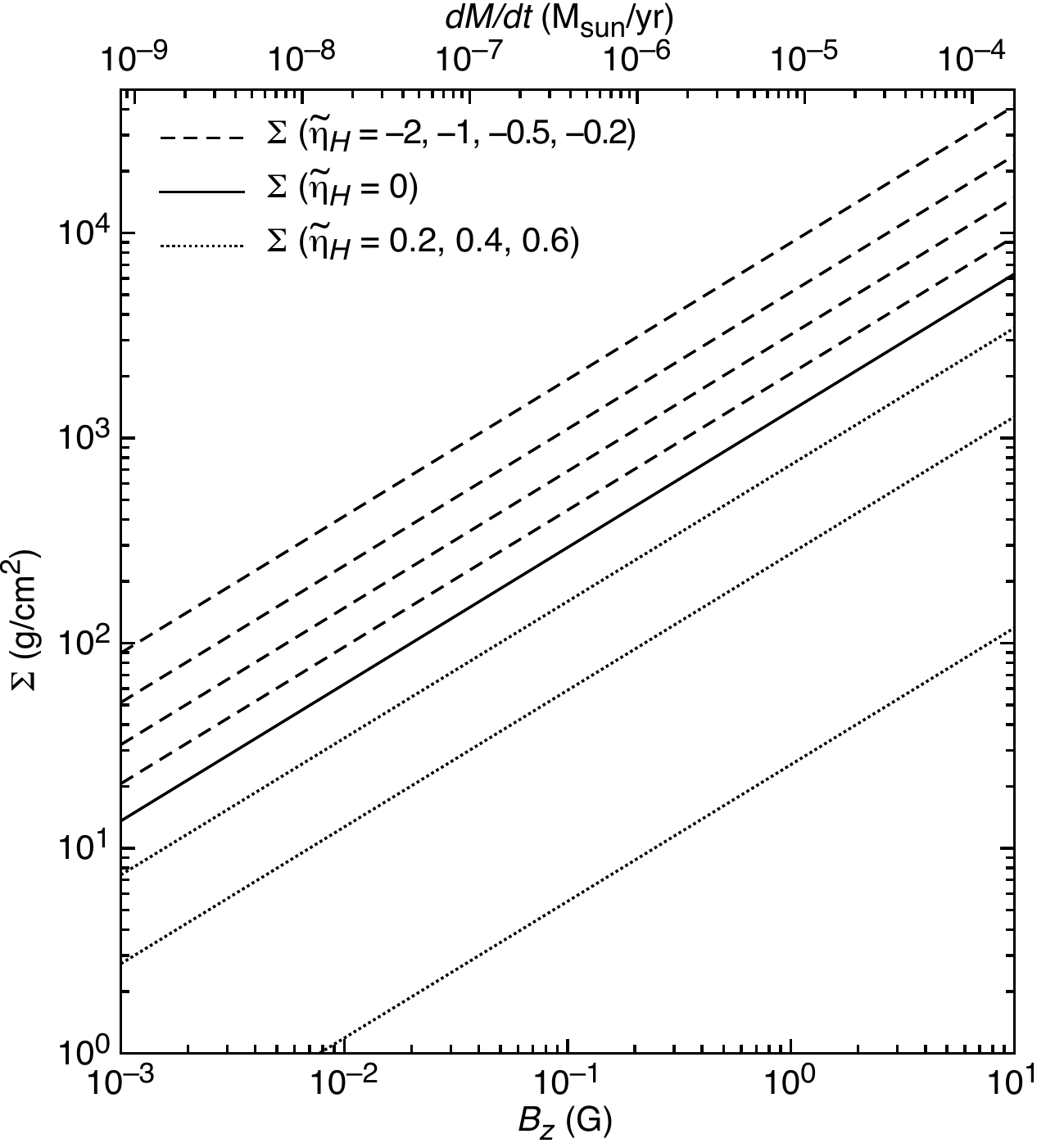}
  \caption[$\Sigma$ vs $B_z$ and $\dot{M}$ with varying $\tilde{\eta}_H$ for
the Keplerian disc solution]{Plot of $\Sigma$ against $B_z$ and $\dot{M}$ for
different values of $\tilde{\eta}_H$ in the Keplerian disc solution, where
$\tilde{\eta}_A = \delta = 1.0$ and $c = 0.19$ km s$^{-1}$, at $r = 1$ AU and
$t = 10000$ yr. The solid line presents the behaviour of $\Sigma$ in the case
that $\tilde{\eta}_H = 0$; the dashed lines above this are the cases where 
$\tilde{\eta}_H = -0.2, -0.5, -1.0$ and $-2$ respectively; and the dotted
lower lines are $\tilde{\eta}_H = 0.2, 0.4$ and $0.6$.} 
\label{SigmavsB}
\end{figure}
As an experiment, the equations are solved to obtain a set of variables that
are functions of $B_z$, $r$, $t$, and the core parameters to get the
relationships modelled in Figure \ref{SigmavsB}. These functions are outlined
below: 
\begin{align}
    \dot{M} &= B_z^{4/3}r^{5/3}(2\delta)^{2/3}(Gt)^{-1/3},\\
    M &= B_z^{4/3}r^{5/3}(2\delta{t})^{2/3}G^{-1/3},\\
    \Sigma &= \frac{c_sf\sqrt{M}}{2\delta\pi\sqrt{2Gr^3[(2\delta/f)^2+1]}}\,,
       \label{b_zsetsig}\\
    H &= \frac{\sqrt{2}r^{3/2}c_s}{\sqrt{GM[1+(f/2\delta)^2]}}\,, \label{b_zseth}\\
    J &= (2\delta{G}B_z^2r^4t)^{1/3},\\
    V_{\phi} &= (2\delta{G}B_z^2rt)^{1/3}\\
    \text{and }V_r &= -\frac{\delta{B_z^2r}}{\pi{V_\phi}\Sigma};
    \label{b_zset}
\end{align}
the other field components were defined as being of order $B_z$ in Equations
\ref{kepeqnsbrs}--\ref{kepeqns}. This model emphasizes the build up of the
central mass through accretion, and Figure \ref{SigmavsB} shows the effect of
Hall diffusion and the magnetic field strength on the surface density at $r =
1$ AU for a system with $\tilde{\eta}_A = \delta = 1$. 

Alternatively, the disc variables can be thought of as functions of $M$ and
$\dot{M}$, treating $t$ as the ratio between $M$ and $\dot{M}$ rather than the
actual age of the system. This makes it possible to regard the system as a
steady state disc, where the other variables are defined by 
\begin{align}
    B_z &= \left(\frac{GM\dot{M}^2}{r^5(2\delta)^2}\right)^{1/4},\\
    J &= \sqrt{GMr}\,,\\
    V_{\phi} &= V_K = \sqrt{\frac{GM}{r}}\,\\
    \text{and }V_r &= -\frac{\dot{M}}{2\pi{r}\Sigma}\,;
\label{mset}
\end{align}
$\Sigma$ and $H$ are unchanged from Equations \ref{b_zsetsig} and
\ref{b_zseth}, as are the field components from Equation
\ref{kepeqnsbrs}--\ref{kepeqns}. The relationship between $B_z$ and $\dot{M}$
is shown on the horizontal axes of Figure \ref{SigmavsB}. 

Figure \ref{SigmavsB} shows the influence of Hall diffusion on the surface 
density, plotted against $B_z$ and $\dot{M}$ for constant $\tilde{\eta}_A =
\delta = 1$. When the Hall diffusion term is negative, the surface density
increases, creating a larger and heavier disc, while a positive Hall term,
corresponding to a reversal of the field with respect to the axis of rotation,
causes a decrease in $\Sigma$ which drops off dramatically as the Hall term
approaches the lower limit implied by the inequality $f \ge 0$.

The Keplerian disc solution says nothing about the angular momentum problem of
star formation described in Section \ref{lr:amp}, as the accretion through the
disc is slow and the model for the angular momentum transport by magnetic
braking is oversimplified by the cap that is placed on the azimuthal field
component. Magnetic diffusion is key to the magnetic flux problem outlined in
Section \ref{lr:mfp} and the addition of Hall diffusion to the self-similar
collapse model has been shown to assist in solving this, or to complicate it
further, depending upon the orientation of the field. 

\section{Free Fall Solution}\label{ff}

The second similarity solution, derived in Section \ref{sub:AD} and located in
the blue lower left region AD of the $pq$-plane (Figure \ref{pq-plane}),
describes the behaviour of the infall when the magnetic braking is very
efficient at removing angular momentum from the flow. In this case there is
very little angular momentum remaining and the reduced centrifugal support
inhibits disc formation, so that the collapsing flow becomes a supersonic
magnetically-diluted free fall onto the central protostellar mass. This
solution is representative of the magnetic braking catastrophe that affects
many numerical simulations of gravitational collapse. 

The nondimensional form of this similarity solution is given by the complete
set of fluid variables: 
\begin{align}
    m &= m_c,\label{ff-ssimm}\\
    \sigma &= \sqrt{\frac{m_c}{2x}},\\
    h &= h_{1,2}\,x^{3/2},\\
    u &= -\sqrt{\frac{2m_c}{x}}, \\
    v &= j_{1,2},\\
    j &= j_{1,2}\,x, \label{ff-ssimj}\\
    \psi &= b_{z1,2}\,x, \\
    b_z &= b_{z1,2}\,x^{-1},\\
    b_{r,s} &= b_z,\\
    \text{and }b_{\phi,s} &= -\text{min}\left[\frac{|\tilde{\eta}_H|}
	{\tilde{\eta}_A}\,b\,;\,{\delta}b_z\right].
    \label{ff-ssim}
\end{align}
In the limit where there is only ambipolar diffusion, this similarity solution
reduces to the asymptotic inner solution to the ``strong braking'' collapse,
which was presented without explicit derivation by \citet[][equations
66--71]{kk2002}. In their solutions the angular momentum and the azimuthal
field component are reduced to a small plateau value as the strong magnetic
braking has removed almost all of the angular momentum early in the collapse
(see Figure \ref{fig-kk10} and Section \ref{catastrophe}, where this is
discussed in more detail). Their similarity solutions tended towards this
behaviour when the magnetic braking parameter, $\alpha$, was large, even
though it does not appear in Equations \ref{ff-ssimm}--\ref{ff-ssim}. 

Given that there are two possible values for $b_{\phi,s}$, depending upon the 
chosen values of the parameters $\tilde{\eta}_{H,A}$ and $\delta$, there are
two possible sets of coefficients for $h, j$ and $b_{z}$. These are derived in
Section \ref{sub:BC} in region AD of the $pq$-plane (see Figure
\ref{pq-plane}), and are referenced by the subscripts 1 and 2 depending on 
which term in Equation \ref{ff-ssim} is the minimum that defines $b_{\phi,s}$.

The first set of coefficients occur when 
\begin{equation}
    \sqrt{\frac{2\tilde{\eta}_H^2}{\tilde{\eta}_A^2 - \tilde{\eta}_H^2}} 
	< \delta,
\label{b_phisineq}
\end{equation}
corresponding to the left hand side of the minimum choice for $b_{\phi,s}$ in
Equation \ref{ff-ssim}. The coefficient $b_{z1}$ is then the single real root
of the polynomial 
\begin{equation}
    b_{z1}^8 - \frac{m_c^2}{2\tilde{\eta}_A^2f_1} \,b_{z1}^6 - 
      \frac{m_c^6}{4\tilde{\eta}_A^4f_1^4} = 0, 
    \label{1-bz1}
\end{equation}
where $f_1$ is a constant defined by the magnetic diffusion parameters:
\begin{equation}
    f_1 = \frac{\tilde{\eta}_A^2 + \tilde{\eta}_H^2}
	{\tilde{\eta}_A^2 - \tilde{\eta}_H^2}.
    \label{1-f}
\end{equation}
The coefficients for $j$ and $h$ are given by
\begin{equation}
    j_1 = -\frac{b_{z1}^2}{m_c} \,\frac{\sqrt{2}\,\tilde{\eta}_H}
	{\sqrt{\tilde{\eta}_A^2 - \tilde{\eta}_H^2}}
    \label{1-j1}
\end{equation}
and
\begin{equation}
    h_1 = \frac{f_1b_{z1}^2}{\sqrt{2m_c^3}}
      \left[-1+\sqrt{1-\frac{4m_c^2}{f_1^2b_{z1}^4}}\,\right].
    \label{1-h1}
\end{equation}
This solution is the asymptotic inner solution of \citet{kk2002} for their
strong braking similarity solutions. In their solution, $\tilde{\eta}_H = 0$,
$f_1 = 1$, and the angular momentum coefficient $j = 0$, so the angular
momentum tends towards zero as $x \to 0$.

The second set of coefficients are adopted when the inequality
\begin{equation}
    \sqrt{\frac{2\tilde{\eta}_H^2}
	{\tilde{\eta}_A^2 - \tilde{\eta}_H^2}} > \delta
\label{b_phisineq2}
\end{equation}
is satisfied, so that $b_{\phi,s}$ takes on the second value in \ref{ff-ssim},
$b_{\phi,s} = -\delta b_z$. The coefficient $b_{z2}$ is given by the single 
positive real root of the equation 
\begin{equation}
    b_{z2}^8 - \frac{m_c^2(1+\delta^2)}{2f_2^2}\,b_{z2}^6 
	- \frac{m_c^6}{4f_2^4} = 0
    \label{2-bz2}
\end{equation}
where $f_2$ is given by
\begin{equation}
   f_2 = \tilde{\eta}_A - \tilde{\eta}_H\delta\sqrt{2+\delta^2}.
   \label{2-f}
\end{equation}
This then gives the other coefficients as
\begin{equation}
    j_2 = \frac{\delta{b_{z2}^2}}{m_c}
    \label{2-j2}
\end{equation}
and 
\begin{equation}
    h_2 = \frac{(1+\delta^2)b_{z2}^2}{\sqrt{2m_c^3}}
	\left[-1+\sqrt{1+\frac{4m_c^2}{(1+\delta^2)^2b_{z2}^4}}\right].
    \label{1-h_2}
\end{equation}
This solution was not explored in \citet{kk2002}, as without the Hall term the
left hand side of the minimum condition in Equation \ref{ff-ssim} is always
zero and can never exceed the right hand side.

As in the Keplerian disc solution, this similarity solution can be converted
to dimensional form using Equations \ref{sssigmag}--\ref{sseta}, to better
understand the physics behind the free fall collapse. In this case the
solution becomes: 
\begin{align}
    M&=\frac{c_s^3m_c}{G}\,t, \\
    \dot{M} &= \frac{c_s^2m_c}{G},\\
    \Sigma &= \frac{1}{\pi{G}}\sqrt{\frac{c_s^3m_c}{2^3rt}},\\
    H &= h_{1,2}\,\frac{r^{3/2}}{\sqrt{c_st}},\\
    V_r &=-\sqrt{\frac{2m_cc_s^3t}{r}},\\
    V_\phi &= j_{1,2}\,c_s,\\
    J &=j_{1,2}\,c_sr,\label{ff-dimsJ}\\
    \Psi &= \frac{2\pi{}c_s^2b_{z1,2}}{G^{1/2}}\,r,\\
    B_z &= \frac{c_s^2b_{z1,2}}{G^{1/2}}\,r^{-1},\\
    B_{r,s} &= B_z, \\
    \text{and }B_{\phi,s} &= -\text{min}\left[
	\frac{|\tilde{\eta}_H|}{\tilde{\eta}_A}B\,;\, \delta B_z\right].
\label{ff-dimension}
\end{align}
By taking the absolute value of $\tilde{\eta}_H$ the sign of $B_{\phi,s}$ is
held constant. Were the sign of $B_{\phi,s}$ to change, then the signs of $J$
and $B_z$ would also change, resulting in a set of flow variables that are 
identical but for a reversed direction of $\mathbf{\hat{z}}$. The absolute
value of the Hall diffusion coefficient is adopted merely to simplify the
calculations.  

The power laws are solutions to the simplified equations:
\begin{align}
    V_r\frac{\partial{V_r}}{\partial{r}} &= g_r,\label{ff-eqnsvr}\\
    \frac{\partial{J}}{\partial{r}}
      &= \frac{rB_{\phi,s}B_z}{2\pi\Sigma{V_r}},\\
    -V_rH &= \frac{\eta_H}{B}B_{\phi,s} + \frac{\eta_A}{B^2}B_z^2,\\
    \frac{GM_c}{2r^3}H^2 &+ \frac{(B_{\phi,s}^2 + B_{r,s}^2)}{4\pi\Sigma}\,H
      -c_s^2 = 0,\\
    \dot{M} &= \text{constant}\\
    J &= rV_\phi,\label{ff-eqnsJ}\\
    B_{r,s} &= B_z\label{ff-eqnsbrs}\\
    B_{\phi,s} &= -\text{min}\left[\frac{|\eta_H|}{\eta_A}\,B\,;\,
      {\delta}B_z\right]\label{ff-eqnsbphis}\\
    \text{and } \Psi &= 2\pi r^2B_z;
\label{ff-eqns}
\end{align}
as in the Keplerian disc solution the induction equation takes the simplified
form $V_r + V_{Br} = 0$. In this similarity solution any remaining rotation of
the flow is that induced by the magnetic ``braking'', which can cause rotation
by Hall diffusion in the azimuthal direction of the field lines tied to the
electrons, which creates a rotational torque on the neutrals and grains as
they fall inward rapidly.

Equations \ref{ff-eqnsvr}--\ref{ff-eqnsJ} may be solved to give the disc
variables as functions of the surface density and magnetic field:
\begin{align}
    V_r &= -\sqrt{\frac{2GM}{r}}\,,\\
    M &= -2{\pi}r{\Sigma}V_rt\,\\
    \dot{M} &= -2{\pi}r{\Sigma}V_r\\
    \text{and }J &= \frac{B_{\phi,s}B_z}{2{\pi}{\Sigma}V_r} \,r^2,
\label{ff-sigmaeqns}
\end{align}
where the field components are given by Equations
\ref{ff-eqnsbrs}--\ref{ff-eqns} and the scale height of the disc is given by
the equation 
\begin{align}
    H_1 &= \frac{f_1B_z^2r^{7/2}t}{\sqrt{2GM^3}}
      \left[-1+\sqrt{1-\frac{4c_s^2M^2}{f_1^2B_z^4r^4t^2}}\, \right]\\
    \text{or } H_2 &= \frac{(1+\delta^2)B_z^2r^{7/2}t}{\sqrt{2GM^3}}
      \left[-1+\sqrt{1-\frac{4c_s^2M^2}{(1+\delta^2)^2B_z^4r^4t^2}} \,\right]
    \label{2-H2}
\end{align}
depending on which value of the azimuthal field component is adopted.

This similarity solution is an example of the magnetic braking catastrophe
that occurs in numerical simulations of star formation when the magnetic
braking of a collapsing core is so strong that all angular momentum is removed
from the gas and it is impossible to form a rotationally-supported disc
\citep[e.g.][]{ml2008, ml2009}. It is clear from Equations \ref{ff-ssimj},
\ref{ff-ssim} and \ref{1-j1} that when there is no Hall diffusion the magnetic
braking in this solution causes $B_{\phi,s} = J = 0$, which is indicative of
the catastrophe and prevents disc formation. This behaviour was demonstrated
in the ``strong braking'' solution of \citet{kk2002} where $\delta = \alpha =
10$ and the fluid was able to fall onto the protostar with very little angular
momentum remaining. This particular solution, and the magnetic braking
catastrophe more generally, are discussed in more detail in Section
\ref{catastrophe}. 

The introduction of Hall diffusion to the similarity solution can cause
additional twisting of the field lines and magnetic braking, or it can cause a
reduction in these by twisting the magnetic field lines in the opposite
direction, in effect spinning up the collapse. The direction of the Hall
diffusion depends upon the orientation of the field with respect to the axis
of rotation, and it is obvious that this directionality has an important
effect on the magnetic braking catastrophe. The linear scaling of the angular
momentum with radius (Equation \ref{ff-dimsJ}) suggests that the point mass at
the origin has no angular momentum, however, Hall diffusion will likely ensure
that in the innermost regions of the collapse the angular momentum shall reach
a plateau value similar to that in Figure \ref{fig-kk10} for the ambipolar
diffusion-only collapse. No similarity solutions to the full collapse problem
were found that matched onto the free fall inner asymptotic solution discussed
in this section, however, work is underway to find similarity solutions with
strong $\alpha$ and $\delta$ that demonstrate this asymptotic behaviour and
illustrate how the Hall effect influences the magnetic braking catastrophe
directly.

\section{Summary}

This chapter saw the derivation of two distinct power law similarity solutions
to the fluid equations in the innermost regions of the collapse as $x \to 0$.
The first of these was a rotationally-supported disc through which the gas is
slowly accreted. The surface density and scale height of the disc are
determined by the magnetic diffusion coefficients, with Hall diffusion either
adding to or reducing the surface density. The magnetic field and surface
density scale with radius in the same way as in previous protostellar disc
models and any accretion through the disc is regulated by the outward
diffusion of the magnetic field against the flow. 

The diffusion coefficients also placed limits upon the flow. In order to form
a Keplerian disc as described by the first asymptotic solution, certain
restrictions are placed upon the diffusion coefficients. Hall diffusion can
counteract the ambipolar diffusion and prevent the gas from falling in; a disc
may only form when the nondimensional Hall parameter $\tilde{\eta}_H$ is no
larger than $\tilde{\eta}_A \sqrt{\frac{8}{17}}$ (for $\delta = 1$). There is
no lower limit on the size of the Hall diffusion parameter, as when
$\tilde{\eta}_H$ is negative it aids the outward diffusion of the field,
reducing the amount of magnetic flux that is accreted onto the central
protostar. The orientation of the field with respect to the axis of rotation
determines the direction of the Hall diffusion and whether it resolves or
furthers the magnetic flux problem outlined in Section \ref{lr:mfp}.

Similarly, the Hall effect can increase or reduce the angular momentum problem
in the second asymptotic solution, where the matter is free falling onto the
central protostar and the only rotational velocity is that induced by Hall
diffusion in the azimuthal direction. No disc may form in this solution, as
the strong magnetic braking prevents the centrifugal force from becoming large
enough to support the infalling gas against gravity. The amount of magnetic
braking affecting the flow depends upon the Hall parameter as well as the
values of the magnetic braking parameter, $\alpha$ (Equation \ref{alpha}), and
the cap on $b_{\phi,s}$, $\delta$ (see Equation \ref{ssb_phis}). This is
expected to be the asymptotic inner solution for the collapse when the
parameters $\alpha$ and $\delta$ are large, as in the ``strong braking''
similarity solution of \citet{kk2002}, which is discussed further in Section
\ref{catastrophe}. 

The two similarity solutions represent both sides of the disc formation
problem that is referred to in the literature as the magnetic braking
catastrophe. In the first, the magnetic braking is limited, and a
rotationally-supported disc such as those in the simulations of \cite{mim2010}
forms, while in the second no disc forms as the catastrophic magnetic braking
removes almost all of the angular momentum so that the matter is falling
rapidly onto the central protostar as in the simulations of \citet{ml2009} and
others. This problem has yet to be fully resolved, however the magnetic
diffusion is clearly important in determining whether or not a disc forms and
which of the two asymptotic solutions shall describe the inner region of any
given collapsing molecular cloud core.

All of the similarity solutions presented in Chapters \ref{ch:nohall} and 
\ref{ch:hall} form rotationally-supported discs, and those with Hall and
ambipolar diffusion match specifically onto the Keplerian disc solution
described here. The asymptotic similarity solution is used as the inner
boundary condition for these models, enforcing disc formation to show how the
Hall effect influences the properties of disc-forming similarity solutions.
The free fall solution is not studied further as part of the full collapse
models, but it shall be discussed in Section \ref{catastrophe} and it is
expected that similarity solutions matching onto this asymptotic solution will
be found in future. 

\cleardoublepage

\chapter{Collapse without the Hall Effect}\label{ch:nohall}

The construction of the self-similar model for studying gravitational collapse
with the Hall effect was a gradual affair, taking place in stages of
increasing physical complexity. The reasons for this were twofold: firstly, a
good initial guess of the variables was required in order to solve the
equations with Hall diffusion --- the calculation of similarity solutions
without Hall diffusion would provide this guess. Secondly, by reproducing the
results of previous self-similar collapse models it is possible to check the
calculation code to ensure that it is both numerically robust and physically
sound, and to be certain that the changes brought on by the introduction of
Hall diffusion are properly contrasted with similarity solutions that contain
less complex physics. The three models calculated in this chapter belong to
different families of similarity solutions: those that are nonmagnetic, those
with ideal magnetohydrodynamics, and those with non-ideal MHD and ambipolar
diffusion. 

This chapter is dedicated to outlining the construction, testing and physical
behaviour of these models, and the derivation of the necessary inner
asymptotic solutions and jump conditions that determine the collapse behaviour
as $x \to 0$. These are then compared to models found in the literature,
adopting their boundary conditions and reproduce their results. As may be
expected, the primary targets of these comparisons are the results of
\citet{kk2002}, although the work of \citet{sh1998} shall also be discussed.
The similarity solutions presented in this chapter provide a baseline for the
discussion of the physics affecting the solutions presented in Chapter
\ref{ch:hall} once the Hall term is activated. 

All of the similarity solutions presented in this chapter include the effects
of rotation, and all of them form rotationally-supported discs. Although it
was shown by \citet{kk2002} that there exist similarity solutions for which
the magnetic braking is so strong that no disc forms (the free fall asymptotic
solution presented in Section \ref{ff} demonstrates the inner boundary
conditions for such a solution in the Hall regime), those solutions are not
reproduced in this work. This modified angular momentum problem, in which the
magnetic braking is so strong that it removes all of the angular momentum from
the collapsing flow has been seen in many numerical simulations
\citep[e.g.][]{ml2008, ml2009, hc2009}, and has yet to be fully resolved. The
inclusion of Ohmic diffusion \citep{db2010} and Hall diffusion
\citep{kls2010-2} has been shown to reduce the magnetic braking catastrophe,
which is discussed further in Section \ref{catastrophe}.

Discs form in the collapse solutions presented in this chapter because the
magnetic braking is artificially capped in a manner that could be a reasonable
substitute for nonaxisymmetric effects, or a disc wind, that could change the
transport of angular momentum above the pseudodisc. By studying those
solutions with discs the importance of the magnetic field diffusion in driving
and controlling gravitational collapse will be demonstrated. 

This chapter outlines the construction and results of collapse models that are
nonmagnetic (Section \ref{nonmag}), that are magnetic under ideal MHD (Section
\ref{imhd}), and that are magnetic with ambipolar diffusion in Section
\ref{ad}. For each of these collapse simulations the inner asymptotic solution
must be derived, as well as the jump conditions at the centrifugal shock and
an estimation of the shock position. The calculation procedure is then
described and the results of each model examined. The impact of rotation and
magnetic fields on the structural features of the similarity solutions, 
particularly the centrifugal shock and the size of the rotationally-supported
disc, are analysed with reference to previous models of collapse in order to
provide a solid foundation for discussing the influence of the Hall effect on
the solutions in the following chapter. 

\section{Nonmagnetic Solutions}\label{nonmag}

The simplest model constructed here is that of a nonmagnetic rotating
collapse, which has been examined by \citet{sh1998} and \citet{kk2002}. This
model is characterised by the equations 
\begin{align}
    \frac{\partial\Sigma}{\partial{t}} 
        &+ \frac{1}{r}\frac{\partial}{\partial{r}}(r\Sigma V_r) = 0,
        \label{nmCM}\\
    \frac{\partial V_r}{\partial t} + V_r\frac{\partial V_r}{\partial r} 
         &= g_r - \frac{c_s^2}{\Sigma} \frac{\partial\Sigma}{\partial r}
         + \frac{J^2}{r^3}, \label{nmRM}\\
    \frac{\partial J}{\partial t} &+ V_r\frac{\partial J}{\partial r} = 0
         \label{nmAM}\\
    \text{and } \frac{c_s^2\Sigma}{2H} &= \frac{\pi}{2}G\Sigma^2,
\label{nmVE}
\end{align}
which are the nonmagnetic form of Equations \ref{cmfinal}--\ref{hefinal}.
Because there is no mechanism for braking the angular momentum, no central
mass forms and the only term left to control the vertical squeezing of the
pseudodisc is its self-gravity.

The nondimensional form of Equations \ref{nmCM}, \ref{nmRM} and \ref{nmVE}
is
\begin{align}
    \frac{dm}{dx} &= x\sigma,\label{nmm}\\
    (1-w^2)\frac{1}{\sigma}\frac{d\sigma}{dx} &=-\frac{m}{x^2} 
     + \frac{j^2}{x^3} + \frac{w^2}{x}, \label{nmsigma}\\
    \text{and } h &= \frac{2}{\sigma};
\end{align}
while the simplified self-similar angular momentum equation, 
\begin{equation}
    \frac{dj}{dx} = \frac{j}{w},
\label{nmdj}
\end{equation}
is integrated to give 
\begin{equation}
    j = \Phi m
\label{nmj}
\end{equation}
where $\Phi$ is a constant. Its value is the initial ($x\to\infty$) ratio of
the angular momentum to the mass at the outer boundary, given by the
expression 
\begin{equation}
    \Phi = \frac{v_0}{A},
\label{nmphi}
\end{equation}
where the constants $v_0$ and $A$ characterise the outer boundary conditions
$j = v_0x$ and $m = Ax$ (see Section \ref{eqouter}). This ratio $\Phi$ is
denoted $\omega$ by \citet{sh1998}; their outer boundary conditions are
described by the same equations as in this work, however the values of the
parameters characterising any given similarity solution are subtly different
from those chosen in this work. Their $\omega = 0.3$ solution corresponds
quite well to the fast rotation solution presented in subsection
\ref{nm:sols}. 

Equation \ref{nmj}, which takes the dimensional form
\begin{equation}
    J = \frac{\Phi GM}{c_s},
\label{nmJ}
\end{equation}
holds true in the outer (large $x$) regions of all of the solutions presented
in this thesis, although it breaks down in the magnetic solutions at lower $x$
as the flux builds up and magnetic braking starts to reduce the angular
momentum. \citet{mhn1997} showed that Equation \ref{nmJ} held true across many
orders of magnitude in their nonmagnetic two-dimensional numerical solutions
of the collapse of a rotating cloud into a rotating disc. Clearly this
relation holds when the rotational velocity and the density are uniform, and
if the initial cloud has constant density and rotational velocity before it is
centrally condensed Equation \ref{nmJ} will hold true for the nonmagnetic
collapse so long as the collapse is axisymmetric. 

From Equations \ref{nmm}--\ref{nmphi} it is possible to derive the asymptotic
similarity solution describing the behaviour of the flow inwards of the
centrifugal shock, the jump conditions and the position of the shock itself.

\subsection{Inner solution}\label{nm:inner}

As in Chapter \ref{ch:asymptotic} for the full model with Hall diffusion, the
inner asymptotic solution is found by assuming that the surface density takes
the form of a power law in $x$, 
\begin{equation}
    \sigma = \sigma_1x^{-p},
\label{nminsig-p}
\end{equation}
where $\sigma_1$ and $p$ are both real constants. The conservation of mass
equation (\ref{nmm}), 
\begin{equation}
    \frac{dm}{dx} = \sigma_{1}x^{1-p},
\label{nminmstuff}
\end{equation}
is integrated to give
\begin{equation}
    m = \frac{\sigma_{1}x^{2-p}}{2-p}.
\label{nminm-p}
\end{equation}
The requirement that the rotational velocity must not diverge as $x\to0$ means
that the angular momentum must vanish at the origin, and that no central point
mass can form. Using Equation \ref{nmj} the angular momentum of the fluid is
then given by
\begin{equation}
    j = \frac{\Phi\sigma_1x^{2-p}}{2-p}.
\label{nminj-p}
\end{equation}

Equations \ref{nminsig-p}, \ref{nminm-p} and \ref{nminj-p} are then
substituted into the conservation of radial momentum equation (\ref{nmsigma}),
so that it is then expressed in terms of $x$ and the constants,
\begin{equation}
    -p\left(1 - \frac{x^2}{(2-p)^2}\right) = - \frac{\sigma_1x^{1-p}}{2-p} 
    + \frac{\Phi^2\sigma_1^2x^{2-2p}}{(2-p)^2} + \frac{x^2}{(2-p)^2}.
\label{nmin-rm2}
\end{equation}
Taking the limit as $x \to 0$, the ${\cal O}(x^2)$ terms are clearly much
smaller than those terms of ${\cal O}(x^0)$, and so they may be dropped from 
the equation. This then becomes
\begin{equation}
    \frac{\Phi^2\sigma_1^2}{(2-p)^2}x^{2-2p} - \frac{\sigma_1}{2-p}x^{1-p} + p
    = 0,
\label{nmin-p}
\end{equation}
and when each pair of exponents are equated it becomes obvious that the only
solution is $p = 1$. Substituting this into Equation \ref{nmin-p} reduces it to a
quadratic in $\sigma_1$,
\begin{equation}
    \Phi^2\sigma_1^2 - \sigma_1 + 1 = 0.
\label{nmin-sigma1}
\end{equation}

The inner asymptotic solution is thus defined by the equations: 
\begin{align}
    \sigma = \sigma_1x^{-1}
     &= \left(\frac{1 + \sqrt{1 - 4\Phi^2}}{2\Phi^2}\right)x^{-1}\label{nmasigma},\\
    m &= \sigma_1x \label{nmam},\\
    j &= \Phi\sigma_1x \label{nmaj},\\
    v_\phi &= \Phi\sigma_1\label{nmavphi},\\
    h &= \frac{2}{\sigma_1}x\label{nmah},\\
    u &= 0\label{nmau}\\
    \text{and }\dot{m} &= 0.\label{nmamdot}
\end{align}
These equations are identical to the inner solution that was presented without
explicit derivation in \S3.1 of \citet{kk2002}. This inner solution ceases to
exist when $\Phi > 0.5$, as in this regime $\sigma_1$ becomes complex; and a
second unstable solution exists when the coefficient of $\sigma$ is the second
root of Equation \ref{nmin-sigma1}. Any increase in $\Phi$ corresponds to a
reduction in the gravitational force, so that it is no longer able to balance
the centrifugal and pressure forces in the disc. If the initial ratio of the
centrifugal to gravitational forces is too high then no disc can form, as
there is no mechanism for reducing the centrifugal forces to create a stable
disc. 

In order to better understand the behaviour of this disc, it is worthwhile to
convert the asymptotic similarity solution back to dimensional form:
\begin{align}
    \Sigma &= \frac{c_s^2\sigma_1}{2\pi G} \,r^{-1}\label{nmaSigma},\\
    M &= \frac{c_s^2\sigma_1}{G} \,r\label{nmaM},\\
    J &= \Phi c_s\sigma_1\, r\label{nmaJ},\\
    V_\phi &= \Phi c_s\sigma_1\label{nmaVphi},\\
    H &= \frac{2}{\sigma_1} \,r\label{nmaH},\\
    V_r &= 0\label{nmaVr}\\
    \text{and }\dot{M} &= 0,\label{nmaMdot}
\end{align}
where the constant $\sigma_1$ retains its definition from Equation
\ref{nmasigma}. At any given radius the enclosed mass and surface density are
constant with time, and the material orbits with a stable rotational velocity.
These relations may also be written as a set of variables that are functions
of 
$\Sigma$ and $M$: 
\begin{align}
    M &= 2\pi r^2 \Sigma,\label{nmaMSigma}\\
    H &= \frac{c_s^2}{\pi G\Sigma},\label{nmaHSigma}\\
    J &= \Phi c_sGM\label{nmaJSigma}\\
    \text{and }V_\phi &= \frac{\Phi c_sGM}{r}.\label{nmaVphiSigma}
\end{align}
These are all steady state equations, as the material joining the disc quickly
loses its radial momentum and stops moving inward. The gas is unable to move
inwards after this point as there is no way to change its angular momentum,
and so the material remains in orbit, unable to fall to the origin to form a
point mass. 

These inner asymptotic solutions match quite well onto the inner regions of
the similarity solutions close to the origin. The outer edge of this region is
marked by a steep shock, which separates the inner centrifugally-supported
disc from the outer dynamic collapse. The properties of this shock are
outlined in the following subsection.

\subsection{Shock position and jump conditions}\label{nm:shock}

The transition between the outer supersonic collapse and the inner steady
state disc takes the form of an abrupt change in both the surface density
$\sigma$ and the radial velocity $u$. Inward of this shock, the density
increases dramatically, and the infall velocity is slowed to a very low value,
which rapidly drops to zero in the post-shock region. In solutions where the
matter is initially rapidly rotating, the radial velocity can change sign as
the shock front overtakes it, creating a region of shocked backflow that
follows behind the shock front until the radial velocity decreases enough that
the gas settles into the asymptotic steady state disc. 

Although the form of the inner asymptotic solution changes with the addition
of magnetic fields and then magnetic diffusion, this shock is a common feature
of all the similarity solutions presented in this work. Its position is
estimated to occur at the \textit{centrifugal radius}, the point where gravity
and the centrifugal forces are first in approximate balance (which is near to
the boundary of the rotationally-supported disc), and this radius will be
shown to be a reasonable approximation to the shock position. It is calculated
by equating the gravitational and centrifugal forces in the radial momentum
equation (\ref{nmsigma}): 
\begin{equation}
    \frac{m(x_c)}{x_c^2} = \frac{j(x_c)^2}{x_c^3}
\label{nmeqforces}
\end{equation}
and then solving for the centrifugal radius, $x_c$:
\begin{equation}
    x_c = \frac{j(x_c)^2}{m(x_c)}.
\label{xc}
\end{equation}
This calculation requires a deep understanding of the behaviour of both the
enclosed mass and the angular momentum in the region of the shock; these
become more difficult to estimate with the increasing complexity of the 
models. However, in the nonmagnetic case this is clearly given by
\begin{equation}
    x_c = \Phi^2m(x_c);
\label{nmxcm}
\end{equation}
where an approximation to the mass is still needed in order to determine the
centrifugal radius. 

The initial (outer) conditions have the gas infalling at a velocity $|u_0|\ll
x$, however, as the mass of the inner disc grows more matter is pulled inwards
faster, so that in the slower rotation cases, the gas close to the centrifugal
shock is falling in at free fall speeds. When the radial velocity becomes much
larger than $x$, then $w \approx |u|$ and $m \approx \dot{m}$, and the
continuity equation becomes
\begin{equation} 
    \frac{dm}{dx} = x\sigma = \frac{m}{w} \ll \frac{m}{x}
\label{nmcon}
\end{equation}
(using $m = xw\sigma$); this causes the value of the enclosed mass to drop to
a plateau value that remains near-constant until the centrifugal shock
position (as can be seen in the results in the following subsection). The
outer edge of this plateau is estimated to occur at 
\begin{equation}
    x_{pl} \approx |u_0|,
\label{xpl}
\end{equation}
which is something of an overestimate as $|u|$ is typically larger than
$|u_0|$ when $m$ reaches the plateau value. At this point the outer asymptotic
solution is still a good approximation to $\sigma$, and so it and Equation
\ref{xpl} are substituted into $m = xw\sigma$ to give an approximation to the
plateau mass: 
\begin{equation}
    m_{pl} \approx x_{pl}(x_{pl} - u_0)\sigma_{pl} 
    = x_{pl}(|u_0| - u_0)\left(\frac{A}{x_{pl}}\right)
    = 2|u_0|A.
\label{nmmpl}
\end{equation}

The value of the mass in the plateau changes very little between $x_{pl}$ and
$x_c$, and so this calculated value can be substituted into Equation
\ref{nmxcm} to give the approximate centrifugal shock position,
\begin{equation}
    x_c \approx \frac{2|u_0|v_0^2}{A}.
\label{nmxc}
\end{equation}
This equation \citep[derived in equations 34--35 of][]{kk2002} only applies if
$x_c \lesssim x_{pl}$, which is true whenever $v_0 \lesssim (A/2)^{1/2}$.
Given the limit on $\Phi$ established in \ref{nm:inner}, this weak inequality
is not violated in the similarity solutions explored in this thesis. There is
no solution if the mass has yet to reach its plateau value. 

\begin{table}[t]
\begin{center}
\begin{tabular}{cccc}
\toprule
 model & $\Phi$ & estimated $x_c$ & actual $x_c$\\
 \midrule
 $v_0 = 0.1$ & $0.0\dot{3}$ & $0.00\dot{6}$ & 0.0077\\
 $v_0 = 1.0$ & $0.\dot{3} $ & $0.\dot{6}  $ & 0.64\\
 \bottomrule
 \end{tabular}
\end{center}
\vspace{-5mm}
 \caption[Estimated vs actual values of $x_c$ in the nonmagnetic model]
{Comparison between the estimated and actual values of the centrifugal shock
position, $x_c$.} 
\label{tab-nmxc}
\end{table}
For the models presented in the next subsection, the actual centrifugal shock
position is close to the one calculated by Equation \ref{nmxc}, and the values
of these are presented for comparison in Table \ref{tab-nmxc}. As they are so
close, Equation \ref{nmxc} is an acceptable initial guess to use when finding
the shock position by the iterative process described in Section
\ref{num:iter} for the Hall similarity solutions. While this process is used
with few modifications in all of the models presented in this work, its
description is left to the following chapter for the reader interested only in
the model with Hall diffusion. 

Because the centrifugal shock manifests as a discontinuity in $\sigma$ and
$u$, the jump conditions must be calculated explicitly at the shock position.
At the shock, the radial momentum equation (\ref{nmsigma}) can be written as 
\begin{equation}
    (1-w^2)\frac{d\sigma}{dx} = 0,
\label{nmrmdots}
\end{equation}
where the terms on the right hand side of Equation \ref{nmsigma} are all small
at the shock in comparison to the steep derivative term on the left. The shock
occurs because of the singularity inherent to the equation, which occurs at
the locus of points where $(1-w^2) = 0$, that is, where 
\begin{equation}
    (x-u)^2 = 1.
\label{nmsing}
\end{equation}
In dimensional form the singularity takes the form of the sonic line,
\begin{equation}
    \left(\frac{r}{t} - V_r\right)^2 = c_s^2.
\label{nmdimsing}
\end{equation}
The shock occurs when the curve describing the flow crosses the singular line
in the $xu$-plane. The shock propagates outwards at the speed of sound, and
occurs near the centrifugal radius defined above. 

The jump conditions are found by solving the continuity and radial momentum
equations at the position of the shock. At the shock, the derivatives of
$\sigma$ and $u$ become large with respect to the other terms in the
equations, which may then be disregarded. The equations are then integrated at
$x_c$ to define the jump conditions.

Starting by examining the conservation of mass across the shock, $m =
xw\sigma$ is substituted into Equation \ref{nmm} to give
\begin{equation}
    \frac{dm}{dx} = w\sigma + x\frac{d(w\sigma)}{dx} = x\sigma,
\label{nmcmsing1}
\end{equation}
which can be simplified to 
\begin{equation}
    \frac{d(w\sigma)}{dx} = \frac{u\sigma}{x}.
\label{nmcmsing2}
\end{equation}
As mentioned above, the terms on the right hand side of this equation are
small compared to the large derivatives on the left, and so they are dropped.
The derivative is then integrated across the shock front to give the first of
the jump conditions: 
\begin{equation}
    w\sigma = \text{constant},
\label{nmwsig}
\end{equation}
which ensures that the mass is conserved across the shock. 

Similarly, the terms on the right hand side of the radial momentum equation
can be dropped, and it is then rearranged into the form
\begin{equation}
    \frac{d\sigma}{dx} = w^2\frac{d\sigma}{dx}.
\label{nmrmsing1}
\end{equation}
This term is integrated across the shock front to become
\begin{equation}
    \sigma = w^2\sigma - \int\limits_{\substack{\text{shock}\\ \text{front}}}
	\sigma\frac{d(w^2)}{dx}\,dx + \text{constant};
\label{nmrmsing2}
\end{equation}
taking advantage of Equation \ref{nmwsig}, a factor of $w\sigma$ can be
removed from the integral in \ref{nmrmsing2}: 
\begin{equation}
    \sigma = w^2\sigma - 2w\sigma\int\limits_{\substack{\text{shock}\\\text{front}}}
	\frac{dw}{dx}\,dx  + \text{constant};
\label{nmrmsing3}
\end{equation}
and the fully integrated radial momentum equation is rearranged into the form
of the second jump condition:
\begin{equation}
    \sigma(1+w^2) = \text{constant}.
\label{nmsigw2}
\end{equation}

The jump conditions presented in Equations \ref{nmwsig} and \ref{nmsigw2} are
solved simultaneously by denoting the upstream and downstream sides of the
shock by the subscripts $u$ and $d$ and rewriting them in the form
\begin{align}
    \sigma_uw_u &= \sigma_dw_d\label{nmshock1}\\
    \text{and }\sigma_u(1+w_u^2) &= \sigma_d(1+w_d^2)\label{nmshock2}.
\end{align}
The first of these is substituted into the second, and this is factorised to
give the equation
\begin{equation}
    \left(w_u - \frac{1}{w_d}\right)(w_u-w_d) = 0,
\label{nmsols}
\end{equation}
which has the non-trivial solution:
\begin{align}
    w_d &= \frac{1}{w_u},\label{nmwsol1}\\
    \text{and }\sigma_d &= \sigma_uw_u^2.
\label{nmwsol2}
\end{align}

These jump conditions, presented with minimal derivation in appendix B1 of
\citet{kk2002}, are applicable to all of the collapse calculations in which
there is no mechanism by which the magnetic field can be changed by the
passage of the shock front. Even in those solutions with magnetic diffusion, 
the magnetic pressure and tension terms are never large enough to influence
the behaviour of the field in the centrifugal shock; the magnetic field is not
affected directly by the shock, although the field behaviour quickly changes
in the post-shock region, and so these jump conditions are used in those
solutions. 

The only similarity solutions with different jump conditions are those for
ideal MHD; these conditions are derived in subsection \ref{imhd:shock}.
\citet{kk2002} also derived jump conditions for a shock in which the scale
height $h$ was unaffected by the change in the density at the shock because
the thickness of the disc is determined by the gravity of the central point
mass. These jump conditions are not required here, as the similarity solutions
studied in this text occur in regions of parameter space where the gravity of
the central mass does not yet control the vertical squeezing of the collapsing
flow in the area of the centrifugal shock. 

The jump conditions in Equations \ref{nmwsol1} and \ref{nmwsol2} make it
possible to calculate the similarity solutions to the nonmagnetic rotational
collapse equations. 

\subsection{Similarity solutions}\label{nm:sols}

\begin{figure}[htp]
  \centering
  \includegraphics[width=5.5in]{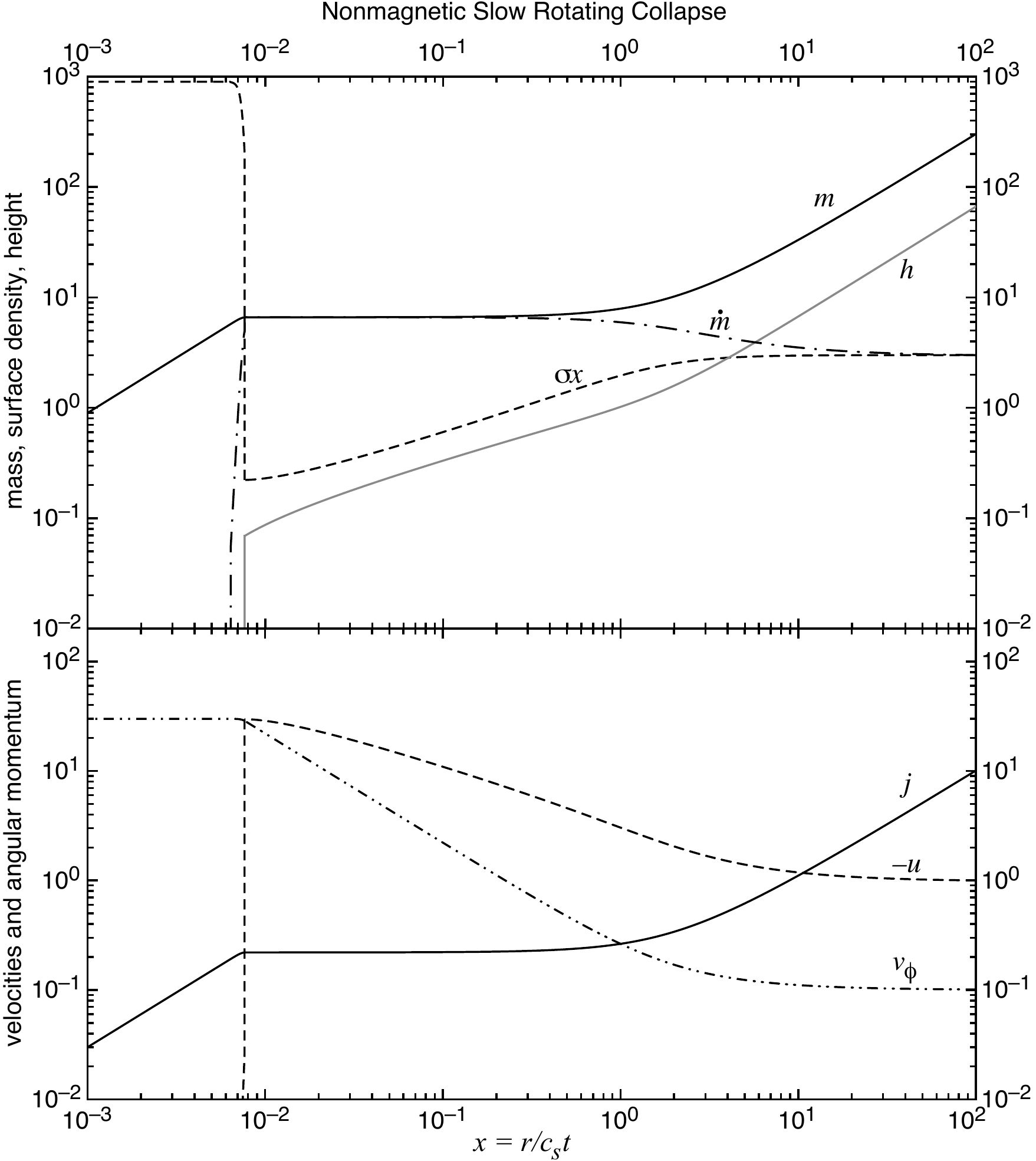}
  \caption[Nonmagnetic slow rotational collapse]{Similarity solution for the
slowly rotating nonmagnetic collapse, with outer asymptotic boundary
conditions $A = 3$, $u_0 = -1$, and $v_0 = 0.1$. The top panel shows the
nondimensional enclosed mass $m$, accretion rate $\dot{m}$, surface density
$\sigma$ and scale height $h$, while the lower panel displays the
nondimensional angular momentum $j$, radial infall speed $-u$ and angular
velocity $v_\phi$ all as functions of the similarity variable $x$. Inwards of
the centrifugal shock (located at $x_c = 7.7\times10^{-3}$) the radial velocity
and accretion rate drop rapidly to zero, and no central mass forms. In this
solution, as the initial rotation rate is slow, it takes longer for the
centrifugal forces to build up and balance gravity, allowing the infall
velocity to achieve free fall speeds.} 
\label{fig-nmslow}
\end{figure}
The similarity solutions are found by integrating the fluid equations from the
outer boundary to the inner one using the \texttt{stifbs} integrator routine
for integrating stiff sets of equations from \textit{Numerical Recipes in
Fortran 77} \citep{NR}, and checked against the output from the fifth order
Runge--Kutter integrator \texttt{rkqs} from the same source. Stiff equations
are those for which the normal numerical methods used to integrate them are
numerically unstable unless the step size is very small; the fluid equations
describing gravitational collapse are often stiff, particularly in the
innermost regions of the collapse. These routines (and their dependencies)
were modified from their original form to use double precision floating point
variables, and produce output that in general is identical to the seventh
significant figure. While either routine may be used to calculate this model,
the more complicated models following on from this are sometimes unable to be
completed using the \texttt{stifbs} routine as it uses a finite differences
method for calculating the Jacobian of the derivatives which operates poorly 
in those regions near to the shocks (it was too complicated to derive a set of
analytic expressions for the Jacobian for the later collapse models). In those
instances the \texttt{rkqs} routine is used; this change is not expected to
introduce significant errors. 

As the code integrates the variables inwards from the outer boundary towards
the inner boundary, the jump conditions derived in the previous subsection are
applied at the approximate position of the shock. The behaviour of the
variables inwards of the shock allows for a better estimate of the shock
position to be determined, and by a process of iteration the true value of the
shock is found. This routine is described in full in Section \ref{numerics}
for the collapse with Hall diffusion; qualitatively, the same downstream
behaviour is observed in each of the earlier collapse models. If the jump
conditions are applied at the true shock position, then as the variables are
integrated inwards they tend asymptotically to the inner power law behaviour
derived in subsection \ref{nm:inner}. 

\begin{figure}[htp]
  \centering
  \includegraphics[width=5.5in]{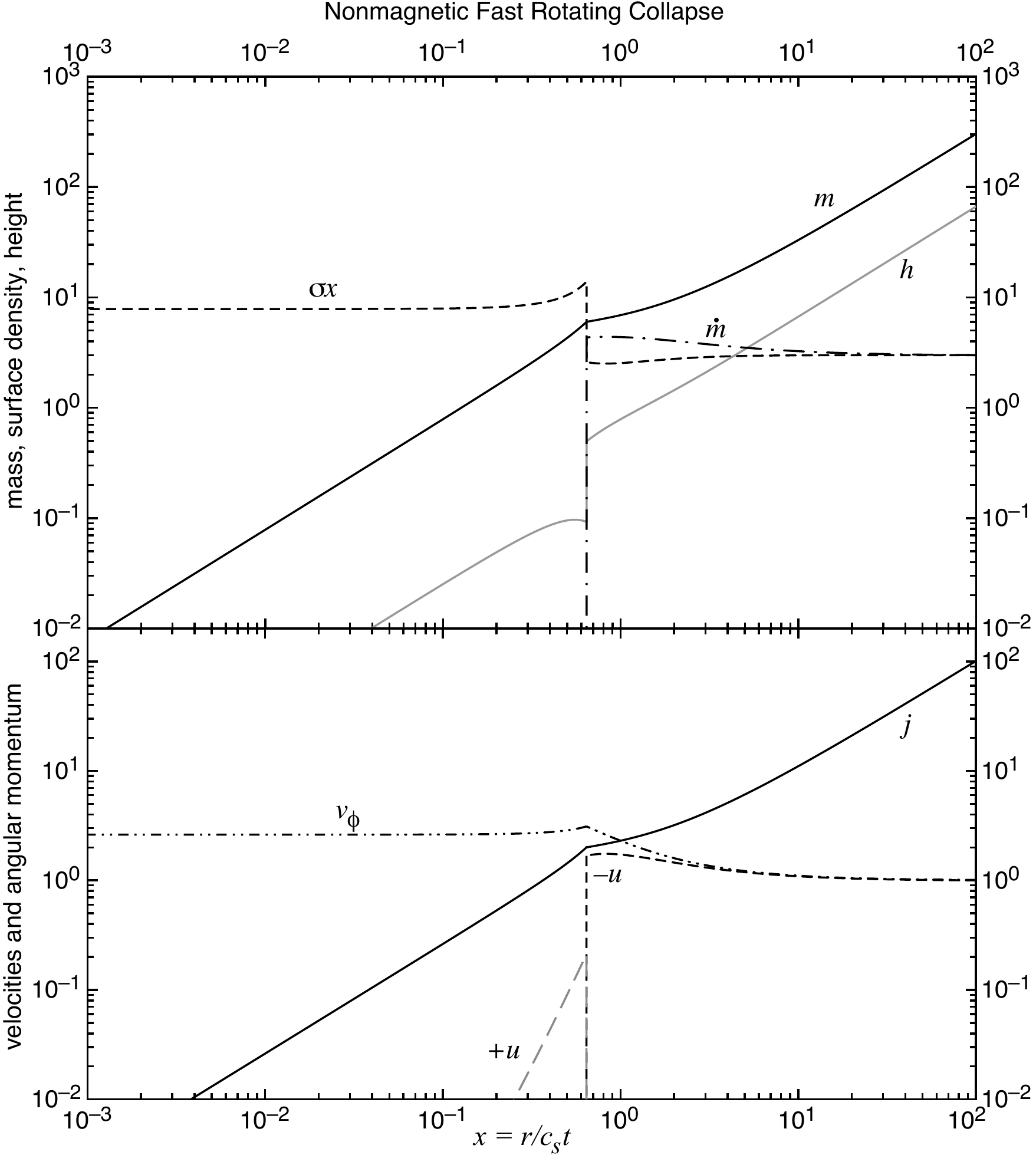}
  \caption[Nonmagnetic fast rotational collapse]{Similarity solution for the
rapidly rotating nonmagnetic collapse. The outer asymptotic boundary
conditions are as in Figure \ref{fig-nmslow}, however the initial (outer)
rotational velocity is increased to $v_0 = 1.0$. The faster initial rotation
causes the centrifugal force to balance gravity earlier in the collapse, so
that the centrifugal shock occurs at $x_c = 0.64$. Although the radial
velocity is much lower in this case than in the slow rotation calculation, the
shock is intense enough to change the sign of the radial velocity immediately
after the shock (indicated by the long-dash curve, $+u$), creating a region of
shocked backflow that moves outwards in physical space.}  
\label{fig-nmfast}
\end{figure}

Two similarity solutions are presented in Figures \ref{fig-nmslow} and
\ref{fig-nmfast}, the first showing a slow rotation case where the initial
rotational velocity is $V_\phi = 0.1c_s$, and the second a faster rotation
case characterised by an initial $V_\phi = c_s$. These values were chosen to
match the solutions in figures 1 and 2 of \citet{kk2002}, both to test the
modelling code and to duplicate their results. The fast rotation similarity
solution is also close to the $\omega = 0.3$ solution from \citet{sh1998},
however their initial (outer) boundary conditions are very slightly different,
as the values $v_0 = 1.05$ and $A = 3.5$ (compared to $A = 3.0$ in these
solutions) were chosen to match onto their solutions for runaway dynamic
collapse. Despite these differences, as the inner solution depends only upon
$\Phi = \omega$ their solution is identical to that in Figure \ref{fig-nmfast}
in the regions interior to the centrifugal shock. Similarly, their centrifugal
shock position is close to the value for the fast solution of $x_c = 0.643$.

As outlined in Section \ref{eqouter}, the outer region of the collapse
corresponds to a collapsing core just before point mass formation. For the two
solutions presented here, the outer collapse is similar, as the difference in
the initial angular momentum between the two only really becomes important in
the region close to the centrifugal shock. The material initially falls from
the outer edge at the constant radial velocity $V_r = -c_s$, with the infall
speed gradually increasing as the matter nears the disc, so that in the slow
rotation solution the gas is falling onto the disc at free fall speeds. In the
fast rotation similarity solution the centrifugal shock occurs before the mass
plateau is fully established; because of this the fluid is infalling slowly
when it hits the shock.

The mass plateau forms because the material is moving inward rapidly, with the
radial pressure terms in the radial momentum equation becoming less important
as gravity and the centrifugal forces start to dominate. The height of the
collapsing flow continues to decrease as the surface density does; because all
of the matter in this region is infalling rapidly, the pseudodisc self-gravity
pulls what remains in the thin disc towards the equator. 

The increase in the gravitational and centrifugal forces causes the formation
of a centrifugal barrier occurring at approximately the position predicted by
the estimation in the previous subsection, as was shown in Table
\ref{tab-nmxc}. The centrifugal shock moves outwards in physical space at a
velocity equal to the sound speed multiplied by the nondimensional shock
position, so that it is propagating slowly outwards in the slow rotation case
where it is located close to the origin in self-similar space, and rapidly
outwards for the fast rotation solution. If the material is initially
rotating rapidly, it may be so shocked that it is pulled along after the shock
in a backflow, before losing its outward momentum and settling to become part
of the disc. Using the jump condition \ref{nmwsol1}, it is possible to show
that a backflow will occur when
\begin{equation}
    w_u > \frac{1}{x_c},
\label{nmback1}
\end{equation}
that is, when the upstream radial velocity satisfies the inequality
\begin{equation}
    u_u < x_c - \frac{1}{x_c}.
\label{nmback2}
\end{equation}
The region of backflow is usually thin, and the material is quickly
decelerated by the shocked increase in density to a value near to zero.

The passage of the centrifugal shock causes the material to be abruptly
slowed, creating a large increase in the surface density and a flattening of
the centrifugally-supported disc that forms interior to the shock. The
variables quickly settle to their asymptotic values, taking the form of a
rotationally-supported disc of material orbiting around the origin without
ever falling in, as there is no way to brake its angular momentum. The
enclosed mass in the disc decreases linearly, as does the angular momentum. It
is clear that introducing a mechanism for braking the angular momentum will
allow the material to move inwards through the Keplerian disc and fall onto a
central point mass, resulting in the formation of a protostar at the centre of
the disc. The most obvious way to achieve this is by the introduction of
magnetic braking, which will be explored in the following section. 

\section{Ideal Magnetohydrodynamics Solutions}\label{imhd}

The most elementary form that the magnetic field behaviour can take is that
dictated by ideal magnetohydrodynamics (IMHD), where there is no magnetic
diffusion and the gas and magnetic field are tied by the insistence that the 
mass-to-flux ratio is constant. The field is frozen into the gas and moves
with it throughout the collapse, with the constant nondimensional mass-to-flux
ratio given by its initial value $\mu = \mu_0$. In this situation the
collapse is described by the equations:
\begin{align}
    \frac{\partial\Sigma}{\partial{t}} 
        + \frac{1}{r}\frac{\partial}{\partial{r}}&(r\Sigma V_r) = 0,
        \label{imhdCM}\\
    \frac{\partial V_r}{\partial t} + V_r\frac{\partial V_r}{\partial r} 
        = g_r - \frac{c_s^2}{\Sigma} \frac{\partial\Sigma}{\partial r}
        &+ \frac{B_z}{2\pi\Sigma}
         \left(B_{r,s} - H\frac{\partial B_z}{\partial r}\right)
        + \frac{J^2}{r^3}, \label{imhdRM}\\
    \frac{\partial J}{\partial t} + V_r\frac{\partial J}{\partial r} &= 
        \frac{rB_zB_{\phi,s}}{2\pi\Sigma}\label{imhdAM}\\
    \text{and } \frac{\Sigma c_s^2}{2H} = \frac{\pi}{2}G\Sigma^2
        + \frac{GM_c\rho H^2}{2r^3} &+ \frac{1}{8\pi}\left(B_{r,s}^2
         + B_{\phi,s}^2 - B_{r,s}H\frac{\partial B_z}{\partial r}\right);
\label{imhdVE}
\end{align}
where the flux and field are tied to the matter and defined by the relations
\begin{align}
    \Psi &= \frac{2\pi G^{1/2}M}{\mu_0}\label{imhdPsi},\\
    B_z &= \frac{2\pi G^{1/2}\Sigma}{\mu_0}\label{imhdBz},\\
    B_{r,s} &= \frac{\Psi}{2\pi r^2}\label{imhdBrs}\\
    \text{and } B_{\phi,s} &= -\text{min}
      \left[\frac{\Psi}{2\pi r}\frac{V_\phi}{V_{A,ext}}; \delta B_z\right].
\end{align}
In this model, the presence of a magnetic braking term in the angular momentum
equation allows for the transfer of angular momentum from the pseudodisc to
the envelope, so the gas may accrete onto a central mass (with mass $M_c$).
The equation set is essentially the full set from Chapter \ref{ch:derivs} with
$\tilde{\eta}_H = \tilde{\eta}_A = 0$, save for the induction and flux
conservation equations, which have been reduced to the flux freezing
descriptions in Equations \ref{imhdPsi} and \ref{imhdBz}. 

The dimensionless form of these equations is 
\begin{align}
    \frac{dm}{dx} &= x\sigma,\label{imhdcm}\\
    (1-w^2)\frac{1}{\sigma}\frac{d\sigma}{dx} =-\frac{m}{x^2} 
     + \frac{b_z}{\sigma}&\left(b_{r,s} - h\frac{db_z}{dx}\right) 
     + \frac{j^2}{x^3} + \frac{w^2}{x}, \label{imhdrm}\\
    \frac{dj}{dx} = \frac{1}{w}
     &\left(j - \frac{xb_zb_{\phi,s}}{\sigma}\right)\label{imhdam}\\
    \text{and } \left(\frac{\sigma m_c}{x^3} - b_{r,s}\frac{db_z}{dx}\right)h^2 
     + &\left(b_{r,s}^2 + b_{\phi,s}^2 + \sigma^2\right)h - 2\sigma = 0;
     \label{imhdve}
\end{align}
with the magnetic field terms given by 
\begin{align}
    \psi &= \frac{m}{\mu_0}\label{imhdpsi}\\
    b_z &= \frac{\sigma}{\mu_0}\label{imhdbz}\\
    b_{r,s} &= \frac{\psi}{x^2}\label{imhdbrs}\\
    \text{and }b_{\phi,s} &= 
     -\text{min}\left[\frac{2\alpha\psi j}{x^3};\delta b_z\right].
\label{imhdbphis}
\end{align}

In order to calculate the solutions to these equations, a set of inner
boundary conditions must be derived, as the formation of a protostellar mass
at the origin changes the dynamics of the inner accretion disc. The jump
conditions for the centrifugal shock that separates the dynamic outer collapse
from the inner slowly-accreting Keplerian disc must also be derived anew, as
the requirements of flux freezing demand that the centrifugal shock force a
change in the strength of the magnetic field as well as the density at the
boundary of the inner disc. The derivations of the conditions describing each
of these phenomena are presented in the following subsections; once their
behaviour is understood it is then possible to calculate the similarity
solutions presented in subsection \ref{imhd:sols}. The similarity solutions
show how the addition of a magnetic field changes the collapse behaviour,
particularly in the innermost regions around the newly-formed protostar. 

\subsection{Inner solution}\label{imhd:inner}

The addition of a magnetic field changes the form of the inner asymptotic
solution, allowing for the formation of a semi-Keplerian, magnetically-diluted
disc around a central point mass. It is not possible to obtain this solution
by setting the diffusion terms in the asymptotic similarity solution derived
in Chapter \ref{ch:asymptotic} to zero --- doing this causes the density to
vanish --- so the derivation of a new inner similarity solution is performed
here, using the same methodology as in the nonmagnetic and diffusive cases. 

As before, the inner asymptotic solution is assumed to take the form of a
series of power laws in $x$: 
\begin{align}
    \sigma &= \sigma_1x^{-p}\label{imhdinsig-p}\\
    \text{and }j &= j_1x^{-r}\label{imhdinj-r}.
\end{align}
In this work, only those solutions in which a central mass forms are sought,
and so the domain of $p$ is limited such that $p < 2$. There exists a solution
in which all of the angular momentum is removed from the collapse and no
central mass may form, however no such collapse calculations are presented in
this work. For further information about this ``strong braking'' solution the
reader is directed to \S3.2.3 of \citet{kk2002}. 

Adopting this limit on $p$, the continuity equation then integrates to
\begin{equation}
    m = m_c
\label{imhdinm}
\end{equation}
(as in Sections \ref{sub:AC}, \ref{sub:AD} and \ref{sub:BD}). This is
substituted into Equations \ref{imhdpsi} and \ref{imhdbrs}: 
\begin{align}
    \psi &= \frac{m_c}{\mu_0}\label{imhdinpsi}\\
    \text{and }b_{r,s} &= \frac{m_c}{\mu_0}\,x^{-2}.\label{imhdinbrs}
\end{align} 
These scalings for $m$ and $\sigma$ are then used to define
\begin{align}
    w &= \frac{m_c}{\sigma_1}\,x^{p-1},\label{imhdinh}\\
    \text{and }b_z &= \frac{\sigma_1}{\mu_0}\,x^{-p}.\label{imhdinbz}
\end{align}

The behaviour of the azimuthal component of the magnetic field is examined by
substituting the above scalings into Equation \ref{imhdbphis} to obtain
\begin{equation}
    b_{\phi,s} = -\text{min}\left[\frac{2\alpha j_1m_c}{\mu_0}\,x^{-r-3};
     \frac{\delta\sigma_1}{\mu_0}\,x^{-p}\right];
\label{imhdinbphisscale}
\end{equation}
of the two terms inside the brackets the term with the higher exponent is
sought, in order to satisfy the requirements of the cap on the magnetic
braking (for it is the smaller term as $x \to 0$). There are then two possible
values of $b_{\phi,s}$:
\begin{equation} 
    b_{\phi,s} = -\frac{2\alpha j_1m_c}{\mu_0}\,x^{-r-3}
\label{imhdinbphis1}
\end{equation}
which implies that $-r-3>-p$ (this, together with $p<2$, further requires that
$r<-1$); or 
\begin{equation}
    b_{\phi,s} = -\frac{\delta\sigma_1}{\mu_0}\,x^{-p}
\label{imhdinbphis2}
\end{equation}
in the case that $r \ge -1$. Using the knowledge that a rotationally-supported
disc is sought, the second of these values is chosen (the first leads to a
solution in which the surface density $\sigma$ is constant with respect to
$x$, and the angular momentum of the fluid is too low to support it against
gravity, preventing the formation of a rotationally-supported disc). This
implies that the magnetic braking is strong enough in the disc that the
artificial cap on $b_{\phi,s}$ must be invoked in order to prevent the removal
of all angular momentum from the gas. While the cap is a simplistic way of
limiting the transferral of angular momentum, it is not an unreasonable
assumption, as it is expected that other effects such as a disc wind or the
MRI will prevent the field lines from twisting too much and removing all
chance of disc formation \citep{kk2002}. 

The scalings for all of the terms are then substituted into the angular
momentum equation (\ref{imhdam}), which becomes
\begin{equation}
    -rj_1x^{-r-1} = \frac{j_1\sigma_1}{m_c}x^{1-r-p} 
    + \frac{\delta\sigma_1^2}{\mu_0^2m_c}x^{2-2p};\label{heythere}
\end{equation}
the exponent of the left hand term is then compared to the exponent of each of
the terms on the right hand side in order to determine which is the dominant
term. There are two possible solutions:
\begin{align}
-r-1 &= 1-r-p    & \text{or}&    &-r-1&= 2-2p,
\end{align}
which are simplified to give
\begin{align}
p &= 2           & \text{or}&    & r  &= 2p-3.
\end{align}
As the domain of $p$ has been limited to that where $p < 2$, the relevant
solution is clearly $r = 2p -3$, so that in this inner limit the angular
momentum evolution is dominated by the magnetic braking. Furthermore, the
condition $r \ge -1$ may be used to find a lower limit on $p$ so that $1 \le p
< 2$. Having discarded the smaller term on the right hand side of Equation
\ref{heythere}, the constant terms in the angular momentum equation are then
rearranged to give the constant coefficient $j_1$ in terms of $\sigma_1$: 
\begin{equation}
    j_1 = \frac{\delta\sigma_1^2}{(3-2p)\mu_0^2m_c}.
\label{imhdamshort}
\end{equation}

The vertical momentum equation (\ref{imhdve}), upon substitution of the above
power laws, becomes
\begin{equation}
    \left(\sigma_1m_cx^{-3-p} + \frac{pm_c\sigma_1}{\mu_0^2}x^{-3-p}\right)h^2
      + \left(\frac{m_c^2}{\mu_0^2}x^{-4} 
        + \frac{\delta^2\sigma_1^2}{\mu_0^2}x^{-2p} + \sigma_1^2x^{-2p}\right)h
      -2\sigma_1x^{-p} = 0.
\label{imhdhmess1}
\end{equation}
Once more, using the restriction that $p < 2$ it is easy to show that $-4 <
-2p$, so that the second and third term in the second set of brackets are
small and may be dropped:
\begin{equation} 
    \left(\sigma_1m_c + \frac{pm_c\sigma_1}{\mu_0^2}\right)h^2x^{-3-p}
      + \left(\frac{m_c^2}{\mu_0^2}x^{-4}\right)h -2\sigma_1x^{-p} = 0;
\label{imhdhmess2}
\end{equation}
this equation is then rearranged into the neater form
\begin{equation} 
    \left(1 + \frac{p}{\mu_0^2}\right)h^2 + \frac{m_c}{\sigma_1\mu_0^2}x^{p-1}h 
    -\frac{2}{m_c}x^{3} = 0.
\label{imhdhmess3}
\end{equation}
The positive solution to this quadratic is 
\begin{equation}
    h = \frac{1}{2}\left(1 + \frac{p}{\mu_0^2}\right)^{-1}
    \left[-\frac{m_c}{\sigma_1\mu_0^2}x^{p-1} 
     + \sqrt{\frac{m_c^2}{\sigma_1^2\mu_0^4}x^{2p-2} 
       + \frac{8}{m_c}\left(1+\frac{p}{\mu_0^2}\right)x^3}\right];
\label{imhdhmess4}
\end{equation}
clearly as $2p - 2 < 2$, then the second term in the square root is small and
should be disregarded, however, if this is the case then the solution to
Equation \ref{imhdhmess4} is $h = 0$, which is unrealistic. Equation
\ref{imhdhmess4} can be rewritten as 
\begin{equation} 
    h = 4\sigma_1x^{-p}\left[\frac{m_c}{\sigma_1\mu_0^2}x^{p-1} 
     + \sqrt{\frac{m_c^2}{\sigma_1^2\mu_0^4}x^{2p-2} 
       + \frac{8}{m_c}\left(1+\frac{p}{\mu_0^2}\right)x^3}\right]^{-1},
\label{imhdhmess5}
\end{equation} 
which becomes 
\begin{equation}
    h = \frac{2\sigma_1\mu_0^2}{m_c^2}x^{4-p}
\label{imhdhsigma}
\end{equation}
when the second term in the square root is small. The vertical pressure in the
disc is therefore dominated by magnetic squeezing from the radial field term
in the IMHD limit, rather than the gravitational field of the central mass.
This result is to be expected from the flux freezing condition --- as the
density of the disc increases in the innermost region the flux is increased
proportionately so that it quickly comes to be the dominant force in
determining the scale height of the disc.  

All of the above power law scalings for the variables are then directly
substituted into the radial momentum equation (\ref{imhdrm}) so that it
becomes
\begin{align}
    \left(1 - \frac{m_c^2}{\sigma_1^2}x^{2p-2}\right)
      \frac{1}{\sigma_1x^{-p}}\frac{d}{dx}(\sigma_1x^{-p}) 
    = &- \frac{m_c}{x^2} + \frac{1}{\mu_0}\left(\frac{m_c}{\mu_0x^2}
     - \frac{2\sigma_1\mu_0}{m_c^2}x^{4-p}\frac{d}{dx}(\sigma_1x^{-p})\right)\nonumber\\
    &+ \left(\frac{\delta\sigma_1^2}{(3-2p)\mu_0^2m_c}\right)^2x^{3-4p} 
     + \frac{m_c^2}{\sigma_1^2}x^{2p-3},
\label{imhdrmmess1}
\end{align}
which is rearranged and simplified into
\begin{align}
    -\left(1 - \frac{m_c^2}{\sigma_1^2}x^{2p-2}\right)\frac{p}{x}
    &= -\frac{m_c}{x^2}\left(1 - \frac{1}{\mu_0^2}\right)
     + \frac{2p\sigma_1^2}{m_c^2}x^{3-2p}\nonumber\\
    &+ \left(\frac{\delta\sigma_1^2}{(3-2p)\mu_0^2m_c}\right)^2x^{3-4p} 
     + \frac{m_c^2}{\sigma_1^2}x^{2p-3}.
\label{imhdrmmess2}
\end{align}
As $p \ge 1$, then $2p -2 \ge 0$ and the second term in the first set of
brackets is smaller than its predecessor and can be disregarded. The first
term in those brackets scales as $\sim x^{-1}$ and as such becomes smaller
than the gravitational term as $x \to 0$, so that the entire left hand side of
Equation \ref{imhdrmmess2} is effectively zero. Because $3 - 2p > -1$ the
$h(db_z/dx)$ term may also be dropped, and as $r = 2p-3 \ge -1$ the final term
is also smaller than the gravitational force and becomes negligible as $x \to
0$. Thus the radial momentum equation may be simplified into the form
\begin{equation}
    m_c\left(1 - \frac{1}{\mu_0^2}\right)x^{-2} = 
    \left(\frac{\delta\sigma_1^2}{(3-2p)\mu_0^2m_c}\right)^2x^{3-4p}, 
\label{imhdrmmess3}
\end{equation}
which can be solved to give the exponents of the density and angular momentum:
\begin{align}
    p &= \frac{5}{4}\\
    \text{and }r &= -\frac{1}{2}.
\label{imhdscalings}
\end{align}
The coefficients in Equation \ref{imhdrmmess3} are then solved for the
constant $\sigma_1$:
\begin{equation} 
    \sigma_1 = \frac{\mu_0m_c^{3/4}}{\sqrt{2\delta}}
     \left(1-\frac{1}{\mu_0^2}\right)^{\!\frac{1}{4}},
\label{imhdsigma1}
\end{equation}
and this is substituted into Equations \ref{imhdamshort} and \ref{imhdhsigma}
to calculate the other coefficients:
\begin{align}
    j_1 &= \sqrt{m_c\left(1-\frac{1}{\mu_0^2}\right)}\\
    \text{and } h_1 &= \sqrt{\frac{2}{\delta}}\,\frac{\mu_0^3}{m_c^{5/4}}
       \left(1-\frac{1}{\mu_0^2}\right)^{\!\frac{1}{4}}.
\end{align}

The full inner asymptotic solution is then given by the set of power law
relations:
\begin{align}
    \sigma = \sigma_1x^{-5/4} &= \frac{\mu_0m_c^{3/4}}{\sqrt{2\delta}}
      (1- \mu_0^{-2})^{1/4}\:x^{-5/4}, \label{imhdin-sig}\\
    m &= m_c, \label{imhdin-m}\\
    v &= \sqrt{m_c(1 - \mu_0^{-2})}\: x^{-1/2}, \label{imhdin-v}\\
    j &= \sqrt{m_c(1 - \mu_0^{-2})}\: x^{1/2}, \label{imhdin-j}\\
    u &= -\frac{m_c}{\sigma_1}\:x^{1/4}, \label{imhdin-u}\\
    h &= \frac{2\mu_0^2\sigma_1}{m_c^2}\:x^{11/4}, \label{imhdin-h}\\
    \psi &= \frac{m_c}{\mu_0}, \label{imhdin-psi}\\
    b_z &= \frac{\sigma_1}{\mu_0}\:x^{-5/4}, \label{imhdin-bz}\\
    b_{r,s} &= \frac{m_c}{\mu_0}\:x^{-2}, \label{imhdin-brs}\\
    b_{\phi,s} &= -\frac{\delta\sigma_1}{\mu_0}\:x^{-5/4} \label{imhdin-bphis}\\
    \text{and }\dot{m} &= m_c; \label{imhdin-mdot}
\end{align}
these are the same as those presented without explicit derivation in \S3.2 of
\citet{kk2002}. In dimensional form the similarity solution becomes: 
\begin{align}
    \Sigma &= \frac{c_s^{9/4}\sigma_1}{2\pi G}\,\frac{t^{1/4}}{r^{5/4}}, 
     \label{imhdin-Sig}\\
    M &= \frac{c_s^3m_c}{G}\,t, \label{imhdin-M}\\
    V_\phi &= \sqrt{c_s^3m_c(1-\mu_0^{-2})\frac{t}{r}} 
     \,= \sqrt{\frac{GM}{r}(1-\mu_0^{-2})} \label{imhdin-Vphi}\\
    J &= \sqrt{c_s^3m_c(1-\mu_0^{-2})rt} 
      \,= \sqrt{GMr(1-\mu_0^{-2})}, \label{imhdin-J}\\
    V_r &= -\,\frac{c_s^{3/4}m_c}{\sigma_1} \left(\frac{r}{t}\right)^{1/4},
     \label{imhdin-Vr}\\
    H &= \frac{2\mu_0^2\sigma_1}{m_c^2}\,\frac{r^{11/4}}{(c_st)^{7/4}},
     \label{imhdin-H}\\
    \Psi &= \frac{2\pi c_s^3m_ct}{\mu_0G^{1/2}} = \frac{2\pi\sqrt{G}}{\mu_0}\,M,
     \label{imhdin-Psi}\\
    B_z &= \frac{\sigma_1c_s^{9/4}t^{1/4}}{\mu_0G^{1/2}r^{5/4}}
     = \frac{2\pi\sqrt{G}}{\mu_0}\,\Sigma,\label{imhdin-Bz}\\
    B_{r,s} &= \frac{\Psi}{2\pi r^2} = \frac{\sqrt{G}}{\mu_0r^2}\,M,
     \label{imhdin-Brs}\\
    B_{\phi,s} &= -\delta B_z = -\,\frac{2\pi\sqrt{G}\delta}{\mu_0}\,\Sigma
     \label{imhdin-Bphis}\\
    \text{and }\dot{M} &= \frac{c_s^3m_c}{G} = -2\pi rV_r\Sigma.
     \label{imhdin-Mdot}
\end{align}
The disc is in near-Keplerian rotation, with the deviation from Keplerian
determined by the magnetic ``dilution'' factor $(1-\mu_0^{-2})^{1/2}$. The
larger the mass-to-flux ratio $\mu_0$ (i.e.\ the less flux there is in the
initial molecular cloud) the closer the rotational speed is to that of the
nonmagnetic Keplerian disc solution in the preceding section.  It is the
magnetic braking that causes this effect, as it transports the angular
momentum of the disc material to the envelope, allowing the fluid to spiral 
inwards towards the central mass. For the similarity solutions presented in
subsection \ref{imhd:sols} this magnetic dilution factor is close to one
(0.939), so the discs are in near-Keplerian rotation. 

Equation \ref{imhdin-sig} shows that the magnetic dilution factor also reduces
the surface density, however the presence of the mass-to-flux ratio in the
definition of the coefficient $\sigma_1$ typically has more of an effect on
the density of the inner disc, enhancing it while also reducing the equivalent
constant in the radial velocity power law relation (Equation \ref{imhdin-u}).
The magnetic forces determine the radial dependence of these terms and the
scale height, which becomes very small in the disc as expected. The low infall
speed ensures that the disc remains in a near-exact dynamical equilibrium. 

The magnetic field takes on a split monopole form, with the field lines
strongly inclined due to the domination of the radial component over the
vertical and azimuthal components (see Equations \ref{imhdin-bz} and
\ref{imhdin-brs}). The strong magnetic field changes the dynamics of the disc
from being strictly Keplerian, and its increasing strength in the innermost
regions demonstrates the magnetic flux problem that occurs in simulations of
star formation \citep[][also Section \ref{lr:mfp}]{l1998}. As the angle
between the field and the disc surface is $<60^\circ$ ($B_{r,s}/B_z >
1/\sqrt{3}$) the disc is able to drive a centrifugal wind from its surface,
which would reduce the amount of matter that reaches the origin and carry away
excess flux \citep{bp1982}. Although this is not explored further in this
work, it has been shown that the presence of a disc wind could assist in
solving the magnetic flux problem \citep[appendix C]{kk2002}. 

As in the nonmagnetic case, the transition between the collapsing flow and the
near-Keplerian inner disc is marked by a sharp transition in the radial
velocity and surface density. The constraint of flux freezing means that the
magnetic field must also change in the shock transition to ensure continuity.
These new jump conditions are derived in the following subsection. 
\raggedbottom

\subsection{Shock position and jump conditions}\label{imhd:shock}

Similarly to the nonmagnetic case, the position of the centrifugal shock is
found iteratively, using the behaviour of the downstream variables to refine
the shock position, $x_c$, until a convergence is reached. Because the ratio
$j/m$ does not change greatly from its initial value before it encounters the
centrifugal shock, the initial guess used to find the position of $x_c$ is
that defined by Equation \ref{nmxc} from the nonmagnetic case. Table
\ref{tab-imhdxc}, which lists both the estimated and true shock positions for
the similarity solutions calculated in the following subsection, shows that
this is still an acceptable approximation to the shock position; the
difference between the estimated and actual shock positions is typically
$\lesssim 10\%$.

\flushbottom
There are two different sets of jump conditions used in the ideal
magnetohydrodynamics models, and the choice of which of these to use is
determined by the dominant vertical forces in the disc in the region of the
shock. For both solutions the continuity equation gives the first jump
condition to be
\begin{equation}
    w\sigma = \text{constant}
\label{imhdwsig}
\end{equation}
as in the nonmagnetic case. Flux freezing, which takes the form of $b_z =
\sigma/\mu_0$ from the induction equation, allows for the vertical field jump
condition to be given by the similar equation
\begin{equation}
    wb_z = \text{constant}.
\label{imhdwbz}
\end{equation}
As before, the radial momentum equation takes the form
\begin{equation}
    (1-w^2)\frac{d\sigma}{dx} = -b_zh\frac{db_z}{dx} 
\label{imhdxcrm11}
\end{equation}
where the other terms in Equation \ref{imhdrm} are small at the shock position
compared to the derivatives of the rapidly changing surface density and
magnetic field. The flux freezing condition is substituted into the right hand
side of this equation so that it becomes  
\begin{equation}
    (1-w^2)\frac{d\sigma}{dx} = -\frac{h\sigma}{\mu_0^2}\frac{d\sigma}{dx};
\label{imhdxcrm12}
\end{equation}
and a good understanding of how the disc scale height behaves in the shock
region is required in order to calculate the integral of the right hand side
of this equation. 

If magnetic squeezing due to the radial field component dominates the scale
height (as in the asymptotic inner solution) then $h \approx
2\sigma/b_{r,s}^2$ and the right hand side of Equation \ref{imhdxcrm12} becomes
\begin{equation}
    -\frac{h\sigma}{\mu_0^2}\frac{d\sigma}{dx} 
    = -\frac{2\sigma^2}{\mu_0^2b_{r,s}^2}\frac{d\sigma}{dx}
    = -\frac{2}{3\mu_0b_{r,s}^2}\frac{d(\sigma^3)}{dx}.
\label{imhdxcrm13}
\end{equation}
As $b_{r,s} = m/\mu_0x^2$ is constant across the shock due to the conservation
of mass, the integral of Equation \ref{imhdxcrm12} is then the final jump
condition 
\begin{equation}
    \sigma(1 + w^2) = -\frac{2\sigma^3}{3\mu_0b_{r,s}^2} + \text{constant}
\label{imhdrmjump1}
\end{equation}
using the result from Section \ref{nm:shock} for the left hand side. Denoting
the upstream and downstream variables by the subscripts $u$ and $d$ as in the
nonmagnetic case, it can be shown that these jump conditions have one real
solution that is given by 
\begin{equation}
    \sigma_d = -\frac{\sigma_u}{3} + q_+ + q_-,
\label{imhdjump1}
\end{equation}
where
\begin{align}
    q_\pm &= \left(-q/2 \pm D^{1/2}\right)^{1/3},\label{imhdjump1qpm}\\
    q &= -\frac{3}{2}\mu_0^2b_{r,s}^2\sigma_uw_u^2
     - \frac{\sigma_u}{3}\left(\sigma_u^2 
     + \frac{3}{2}\mu_0^2b_{r,s}^2\right),\label{imhdjump1q}\\
    D &= (p/3)^3 + (q/2)^2\label{imhdjump1D}\\
    \text{and }p &= \frac{2}{3}\sigma_u^2 + \frac{3}{2}\mu_0^2b_{r,s}^2.
\label{imhdjump1p}
\end{align}
The greater details of this derivation are provided in appendix B3 of
\citet{kk2002}, as this set of jump conditions is the same as those for their
``magnetically squeezed shock''. These jump conditions are used in the slowly
rotating collapse solution shown in Figure \ref{fig-imhdslow} where the radial
field component is already large in the region of the shock. 
\begin{table}[t]
\begin{center}
\begin{tabular}{ccc}
\toprule
 model & estimated $x_c$ & actual $x_c$\\
 \midrule
 $v_0 = 0.1$, $\alpha = 0.1$  & $0.00\dot{6}$ & $0.0049 $\\
 $v_0 = 1.5$, $\alpha = 0.1$  & $1.5$         & $1.48   $\\
 $v_0 = 1.5$, $\alpha = 0.01$ & $1.5$         & $1.59   $\\
 \bottomrule
 \end{tabular}
\end{center}
\vspace{-5mm}
 \caption[Estimated vs actual values of $x_c$ in the IMHD model]
{Comparison between the estimated and actual values of the centrifugal shock
position, $x_c$, for the models calculated in the following subsection. The
estimated values are typically larger than the true shock position, save for
when the magnetic braking parameter $\alpha$ is particularly weak, however
they are close enough that they provide a reasonable first approximation for
the iterative routine.} 
\label{tab-imhdxc}
\end{table}

The second set of possible jump conditions are those that apply when the shock
occurs in the region where the disc self-gravity still dominates the vertical
forces in the disc, so that $h \approx 2/\sigma$. In this case the right hand
side of the radial momentum equation at the shock (\ref{imhdxcrm12}) is then 
\begin{equation}
    -b_zh\frac{db_z}{dx} = -\frac{2}{\mu_0^2}\frac{d\sigma}{dx};
\label{imhdxcrm23}
\end{equation}
this is integrated at the shock position to give the jump condition
\begin{equation}
    \sigma(1+w^2) = -\frac{2\sigma}{\mu_0^2} + \text{constant}.
\label{imhdrmjump2}
\end{equation}
Combining this with the other jump conditions in Equations \ref{imhdwsig} and
\ref{imhdwbz} gives the nontrivial solution to the jump conditions:
\begin{align}
    w_d &= \left(1 + \frac{2}{\mu_0^2}\right)\frac{1}{w_u},\label{imhdjump2w}\\
    \sigma_d &= \sigma_uw_u^2
      \left(1 + \frac{2}{\mu_0^2}\right)^{-1}\label{imhdjump2sig}\\
    \text{and }b_{zd} &= b_{zu}w_u^2
      \left(1 + \frac{2}{\mu_0^2}\right)^{-1}\label{imhdjump2bz},
\end{align}
which is the solution for the nonmagnetic case with the additional magnetic
factor $(1 + 2/\mu_0^2)$. These jump conditions are used in the rapidly
rotating solutions shown in Figures \ref{fig-imhdfast} and
\ref{fig-imhdfastalpha}, in which the shock occurs much earlier in the
collapse process before the radial field has built up enough to dominate the
vertical squeezing. 

\subsection{Similarity solutions}\label{imhd:sols}

The IMHD similarity solutions are calculated in a similar manner to those for
the nonmagnetic model, by integrating the equations (in this case Equations
\ref{imhdcm}--\ref{imhdam}) from the outer boundary to the centrifugal shock
position, performing the jump calculations and then continuing the integration
to the inner boundary. The exact location of the shock is found iteratively by
performing the integration using an estimated shock position (starting at the
value of $x_c$ derived in subsection \ref{nm:shock}) and using the post-shock
behaviour to refine the estimate of the shock position until the integrated
variables at the inner boundary match onto the asymptotic inner solution as
expected. The full details of this routine are described in Section
\ref{numerics} for the Hall similarity solutions.  

The addition of magnetic braking to the calculations causes the formation of a
central point mass, parameterised by the nondimensional mass and accretion
rate $m_c$. The value of $m_c$ is initially unknown, although it is first
approximated by the value of the plateau mass $m_{pl}$ given by Equation
\ref{nmmpl}. The true value of the central mass is determined by iteration and
is typically easier to find than the shock position, as the gravity of the
central mass becomes important to the collapse dynamics only in the innermost
regions of the collapsing pseudodisc. While small changes to the assumed value
of $m_c$ may greatly change the integrated values of $m$ and the surface
density at the inner boundary, the outer and mid-regions are only
superficially altered. 

The full calculation, including the convergence on the true values of $x_c$
and $m_c$, typically takes under a minute on a current generation desktop
computer, provided that the influence of an incorrect shock position on the
post-shock variables is well-understood. A discussion of the types of
post-shock behaviour observed while trying to calculate the shock position is
provided in Chapter \ref{ch:hall} for the Hall MHD solutions. 

The three similarity solutions presented here show a slow collapse with
initial conditions matching those in the nonmagnetic slow collapse and a
faster collapse with two different values of the magnetic braking parameter
$\alpha$ (the ratio of the sound speed to the Alfv\'en speed in the external
envelope) defined in Equation \ref{alpha}. As before, the parameters were
chosen to match those used in the solutions in \S3.2 of \citet{kk2002} as a
mechanism for testing the model calculations. The mass-to-flux ratio is held
constant at $\mu_0 = 2.9$; and the other initial parameters match those in the
nonmagnetic solutions, with mass parameter $A = 3$ and infall velocity $u_0 =
-1$. The azimuthal field cap is fixed at $\delta = 1$, as the azimuthal field
component is unlikely to exceed the vertical one dramatically in a real
collapsing flow. These values were chosen to match onto both the numerical
calculations of \citet{ck1998} and observations such as those by
\citet{c1999}; all are within the range of physical parameters believed to be
encountered in collapsing molecular cloud cores. 

All three solutions look similar to the nonmagnetic solutions from Section
\ref{nonmag} in the outer regions of the collapse, where the ratio of the
enclosed mass to the specific angular momentum is constant and the mass scales
with $x$. However, the nonmagnetic and IMHD similarity solutions diverge as
the magnetic field  builds up and the magnetic braking begins transporting the
angular momentum to the envelope, increasing this ratio and breaking the
resemblance. The behaviour of the solutions interior to this outer collapse is
determined by the initial rotation speed $v_0$ and the value of the magnetic
braking parameter $\alpha$. 

\begin{figure}[htp]
  \centering
\vspace{-5mm}
  \includegraphics[width=5.4in]{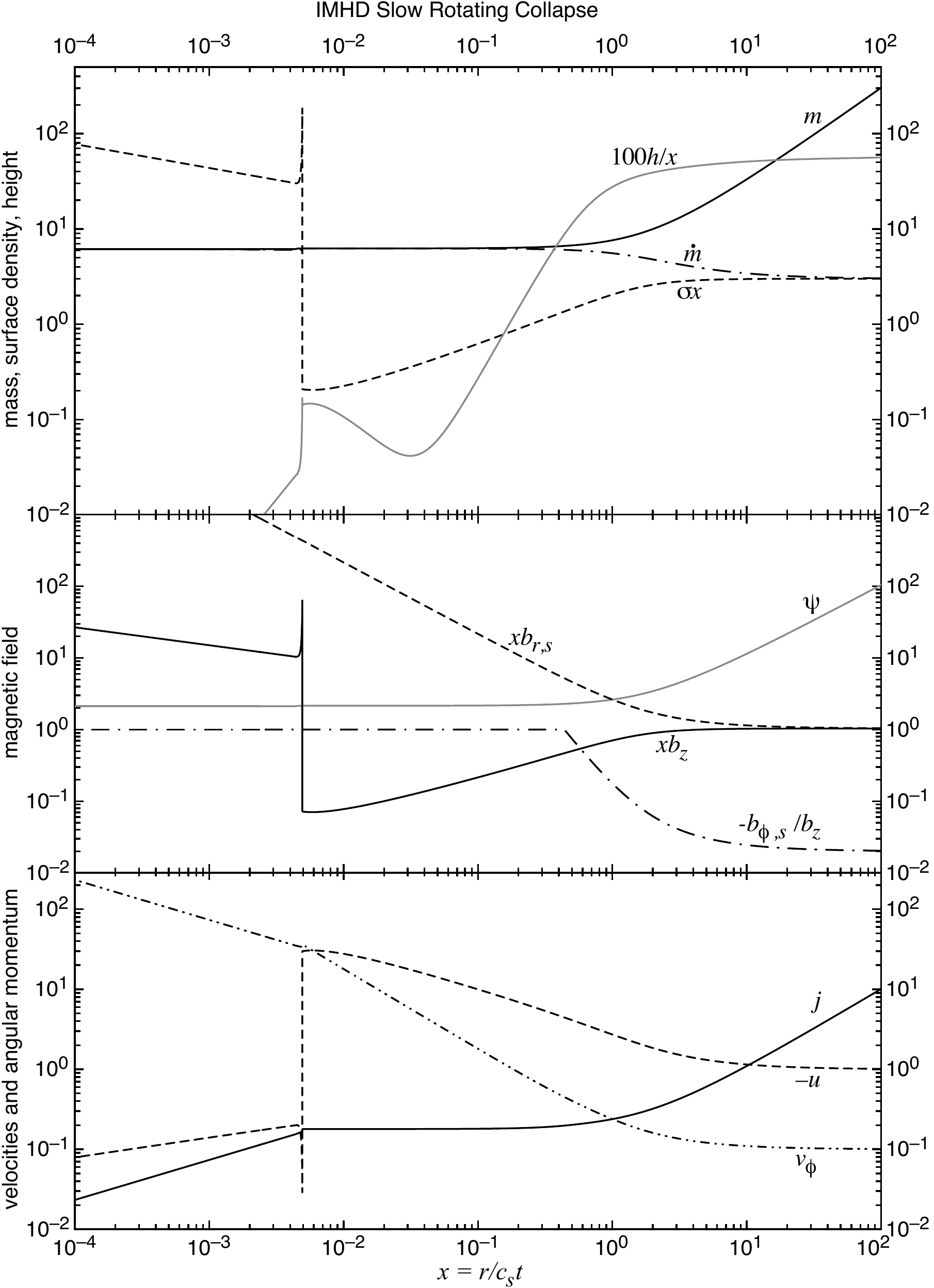}
\vspace{-5mm}
  \caption[IMHD slow rotational collapse]{Similarity solution for the slowly
rotating IMHD collapse, with outer asymptotic boundary conditions $A = 3$,
$u_0 = -1$, $\mu_0 = 2.9$ and $v_0 = 0.1$, matching those in Figure
\ref{fig-nmslow} and figure 3 of \citet{kk2002}. The top and lower panels
display the same variables as in Figure \ref{fig-nmslow}, while the central
panel displays the nondimensional magnetic field components $b_{r,s}$,
$b_{\phi,s}$ and $b_z$, as well as the magnetic flux $\psi$, as functions of
the similarity variable $x$. A magnetically-diluted Keplerian disc forms
inside a centrifugal shock (located at $x_c = 4.93\times10^{-3}$). } 
\label{fig-imhdslow}
\end{figure}

The slow rotation solution presented in Figure \ref{fig-imhdslow} is
characterised by the parameters $V_0 = 0.1c_s$ and $\alpha = 0.1$,
corresponding to moderate rotation and magnetic braking rates. As in the
slowly rotating nonmagnetic solution in Figure \ref{fig-nmslow}, the mass and
angular momentum plateau as the radial velocity of the collapsing matter
increases. The addition of magnetic braking to the calculations causes a
slight reduction of the angular momentum in the plateau region from $j_{pl}
\sim 2.5$ to $\sim 1.8$, which in turn reduces the centrifugal force so that
the shock position drops from $x_c = 7.7 \times 10^{-3}$ in the nonmagnetic
similarity solution to $x_c = 4.9 \times 10^{-3}$ in the IMHD solution.  

The magnetic field builds up in the plateau region until the magnetic
squeezing comes to dominate the vertical compression of the disc, so that it
is much thinner than its nonmagnetic counterpart in Figure \ref{fig-nmslow}
(note that in Figure \ref{fig-imhdslow} and the other IMHD similarity
solutions the scale height is plotted as $100h/x$, while for the nonmagnetic
similarity solutions in Figures \ref{fig-nmslow} and \ref{fig-nmfast} $h$ is
plotted directly). The scale height is therefore approximated by $h \approx
2\sigma /b_{r,s}$ and the jump conditions used at the centrifugal shock are
the first set derived in subsection \ref{imhd:shock} (Equations
\ref{imhdwsig}, \ref{imhdwbz} and \ref{imhdjump1}). 

As in the nonmagnetic case, the matter is falling in at near-free fall speeds
as the centrifugal force builds up until it is able to counter the influence
of the central mass gravity and the centrifugal shock is formed. The fluid
is rapidly decelerated by the encounter with the ring of increased density
that is the shock front. The spike that occurs in the surface density
corresponds to an equivalent increase in the magnetic field strength; this in
turn produces a rapid drop in the angular momentum caused by increased
magnetic braking. 

The post-shock region of the slowly rotating similarity solution is very
narrow, with no backflow region as may occur in rapidly rotating solutions.
After a thin transition region the flow merges into the asymptotic
magnetically-diluted Keplerian disc solution outlined in subsection
\ref{imhd:inner}. Accretion through this disc is slow and driven by the
magnetic braking, which gives $j$ its characteristic near-Keplerian profile
(which scales with $x^{1/2}$), so that the disc is in almost perfect dynamical
equilibrium. The disc mass is $\sim 3\%$ that of the central mass, and the
accretion rate from the  disc onto the point mass is given by $\dot{m} = m_c =
6.0$; this corresponds to a dimensional value of $\dot{M}_c \approx 10^{-5}$
M$_\odot$ yr$^{-1}$, which is at the high end of the range of expected
accretion rates for protostellar cores \citep{lmt2001}.

\begin{figure}[htp]
  \centering
\vspace{-5mm}
  \includegraphics[width=5.3in]{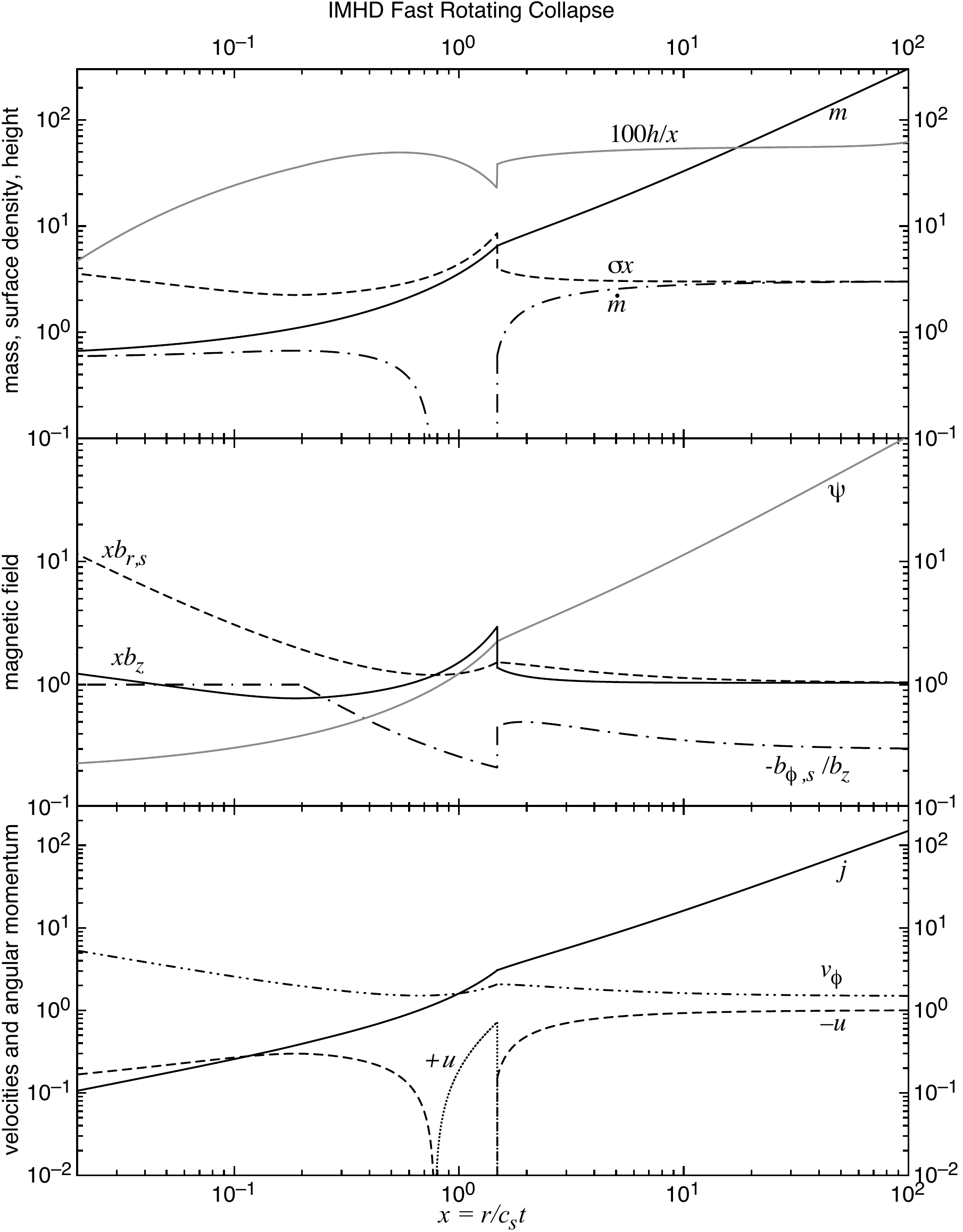}
\vspace{-5mm}
  \caption[IMHD fast rotational collapse]{Similarity solution for the rapidly
rotating IMHD collapse, with outer asymptotic boundary conditions matching
those in Figure \ref{fig-imhdslow}, save for the initial rotational velocity,
$v_0 = 1.5$, which is higher than the equivalent nonmagnetic case in Figure
\ref{fig-nmfast} but matches figure 4 of \citet{kk2002}. Note that the
horizontal scale is different to that in Figure \ref{fig-nmfast}, as the
variables attain their asymptotic forms much earlier in this solution. In this
case the centrifugal shock occurs sooner at $x_c = 1.48$, and as in the
nonmagnetic case the infall velocity changes sign at the shock, creating a
region of backflow. The variables gradually settle to their asymptotic values
once the gas starts collapsing again and the azimuthal field reaches its
capped value; however, the forming central mass is very small ($m_c = 0.57$)
compared to that in the slow rotation solution.} 
\label{fig-imhdfast}
\end{figure}

Figure \ref{fig-imhdfast} shows the similarity solution for a rapidly rotating
IMHD collapse, with the same initial conditions and parameters as those in the
slowly rotating solution in Figure \ref{fig-imhdslow} save for the initial
rotational velocity which has been increased to $V_0 = 1.5 c_s$. This is
higher than the value in the corresponding nonmagnetic similarity solution
presented in Figure \ref{fig-nmfast} ($v_0 = 1.0$), however, the two solutions
are qualitatively similar. The increase in the initial angular momentum
implies that the centrifugal force comes to balance gravity earlier in the
collapse, resulting in a much higher value of the shock position $x_c = 1.48$.
In this region of the flow the enclosed mass is still much higher than $m_c$,
the infall rate is slow and the disc scale height is still dominated by the
disc self-gravity; the jump conditions applied at this shock are the
generalised isothermal shock conditions given by Equations
\ref{imhdjump2w}--\ref{imhdjump2bz}. 

Similarly to the fast nonmagnetic similarity solution, the centrifugal shock
in Figure \ref{fig-imhdfast} is so strong that it creates a region of shocked
backflow in the post-shock annulus. This region has a finite width in $x$,
during which the outflowing gas is first slowed and then begins to inflow once
more as the surface density decreases from its shocked value. The disc of
accreting material is still larger than in the slowly rotating solution, and
contains approximately four times as much mass as the central protostar. It is
only in the inner regions of this disc that the variables attain their
asymptotic magnetically-diluted Keplerian disc values, after an extended
region in which the disc is self-gravitating.

The angular momentum remains high throughout the extended disc that forms
behind the shock as the low field density implies that magnetic braking has
little effect and the outer disc regions are strongly non-Keplerian. Because
the braking is so inefficient, very little mass accumulates at the origin
compared with the slow rotation case; the central mass is $m_c = 0.57$, which
corresponds to a reduced accretion rate of $\dot{M}_c \approx 9 \times
10^{-7}$ M$_\odot$ yr$^{-1}$. It is clear that increasing the angular momentum
in the initial cloud creates an angular momentum problem similar to that
encountered in the nonmagnetic case, in which the inefficiency of magnetic
braking transportation of the angular momentum inhibits the star formation
process, creating a smaller central mass surrounded by an extended disc that
very slowly falls inward. 

\begin{figure}[htp]
  \centering
\vspace{-5mm}
  \includegraphics[width=5.3in]{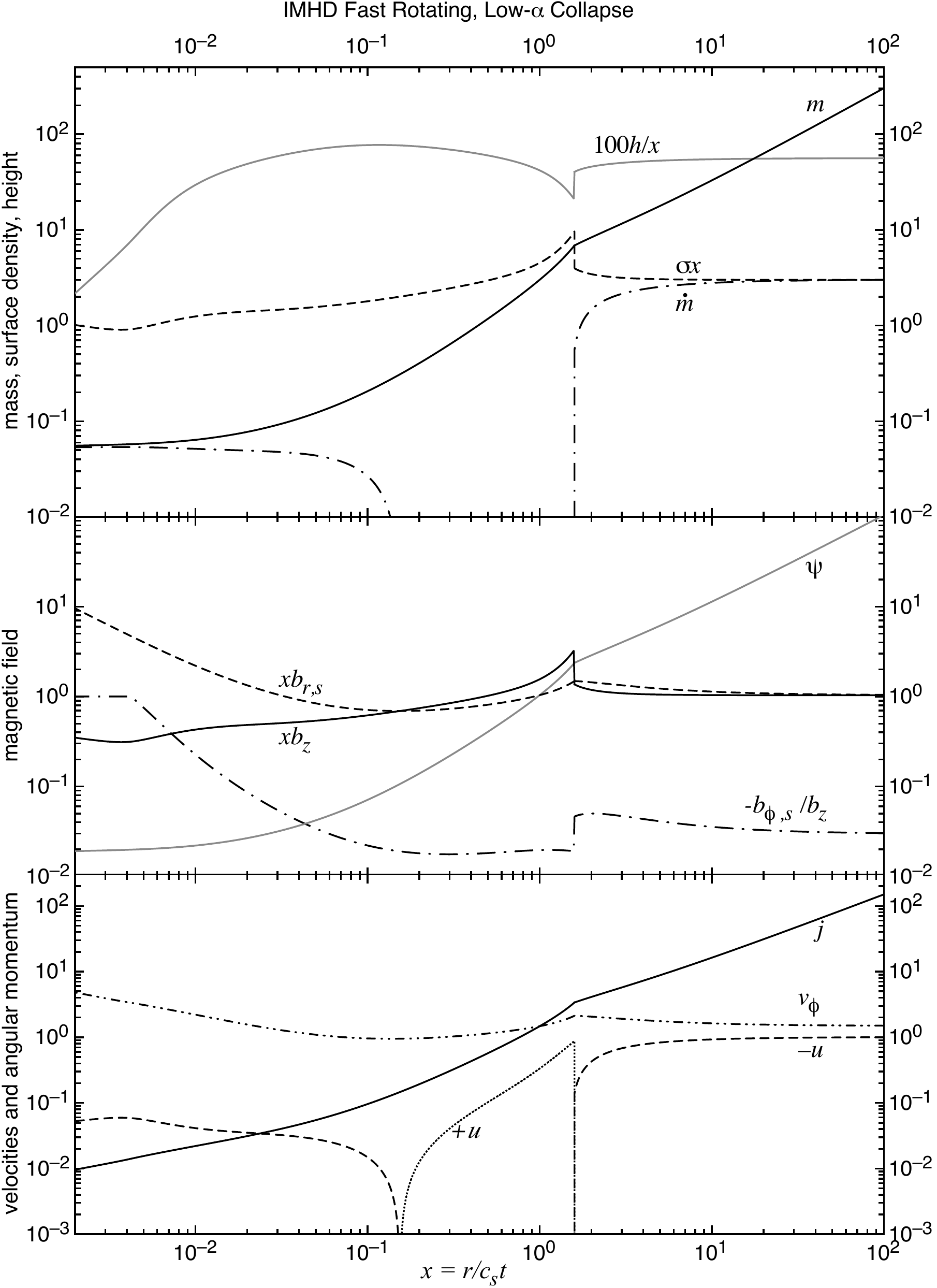}
\vspace{-5mm}
  \caption[IMHD fast rotational collapse with reduced $\alpha$ ]
{Similarity solution for the rapidly rotating IMHD collapse with reduced
magnetic braking parameter $\alpha = 0.01$. The boundary conditions otherwise
match those in Figure \ref{fig-imhdfast} and figure 5 of \citet{kk2002}; the
horizontal scale is again changed to show the inner regions of the collapse.
The reduced magnetic braking parameter causes there to be a wider backflow
region inwards of the centrifugal shock at $x_c = 1.59$, which reduces the
size of the extended accretion disc. The near-Keplerian disc region is much
smaller, beginning around $x \approx 0.004$, and the central mass is reduced
to $m_c = 0.054$.}
\label{fig-imhdfastalpha}
\end{figure}

The final plot, Figure \ref{fig-imhdfastalpha}, shows a second rapidly
rotating similarity solution that has the same initial values as Figure
\ref{fig-imhdfast}, save that the magnetic braking parameter $\alpha$ is
reduced from 0.1 to 0.01. This solution has even less efficient braking than
Figure \ref{fig-imhdfast}, which causes the centrifugal shock to occur even
sooner at $x_c = 1.59$ and extends the region of backflow so that it occurs
over an order of magnitude in self-similar space. The backflow fuels the
shock, but eventually the matter slows enough that it is able to start
infalling once more. 

Even once the fluid is inflowing again, it takes much longer to join the
asymptotic inner solution (see the turning point in $\sigma x$ at around $x =
0.004$) as the low rate of infall prevents the magnetic field from building up
and compressing the disc. It is only once the magnetic field is strong enough
that the azimuthal field parameter attains its capped value that the collapse
starts to behave in a manner similar to the asymptotic solution. As the region
of inflow is reduced so too is the central mass, with $m_c = 0.05$
corresponding to a very low accretion rate of $\dot{M}_c \approx 8 \times
10^{-8}$ M$_\odot$ yr$^{-1}$. Again, the mass of the rotationally-supported
and self-gravitating disc is about four times that at the origin. 

In all of the similarity solutions the magnetic braking acts to transfer the
angular momentum to the external medium, and the reduction of $\alpha$ or an
increase in the initial angular momentum of the cloud leads to a reduction in
the amount of matter that can fall onto the central point mass, resulting in
an enhanced angular momentum problem. As the angular momentum inside of the
centrifugal shock is increased, the inner similarity solutions tend towards a
near-nonmagnetic solution similar to that in Figure \ref{fig-nmfast} in which
no central mass is able to form. 

The azimuthal field component reaches its capped value in all of the solutions
presented here, which limits the amount of magnetic braking that is possible
in the inner regions of the collapse. If this cap were lifted then further
braking could reduce the disc mass and size while increasing accretion onto
the central point mass that forms, perhaps even to the point where no disc is
able to form around the star, as in the numerical solutions of \citet{ml2008}.
It is clear that limiting the magnetic braking is a solution to the problem of
disc formation, yet this needs to be better studied, either by increasing the
resolution of the numerical models so that those mechanisms which could
decrease the magnetic braking (such as nonaxisymmetric turbulence in the thin
inner disc) can be explored, or by adopting a better prescription for the
azimuthal field component in the similarity solutions.

The other major problem in star formation is introduced here in the IMHD
models: the magnetic flux problem \citep[][also Section \ref{lr:mfp}]{s1978},
in which too much magnetic flux is dragged into the central protostar and
disc. This is obvious from the way that $b_{r,s}$, which scales with $x^{-2}$
(and the other magnetic field components to a lesser extent), increase
dramatically in the innermost regions of disc. This problem is exemplified by
the restrictions of IMHD, as the matter is tied to the field lines and any gas
accreted through the disc will bring flux with it, leading to a high
concentration of the magnetic field in the inner disc regions of high density
and the central point mass.

The angular momentum and magnetic flux problems are mitigated by breaking the
assumption of flux freezing in the collapsing gas, which changes the dynamics
of both the angular momentum and magnetic field transport in the pseudodisc.
The first step in achieving this is to introduce ambipolar diffusion, which
causes the decoupling of the field from the neutral particles at the
densities encountered around the beginning of the mass plateau and enhances
the magnetic braking in the forming disc. The influence of ambipolar diffusion
on the collapse process is explored in detail in the following section. 

\section{Ambipolar Diffusion Solutions}\label{ad}

The last of the preliminary models used to test the code is the non-ideal MHD
calculation with ambipolar diffusion. In this model, the field is no longer
strictly tied to the neutral material, for as the density increases the field
is decoupled from the neutral particles and is instead tied to the ions via
the nondimensional ambipolar diffusion parameter $\tilde{\eta}_A$. The
magnetic field is then able to be advected against the flow, reducing the
magnitude of the magnetic flux problem that occurs in the central region of
the IMHD solutions. The ambipolar diffusion term in the induction equation
becomes dominant near the origin; this changes the structure of the inner
Keplerian disc, increasing the density and reducing the angular momentum and
flux carried inwards by the gas. 

The ambipolar diffusion model is described by Equations 
\ref{cmfinal}--\ref{infinal} in the limit that $\tilde{\eta}_H = 0$. For
brevity these shall not be duplicated here, however the corresponding set of 
nondimensional equations (originally stated in Equations
\ref{sspsi}--\ref{ssb_phis}) are reproduced in the interests of clarity, as
the important terms in these equations shall be discussed in the following
subsections. These are 
\begin{align}
   \frac{dm}{dx} &= x\sigma, \label{adcm}\\
   (1-w^2)\frac{1}{\sigma}\frac{d\sigma}{dx} = -\frac{m}{x^2} 
     &+ \frac{b_z}{\sigma}\left(b_{r,s} - h\frac{db_z}{dz}\right)
     + \frac{j^2}{x^3} + \frac{w^2}{x}, \label{adrm}\\
   \frac{dj}{dx} &= \frac{1}{w}\left(j 
     - \frac{xb_zb_{\phi,s}}{\sigma}\right),\label{adam}\\
   \left(\frac{\sigma{m_c}}{x^3} - b_{r,s}\frac{db_z}{dx}\right)h^2
     &+ \left(b_{r,s}^2 + b_{\phi,s}^2 + \sigma^{2}\right)h - 2\sigma 
     = 0,\label{advhe}\\
   \frac{d\psi}{dx} &= xb_z, \label{adflux}\\
   \text{and }\psi - xwb_z 
     + \tilde{\eta}_Axb_z^2&h^{1/2}\sigma^{-3/2}\left(b_{r,s} 
                   - h\frac{db_z}{dx}\right) = 0;\label{adin}
\end{align}
while the accretion rate and other field components are given by
\begin{align}
   \dot{m} &= -xu\sigma, \label{admdot}\\
   b_{r,s} &= \frac{\psi}{x^2}, \label{adb_rs}\\
   \text{and }b_{\phi,s} &= -\min\left[\frac{2\alpha\psi j}{x^3} 
    \left(1 + \frac{2\alpha\tilde{\eta}_Ah^{1/2}\psi{b_z}}{x^2\sigma^{3/2}}
    \right)^{-1}; \delta{b_z}\right]. 
\label{adb_phis}
\end{align}

The ambipolar diffusion calculations were the most advanced of those performed
by \citet{kk2002} and their solutions shall be examined in detail in this
section. In addition to the centrifugal shock that separates the region of
dynamic inflow and the slow-infall Keplerian disc, ambipolar diffusion drives
a continuous shock outwards as it decouples the field from the neutral fluid.
The position and behaviour of these shocks, and the dynamics of the asymptotic 
inner disc solution must both be discussed before the similarity solutions to
the full set of MHD equations with ambipolar diffusion can be presented and
analysed.
\vspace{2mm}

\subsection{Inner solution}\label{ad:inner}

As was mentioned in Section \ref{kepdisc}, the inner asymptotic solution for
the full model with both ambipolar and Hall diffusion reduces to the solution
for a model with just ambipolar diffusion in the limit that the Hall diffusion
parameter $\tilde{\eta}_H = 0$. This solution was presented without explicit
derivation in \S3.3 of \citet{kk2002}, and it is reproduced here for the
purpose of discussion. 

The derivation of the nondimensional inner similarity solution directly
follows that in Section \ref{indev} so that the asymptotic power law relations
take the form: 
\begin{align}
    m &= m_c, \label{adin-m}\\  
    \dot{m} &= m_c, \label{adin-mdot}\\
    \sigma = \sigma_1x^{-3/2} &= 
      \frac{2\tilde{\eta}_A\sqrt{2m_c}}
      {3\delta\sqrt{(3\delta/2\tilde{\eta}_A)^2 + 1}} \,x^{-3/2},\label{adin-sig}\\
    h = h_1x^{3/2} &= \sqrt{\frac{2}{m_c[1+(2\tilde{\eta}_A/3\delta)^2]}}
      \,x^{3/2}, \label{adin-h}\\
    u &= -\frac{m_c}{\sigma_1}\,x^{1/2}, \label{adin-u}\\
    v &= \sqrt{\frac{m_c}{x}}, \label{adin-v}\\
    j &= \sqrt{m_cx}, \label{adin-j}\\
    \psi &= \frac{4}{3}b_zx^2, \label{adin-psi}\\
    b_z &= \frac{m_c^{3/4}}{\sqrt{2\delta}}\,x^{-5/4}, \label{adin-bz}\\
    b_{r,s} &= \frac{4}{3}\,b_z, \label{adin-brs}\\
    \text{and } b_{\phi,s} &= -\delta\,{b_z}; \label{adin-bphis}
\end{align}
$m_c$ is the constant nondimensional mass infall rate. The dimensional form of
these variables are given by substituting the above definitions of the
constants $\sigma_1$ and $h_1$ into Equations \ref{kepfakeM}--\ref{kepfake}; 
the other coefficients and the dimensional scaling of the variables are
unchanged from the inner asymptotic solution with both ambipolar and Hall
diffusion. 

The inner solution represents a disc in Keplerian rotation; it is supported
against gravity by the angular momentum and has a low accretion rate onto the
central protostar (which has mass $M_c = m_cc_s^3t/G$, where $c_s$ is the
thermal sound speed and $G$ the gravitational constant). As in the full
Keplerian disc solution discussed in Section \ref{kepdisc}, the scale height
and the surface density of the centrifugally-supported disc depend upon the
ambipolar diffusion coefficient, while the magnetic field strength in the disc
depends only on the nondimensional mass infall rate and the cap on the
azimuthal component of the magnetic field. 

The dimensional value of the surface density $\Sigma$ similarly depends on the
ambipolar diffusion parameter $\tilde{\eta}_A$. For the position $r = 1$ AU at
a time $t = 10,000$ yr in a disc with sound speed $c_s = 0.19$ km s$^{-1}$,
azimuthal field cap $\delta = 1$ and accretion rate $\dot{M}_c = 10^{-5}$
M$_\odot$ yr$^{-1}$ (which corresponds to a vertical magnetic field component
$B_z = 1.15$ G), the surface density is given by 
\begin{equation} 
    \Sigma \simeq \frac{2690\,\tilde{\eta}_A}{\sqrt{1+2.25\tilde{\eta}_A^{-2}}} 
    \text{ g cm}^{-3}.
\label{adin-signo}
\end{equation}
The size of the ambipolar diffusion parameter $\tilde{\eta}_A$ determines the
build-up of material relative to the magnetic field, which moves inward slower
than the neutral particles in the disc. The disc is kept in Keplerian rotation
by the ambipolar diffusion, which holds up the gas and balances its inward
radial velocity with the drift of the field lines against the flow. 

As the only real difference between this solution and the full ambipolar and
Hall diffusion solution is the change in the values of $\Sigma$ and $H$, which
depend upon the ratio of the diffusion parameters, it will not be discussed
further here. The reader is directed to Section \ref{kepdisc} and Chapter
\ref{ch:hall} for further analysis and discussion of the Keplerian disc
behaviour in the diffusive regime. 

\subsection{Shock positions and jump conditions}\label{ad:shock}

Ambipolar diffusion causes the magnetic field and charged particles to accrete 
onto the central point mass more slowly than the neutral particles, and
collisions between the neutrals and ions slow the primarily neutral fluid so
that it accretes even more slowly than in the IMHD solution. Its importance to
the dynamics of the flow depends upon the density, and so in the outermost
regions of the collapsing cloud it has little effect on the infall rate of the
field, which is ruled by IMHD as in the previous model. However, at lower $x$,
as the magnetic field and the density build up, ambipolar diffusion becomes
important and the field is decoupled from the neutral matter, though it
remains attached to the charged particles. The decoupling causes the field to
build up rapidly, and the magnetic forces may become stronger than gravity as
the field lines are forced to diffuse outwards against the accreting neutrals
at a speed almost as high as the accretion speed. The compressed field lines
take the form of an extended shock front that slows the accretion and
compresses the disc in the vertical direction.

This magnetic diffusion shock was first predicted to occur by \citet{lm1996},
and appeared in the numerical and analytic solutions of \citet{ck1998},
\citet{cck1998} and \citet{kk2002}, where it was referred to as the
``ambipolar diffusion shock''. In the model of \citet{lm1996} the shock was
driven by the decoupling of the field from the gas by Ohmic dissipation,
however ambipolar diffusion is known to become important at lower densities
than Ohmic diffusion (see Section \ref{lr:mhd} for an overview) and so it
causes the development of the shock at this early stage of collapse before
Ohmic dissipation is able to decouple the field from the gas. Ohmic
dissipation does become important in the innermost and later stages of
collapse \citep{sglc2006}, when the density is high and the field becomes
entirely decoupled from the gas. 

The magnetic diffusion shock is of C-type \citep{dm1993} and can be resolved
as a continuous transition in the flow variables, so that no explicit jump
conditions need to be imposed. \citet{lm1996} showed that their shock creates
a region of downstream turbulent inflow that may be subject to interchange
instabilities \citep{ssp1995} and the Wardle instability \citep{w1990}; it is
not possible to observe such instabilities in this model, as the turbulent
region is smoothed by the various approximations adopted, in particular the
assumption of axisymmetry. Inwards of the shock \citet{cck1998} observed that
the gas establishes a laminar free fall collapse, as their simulations were
nonrotating; it is the presence of rotation that causes the formation of the
Keplerian disc in this solution, while the size of the disc is determined by
the amount of magnetic diffusion and braking. 

The position of the magnetic diffusion shock, $x_d$, can be estimated by
examining the induction equation (\ref{adin}). The inequality $b_{r,s} \ll
hdb_z/dx$ holds true everywhere except during the shocks, and as this term is
otherwise smaller than any others it can be disregarded in order to simplify
and solve the induction equation. This can then be written as a quadratic
equation in $b_z$: 
\begin{equation}
    xh^{1/2}\sigma^{-3/2}\tilde{\eta}_Ab_{r,s}b_z^2 - xwb_z + \psi = 0.
\label{adxdin}
\end{equation}
The two regimes of flux behaviour, described by ideal MHD in the outer
asymptotic solution and ambipolar diffusion in the inner, can be approximated
by the two roots of this equation, which are usually well-separated.  

In the large $x$ limit, the quadratic term in Equation \ref{adxdin} becomes
small and IMHD is dominant. The smaller of the two roots is then a good
approximation to the vertical field component, and is given by 
\begin{equation}
    b_{z,low} \approx \frac{\psi}{xw} = \frac{\psi\sigma}{m}.
\label{adbzlow1}
\end{equation}
As this holds true during the initial dynamic collapse when the mass-to-flux
ratio is constant and given by the initial value $\mu = \mu_0$, Equation
\ref{adbzlow1} can be simplified into 
\begin{equation}
    b_{z,low} \approx \frac{\sigma}{\mu_0}
\label{adbzlow2}
\end{equation}
which is the initial (outer) boundary condition for the field derived in
Section \ref{eqouter}.

The larger root of Equation \ref{adxdin} gives the vertical field component in
the ambipolar diffusion regime, where it is approximated by dropping the
now-small constant (with respect to $b_z$) term in the quadratic (Equation 
\ref{adxdin}). In this case the equation is solved to give
\begin{equation}
    b_{z,high} 
      \approx \frac{xm}{\tilde{\eta}_A\psi}\left(\frac{\sigma}{h}\right)^{1/2},
\label{adbzhigh1}
\end{equation}
which is equation 50 of \citet{kk2002} --- the smaller root is their equation
49. Following their derivation further, inwards of the diffusive shock
position, the vertical compression of the disc is controlled by the magnetic
squeezing induced by $b_{r,s}$ (as $b_{\phi,s}$ is still small). The scale
height is then given by  
\begin{equation}
    h \approx \frac{2\sigma x^4}{\psi^2};
\label{adbzhighh}
\end{equation}
and this is then substituted into the approximation to the vertical field
component so that
\begin{equation}
    b_{z,high} \approx \frac{xm}{\tilde{\eta}_A\psi}
     \left(\frac{\psi^2}{2x^4}\right)^{1/2}
     = \frac{m}{\sqrt{2}\tilde{\eta}_Ax}.
\label{adbzhigh2}
\end{equation}

The magnetic diffusion shock is smooth, even though the $db_z/dx$ term is
large in the shock itself, and its position may be approximated by recognising
that just inwards of the shock the radial field component may be approximated
by $b_{r,s} \approx b_z \approx b_{z,high}$. This then gives the relation
\begin{equation}
    \frac{\psi}{x_d^2} \approx \frac{m}{\sqrt{2}\tilde{\eta}_Ax_d},
\label{adxd1}
\end{equation}
and as flux freezing is still approximately valid in this region, then $\psi =
m/\mu_0$ and Equation \ref{adxd1} may be solved for the position of the
magnetic diffusion shock:
\begin{equation}
    x_d \approx \frac{\sqrt{2}\tilde{\eta}_A}{\mu_0}
\label{adxd}
\end{equation}
\citep[equation 58 of][]{kk2002}, which depends only on the initial conditions
of the collapsing molecular cloud. This expression is generally a good
approximation to the shock position for all of the similarity solutions
explored in this section, with the estimated and actual positions of the 
magnetic diffusion shocks for each solution listed in Table \ref{tab-adxd} for
the purpose of comparison. Typically, Equation \ref{adxd} gives the position
of $x_d$ to $\sim 10\%$, although it is much closer for those solutions in
which the initial rotation rate is slow, as the centrifugal force is not yet
significant, and this affects the amount of magnetic braking and the behaviour
of the scale height in this approximation. 
\begin{table}[t]
\begin{center}
\begin{tabular}{ccc}
\toprule
 model & estimated $x_d$ & actual $x_d$\\
 \midrule
 $v_0 = 0.73$, $\tilde{\eta}_A = 1.0$ & 0.49 & 0.41\\
 $v_0 = 0.18$, $\tilde{\eta}_A = 1.0$ & 0.49 & 0.46\\
 $v_0 = 0.18$, $\tilde{\eta}_A = 0.7$  & 0.34 & 0.33\\
 \bottomrule
 \end{tabular}
\end{center}
\vspace{-5mm}
 \caption[Estimated vs actual values of $x_d$ in the AD model]
{Comparison between the estimated and actual values of the magnetic diffusion
shock position, $x_d$, for the similarity solutions in subsection \ref{ad:sols}.
The estimated position of the shock is typically accurate to $10\%$.} 
\label{tab-adxd}
\end{table}

Inwards of the magnetic diffusion shock and its associated turbulent
post-shock region, the slowed gas is accelerated inwards by the gravity of the
central point mass until it is falling inwards at near-free fall speeds. As in
the previous solutions, the centrifugal force builds up as the matter falls in
and triggers the formation of the centrifugal shock. In order to estimate the
position of the centrifugal shock the behaviour of the angular momentum and
magnetic braking in the free fall region must be well understood. 

During the free fall collapse region, the angular momentum is reduced by the
magnetic braking in an exponential manner, and if the region is wide enough
then the angular momentum may be reduced to an essentially constant value $j =
j_{pl2}$ (which is much smaller than the first angular momentum plateau value
$j_{pl}$ that is attained at the inner edge of the IMHD region). The
amplification of the magnetic field in the magnetic diffusion shock causes an
increase in the amount of magnetic braking that is strongly dependent on
$\tilde{\eta}_A$, and so the value of the secondary angular momentum plateau
can not be easily approximated from the initial conditions of the collapse.

Angular momentum is transported from the disc to the envelope by the twisting
of the field lines in the azimuthal direction, and so it is the azimuthal
field component that must be examined in order to determine the degree of
magnetic braking affecting the disc. As mentioned above, inwards of the
magnetic diffusion shock ambipolar diffusion dominates the behaviour of the
vertical field in the collapse region. Therefore, substituting $b_{r,s}
\approx b_z \approx b_{z,high}$ (given by Equation \ref{adbzhigh2}) into
Equation \ref{adb_phis} for $b_{\phi,s}$ gives an approximation to the
azimuthal field in this region,
\begin{equation}
    b_{\phi,s} \approx -\frac{2\alpha\psi j}{x^3}(1 + 2\alpha w)^{-1};
\label{adxcbphis1}
\end{equation}
this does not strictly hold true, as $b_{\phi,s} = -\delta b_z$ in the
innermost area of the free fall collapse region. Equation \ref{adxcbphis1} is
then an overestimate of the azimuthal field near the centrifugal shock, but is
an acceptable approximation for the purpose of estimating $j_{pl2}$ and the
position of the centrifugal shock. In this region the angular momentum
equation (\ref{adam}) can be simplified into 
\begin{equation}
    \frac{dj}{dx} \approx -\frac{x^2}{m}\,b_zb_{\phi,s}.
\label{adxcam1}
\end{equation}
Substituting Equations \ref{adbzhigh2} for $b_z$ and \ref{adxcbphis1} into
this equation yields
\begin{equation}
    \frac{dj}{dx} \approx \frac{\alpha m}{\tilde{\eta}_A^2x}\,j(1+2\alpha w)^{-1}.
\label{adxcam2}
\end{equation}
In this free fall region the mass is well approximated by its plateau value $m
= m_{pl}$ (see Section \ref{nm:shock}), which is also a good first
approximation to the central mass $m_c$. Similarly, as the matter is falling
in under gravity, $w \approx \sqrt{2m/x}$, and this and $m \approx m_c$ are
substituted into Equation \ref{adxcam2} to give
\begin{equation}
    \frac{dj}{dx} \approx \frac{\alpha m_c}{\tilde{\eta}_A^2x} \,j 
     \left(1 + \alpha\sqrt{\frac{8m_c}{x}}\right)^{-1}.
\label{adxcam3}
\end{equation}

As $x$ is small, then $2\alpha w \gg 1$ and the angular momentum equation may
then be simplified as 
\begin{equation}
    \frac{dj}{dx} \approx \frac{j}{\tilde{\eta}_A^2} \sqrt{\frac{m_c}{8x}}.
\label{adxcam4}
\end{equation}
This ordinary differential equation is separable and is integrated to give
\begin{equation}
    j \approx j_1 
      \exp\Biggl[\frac{1}{\tilde{\eta}_A^2} \sqrt{\frac{m_cx}{2}}\Biggr],
\label{adxcj}
\end{equation}
where $j_1$ is a constant. This exponential is a good approximation to $j$
between the two shocks, and may be used to estimate the position of the
centrifugal shock once the boundary condition is solved. The constant, $j_1$,
is the value of the second plateau in the angular momentum, $j_{pl2}$, which
is used to approximate the position of the centrifugal shock. It is calculated
by evaluating Equation \ref{adxcj} at $x_d$ where the angular momentum is
given by the first plateau value 
\begin{equation}
    j_{pl} \approx \frac{v_0}{A}\,m_{pl} \approx \frac{v_0}{A}\,m_c,
\label{adxcjpl1}
\end{equation}
assuming that there is little magnetic braking before the formation of the
magnetic diffusion shock (so that the ratio of the mass to the angular
momentum is equal to its initial value), and that the value of the mass plateau
is approximately equal to that of the central mass. The second plateau value
of the angular momentum is then given by
\begin{equation}
    j_{pl2} \approx \frac{v_0}{A}\,m_c
     \exp\left[-\sqrt{\frac{m_c}{\mu_02^{1/2}\tilde{\eta}_A^3}}\right]
\label{adxcjpl2}
\end{equation}
\citep[equation 64 of][]{kk2002}; and using Equation \ref{xc}, which defines
the centrifugal radius, the centrifugal shock position is approximately: 
\begin{equation}
    x_c \approx \frac{v_0^2}{A^2}m_c 
     \exp\left[-\sqrt{\frac{2^{3/2}m_c}{\mu_0\tilde{\eta}_A^3}}\right]
\label{adxc}
\end{equation}
\citep[equation 65 of][]{kk2002}.

\begin{table}[tb]
\begin{center}
\begin{tabular}{ccccc}
\toprule
 model & $j_{pl2}$ & $j_{x_c}$ & estimated $x_c$ & actual $x_c$\\
 \midrule
 $v_0 = 0.73$, $\tilde{\eta}_A = 1.0$   & 0.39 & 0.20      & 
   $3.3 \times 10^{-2}$ & $1.32 \times 10^{-2}$\\
 $v_0 = 0.18$, $\tilde{\eta}_A = 1.0$   & 0.11 & $\sim0.1$  & 
   $1.9 \times 10^{-3}$ & $1.7 \times 10^{-3}$\\
 $v_0 = 0.18$, $\tilde{\eta}_A = 0.7$   & 0.05 & $\sim0.04$ & 
   $3.5 \times 10^{-4}$ & $2.6 \times 10^{-4}$\\ 
 $v_0 = 1.50$, $\tilde{\eta}_A = 1.0^*$ &       &            & 
   $1.5$                 & $1.5$               \\
 \bottomrule
 \end{tabular}
\end{center}
\vspace{-5mm}
 \caption[Estimated vs actual values of $j_{pl2}$ and $x_c$ in the AD model]
{Comparison between the estimated and actual values of the angular momentum
$j$ at the centrifugal shock and the shock position $x_c$ for the similarity
solutions in subsection \ref{ad:sols}. The first similarity solution has poor
matches between estimated and actual values, as the poor approximation of the
magnetic diffusion shock position carries through to give a poor estimate of
the angular momentum plateau and the centrifugal shock position. 

\vspace{0.3em}$^*$Note: The $v_0 = 1.50$, $\tilde{\eta}_A = 1.0$ model (which also has
$\alpha = 0.1$ instead of $0.08$) is a rapidly rotating similarity solution in
which the angular momentum does not form a plateau before the centrifugal
shock. In this instance the centrifugal shock position is estimated using
Equation \ref{nmxc} as in the nonmagnetic and IMHD solutions.} 
\label{tab-adxc}
\end{table}
This estimate of the shock position is only valid if the centrifugal radius
occurs once the angular momentum has attained its secondary plateau value,
which makes it a poor estimate for the moderately rotating similarity solution
calculated in subsection \ref{ad:sols}. It is, however, a good fit to the
position of the shock for the other solutions of \citet{kk2002}, which are 
reproduced in this work for discussion purposes. The estimated value of the
angular momentum plateau, the actual value of $j$ at $x_c$ and the estimated
and converged centrifugal shock positions of each of the similarity solutions
in subsection \ref{ad:sols} are presented in Table \ref{tab-adxc} for the
purpose of comparison. The two estimations are intimately tied; when the
angular momentum plateau is approximated to a value close to the actual value
of $j$ at the shock, the position of the centrifugal shock is also estimated
to a high precision. The discrepancy in the calculations for the first
similarity solution is likely due to the overestimate of the diffusion shock
position (see Table \ref{tab-adxd}), which affects both the value of $j_{pl2}$
and the centrifugal shock position in turn.

The centrifugal shock is again treated as a discontinuity in the radial
velocity and surface density in which the flow changes from being in near-free
fall collapse to a subsonic accretion disc in Keplerian orbit. The shock is
calculated explicitly using the jump conditions derived for the nonmagnetic
model in subsection \ref{nm:shock} (Equations \ref{nmwsol1} and \ref{nmwsol2}),
as the magnetic field is decoupled from the neutrals in this region and so
does not change across the shock. The shock is followed by a thin numerically
resolvable post-shock layer in which $b_z$ increases, and the flow then
settles into its asymptotic disc solution, as will be shown in the following 
subsections. 

\subsection{Model construction}\label{ad:model}

\begin{figure}[!t]
  \centering
  \vspace{-2mm}
  \includegraphics[width=4in]{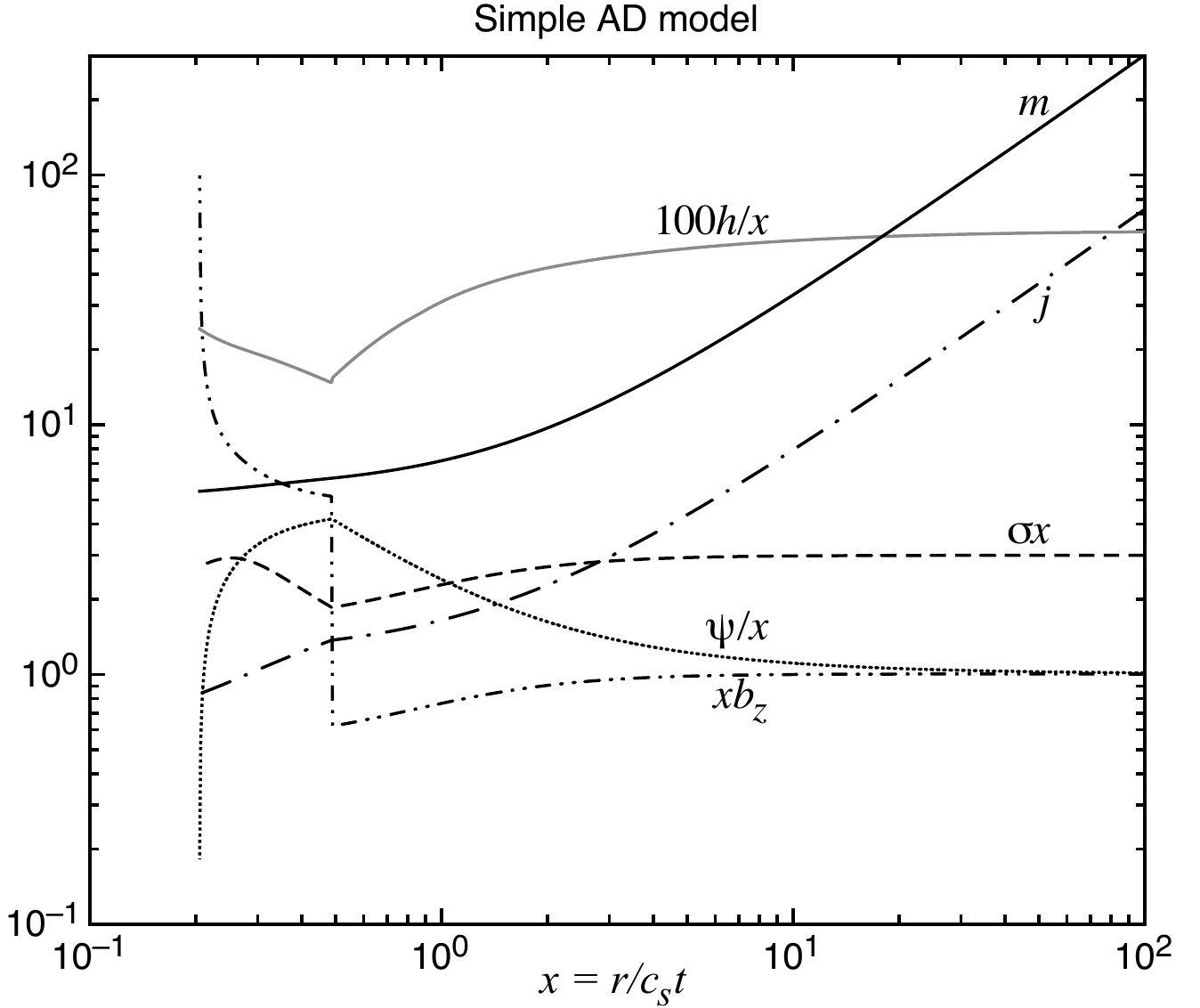}
  \vspace{-4mm}
  \caption[Simple ambipolar diffusion model]{Simple model for the ambipolar
diffusion collapse, with $x_d = 0.487$ (the estimated value calculated in
Table \ref{tab-adxd}) and outer asymptotic boundary conditions matching those
in Figure \ref{fig-admod}. Note that $h$, $\psi$ and $b_z$ are unstable and
veer away from a singularity in the inner regions, so the matching point is
chosen to be $x_m = 0.3$, before the solution has been much affected by these
instabilities.}\label{fig-adsimp} 
\end{figure}

The addition of two further equations to the set to be integrated complicates
the numerical routine so that it is no longer possible to integrate inwards
from the outer boundary, as small numerical errors in the calculation of the
derivatives can compound and cause the integration to veer unphysically from
the expected asymptotic collapse solution (see the inner region of Figure
\ref{fig-adsimp} for an example of this behaviour). Similarly, it is not
possible to integrate out from the inner boundary, as the calculation rapidly
breaks down in the outward direction as well. 

The solution to this problem is to treat the integration as a two-point
boundary value problem in which the values of the variables $m$, $\sigma$,
$j$, $\psi$ and $b_z$ and the constant boundary condition $m_c$ are known at
the inner and outer boundary, but unknown at an intermediate value of $x$ (the
``matching point'', denoted $x_m$). The values of the variables at this
matching point are treated as free parameters that are initially ``guessed''
and refined by iteration. The integration of the equations is performed in
both directions from the matching point, with the shock position and
conditions calculated on the inwards path, until they reach the boundaries.
Here the discrepancies between the integrated variables and the boundary
conditions are evaluated --- these are initially nonzero. The integration
process can then be treated as a multidimensional root-finding problem, which
can be solved using a globally convergent Newton-Raphson procedure
\citep[\texttt{newt} and its dependencies from \textit{Numerical
Recipes};][]{NR}. As the differential equations are nonlinear, zeroing the 
discrepancies between the integrated variables and the boundary conditions by
varying the variables at $x_m$ is a time-consuming process requiring many
iterations of the integration. This technique for solving two-point boundary
value problems is known as the ``shooting method'', and is discussed in more
detail in chapter 17 of \citet{NR}. 

The shooting method for solving the boundary condition problem requires a good
initial estimate of the variables at the matching point, which is chosen such
that it lies between the centrifugal and magnetic diffusion shocks (i.e.\ $x_c
< x_m < x_d$). This estimate is found by calculating a simplified model of the
collapse with ambipolar diffusion in which the derivative of the vertical
magnetic field component with respect to $x$ is assumed to be small everywhere
and may be disregarded. As explained in Section \ref{eqsimple}, this is
generally valid because $hdb_z/dx$ is small everywhere (except in the magnetic
diffusion shock) due to the thin disc approximation which requires that $h \ll
x$. The induction equation is then replaced by the approximations to $b_z$
that were used in estimating the shock positions: $b_z = b_{z,low}$ (Equation
\ref{adbzlow2}) when $x > x_d$, and $b_z = b_{z,high}$ (Equation
\ref{adbzhigh2}) when $x < x_d$; the position of the magnetic diffusion shock
is assumed to be that given by Equation \ref{adxd}. The other variables ($m$,
$\sigma$, $j$, and $\psi$) are integrated from the upper boundary to the
matching point, creating a solution that is close enough to the expected full
solution that the values of the variables at the matching point may be used as
an initial guess for the shooting routine. 

\begin{figure}[ht]
  \centering
  \vspace{-2mm}
  \includegraphics[width=4.3in]{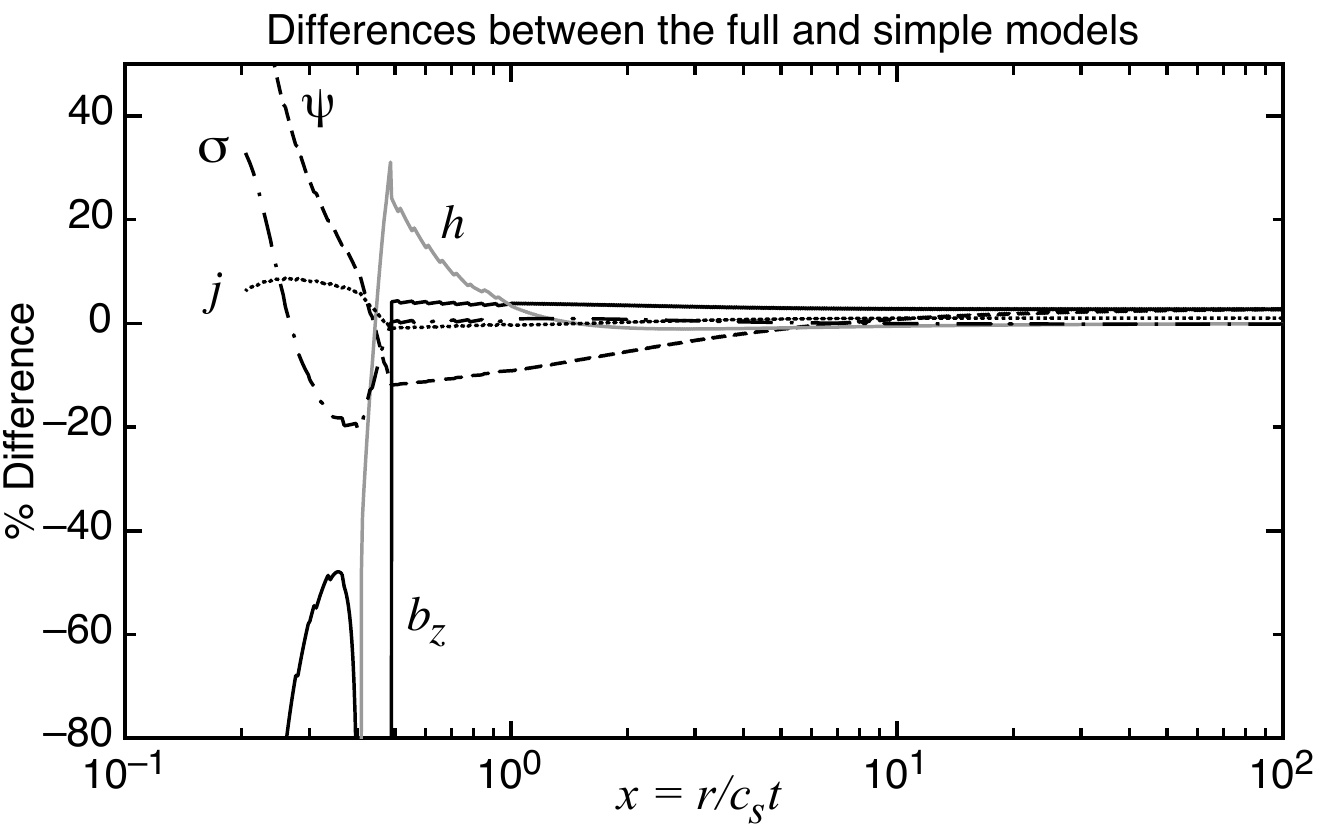}
  \vspace{-4mm}
  \caption[Differences between the full and simple AD models]{The difference
between the values of $\sigma$, $j$, $\psi$, $b_z$ and $h$ in the simple model
from Figure \ref{fig-adsimp} and the full similarity solution in Figure
\ref{fig-admod} as percentages of the true value. In the outer regions the
variables match very well save for some minor constant discrepancies, however,
near the magnetic diffusion shock the magnetic flux and the scale height drift
from their expected values. The biggest discrepancy between the two solutions
is the position of the diffusion shock (see Table \ref{tab-adxd}), which
causes the large drop in $b_z$ at the simple shock position (to $-5$x that of
the full model) and then the peak downstream at the true shock. It is possible
that this discrepancy causes the singularity inwards of the shock shown in
Figure \ref{fig-adsimp}.}  
\label{fig-adsimpdiff}
\end{figure}
Figure \ref{fig-adsimp} shows such a simple calculation for the same
parameters as the full solution in Figure \ref{fig-admod} (the matching point
for that solution is chosen to be $x_m = 0.3$). This similarity solution is
qualitatively similar to the IMHD similarity solutions in the outer region of
the collapse where IMHD is dominant, however inwards of the magnetic diffusion
shock the vertical field component and the flux diverge from their expected
behaviour: the field becomes infinitely large as the enclosed flux decreases
rapidly, causing the integrator to fail. The percentage error between the
variables in the full solution and their values in this simple model is shown
in Figure \ref{fig-adsimpdiff} --- the biggest discrepancy between the two
solutions is in the position of the magnetic diffusion shock, which is located
at $x_d = 0.406$ in the full similarity solution and at $x_d = 0.487$ in the
simple model. This incorrect shock position fuels the divergence of the simple
solution; if a more accurate estimate of the shock location were adopted then
the simple model would be able to produce variables at the matching point that
are closer to the true values. 

Divergences from the expected behaviour such as those undergone by $\psi/x$
and $xb_z$ in Figure \ref{fig-adsimp} can be so strong that using such values at
the matching point (even after choosing $x_m$ as close to $x_d$ as possible)
may cause the full integration to fail. Typically, adopting an initial guess
that is too far from the true values causes the integration to either
encounter a spontaneous singularity and diverge (as in Figure
\ref{fig-adsimp}) or to score so badly against the boundary conditions that
the routine will never converge on the true solution. In these cases it is
prudent to adjust $\psi$ and $b_z$ at the matching point by hand until the
code is able to integrate to both boundaries while matching the boundary
conditions to $\lesssim10\%$. Such tweaking is not always needed, however it
can be a significant additional source of overhead in an already much-slowed
process. 
\begin{figure}[t]
  \centering
  \vspace{-2mm}
  \includegraphics[width=4.5in]{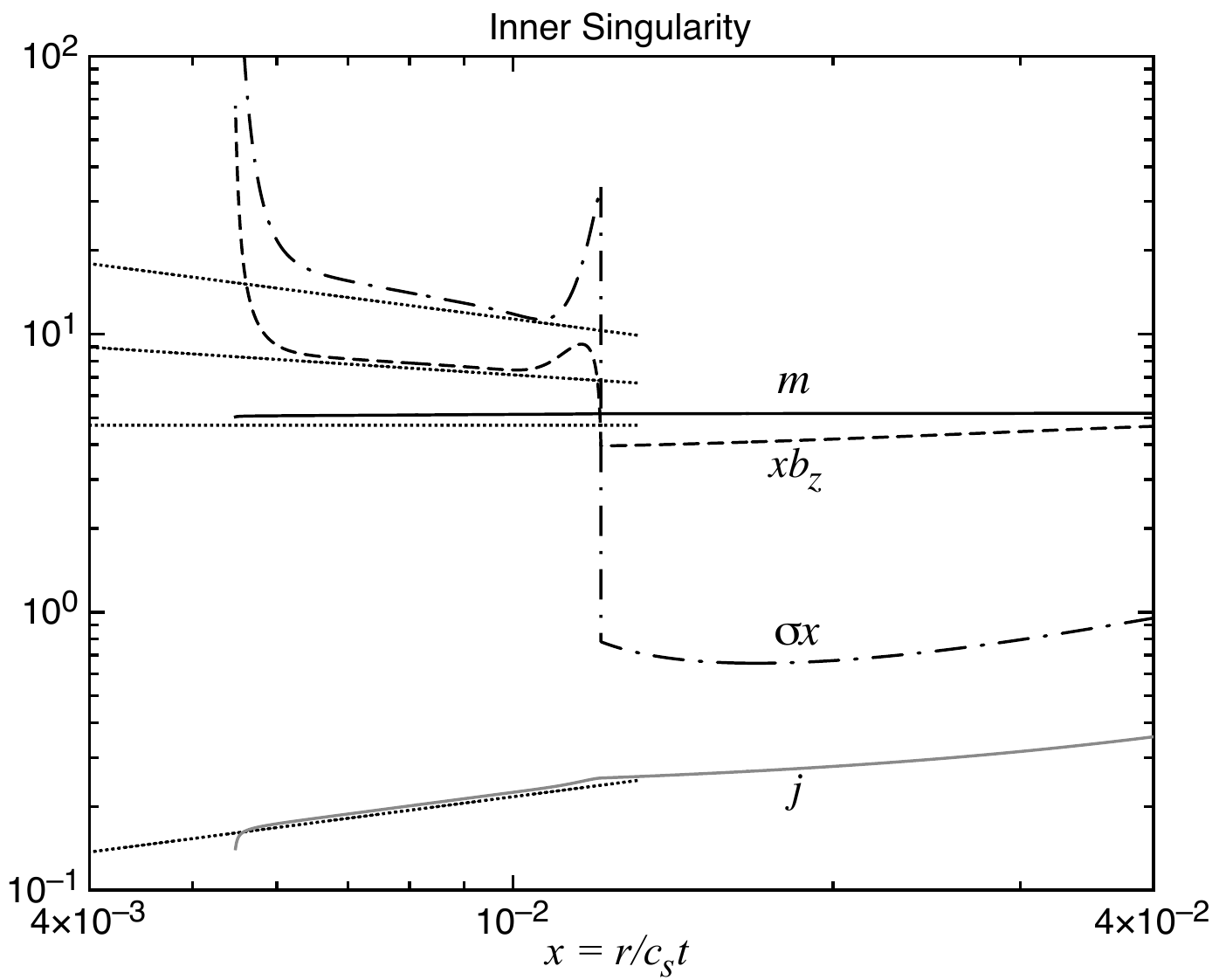}
  \vspace{-5mm}
  \caption[The inner singularity]{Close-up on the centrifugal shock region for
a collapse calculation with the same boundary conditions and parameters as
Figure \ref{fig-admod}. Inwards of the centrifugal shock at $x_c = 1.32 \times
10^{-2}$ the variables tend towards the asymptotic solutions (the dotted
lines), before encountering a singularity at $\sim 5.5 \times 10^{-3}$ and
veering away. Note that this particular simulation has yet to converge on the
true value of $m_c$.} 
\label{fig-adzderivs}
\end{figure}

One other simplification is required in order to integrate inwards to the
inner boundary. Even though the position of $x_c$ can be calculated to the
maximum possible precision, a spontaneous singularity may occur at some point
$0< x< x_c$ after the solution has seemingly matched onto the inner asymptotic
solution. An example of this behaviour is shown graphically in Figure
\ref{fig-adzderivs}, in which the asymptotic solutions are shown as dotted
lines that run parallel to the similarity solution for some length of
similarity space before diverging (note that this solution has yet to converge
on the true value of $m_c$). Rather than choosing a point just outwards of
this singularity to be the inner boundary, it is more productive to switch the
integration to a simpler set of equations while the variables are still
close to their asymptotic values. 

The inner singularity occurs because the derivatives for the surface density
and the vertical field component become large in the inner regions, and in
calculating these (as is required to numerically integrate the ordinary
differential equations) numerical errors compound and cause the code to fail.
In order to counteract this effect a simplified model is used to perform the
innermost integration; in this model the problematic derivatives are given by
the values from the inner asymptotic solution in subsection \ref{ad:inner}: 
\begin{align}
    \frac{d\sigma}{dx} &= -\frac{3}{2}\,\sigma_1\,x^{-5/2} \label{ad-zdsigma}\\
    \text{and }\frac{db_z}{dx} &= -\frac{5m_c^{3/4}}{4\sqrt{2\delta}}\,x^{-9/4}
\label{ad-zdbz},
\end{align}
where $\sigma_1$ is the constant coefficient defined in Equation
\ref{adin-sig}. These values are substituted into the simplified equation set:
\begin{align}
    \frac{\sigma m_c}{x^3}\,h^2 + \left(b_{r,s}^2 + b_{\phi,s}^2 
     + \sigma^2\right)h -2\sigma &= 0, \label{ad-zh}\\
    g + \frac{b_zb_{r,s}}{\sigma} + \frac{j^2}{x^3} 
     - \frac{1}{\sigma}\frac{d\sigma}{dx} 
     - \frac{b_zh}{\sigma}\,\frac{db_z}{dx} &= 0 \label{ad-zrm}\\
    \text{and  }-xwb_z + \tilde{\eta}_Axb_z^2b_{r,s}h^{1/2}\sigma^{-3/2}
     &=0, \label{ad-zin}
\end{align}
which is then solved for $h$, $\sigma$ and $b_z$. The other variables are
determined by integrating the remaining differential equations
\begin{align}
    \frac{dm}{dx} &= x\sigma, \label{ad-zm}\\ 
    \frac{d\psi}{dx} &= xb_z \label{ad-zpsi}\\
    \text{and }
    \frac{dj}{dx} &= \frac{j}{w} - \frac{x^2b_zb_{\phi,s}}{m} \label{ad-zj}
\end{align}
to the inner boundary, whereupon the discrepancy between the boundary
condition $m = m_c$ and the integrated value of $m$ is passed back to the
shooting routine for refinement. 

The simplified model is able to integrate to the inner boundary because of the
assumption that $d\sigma/dx$ and $db_z/dx$ are given by their asymptotic
expressions, which reduces the numerical error of the calculation, and makes
it possible to solve the resulting reduced set of fluid equations. The
validity of this assumption is demonstrated in Figure \ref{fig-adzderivs},
which shows that downstream of the centrifugal shock the variables settle
quickly to their asymptotic forms and the derivatives of $\sigma$ and $b_z$
calculated using the full set of equations are close to the asymptotic values.
Because of this, adopting the simple model is not expected to introduce any
significant errors to the calculations, provided that it matches well to the
full solution.

The transition between the full and simplified models occurs when both the
derivatives of $\sigma$ and $b_z$ match their asymptotic values in Equations
\ref{ad-zdsigma} and \ref{ad-zdbz} to an appropriate degree. In most of the 
solutions calculated in the following subsection and the Hall diffusion
solutions in Chapter \ref{ch:hall}, matching the derivatives to 1\% is
considered acceptable for switching between the two models, although a finer
match may be required if the solution is otherwise unable to converge. The
transition is usually seamless, however if not all of the values of the
variables (particularly $m_c$) at the matching point have converged, or if the
criteria for changing between the models is not vigorous enough, the point of
transition may be visible in the plots, as occurs to $\psi/x$ in Figure
\ref{fig-kk8b} just inwards of the expanded shock region. If the variables
(including $m_c$) at $x_m$ are particularly poorly chosen then the simple
model may also encounter spontaneous singularities as it integrates inwards.

Even with a good estimate of the variables at the matching point as input to
the root-finding routine, and the simplified model for calculating the
innermost integrals, the calculation of the true solution with its
minimized scores (typically of order $10^{-5}\%$) may take up to a day to
compute. The initial step of finding a set of variables at $x_m$ that would
integrate to both boundaries was not fully automatable, as poor guesses would
cause the code to crash in ways that could not always be predicted or
accounted for. Multiple iterations of this procedure were often necessary to
ensure convergence on the value of the central mass. 

\subsection{Similarity solutions}\label{ad:sols}

\begin{figure}[htp]
  \centering
\vspace{-5mm}
  \includegraphics[width=5.1in]{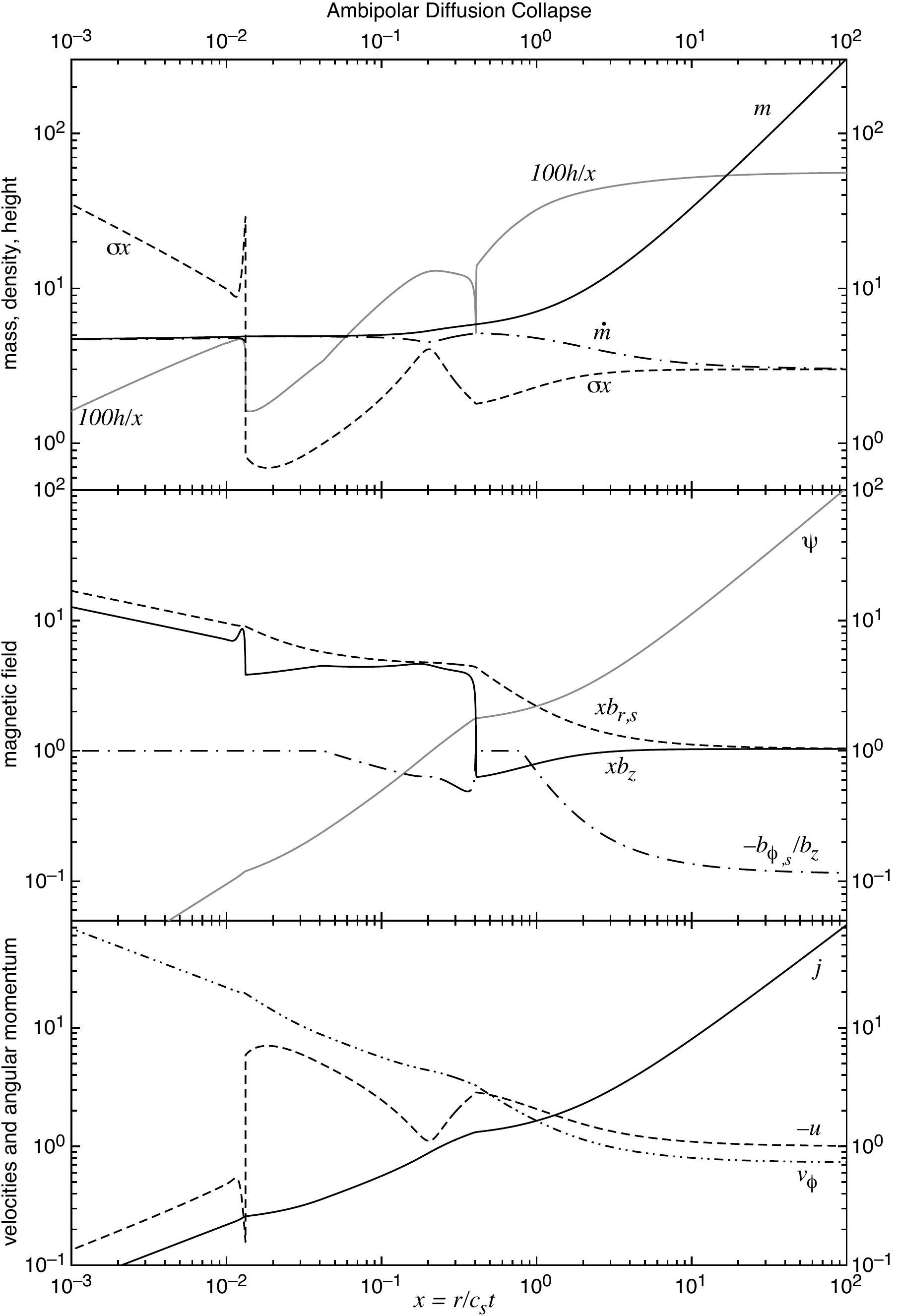}
\vspace{-5mm}
  \caption[Ambipolar diffusion collapse]{Similarity solution for the
moderately rotating ambipolar diffusion collapse, with outer asymptotic
boundary conditions $A = 3$, $u_0 = -1$, $\mu_0 = 2.9$ and $v_0 = 0.73$, and
magnetic parameters $\alpha = 0.08$, $\delta = 1$ and $\tilde{\eta}_A = 1.0$;
these match the parameters in figure 7 of \citet{kk2002}. The displayed
variables are the same as in Figure \ref{fig-imhdslow} for the IMHD case; the
nondimensional central mass is $m_c = 4.67$, and the magnetic diffusion and
centrifugal shocks are located at $x_d = 0.406$ and $x_c = 1.32 \times
10^{-2}$ respectively.} 
\label{fig-admod}
\end{figure}

\begin{table}[t]
\begin{center}
\begin{tabular}{lcccc}
\toprule
 Solution & Figure & $\tilde{\eta}_A$ & $v_0$ & $\alpha$\\
 \midrule
 moderate rotation & \ref{fig-admod} & 1.0 & 0.73 & 0.08 \\
 slow rotation     & \ref{fig-kk8a}  & 1.0 & 0.18 & 0.08 \\ 
 slow rotation, reduced $\tilde{\eta}_A$ & 
                     \ref{fig-kk8b}  & 0.7 & 0.18 & 0.08 \\
 fast rotation     & \ref{fig-kk9}   & 1.0 & 1.50 & 0.10 \\
 \bottomrule
 \end{tabular}
\end{center}
\vspace{-5mm}
 \caption[Parameters of the ambipolar diffusion similarity solutions]
{Ambipolar diffusion, initial rotational velocity and magnetic braking
parameters for the ambipolar diffusion similarity solutions. All of the other
parameters are identical and given by $A = 3$, $u_0 = -1.0$, $\mu_0 = 2.9$ and
$\delta = 1$.} 
\label{tab-adparams}
\vspace{1mm}
\end{table}
The ambipolar diffusion similarity solutions are calculated using the method
outlined above, however only the first of the ambipolar diffusion solutions
from \citet{kk2002} was calculated as a test of the computation code; this
similarity solution is shown in Figure \ref{fig-admod}. Their other ambipolar
diffusion solutions are reproduced in Figures \ref{fig-kk8a}--\ref{fig-kk9} in
order to properly examine and discuss the role of ambipolar diffusion in the
star formation process. The initial conditions and parameters of all of the
similarity solutions presented in this section are given in Table
\ref{tab-adparams}. 

The similarity solutions presented in this section have much in common with
the IMHD solutions, as the basic interplay between the magnetic, gravitational
and centrifugal forces remains the same. However, altering the coupling
between the gas and the magnetic field allows the field lines to be
transported from the innermost regions of the collapse to the outer ones,
reducing the magnetic flux problem so there is less of a magnetic field excess
near the origin. Similarly, the magnetic braking is reduced by the presence of
ambipolar diffusion, so that the magnetic braking catastrophe may be resolved
completely. 

The amount of ambipolar diffusion present in the solutions is determined by
the nondimensional parameter $\tilde{\eta}_A$, which is a constant of order
$\sim1$. The justification for this is that in the outer regions of collapsing
cores at radii $\gtrsim 10^3$ AU, the grains have a typical radius $a = 0.1
\mu$m, the temperature is $10$ K and the cosmic ray ionisation rate is given
by $\xi = 10^{-17} \xi_{-17}$ s$^{-1}$ \citep[where $\xi_{-17} \approx
1$;][]{cm1998,kn2000}. The ambipolar diffusion parameter is given by
$\tilde{\eta}_A \approx 0.2 \xi^{-1/2}_{-17}$ when the scaling of the ion
density is proportional to the square root of the neutral density (see Section
\ref{lr:mhd}) and the molecular ionisation by cosmic rays is balanced by rapid
dissociative recombination of the molecular ions. 

Further inward, at radii $\lesssim 10$ AU where the density is higher, the
temperature in the collapsing core is around $T = 10^2T_2$ K (where $T_2
\approx 1$). The ambipolar diffusion parameter is then given by
$\tilde{\eta}_A \approx 0.07 \xi_{-17}^{-1/2}T_2^{-1/4}(a/5$ \AA$)^{-1/4}$
\citep[see][for the justification of this approximation]{kk2002}, where $a$ is
the average grain size, assumed in this region to be small singly-charged
particles that are PAH-like \citep{nh1994}. This relationship breaks down
during the intermediate regions where Hall diffusion is expected to be
important, however, given that the value of $\tilde{\eta}_A$ only changes by
an order of magnitude over 8 orders of magnitude in density, choosing a
constant $\tilde{\eta}_A$ of order 1 is an acceptable parameterisation for the
ambipolar diffusivity. The value of the diffusion parameter is varied in the
solutions presented in this section in order to properly explore the role
played by ambipolar diffusion in self-similar collapse, and its variation is
discussed further in Section \ref{catastrophe}. 

Figure \ref{fig-admod} shows the only ambipolar diffusion similarity solution
calculated in this work, which was performed in order to test the code for
solving the collapse problem with both Hall and ambipolar diffusion and
confirm the results of \citet{kk2002}. It has the same initial (outer)
conditions as the fiducial solution in \S3.3.1 of their work: the density
parameter is $A = 3$; the initial radial velocity is parameterised by $u_0 =
-1$; the nondimensional mass-to-flux ratio is $\mu_0 = 2.9$ and the initial
azimuthal velocity is $v_0 = 0.73$. This core is initially rotating at a rate
that would be termed fast in the previous models, however the changed magnetic
braking caused by ambipolar diffusion in the azimuthal direction causes the
centrifugal force to remain unimportant until $x$ is small. The magnetic
braking parameter is given by $\alpha = 0.08$, which is slightly reduced from
the value $\alpha = 0.1$ used in the IMHD solutions; the azimuthal field cap
remains at $\delta = 1$ for simplicity (so that the maximum value of the
azimuthal field is $b_{\phi,s} = -b_z$) and the ambipolar diffusion parameter
is chosen to be $\tilde{\eta}_A = 1.0$, which is slightly larger than that
expected. 

The outer regions of the ambipolar diffusion collapse match those from the
IMHD similarity solutions, as the mass-to-flux and the mass-to-angular
momentum ratios remain constant as the material falls in at supersonic speeds.
The radial velocity and scale height are dominated by the self-gravity of the
disc, which pulls the gas towards the equator before it then flows towards the
central mass. The magnetic field gradually builds up as the matter falls
inwards, and it becomes important to the dynamics at around $x \approx 2$ where
the magnetic braking starts to affect the angular momentum transport and the
constant ratio of the mass to the angular momentum breaks down. The azimuthal
field component attains its capped value, and is important to the angular
momentum transport throughout the rest of the collapse. 

The mass and angular momentum tend towards their plateau values as in the
previous solutions, and ambipolar diffusion becomes important to the field
transport as the surface density and magnetic field build up. The gravity of
the central point mass becomes important to the radial velocity --- this pulls
more mass and flux inwards until ambipolar diffusion causes flux freezing to
break down. 

The magnetic diffusion shock at $x_d = 0.41$ (which is much smaller than the 
estimated position of $x_d \approx 0.49$; see Table \ref{tab-adxd}) takes the
form of a sudden increase in the vertical field component as the field lines
diffusing against the flow from the downstream ambipolar diffusion-dominated
regime meet those coming inward with the IMHD collapse. The sudden increase in
the field causes the gas to slow down due to the magnetic pressure terms in
the radial momentum equation (\ref{adrm}) and the magnetic squeezing terms
come to dominate the vertical compression of the disc. The particular dip in
the scale height at the shock suggests that the rapid compression of the disc
causes a breakdown of the vertical hydrostatic equilibrium in the shock; in
reality the magnetic squeezing at the shock should produce a smooth reduction
of the disc thickness over the shocked region. 

IMHD breaks down at around the position of the magnetic diffusion shock where
$m$ (and by extension $\psi$) have fallen to a plateau value. This value is
not affected by the addition of ambipolar diffusion to the calculations, so
that the amount of flux contained within $x_d$ is roughly that trapped at the
origin in the IMHD solutions. Ambipolar diffusion causes the flux to be
redistributed downstream of the magnetic diffusion shock, so that no flux is
contained at the origin.

\begin{figure}[t]
  \centering
  \includegraphics[width=4in]{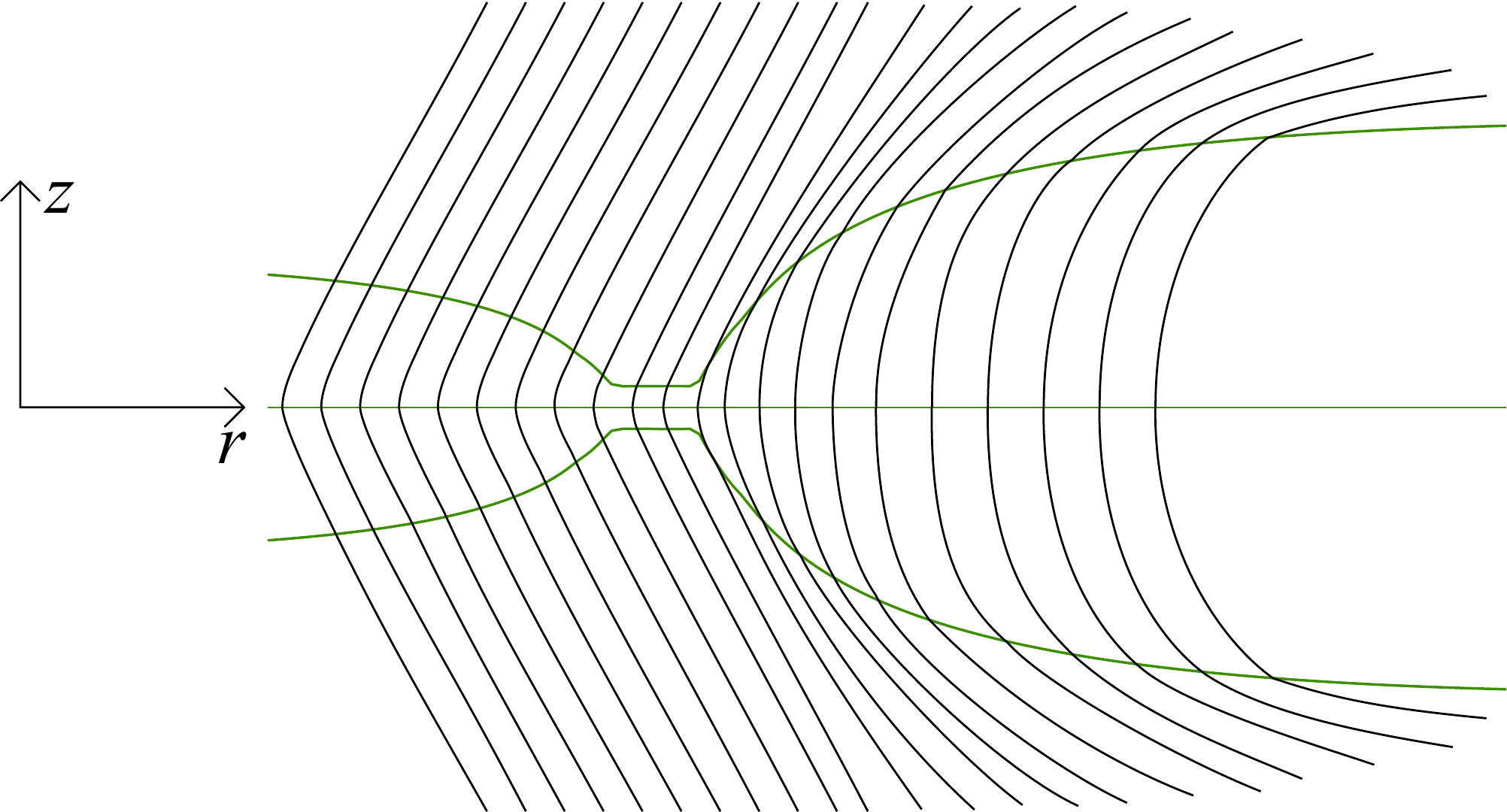}
  \caption[Field behaviour in the magnetic diffusion shock]{Schematic of the
magnetic field behaviour in the magnetic diffusion shock. The disc (green) is
compressed as the vertical field becomes large, causing the field lines at the
surface to straighten from being largely radial upstream of the shock to
having roughly equal values of $b_z$ and $b_{r,s}$ downstream of the shock.
(Not to scale.)} 
\label{fig-adsilverfish}
\end{figure}

\begin{figure}[htp]
  \vspace{-5mm}
  \centering
  \includegraphics[width=5.5in]{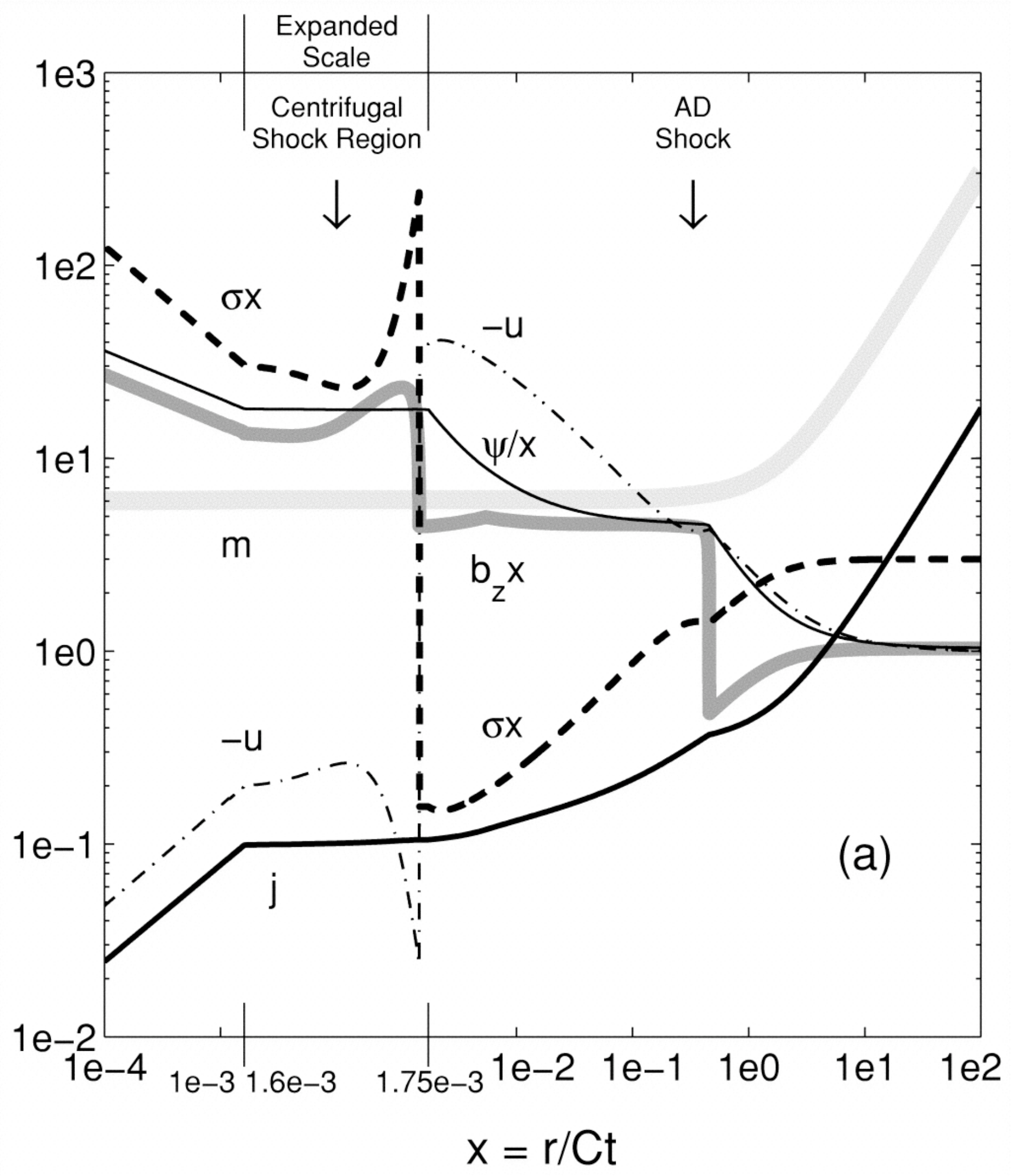}
 \vspace{-2mm}
  \caption[Slow ambipolar diffusion collapse]{Similarity solution for the
slowly rotating ambipolar diffusion collapse \citep[figure 8a of][]{kk2002}.
The initial conditions and parameters are the same as those in Figure
\ref{fig-admod}, save that the initial azimuthal velocity is reduced to $v_0 =
0.18$. The position of the magnetic diffusion shock (labelled ``AD'') is
increased to $x_d = 0.46$, which smooths out the post-magnetic diffusion shock
region before the gas starts accelerating towards the free fall collapse. The
reduction of the initial angular momentum moves the centrifugal shock inwards
(as in the IMHD solutions) to $x_c = 1.7 \times 10^{-3}$, for it takes longer
for the centrifugal force to balance gravity. The central mass is increased to
$m_c = 6.0$. The horizontal scale covers a wider range of $x$ than in the
previous plot, and inwards of the centrifugal shock the scale has been
expanded to better highlight the post-centrifugal shock transition to the
asymptotic solution.}  
\label{fig-kk8a}
\end{figure}

\begin{figure}[htp]
  \centering
  \vspace{-5mm}
  \includegraphics[width=5.5in]{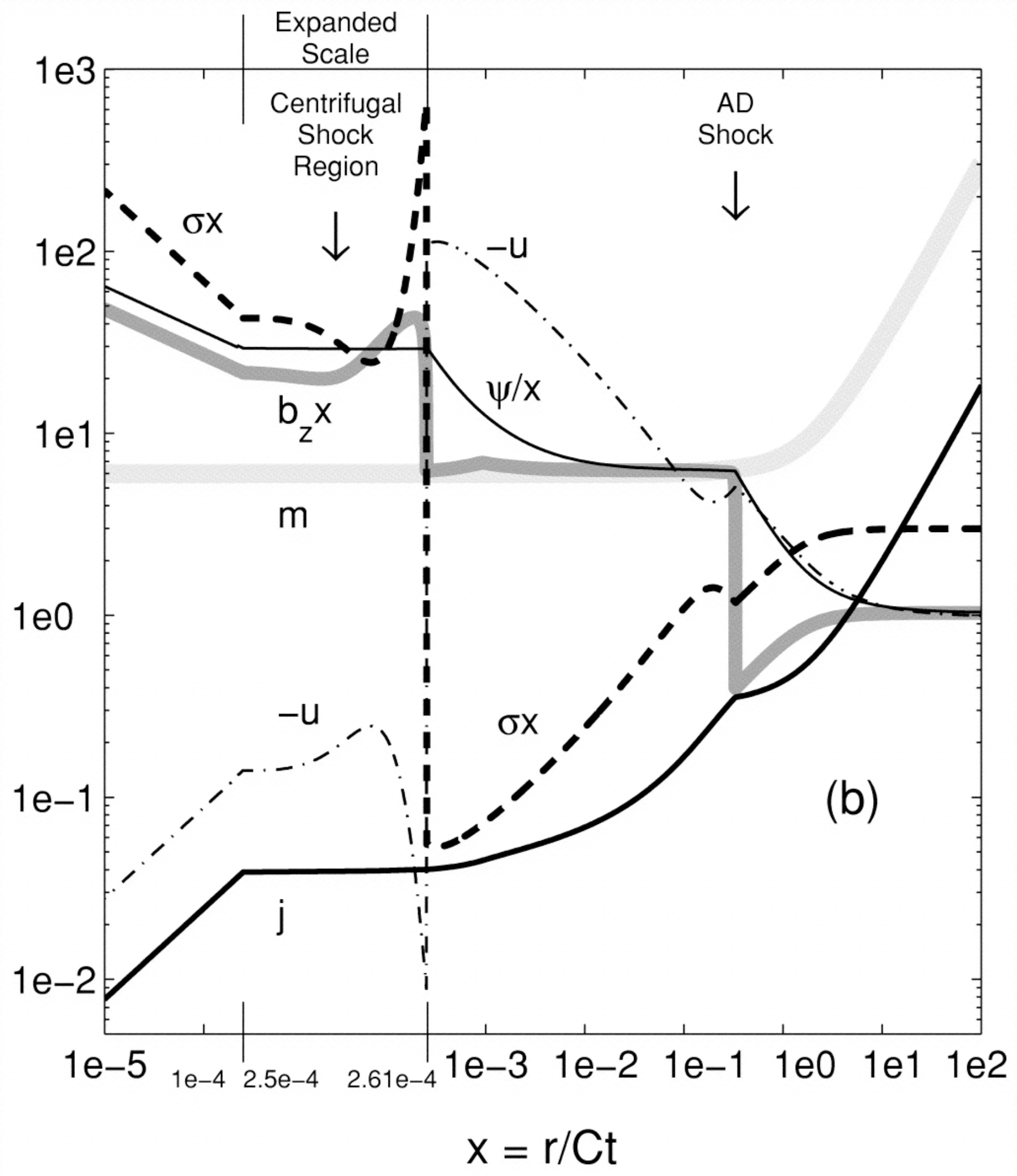}
  \vspace{-2mm}
  \caption[Slow ambipolar diffusion collapse with reduced
$\tilde{\eta}_A$]{Slowly rotating reduced ambipolar diffusion collapse,
reproduced from figure 8b of \citet{kk2002}. This similarity solution has the
same parameters and initial conditions as Figure \ref{fig-kk8a}, with an
initial rotational velocity of $v_0 = 0.18$ and reduced ambipolar diffusion
parameter $\tilde{\eta}_A = 0.7$. The reduced ambipolar diffusion parameter
causes the decoupling front to move inwards to $x_d = 0.33$, while its
influence on the magnetic braking causes the secondary angular momentum
plateau to be reduced further so that the centrifugal shock position is also
closer to the origin at $x_c = 2.6 \times 10^{-4}$. The central point mass is
$m_c = 6.0$ as in the previous solution.} 
\label{fig-kk8b}
\end{figure}

\begin{figure}[htp]
  \centering
  \vspace{-5mm}
  \includegraphics[width=5.5in]{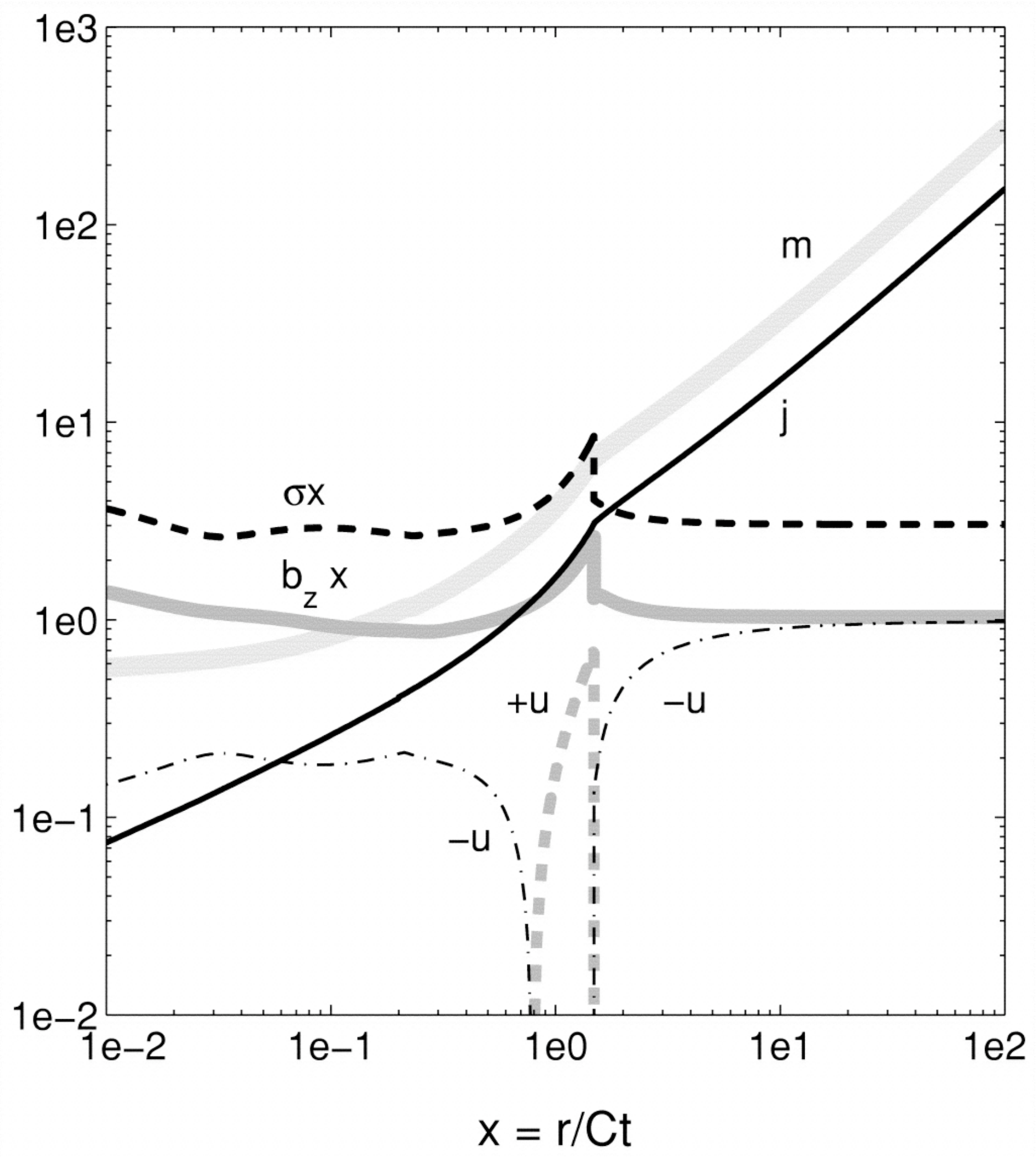}
  \vspace{-2mm}
  \caption[Fast ambipolar diffusion collapse]{Similarity solution for the
rapidly rotating ambipolar diffusion collapse, with outer asymptotic boundary
conditions matching those in Figure \ref{fig-admod} except for the azimuthal
velocity parameter which has been increased from 0.73 to $v_0 = 1.5$, and the
magnetic braking parameter $\alpha$ has been increased to $\alpha = 0.1$ from
0.08 \citep[figure 9 of][]{kk2002}. These match the parameters from the fast
IMHD similarity solution in Figure \ref{fig-imhdfast}. The solution is similar
to that in the IMHD case, with an increased centrifugal shock position $x_c =
1.5$, and reduced central mass $m_c = 0.5$. There is no magnetic diffusion
shock, as the density remains too low for ambipolar diffusion to become
important before the centrifugal shock occurs. After an extended region of
backflow and slow infall the variables settle onto the inner asymptotic
solutions. } 
\label{fig-kk9}
\end{figure}

The decoupling of the field from the neutral particles primarily takes place
at the magnetic diffusion shock, which also changes the disc geometry.
Upstream of the shock, the magnetic field is dominated by the radial
component, which can be an order of magnitude larger than the vertical and
azimuthal field components. During the shock and the transition region that
follows downstream of it, the field lines straighten until the poloidal
components at the disc surface are approximately equal; this field geometry is
illustrated in Figure \ref{fig-adsilverfish}. Relative to the vertical
component $b_{\phi,s}$ drops in the shock, however the decoupling of the field
causes the magnetic braking rate to increase so that the azimuthal field
component increases by a factor of 1.5 during the immediate post-shock region.

Inside the magnetic diffusion shock the poloidal field components scale with
$\sim x^{-1}$ as the surface density and thickness of the disc increase and
the radial velocity and accretion rate drop. The gas is slowed due a weak
outward acceleration caused by the radial magnetic pressure, and this in turn
causes the density to rise. Rotation is not dynamically important in this
region and so the shock has a similar structure to that seen in the
nonrotating similarity solution of \citet{cck1998}. The gravity of the central
mass eventually overcomes the magnetic pressure and the gas starts
accelerating inwards once more. 

As in the slowly rotating similarity solutions obtained from the IMHD and 
nonmagnetic models, rotation remains dynamically unimportant until the
vicinity of the centrifugal shock. For most of the region between the two
shocks the gravity of the central mass dominates the radial acceleration of
the gas until it is in near-free fall collapse, slowed only a little by
ambipolar diffusion of the magnetic flux and the magnetic pressure. The
azimuthal field component increases in this free fall region until it is again
equal to the vertical component, which causes the change in the scale height
behaviour at $x \sim 4.5\times 10^{-2}$ as it contributes more to the magnetic
squeezing forces in the pseudodisc. By this point the enclosed mass and the
accretion rate have flattened and remain near-constant throughout the
remainder of the collapse. The angular momentum has yet to reach its second 
plateau value (see Table \ref{tab-adxc}) when the centrifugal force becomes
large and triggers the formation of the Keplerian disc. 

Eventually the centrifugal force becomes equal to the gravitational force and
the centrifugal shock occurs at $x_c = 1.32 \times 10^{-2}$, which is much
smaller than the estimated value $x_{c} \approx 3 \times 10^{-2}$ (see Table
\ref{tab-adxc}). This discrepancy is likely caused by the overlarge estimate
of the magnetic diffusion shock radius used to approximate the centrifugal
shock position. In this similarity solution the shock is not strong enough
that it can create a region of backflow, however the gas is slowed so that the
infall is now subsonic and the surface density increases by more than an order
of magnitude. 

The centrifugal shock is followed immediately by a very thin layer in which
the azimuthal and vertical magnetic field components increase rapidly. This
increase causes a decrease in the angular momentum to its asymptotic value, as
the surface density and the other variables adjust with a few overshoots
towards their expected rotationally-supported disc behaviour. The transition
between the full model and the simplified set of equations outlined in
subsection \ref{ad:model} occurs at $x \sim 8.6 \times 10^{-3}$, after the
variables have joined onto the asymptotic inner disc described by Equations
\ref{adin-m}--\ref{adin-bphis}. 

The Keplerian disc itself is rather small compared to that in the rapidly
rotating similarity solutions without ambipolar diffusion (as $x_c = 1.32
\times 10^{-2}$ corresponds to a disc radius of $r_c \approx 53$ AU at time $t
= 10^5$ years), and it has a mass $m_d = 0.23$ that is $\sim5\%$ that of the
central point mass. The nondimensional mass at the origin is $m_c = 4.67$,
which corresponds to a moderate accretion rate of $\dot{M}_c = 7.6 \times
10^{-6}$ M$_\odot$ yr$^{-1}$ (so that at time $t = 10^5$ yr, the central mass
is $M_c = 0.76$ M$_\odot$). The surface density of the disc depends upon the
ambipolar diffusion and azimuthal field cap parameters, as does the infall
velocity, which is subsonic and very low. The disc is extremely thin and the
vertical squeezing is dominated by the tidal and self-gravitational forces. 

Within the rotationally-supported disc the angular momentum problem of star
formation is solved as the disc is in Keplerian orbit, and as the flux is
clearly reduced from the constant value in the IMHD solution the magnetic flux
problem is seemingly resolved as well. The amount of magnetic flux in the disc
scales with $x^{3/4}$, so that $\psi \to 0$ as $x \to 0$; clearly the amount
of flux present in the protostar depends upon more detailed flux transport and
destruction mechanisms than are included in this model, such as Ohmic
diffusion \citep[e.g.][]{lm1996} and reconnection \citep[e.g.][]{gs1993b,
l2005}. 

Figure \ref{fig-kk8a}, reproduced from figure 8a of \citet{kk2002}, shows a
slowly rotating similarity solution with the same initial conditions as Figure
\ref{fig-admod}, save for the initial azimuthal velocity which has been
reduced from $v_0 = 0.73$ to $v_0 = 0.18$. This solution is qualitatively
similar to that in Figure \ref{fig-admod} in the outermost region where IMHD
is dominant; the values of the angular momentum and the enclosed mass plateau
as the magnetic braking and ambipolar diffusion start to become important to
the flux and angular momentum transport. In this region the self-gravity of
the disc and the gravity of the central mass dominate the forces on the radial
velocity so that it increases rapidly, and rotation is not yet important to
the dynamics of collapse. 

The magnetic diffusion shock occurs further from the origin in this solution
at $x_d = 0.46$, which is much closer to the estimated position $x_d \approx
0.49$ than in the previous similarity solution. This occurs because the
reduced rotational support allows more gas and flux to fall inwards so that
the ambipolar diffusion term becomes important sooner, triggering the shock. 
Ambipolar diffusion comes to dominate the field transport inwards of the
magnetic diffusion shock as the increase in the vertical field component at
the shock is stronger in this solution than in the previous solution. The
post-magnetic diffusion shock transition region is smoothed by the reduction
in the initial angular momentum --- the radial velocity only decreases very
slightly downstream of the shock before accelerating inwards once more. 

The decrease in the initial azimuthal velocity changes the width of the free
fall region between the two shocks, as there is less angular momentum to
trigger the formation of the centrifugal shock. The azimuthal field component
reaches its capped value at around $x \sim 6\times 10^{-3}$ (where $xb_z$ has
its maximum turning point in Figure \ref{fig-kk8a}), so that it is better able
to transport the angular momentum from the disc than in Figure
\ref{fig-admod}. In this solution $j$ falls to the second plateau value given
by $j_{pl2} \approx 0.1$, which matches the predicted value quite well (see
subsection \ref{ad:shock} for more on the formation of this angular momentum
plateau). 

The centrifugal force becomes large enough to balance gravity and cause the
formation of the centrifugal shock at $x_c \approx 1.75 \times 10^{-3}$, which
is an order of magnitude smaller than in the moderately rotating similarity
solution, and also smaller than the $v_0 = 0.1$ IMHD solution in Figure
\ref{fig-imhdslow}. The horizontal scale in the post-centrifugal shock region
has been expanded to show that there is no overshoot and adjustment period
inwards of the shock; the variables tend directly towards their asymptotic
solutions. The Keplerian disc in this similarity solution is much smaller than
in Figure \ref{fig-admod} (corresponding to only $r_c = 7$ AU at a time $t =
10^5$ yr), and it contains a mass that is only $\sim 2\%$ of the central mass
$m_c = 6.0$  \citep[corresponding to $\dot{M}_c \approx 9.8 \times 10^{-6}$
M$_\odot$ yr$^{-1}$; values from][]{kk2002}. The central point mass has
increased compared with the more rapidly rotating similarity solution; this
also occurred in the slow and fast IMHD solutions, suggesting that the
reduction in the initial azimuthal velocity causes a reduction in the radial
support which allows more matter to accrete in the slower similarity
solutions. 

The third solution, plotted in Figure \ref{fig-kk8b} \citep[figure 8b
of][]{kk2002} is another slowly rotating collapse, with initial conditions
matching those in Figure \ref{fig-kk8a} save that the ambipolar diffusion
parameter is reduced from $1.0$ to $\tilde{\eta}_A = 0.7$. Both shocks move
even further inwards as predicted by the theory outlined in subsection
\ref{ad:shock}, because the decrease in ambipolar diffusion causes the build
up of flux necessary to trigger the magnetic diffusion shock to be a slower 
process. The post-magnetic diffusion shock region in which the density
increases and the infall is slowed is more dynamic than in the previous slowly
rotating solution, however the magnetic field does not increase downstream of
the shock as in the more moderately rotating Figure \ref{fig-admod}. 

The width of the near-free fall region in logarithmic space is again increased
in comparison to the other slow similarity solution, and the degree of
magnetic braking is increased, so that it takes longer for the angular
momentum to reach its second plateau value of $j_{pl2} \approx 0.04$ and 
balance gravity. The centrifugal shock at the boundary of the Keplerian disc
occurs at $x_c = 2.6 \times 10^{-4}$ (which corresponds to $r_c = 1$ AU at
$t = 10^5$ yr), showing that the amount of ambipolar diffusion (as well as the
initial rotation rate) determines the size and mass of the rotationally
supported protostellar disc, which in this similarity solution has only 0.4\%
the mass of the central point mass. The mass at the origin is unchanged by the
reduction of $\tilde{\eta}_A$, remaining constant at $m_c = 6.0$; this implies
that the amount of ambipolar diffusion does not affect the accretion rate onto
the central star. Again, the variables match rapidly onto their asymptotic
values, the surface density within the disc is decreased and the infall
velocity is higher than in the other slow solution in Figure \ref{fig-kk8a}. 

The final ambipolar diffusion solution in Figure \ref{fig-kk9} is a rapidly
rotating collapse with an initial azimuthal velocity $v_0 = 1.5$ and magnetic
braking parameter $\alpha = 0.1$, matching the parameters of the IMHD solution
presented in Figure \ref{fig-imhdfast}. The ambipolar diffusion parameter is
restored to $\tilde{\eta}_A = 1.0$ so that the amount of diffusion is high,
however it is not large enough to generate a magnetic diffusion shock as in
the previous solutions. The infall rate in the outer regions is slow, and is
already dropping towards zero when the gas encounters the centrifugal radius. 
As the gas is rapidly rotating, the centrifugal force becomes large early in
the collapse (as in the fast IMHD solution), and the centrifugal shock causes
the already slow radial velocity to become a backflow, representing a ring of
material downstream of the shock forcing it outwards.

As most of the momentum of the gas is in the azimuthal direction, the radial
velocity is always slow and subsonic (even in the backflow region) in this
solution. Downstream of the centrifugal shock the magnetic braking reduces the
angular momentum as the density and magnetic field (which are still tied by
flux-freezing until around $x \sim 0.2$) increase until ambipolar diffusion
becomes important and the variables settle with a number of overshoots to the
asymptotic values at around $x \sim 0.03$. Note that the disc mass is much
larger (by an order of magnitude) than the central point mass which has $m_c =
0.5$, and that at the inner edge of this plot the enclosed mass has yet to
plateau to its asymptotic value. Although the accretion rate onto $m_c$ is
slow ($\dot{M}_c \approx 8 \times 10^{-7}$ M$_\odot$ yr$^{-1}$), the disc is
very large in comparison, so that by a time of $10^5$ years the disc has
radius $r_c = 6000$ AU and is orbiting a ``protostar'' of mass $M \approx 8
\times 10^{-2}$ M$_\odot$, which is startlingly small. Without the requirement
of axisymmetry, such a disc would experience gravitational instability and
fragment, leading to the formation of a small cluster of stars
\citep[e.g.][]{mh2003}. 

The addition of ambipolar diffusion to the model changes the dynamics of the
gravitational collapse process by decoupling the magnetic field from the
neutral particles at moderate densities so their behaviour is no longer tied
by flux freezing. Because of this, less magnetic flux is carried inwards to
the origin than in the IMHD model, and the magnetic flux problem of star
formation is essentially resolved. The change in the field behaviour also
allows more angular momentum to be transported to the external envelope by
magnetic braking, which helps solve the angular momentum problem. The inner
Keplerian disc is changed by the inclusion of ambipolar diffusion, which
affects the radius of the disc and the surface density and infall velocity of
the gas inwards of the centrifugal shock. These are all expected to be further
changed by the inclusion of Hall diffusion, which alters the magnetic
behaviour of the collapsing core further from that in the simple models. 

\section{Summary}

Three forms of gravitational collapse were described within this chapter: a
nonmagnetic model, an ideal MHD model, and finally one with ambipolar diffusion
as the dominant flux transport mechanism in the collapsing flow once the
density is high enough that flux freezing breaks down. These were intermediate
models constructed as part of the process of building and testing a model of
rotational molecular cloud core collapse that shall demonstrate the influence
of both ambipolar and Hall diffusion, which shall be described in the
following chapter. The solutions here were reconstructions of those by
\citet{kk2002}, and their results were duplicated in order to test the code
calculations. Comparisons were also made to other collapse simulations in the
literature, demonstrating the power of the self-similar formulation as a tool
for exploring the physics of star formation over a wide range of densities and
length and time scalings. 

In order to ensure that the similarity solutions matched onto appropriate
inner boundary conditions, the inner asymptotic power law solutions describing
the collapse behaviour close to the origin had to be derived; for each of the
models these took the form of rotationally-supported accretion discs. The
outer boundary of these discs was marked by a shock discontinuity that
occurred near the centrifugal radius, and the exact position of this shock 
determined the matching onto the inner asymptotic solution. The jump
conditions, and an estimation of the centrifugal shock position were derived,
and for the ambipolar diffusion model an estimate of the magnetic diffusion
shock position was also derived.

The ambipolar diffusion model required the calculation of a simplified model
for the purpose of estimating the variables at a matching point located
between the two shocks, which is used as the initial guess in the shooting
routine. Inside of the centrifugal shock, a second simplified model is used to
perform the innermost integration and ensure convergence on the true
similarity solution. The numerical procedures in each version of the model
code were briefly described, along with a discussion of the numerical
instabilities and complexities that limit the precision of the similarity
models. Finally, similarity solutions of interest in each case were presented
and the physics discussed, in order to explore the effect of the magnetic
field transport, the magnetic braking and the initial rotation rate in the
core on the collapse process. 

The nonmagnetic model (Figures \ref{fig-nmslow} and \ref{fig-nmfast}) has no
way to brake the angular momentum in the gas, so the angular momentum-to-mass
ratio, $\Phi$, is a constant that parameterises the similarity solutions. The
pseudodisc thickness is dominated by its self-gravity, as is the radial
velocity. In these solutions the fluid falls in from the outer asymptotic
solution describing a rotationally-flattened core undergoing supersonic
collapse just before point mass formation. The centrifugal force builds up
until it is able to balance the gravitational force, triggering the formation
of the centrifugal shock. 

The shock represents a sudden deceleration of the infalling matter as it
encounters a wall of increased density at the boundary of the inner disc,
which is supported by the radial pressure and rotation. The variables quickly
settle to their asymptotic disc behaviour, and as there is no way to brake the
angular momentum the gas cannot fall inwards through the disc to the origin,
so that no central mass is able to form. The disc has a constant (with $x$)
azimuthal velocity $v_\phi$, which depends only upon the initial ratio $\Phi$.

The size of the inner disc depends upon the initial azimuthal velocity in the
core relative to the mass. As the ratio $\Phi$ increases, the centrifugal
force becomes important earlier in the collapse, resulting in a larger disc.
If this initial velocity is sufficiently high then the material can be pulled
along after the shock in a backflow for quite some time before the gravity of
the protostar causes it to lose its outward radial momentum and settle onto
the centrifugally-supported disc. This backflow appears in all of the
similarity solutions discussed in this work that have a high initial rotation
rate. 

The ideal MHD model (Figures \ref{fig-imhdslow}--\ref{fig-imhdfastalpha}) saw
the introduction of a magnetic field to the collapsing core. The field is
frozen into the material, moving with it so that the nondimensional
mass-to-flux ratio $\mu$ remains constant throughout the collapse; a magnetic
braking term included in the angular momentum equation allows for the removal
of angular momentum from the thin disc to the external envelope by Alfv\'en
waves. This braking allows matter from the rotationally-supported disc to lose
its angular momentum and spiral down to the origin, creating a central point
mass of nondimensional mass $m_c$. 

In the IMHD solutions the inner disc is in near-Keplerian rotation, as the
magnetic pressure term in the radial momentum equation aids the centrifugal
force in supporting the disc against collapse. As flux is dragged into the
central point mass with the gas (creating a magnetic flux problem) the field
takes on a split monopole form with the field lines strongly inclined in the
radial direction. The dominance of the radial magnetic field component over
the vertical component also appears in the discs of the ambipolar diffusion
model, and is strong enough in both models to drive a centrifugal disc wind
from the surface, although such behaviour is not explored in these models. 

The amount of angular momentum in the initial core continues to affect the
size of the protostellar disc, with rapidly rotating cores forming larger
discs. It also affects the central point mass so that those cores with a
higher initial rotational velocity form larger discs with smaller point
masses, as accretion through the disc is low and subsonic. Reducing the amount
of magnetic braking further reduces the central point mass, and causes the
backflow region outside of the disc, in a core that was initially rotating
rapidly, to be extended so that the magnetic and rotationally-supported disc
is much smaller than in those solutions with higher rates of magnetic braking. 

The inclusion of ambipolar diffusion in the similarity solutions (Figures
\ref{fig-admod}--\ref{fig-kk9}) reduces the magnetic flux problem induced in
the IMHD solutions by redistributing the flux in the central protostar
throughout the inner disc and the surrounding near-free fall collapse region.
The boundary between the ambipolar diffusion-dominated free fall and the
outermost dynamic collapse where IMHD is still dominant is marked by the
magnetic diffusion shock, which is a thin (but numerically resolvable)
discontinuity in the vertical magnetic field strength. The magnetic field is
increased rapidly by the decoupling of the field from the infalling neutral
particles; this causes a compression of the pseudodisc thickness by magnetic
squeezing and an increase in the amount of magnetic braking affecting the
collapsing flow. This additional shock is a feature of the ambipolar diffusion
similarity solutions that is not present in the simpler models; the infalling
gas is slowed as the density increases in a post-shock transition region
before the gravity of the protostar comes to dominate the radial acceleration
and the matter begins to infall at near-free fall speeds. 

The mass of the protostar and the size of the rotationally-supported
protostellar disc both depend on the amount of ambipolar diffusion in the
flow, as the decoupling of the field from the neutral particles causes an
increase in the magnetic force that slows the inflow and reduces the accretion
rate onto the protostar. The size of the disc depends also on the amount of
magnetic braking and the initial rotation of the core: reduced braking or a
larger value of $v_0$ correspond to a larger rotationally-supported disc;
stronger braking or a low initial rotation rate lead to a smaller disc, or the
prevention of disc formation entirely. The density, scale height and infall
velocity within the inner Keplerian disc all depend upon the ratio of the
ambipolar diffusion parameter to the azimuthal field cap, which limits the
amount of magnetic braking in the disc; the surface density scales with
$x^{-3/2}$ as expected in protostellar accretion discs. The disc is
rotationally-supported with a low infall speed, so that the angular momentum
problem is effectively resolved. 

A clear problem with all of the models is the cap on the azimuthal field
parameter, which limits the amount of angular momentum that can be removed
from the collapsing core and ensures disc formation. Although not duplicated
in this work, \citet{kk2002} also produced a similarity solution with $\alpha
= \delta = 10$ and $\tilde{\eta}_A = 0.5$, in which the magnetic braking was
so strong that all angular momentum was removed from the gas and no
rotationally-supported disc could form (see their \S3.3.3 and Section
\ref{catastrophe} for a discussion of this similarity solution). Similarly,
the numerical simulations of \citet{ml2008, ml2009} and others have shown that
magnetic braking is more than capable of suppressing disc formation, which
could have serious consequences for studies of protoplanet formation.
Overstrong magnetic braking may be responsible for observations of slowly
rotating YSOs without discs \citep[e.g.][]{smvmh2001}, as such stars could
have formed in cores that were so strongly braked that not only were they
unable to form a rotationally-supported disc but the angular momentum of the
protostar itself was braked during collapse. 

This magnetic braking catastrophe is unresolved at present, however, Hall
diffusion has been neglected in previous simulations of star formation. Hall
diffusion changes the amount of magnetic braking affecting the collapsing
flow, either increasing or decreasing it depending upon the sign of the Hall
parameter; this changes the dynamics of the collapse by allowing more or less
matter to fall onto the central protostar, and affects the infall rate through
the inner Keplerian disc by providing more or less magnetic support. This
behaviour is explored in the following chapter in which Hall diffusion is
introduced into the self-similar collapse model.

\cleardoublepage

\chapter{Collapse with the Hall Effect}\label{ch:hall}

The introduction of Hall diffusion modifies the behaviour of the collapsing
fluid, changing the size and strength of the magnetic diffusion shock, the
effect of magnetic braking and the radius of the rotationally-supported disc
in the central regions of the collapse. The sign of the Hall parameter, which
corresponds to the orientation of the magnetic field with respect to the axis
of rotation, can reduce the complexity of the similarity solutions when it is
negative or add to it when the Hall parameter is positive. Additional shocks
may form in the post-shock regions downstream of the magnetic diffusion and
centrifugal shocks; these slow the infall and reduce the accretion onto the
central protostar. 

The model that is used to find the similarity solutions is a modified version
of that used in Section \ref{ad} to calculate the solutions with only
ambipolar diffusion. The full set of self-similar equations derived in Chapter
\ref{ch:derivs} are integrated from a matching point across a wide range of
$x$ until they match onto the initial supersonic collapse of a molecular 
cloud core at the outer boundary and the Keplerian disc described in Chapter
\ref{ch:asymptotic} at the inner boundary. Only the induction equation and the
azimuthal field component are changed from their counterparts in the ambipolar
diffusion model, as both of these need to account for Hall diffusion: 
\begin{align}
    \psi - xwb_z + \tilde{\eta}_Hxb_{\phi,s}b_zbh^{1/2}\sigma^{-3/2}
    + \tilde{\eta}_Axb_z^2h^{1/2}\sigma^{-3/2}\left(b_{r,s} 
    - h\frac{db_z}{dx}\right) = 0 \label{in}
\end{align}
and
\begin{align}
   b_{\phi,s} = -\min\!\left[\! \frac{2\alpha\psi}{x^2}\!\left[\frac{j}{x} 
    -\frac{\tilde{\eta}_Hh^{1/2}b}{\sigma^{3/2}}\!\left(\!b_{r,s} 
    -h\frac{db_z}{dx}\!\right)\!\right]\!\!
    \left[1 + \frac{2\alpha\tilde{\eta}_Ah^{1/2}\psi{b_z}}{x^2\sigma^{3/2}}
    \right]^{-1}\!;\delta{b_z}\!\right] \label{ha-b_phis}
\end{align}
(Equations \ref{ssin} and \ref{ssb_phis}). The other equations remain
unchanged from those used in the ambipolar diffusion model; they are
reproduced here from Equations \ref{sspsi}--\ref{ssb_r} for ease of reference:
\begin{align}
   \frac{d\psi}{dx} &= xb_z, \label{cf} \\
   \frac{dm}{dx} &= x\sigma, \label{cm} \\
   (1-w^2)\frac{1}{\sigma}\frac{d\sigma}{dx}
      = g &+ \frac{b_z}{\sigma}\left(b_{r,s} - h\frac{db_z}{dz}\right)
      + \frac{j^2}{x^3} + \frac{w^2}{x}, \label{crm} \\
   \frac{dj}{dx} &= \frac{1}{w}\left(j 
      - \frac{xb_zb_{\phi,s}}{\sigma}\right), \label{cam} \\
   \left(\frac{\sigma{m_c}}{x^3} - b_{r,s}\frac{db_z}{dx}\right)h^2
      &+ \left(b_{r,s}^2 + b_{\phi,s}^2 + \sigma^{2}\right)h - 2\sigma = 0,
      \label{vhe} \\
   m &= xw\sigma, \label{m} \\
   \dot{m} &= -xu\sigma, \label{dotm} \\
   g &= -\frac{m}{x^2} \label{g} \\
   \text{and }b_{r,s} &= \frac{\psi}{x^2}. \label{ha-b_rs}
\end{align}

This chapter aims to show the importance of the Hall effect in gravitational
collapse by outlining the construction and results of the model with Hall
diffusion. The inner asymptotic solution for the Keplerian accretion disc
shall be briefly recapped in Section \ref{ha:inner} before the numerical
procedure for finding the similarity solutions is outlined in Section
\ref{numerics}. This is followed by a discussion of the behaviour of the
shocks and their positions in Section \ref{shocks}, which also contains an
exploration of the importance of the Hall term in determining the size of the
centrifugally-supported disc and the region bounded by the magnetic diffusion
shock. Finally, a series of similarity solutions are presented and discussed
in Section \ref{hall},  demonstrating how the Hall term changes the dynamics
of the flow, and how the orientation of the field with respect to the axis of
rotation influences the size of the accretion disc. Hall diffusion also
introduces additional shocks into the flow and modifies the accretion rate
onto the central protostar; these will be briefly discussed in the context of
the magnetic braking catastrophe, which will be described more fully in
Chapter \ref{ch:discuss}. 

\section{Inner Asymptotic Solution}\label{ha:inner}

The inner asymptotic solutions to the collapse with both Hall and ambipolar
diffusion were the focus of the derivation and discussion contained in Chapter
\ref{ch:asymptotic}. As this work is principally concerned with the influence
of the Hall effect on the disc formation problem in star formation, only those
similarity solutions in which a Keplerian disc forms around the protostar are
calculated in this chapter. These solutions are characterised by the existence
of an accretion disc around the central mass which is described by the set of
power laws with respect to $x$ in Equations \ref{kep-ssimm}--\ref{kep-ssim},
which are reproduced here for clarity:
\begin{align}
    m &= m_c, \label{in-m}\\  
    \dot{m}&= m_c, \label{in-mdot}\\
    \sigma &= \sigma_1\,x^{-3/2} 
     = \frac{\sqrt{2m_c}f}{2\delta\sqrt{(2\delta/f)^2 + 1}}\,x^{-3/2}, 
     \label{in-sigma}\\
    h &= h_1\,x^{3/2} 
     = \left(\frac{2}{m_c[1+(f/2\delta)^2]}\right)^{\!1/2}x^{3/2}, \label{in-h}\\
    u &= -\frac{m_c}{\sigma_1}\,x^{1/2}, \label{in-u}\\
    v &= \sqrt{\frac{m_c}{x}}, \label{in-v}\\
    j &= \sqrt{m_cx}, \label{in-j}\\
    \psi &= \frac{4}{3}b_zx^2, \label{in-psi}\\
    b_z &= \frac{m_c^{3/4}}{\sqrt{2\delta}}\,x^{-5/4}, \label{in-bz}\\
    b_{r,s} &= \frac{4}{3}\,b_z \label{in-brs}\\
    \text{and } b_{\phi,s} &= -\delta\,{b_z}. \label{in-bphis}
\end{align}
The constant $m_c$ is the nondimensional central mass infall rate, and
$\delta$ is the artificial cap placed upon $b_{\phi,s}$ to prevent it from
becoming the dominant field component in the innermost regions of the disc.
The diffusion constant $f$ is a function of the ambipolar and Hall diffusion
parameters, and is given by the equation 
\begin{equation}
    f = \frac{4}{3}\,\tilde{\eta}_A 
      - \delta\tilde{\eta}_H\sqrt{\frac{25}{9} + \delta^2}; \label{in-f}
\end{equation}
this definition shows how the Hall term is able to counteract the ambipolar
diffusion term in determining the surface density of the disc and the
accretion rate onto the central protostar when the nondimensional Hall
parameter $\tilde{\eta}_H$ is positive, and add to the ambipolar diffusion if
the Hall parameter is negative. The characteristic parameter of the disc, $f$,
must be positive lest the surface density be negative; this places limits on
the allowed relative sizes of the two diffusion parameters that must be
satisfied in order to form a Keplerian disc. 

The full physical meaning of this solution is discussed at length in Section
\ref{kepdisc}. To summarise, the inner accretion disc is in Keplerian
rotation, with the centrifugal force balancing the inwards pull of gravity.
The accretion rate onto the protostar is constant and low, and accretion
through the disc is determined by the total amount of magnetic diffusion,
which removes radial support by the magnetic field and allows the gas to fall
inwards. 

Section \ref{ff} described another asymptotic solution in which the magnetic
braking is so strong that all angular momentum is removed from the gas which
free falls onto the protostar without forming a disc. No similarity solutions
matching onto this inner solution are calculated in this work, but the
similarity solution with no Hall diffusion but strong magnetic braking, and
its role in describing the magnetic braking catastrophe are discussed further
in Chapter \ref{ch:discuss}. 

The Keplerian disc equations, in particular Equations \ref{in-m} and
\ref{in-psi}, are the inner boundary conditions that must be evaluated to test
the accuracy of any individual integration and satisfied in the true
similarity solution. The method in which these are employed to ensure
convergence of the shooting routine for finding the similarity solutions shall
be described in the following section. 

\section{Numerical Method}\label{numerics}

The numerical procedures used to calculated the similarity solutions are
relatively unchanged from those used in the ambipolar diffusion model, save
for those changes that were necessary to handle the additional complications
introduced with Hall diffusion. As mentioned previously, sonic points and
subshocks occur downstream of the magnetic diffusion and centrifugal shocks,
which may be the ``viscous subshock'' mentioned in \citet{kk2002}, although
these do not appear in their published solutions. Locating and integrating
through such shocks requires the introduction of new routines that monitor the
integration and perform such adjustments as are necessary. 

Overall the iterative routine for finding the correct similarity solution
remains unchanged, using the globally convergent multidimensional
Newton-Raphson root-finding routine (\texttt{newt} and its dependences,
converted to double precision) from \textit{Numerical Recipes} \citep{NR}.
This routine performs a series of integrations (using the routine
\texttt{shoot}) from the unknown initial values to the boundaries whereupon
the difference between the integrated variables and the boundary conditions
are evaluated. Each of the initial values is perturbed in turn, and the
Jacobian of the shooting routine is calculated; \texttt{newt} then uses the
inverse of the Jacobian to determine the changes to the initial values
necessary to zero the discrepancies. This step is then taken and assessed (a
smaller change in the variables is made if needed) and the process is repeated
until the shooting routine converges on the true solution. Once the initial
values are close enough to the true solution the Newton-Raphson routine will
converge quadratically. 

\begin{figure}[htp]
  \centering
  \includegraphics[width=4in]{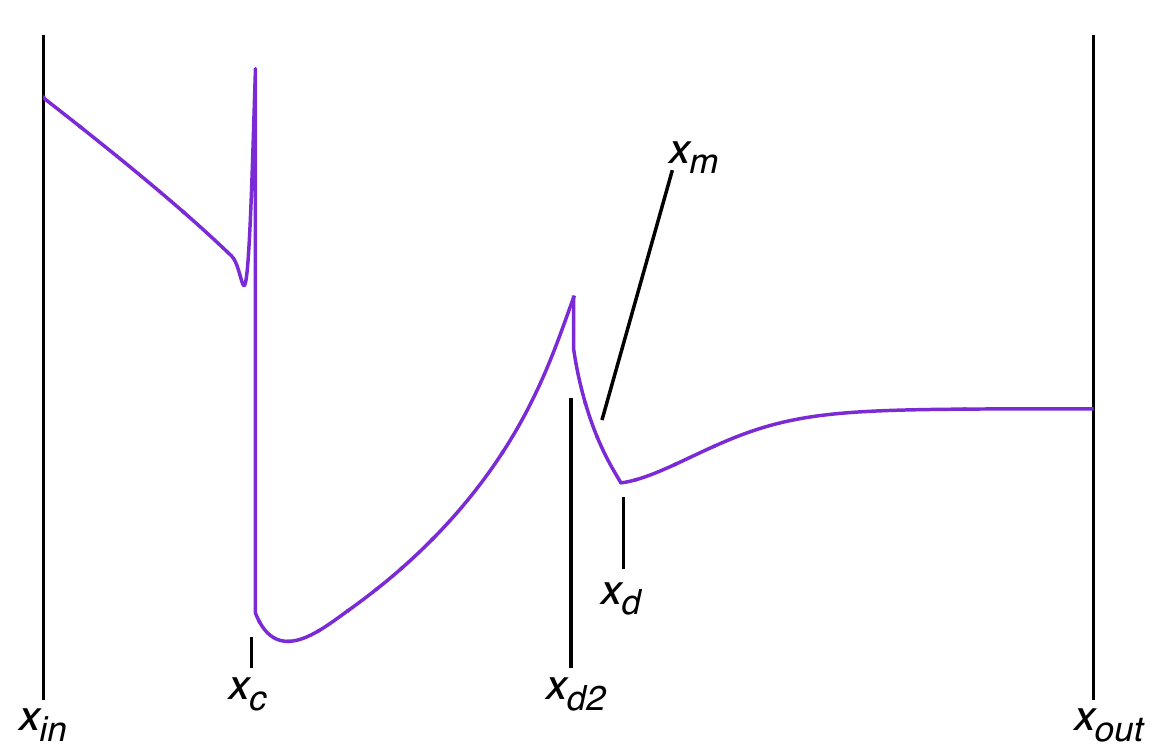}
  \caption[Two-point boundary value problem]{Schematic of the two-point
boundary value problem, showing the integration of a variable (in this case 
$\sigma$) from the matching point $x_m$ to the inner and outer boundaries. In
the outwards direction the integration is slowed only by the continuous
magnetic diffusion shock at $x_d$. Inwards of the matching point the
integration encounters the centrifugal shock at $x_c$; it may also need to
integrate through sonic points and subshocks (e.g.\ $x_{d2}$) downstream of
both the magnetic diffusion and centrifugal shocks.} 
\label{fig-xmhall}
\end{figure}
As in the ambipolar diffusion model, the problem is recast as a two-point
boundary value problem in which the variables are integrated from a matching
point $x_m$ to both the inner and outer boundaries, as depicted in Figure
\ref{fig-xmhall}. The matching point is located at a position $x_c < x_m <
x_d$, and is typically chosen to lie just downstream of the magnetic diffusion
shock and upstream of any subshocks that may occur. The values of the
variables $m$, $\sigma$, $j$, $\psi$ and $b_z$ at $x_m$ are initially unknown
but may be estimated by integrating a simplified set of equations as shall be
outlined in subsection \ref{num:simp}. The initial value of the central mass
parameter $m_c$ is estimated using the plateau value defined in Section
\ref{nonmag}: 
\begin{equation}
    m_c \approx m_{pl} \approx 2|u_0|A. \label{mcguess}
\end{equation}

The integration itself is performed using either a fifth order Runge--Kutta
step (\texttt{rkqs}) or a Bulirsch--Stoer semi-implicit midpoint rule step
(\texttt{stifbs}) as the equations may be stiff, requiring a different method
of integration. The driver routine for both is \texttt{odeint}, which exerts
an adaptive stepsize control to improve the accuracy and efficiency of the
integration; all of these routines and their dependencies were taken from
\textit{Numerical Recipes} \citep{NR} and have been converted to double
precision. 

Integration in the outwards direction is usually performed without difficulty
unless a particularly poor guess of the values of the variables at $x_m$ is
employed. As the magnetic diffusion shock is continuous, both routines for
stepping through the integration are able to integrate through it without
pause. At the outer boundary (which is usually located at $x_{out} = 10^4$ or
$10^5$) the variables are compared to the outer boundary conditions outlined
in Section \ref{eqouter}: 
\begin{align}
    m &= Ax_{out}, \label{outm2}\\
    \sigma &= \frac{A}{x_{out}}, \label{outsigma2}\\
    b_z &= \frac{\sigma}{\mu_0} \label{outbz2}\\
    \text{and } v &= v_0 \label{outv2};
\end{align}
the difference between these variables and their boundary values is passed to
the root-finding routine as a score that must be minimised. Both \texttt{rkqs}
and \texttt{stifbs} give similar results so they may be used interchangeably;
\texttt{stifbs} is typically used for the outer integration, while
\texttt{rkqs} is used for the inwards path, as \texttt{stifbs} occasionally
fails downstream of the centrifugal shock. 

Integrating in the inwards direction is more problematic, as the calculation
is very sensitive both to the initial values of the variables at $x_m$ and the
Hall diffusion parameter $\tilde{\eta}_H$. Close to the magnetic diffusion
shock (downstream of $x_m$) there may occur a subshock in which the supersonic
(but slowing due to the sudden increase in $b_z$ at the magnetic diffusion
shock) inflow is abruptly slowed to a subsonic rate. This only occurs when the
Hall parameter is positive, and is likely caused by the change in the
azimuthal field in the shock affecting the downstream magnetic braking. This
subshock is a sharpening of the post-shock variation of the density and infall
speed in the ambipolar diffusion solution in Figure \ref{fig-admod}, and its
position and jump conditions (discussed in Section \ref{shocks}) must be
calculated explicitly using the method to be described in subsection
\ref{num:iter}. 

The magnetic diffusion subshock's existence is detected by performing a test
integration inwards; if the variables approach a sonic point (where $w^2 = 1$)
then a shock must exist before this point. Downstream of the subshock a sonic
point occurs as the radial velocity becomes supersonic once more; this is
integrated through using the method described in subsection \ref{num:sub}.
Past this point, the variables are integrated without any further interruption
until they reach the centrifugal shock. 

The centrifugal shock is located by integrating to an upper bound on the shock
position and then calling the iterative routine outlined in subsection
\ref{num:iter} which tests the initial guess of the shock position $x_c$ and 
refines this guess based upon the downstream behaviour of the variables. Once
$x_c$ is known as precisely as possible, the variables are integrated to the
shock position and the jump conditions (again described in subsection
\ref{s:jump}) are evaluated. Downstream of this shock, if the Hall parameter
is positive, one or more subshocks may occur. As with the magnetic diffusion
subshock, the centrifugal subshocks are preceded by a sonic point that must
be calculated carefully and then the subshock position is found using the same
iterative process as the centrifugal shock. Inwards of these shocks the
variables approach the expected asymptotic behaviour outlined in Section
\ref{ha:inner}.

In the innermost regions of the collapse the derivatives $d\sigma/dx$ and
$db_z/dx$ become very large in comparison to the other derivatives. These can
cause $\sigma$ and $b_z$ to become large, and small numerical errors in the
calculation of these derivatives and their integrals can cause the appearance
of a spontaneous singularity in which the variables diverge dramatically from
the asymptotic similarity solutions. This behaviour is avoided by simplifying
the calculation of the derivatives in this region using the expected
asymptotic behaviour of the variables; the simplified equations, and the
criteria used to match onto this simple model are described in subsection
\ref{num:insimp}. 

When the inwards integration reaches the inner boundary (typically located
around $x_{in} = 10^{-4}$, depending upon the position of $x_c$) the variables
are compared to the inner boundary conditions 
\begin{align}
    m(x_{in}) &= m_c \label{inm} \\
    \text{and }\psi(x_{in}) &= \frac{4m_c^{3/4}}{3\sqrt{2\delta}}
         \,x_{in}^{3/4}. \label{inpsi}
\end{align}
The difference between the expected boundary value and the integrated variable
are passed back to the shooting routine, which modifies the initial values at
$x_m$ and begins the next iteration of the integration. The similarity
solution is considered to have converged when the integrated variables match
the boundary condition values to at least 0.002\%. Due to the ease of
convergence in the outwards direction, the outer boundary conditions match the
integrated variables to a much higher degree. 

Some of the earlier (low $|\tilde{\eta}_H|$) similarity solutions were
calculated using a modified shooting routine that only integrated in the
outwards direction. This method required the adoption of specific values of
$\psi(x_m)$ and $m_c$, which were held constant as the root-finding routine
minimised the scores at the outer boundary; only after that convergence was
achieved was the inwards integration performed and the values of $\psi(x_m)$
and $m_c$ altered. This process was repeated until the initial values were
properly converged. While this method was less efficient than the full version
of \texttt{shoot} that integrates in both directions, due to the complications
involved in integrating inwards (especially when multiple subshocks exist in
the solution) this was often a more reliable mechanism for solving to the full
set of boundary conditions. 

Both methods require a good initial estimate of the values of the variables at
$x_m$ and take at least a full day to compute on a desktop machine. If a
particularly poor initial guess at the matching point is chosen, or if more
than one sonic point occurs in the collapse, the calculation may require close
monitoring to ensure that integration in the inwards direction proceeds as
expected and convergence on the true solution takes place. 

\subsection{Simplified model}\label{num:simp}

The initial guess of the variables at the matching point is estimated by
calculating a simplified model in which the induction equation is replaced by
an algebraic expression for the vertical field component. As the derivative of
$b_z$ is a negligible term in the equations (see Section \ref{eqsimple}), when
the fluid equations are integrated inward from the outer boundary the
evaluation of $db_z/dx$ causes the integration to become unstable. Removing
this term allows the induction equation to be written as an algebraic
expression for $b_z$ as a function of the other variables. The other equations
are unchanged, and using the expression for $b_z$ it is possible to integrate
these from the outer boundary to the matching point $x_m$. This simplified
model is quite similar to the one constructed for the ambipolar diffusion
collapse in Section \ref{ad}, differing only in the inclusion of the Hall
diffusion term. 

With the exception of during the magnetic diffusion shock where $b_z$ changes
rapidly, the inequality $b_{r,s} \gg h(db_z/dx)$ holds true everywhere during
the collapse; and as this term is always small compared to the other terms in
the induction equation it may be dropped. The magnitude of the magnetic field
$b$ is always of order $b_z$, so that it may be approximated by $b = b_1b_z$
where $b_1$ is a constant that depends on the relative sizes of $b_{r,s}$ and
$b_z$. The induction equation may then be written as a quadratic in $b_z$: 
\begin{equation}
    xh^{1/2}\sigma^{-3/2}
     \left(\tilde{\eta}_Hb_1b_{\phi,s} + \tilde{\eta}_Ab_{r,s}\right)b_z^2
    - xwb_z + \psi = 0.
\label{hall-simpin}
\end{equation}
The two well-separated roots of this equation provide an acceptable
approximation to the behaviour of $b_z$ on either side of the magnetic
diffusion shock, with the prescription 
\begin{equation}
    b_{z,low} \approx \frac{\psi\sigma}{m} \approx \frac{\sigma}{\mu_0}
\label{bzlow}
\end{equation}
(which is unchanged from that in the ambipolar diffusion model) applying in
the large $x$ regime where flux freezing still mostly holds true and the
mass-to-flux ratio is given by its initial value $\mu = \mu_0$. This is
equivalent to the initial condition for the vertical field component on the
outer boundary derived in Section \ref{eqouter}; although IMHD breaks down
before the magnetic diffusion shock this remains a good approximation to the
field in this region. 

The larger root of the two gives the value of the vertical field component in
the magnetic diffusion regime where $x$ is small. It is approximated by
dropping the constant (with respect to $b_z$) term in the quadratic equation
(\ref{hall-simpin}) and solving for $b_z$ to obtain 
\begin{equation}
    b_{z,high} = \frac{m}{x}\left(\frac{\sigma}{h}\right)^{1/2}
     \left(\tilde{\eta}_Hb_1b_{\phi,s} + \tilde{\eta}_Ab_{r,s}\right)^{-1},
\label{bzhigh}
\end{equation}
which reduces to Equation \ref{adbzhigh1} in the ambipolar diffusion limit.

The transition between the two approximations to $b_z$ occurs at the magnetic
diffusion shock, $x_d$, the position of which is estimated in Section
\ref{shocks}. The matching point is usually taken to be $x_m \sim 0.3$
depending slightly upon the position of the magnetic diffusion shock and its
potential subshocks, and $m_c$ is estimated by the plateau value in Equation
\ref{mcguess}. Figure \ref{fig-adsimp} showed such a simple model for the
ambipolar diffusion calculations --- inwards of the matching point the
variables diverged from the expected values due to the poor guess of $x_d$
used. 

The values of the variables at the matching point from this simple model are
often a good enough guess to the true values that the code is able to use them
as the initial values for the iterative root-finding routine that solves the
full set of equations. In those cases where this is not a good estimate
(typically when $|\tilde{\eta}_H| \gtrsim 0.1$), the values of $\psi$ and
$b_z$ are adjusted by hand until the shooting routine is able to integrate in
both directions and match the boundary conditions to approximately 10\%. Once
this is achieved the root-finding routine is typically able to converge on the
similarity solution.

\subsection{Iterative routine for locating the shock position}\label{num:iter}

The position of the centrifugal shock, $x_c$, is found by employing a simple
binary search over an appropriate interval. The upper and lower bounds on this
position are initially described by $x_{c0} \pm 0.2 x_{c0}$ where $x_{c0}$ is
the estimated value that is derived in Section \ref{shocks}. Once the position
of the shock has been found for the initial guess of the variables, a smaller
pair of limits and the previous value of the shock position may be used in
later integrations, as $x_c$ does not vary greatly in its position with
changing initial values of the variables at $x_m$. The variables are 
integrated down to the estimated position of $x_c$, where the jump conditions
from subsection \ref{s:jump} are applied, and then integrated inwards towards
the inner boundary. Unless the shock position is known very precisely, the
variables will approach their asymptotic values and then veer off course.  

\begin{figure}[p]
  \centering
  \vspace{-3mm}
  \includegraphics[width=4.5in]{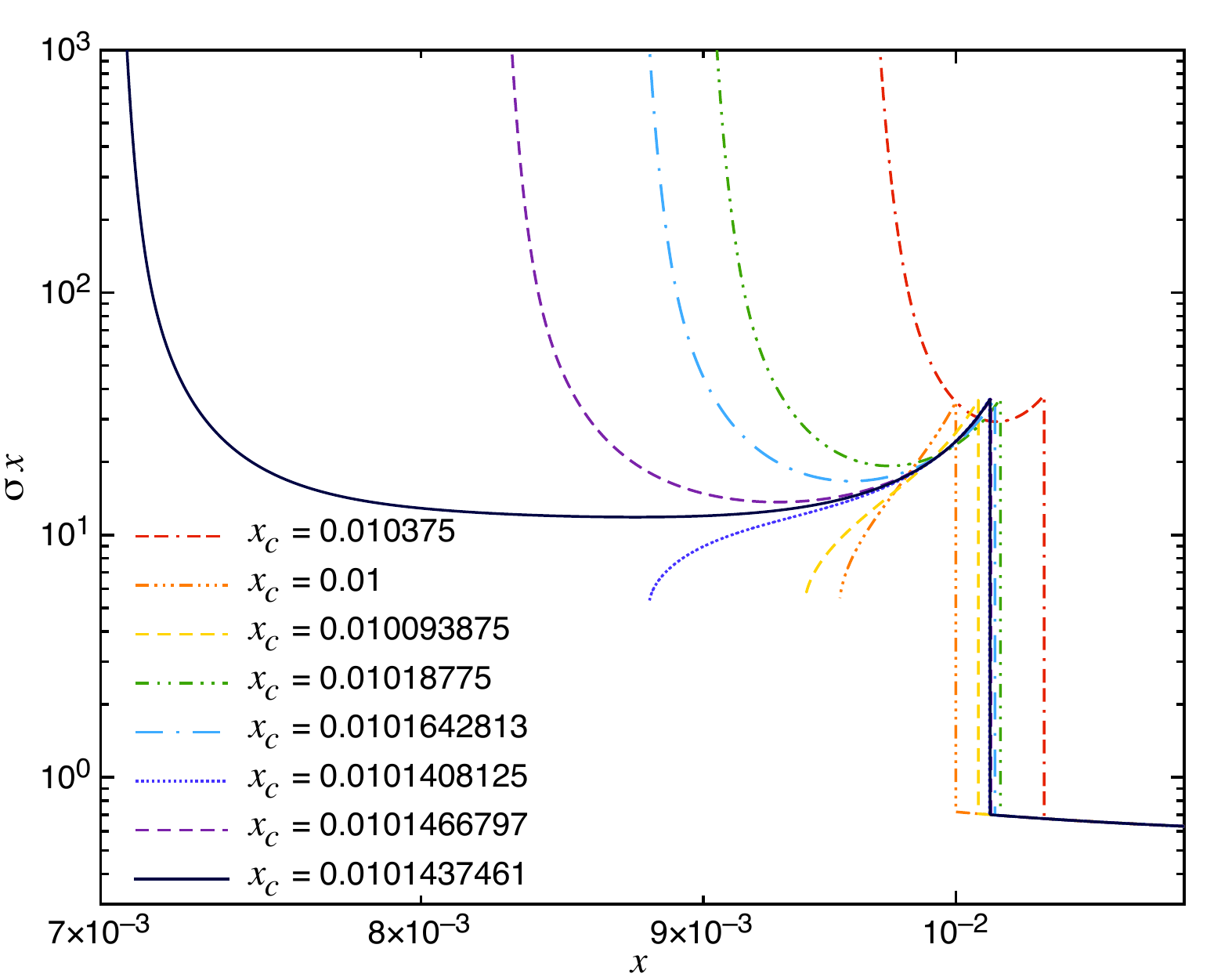}
  \vspace{-5mm}
  \caption[Locating the centrifugal shock]{Locating the centrifugal shock
position for a non-converged solution by integrating inwards and observing the
behaviour of $\sigma x$. When the estimated value of $x_c$ is too high, the
surface density diverges from its expected behaviour and becomes very large;
when it is too low, the surface density becomes small. The more accurate the
value of $x_c$ the longer the variables follow the expected asymptotic
behaviour before diverging.}  
\label{fig-sbhave}
\end{figure}

This behaviour is most clearly seen in the surface density $\sigma$, which
increases rapidly downstream of the shock if $x_c$ is an overestimate to the
true shock position, and decreases dramatically if $x_c$ is an underestimate.
The incorrect estimate is then assigned to be the new upper or lower boundary
on the shock position as appropriate, and a new estimate of the shock position
is chosen at the midpoint between the boundaries. As the position of the shock
is more precisely known, the variables follow the expected asymptotic
behaviour for longer, as can be seen in Figure \ref{fig-sbhave}. 

When the shock position is known to approximately ten decimal places, it is
considered to be known to the precision of the full calculation, as the
precision of the integration itself (typically to around $10^{-10}$) limits
the accuracy of all other calculations. Note that even when the position of
the shock is known ``precisely'' (as is the case with the solid black plot in
Figure \ref{fig-sbhave}), the integrating routine is unable to complete the
integration all the way to the inner boundary. As was the case in the
ambipolar diffusion model, a simplified set of equations need to be integrated
from a point far from the centrifugal shock, where the variables follow their
asymptotic behaviour, to the inner boundary. The simplified set of equations
with Hall and ambipolar diffusion and the matching criteria is outlined in
subsection \ref{num:insimp}. 

\begin{figure}[htp]
  \centering
  \vspace{-3mm}
  \includegraphics[width=5.5in]{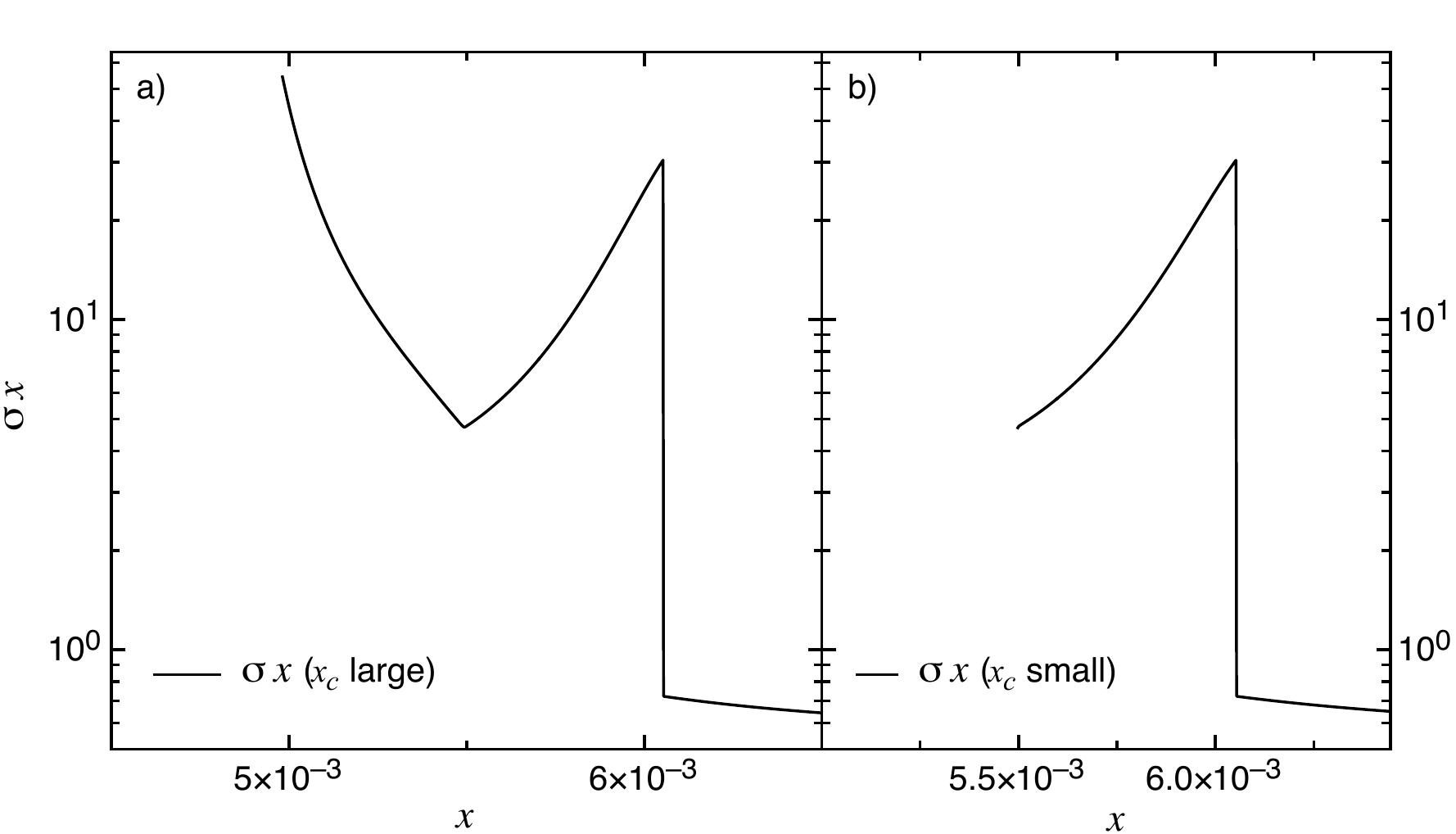}
  \vspace{-5mm}
  \caption[Locating the centrifugal shock in the presence of subshocks
]{Locating the centrifugal shock in the presence of one or more subshocks.
When the estimated value of $x_c$ is too high (panel a) the surface density
diverges from the expected behaviour at the sonic point and becomes very
large. When the estimate of $x_c$ is too low (panel b) the integration simply
fails at the sonic point. This contrasting behaviour is used to refine the
shock position.} 
\label{fig-ssbhave}
\end{figure}

The potential presence of a sonic point and subshock does not particularly
interfere with the initial iterative routine for finding the shock position.
When there exists a sonic point downstream of the shock, if $x_c$ is too high
then $\sigma x$ bounces upwards at the sonic point, as seen in Figure
\ref{fig-ssbhave}a, while if $x_c$ is too low then the integration fails at
the sonic point as demonstrated in Figure \ref{fig-ssbhave}b. The same
iterative routine may then be used to find the shock position even when the
shock is followed by one or more subshocks. 

Downstream of any of the sonic points the iterative routine must again be
employed to find the position of the associated shock front that follows. In
this instance the upper boundary of the search ($x_{up}$) is initially chosen
to be the sonic point, and the lower boundary is located at a value of
$x_{down} = 0.1x_{up}$; these typically enclose the shock. The behaviour of
the variables downstream of the subshock is the same as that downstream of the
principal shock, so the binary search described above is able to locate the
true shock position without alteration, and once $x_{c2}$ has been found the 
variables typically match onto the simplified set of equations used to
integrate to the inner boundary without difficulty. 

\subsection{Subshocks}\label{num:sub}

Perhaps the most obvious difference between those solutions in which the Hall
parameter is positive and those with no or negative Hall diffusion is the
presence of subshocks downstream of either or both of the magnetic diffusion
and centrifugal shocks as described in the previous subsection. To date only
one magnetic diffusion subshock has been found in the similarity solutions,
however many subshocks may occur downstream of the centrifugal shock. These
subshocks occur after the density increase in the principal shock causes the
gas to be slowed to a subsonic speed. The gas is abruptly halted as the
magnetic or centrifugal forces overcome the radial gravity, causing another
steep density increase. 

\begin{figure}[htp]
  \centering
  \vspace{-4mm}
  \includegraphics[width=3.6in]{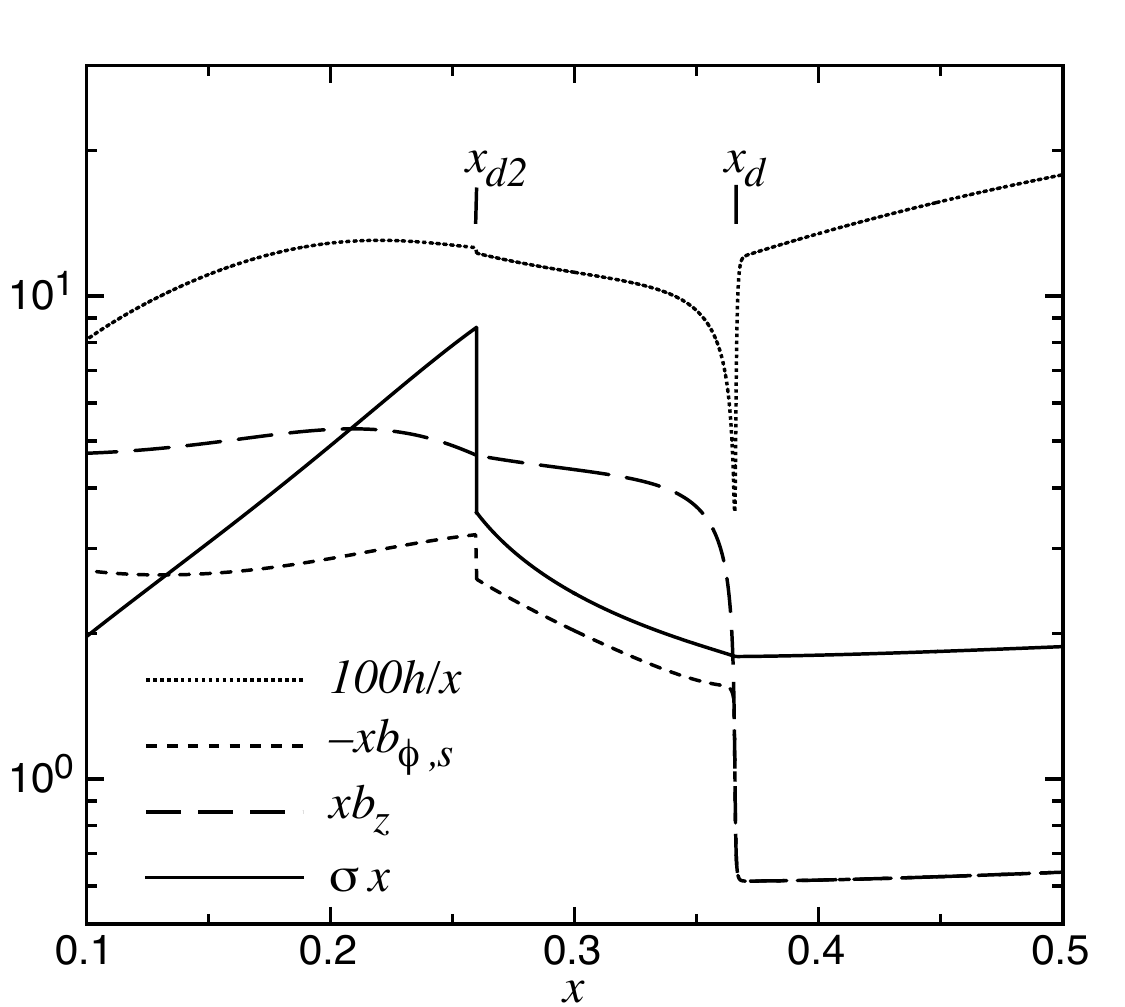}
  \vspace{-4mm}
  \caption[Magnetic diffusion subshock when $\tilde{\eta}_H = 0.2$]{Close
view of the magnetic diffusion shock at $x_d = 0.366$ and associated subshock
at $x_{d2} = 0.260$ for the similarity solution with $\tilde{\eta}_H = 0.2$
and $\tilde{\eta}_A = \delta = 1.0$. The scale height $h$ and the vertical and
azimuthal field components ($b_z$ and $b_{\phi,s}$) change at the principal
shock, while it is the scale height, the azimuthal field component and the
surface density $\sigma$ (as well as the infall speed $-u$, not pictured here)
that change values at the subshock, while the vertical field component remains
constant.} 
\label{fig-etah0.2-xd2}
\vspace{4mm}
  \includegraphics[width=3.6in]{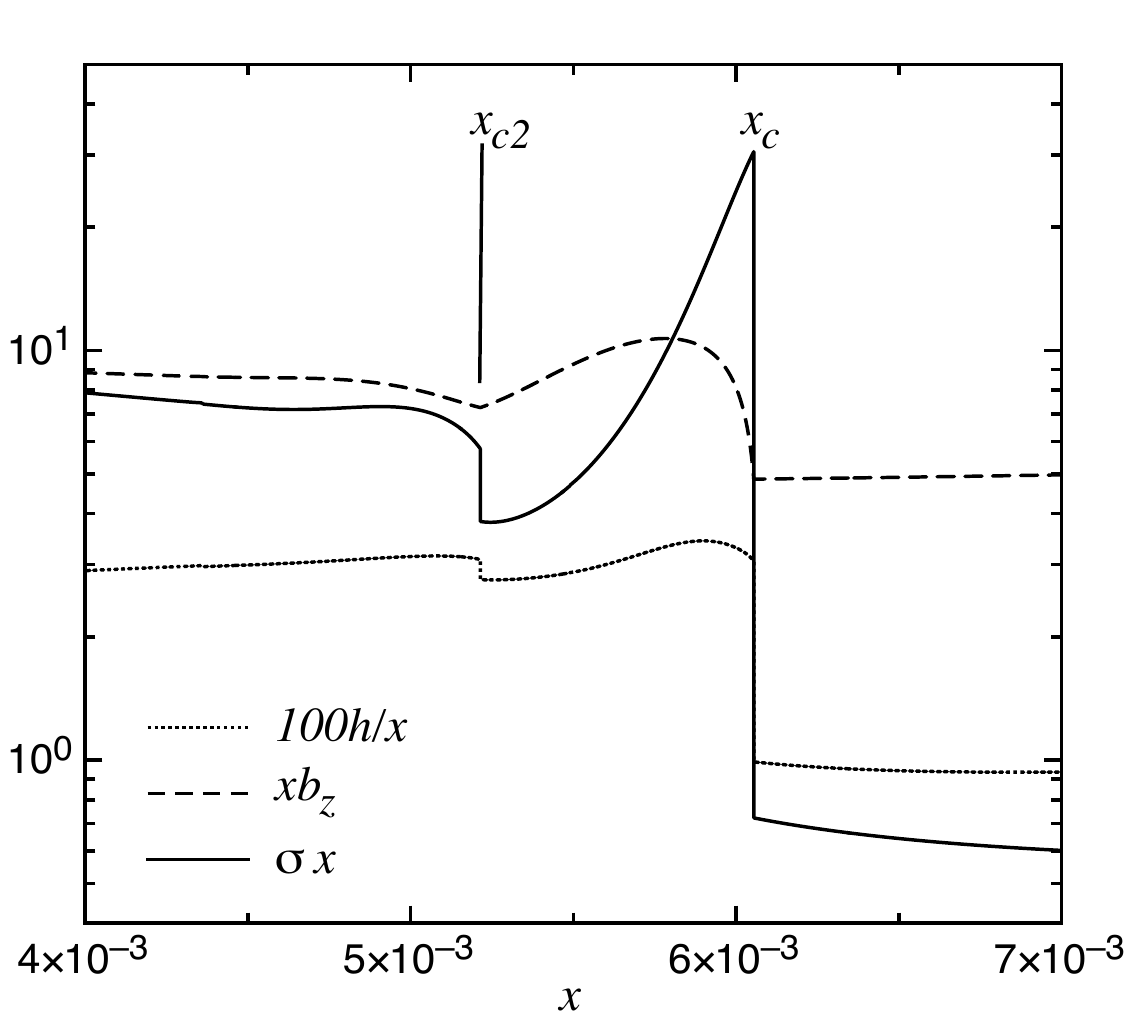}
  \vspace{-4mm}
  \caption[Centrifugal subshock when $\tilde{\eta}_H = 0.2$]{Close
view of the centrifugal shock at $x_c = 6.05 \times 10^{-3}$ and 
associated subshock at $x_{c2} = 5.21 \times 10^{-3}$ for the similarity
solution with $\tilde{\eta}_H = 0.2$ and $\tilde{\eta}_A = \delta = 1.0$. The
scale height $h$ and the surface density $\sigma$ (and the infall speed $-u$,
not pictured here) increase at each of the shocks, while the vertical field
component $b_z$ changes only in the post-shock region. By $4 \times 10^{-3}$
the variables have matched to their asymptotic behaviour.} 
\label{fig-etah0.2-xc2}
\vspace{-4mm}
\end{figure}

The first of these subshocks, downstream of the magnetic diffusion shock,
occurs when the post-shock density increase is so steep that the gas is slowed
dramatically, eventually forming a shock front. This shock can be seen in
Figure \ref{fig-etah0.2-xd2}, which is a close view of the magnetic diffusion
subshock in the $\tilde{\eta}_H = 0.2$, $\tilde{\eta}_A = 1.0$ similarity
solution presented in Section \ref{hall}. At this shock the azimuthal field
component is increased suddenly, and this straightening of the field at the
pseudodisc surface slows the fluid. There are discontinuities in the scale
height and the vertical field derivative, and the corresponding increase in
$b_{\phi,s}$ also increases the amount of magnetic braking downstream of the
shock. 

As described in the introduction to this section, the magnetic diffusion
subshock occurs when the integration tends towards a numerical sonic point
downstream of the principal shock; this sonic point would see the radial
inflow become subsonic gradually, while the code rapidly encounters an
unphysical singularity and fails. This is clearly undesirable and the solution
to this behaviour is to use the iterative routine described in the previous
subsection to find the subshock position, using the magnetic diffusion shock
position as the upper boundary and the sonic point position as the lower
boundary. Once the subshock position is known it is possible to integrate from
the magnetic diffusion shock to the subshock, and from there towards the
centrifugal shock. 

Downstream of the magnetic diffusion subshock the radial velocity increases
gradually under the influence of gravity; it passes from subsonic to supersonic
at a sonic point that must be calculated explicitly. The singularity in the
radial momentum equation (\ref{crm}), occurring when $(x-u)^2 = w^2 = 1$,
prevents the smooth integration through this point as $d\sigma/dx$ diverges;
this behaviour is prevented by checking the value of $w$ at each intermediate
point between the magnetic diffusion subshock and the sonic point. When $|w^2
- 1| < 0.01$ then a small manual step (or series of steps) is taken by
evaluating the derivatives and multiplying these by the stepsize $\Delta x <
0.001$ to give a small change to the variables. This change is added to the
variables to give the value at the new position, $x - \Delta x$, which is
written as 
\begin{equation}
    f(x - \Delta x) = f(x) - \Delta x f'(x)
\label{step}
\end{equation}
for each of the variables $f$. If $w$ has not passed through the sonic point
this manual step is repeated as necessary until it has. Once past the sonic
point the gas flow tends towards the free fall collapse behaviour as in the
ambipolar diffusion solutions. 

Similarly, those subshocks that occur downstream of the centrifugal shock
form because the density increase in the shock is narrow, and once past the
sharp peak of the shock the matter starts falling inward again at a rapid
rate. This behaviour can be seen in Figure \ref{fig-etah0.2-xc2}, which again
is a close view of the similarity solution with $\tilde{\eta}_H = 0.2$ and
$\tilde{\eta}_A = 1.0$ focussing on the centrifugal subshock; the surface
density and other variables overshoot their asymptotic behaviours and the
infall quickly becomes supersonic once more. The centrifugal force remains
important, and once the infalling material has overcome the increased magnetic
pressure caused by the increase in the field downstream of the shock it starts
to fall in under gravity once more, so that the centrifugal force causes the
gas to be shocked once more. Multiple subshocks occur as the Hall parameter
becomes large --- the increasing number of subshocks make it computationally
difficult for the iterative root-finding routine to converge on the true
similarity solution, preventing the publication of larger positive Hall
parameter solutions in this work. 

The presence of the centrifugal subshock is indicated by the presence of a
sonic point downstream of the principal centrifugal shock, in which the infall
speed seemingly passes from being subsonic to supersonic in a smooth manner.
As with the magnetic diffusion subshock, this sonic point is caused by the
singularity in the radial momentum equation and is manually integrated through
using Equation \ref{step}. The position of the subshock downstream of the
sonic point is found iteratively using the routine outlined in subsection
\ref{num:iter}, using the sonic point $x_{sp}$ as the upper bound and a
position 0.6$x_{sp}$ as the lower bound. If multiple subshocks exist (caused
by the steep increase in the magnetic field and correspondingly the magnetic
pressure) then this behaviour will be repeated, with the first subshock being
followed by an additional sonic point and then a second subshock. This
behaviour is repeated until the variables match onto the inner asymptotic
solution and can be integrated without further intervention to the inner
boundary. 

The jump conditions calculated at all of the subshocks are those jump
conditions used for the centrifugal shock that shall be outlined in Section
\ref{shocks}. It is possible that as the positive Hall parameter becomes
larger the magnetic diffusion subshock may require different jump conditions 
that take the twisting of the magnetic field lines due to the Hall effect into
account; for the solutions presented in this work such changes were
unnecessary.

Subshocks do not occur in those similarity solutions where the Hall parameter
is negative, as the reduced radial magnetic field diffusion causes the
post-shock regions to be smoothed and less dynamic. The variables match onto
the expected behaviour downstream from the shocks with smaller overshoots, and
the shocks themselves are weaker. Making the Hall parameter increasingly
negative (while keeping the ambipolar diffusion parameter constant) reduces
the severity of the magnetic diffusion shock (much as reducing
$\tilde{\eta}_A$ does). Furthermore, tweaking $\tilde{\eta}_H$ changes the
amount of magnetic braking affecting the disc, as larger negative values of
$\tilde{\eta}_H$ lead to $b_{\phi,s}$ attaining its capped value faster; the
reduced flux in the inner regions of the pseudodisc slows the rate of magnetic
braking so that the centrifugal shock occurs sooner in the collapse process. 
\raggedbottom

\subsection{Simplified inner integration}\label{num:insimp}

As can be seen in Figure \ref{fig-sbhave}, the shooting routine is sometimes
unable to integrate the equations all the way to the inner boundary, even when
the position of the centrifugal shock is known as accurately as possible. The
divergence from the true similarity solution occurs because the derivatives
$db_z/dx$ and $d\sigma/dx$ become very large downstream of the centrifugal
shock and associated subshocks; the accumulation of small numerical errors and
the contributions of small terms in the equations leads to a situation in
which the integrated variables veer away from the expected asymptotic
solutions. 

In order to avert this behaviour and integrate to the inner boundary, the
method used in the ambipolar diffusion model (described in Section \ref{ad}) 
is reproduced here for the full model with both Hall and ambipolar diffusion.
This simplification sees the MHD equations replaced by a simplified set that
can be integrated all the way to the inner boundary. In this calculation the
problematic derivatives that become large are replaced by approximations that
are derived from the expected inner asymptotic solution that was outlined in
Section \ref{ha:inner}, that is, 
\begin{align}
    \frac{db_z}{dx} &= -\frac{5m_c^{3/4}}{4\sqrt{2\delta}}\,x^{-9/4} 
     \label{simpdbz}\\
    \text{and }\frac{d\sigma}{dx} &= -\frac{3}{2}\,
     \frac{\sqrt{2m_c}f}{2\delta\sqrt{(2\delta/f)^2 + 1}}\,x^{-5/2}.
      \label{simpdsigma}
\end{align}
These are then substituted into the appropriate MHD equations so that
$\sigma$, $b_z$, $b_{\phi,s}$ and $h$ are then found by solving Equations
\ref{in}, \ref{ha-b_phis}, \ref{crm} and \ref{vhe} simultaneously, leaving
only Equations \ref{cf}, \ref{cm} and \ref{cam} to be integrated. 
\flushbottom

Switching between the full and simplified models takes place after the minimum
turning point in the surface density when the variables have joined the inner
asymptotic solutions, but before they diverge from the expected behaviour.
Typically this switch occurs when the old and new values of $\sigma$ match to
within 0.01/$x$ (which is $\sim 0.1\%$ of the value from the full set of
equations) and $d\sigma/dx$ calculated using both methods matches less well to
around 200/$x$ ($\sim 7\%$). As in the ambipolar diffusion model this means
that, in general, the transition between equation sets is visually seamless;
the required precision of the match is raised from those stated here when the
transition is visible and the change in the derivatives is apparent --- such a
slope change would clearly influence the accuracy of the calculations. When
the precision of the match is as high as possible the change between the two
equation sets is smooth. 

The simplified set of equations is not subject to the same numerical
instabilities as the full set unless the initial values of the variables at
the matching point are a particularly poor guess of the true values. Should
such a guess be adopted then the routine is unable to match onto the
simplified model, and the failure of the inwards integration is used to refine
the values at the matching point. Using a simplified set of equations to
integrate to the inner boundary is not expected to introduce significant
errors into the calculation.

The shooting routine can then continue the integration to the inner boundary
where the difference between the integrated variables and the boundary
conditions are evaluated and the values of the variables at the matching point
are adjusted.

\section{Shocks}\label{shocks}

The inclusion of Hall diffusion in the collapse changes the position and
behaviour of the magnetic diffusion and centrifugal shocks, and can cause
there to be subshocks associated with these. As mentioned previously, when the
Hall diffusion parameter $\tilde{\eta}_H$ is positive the increased magnetic
forces and braking may cause the formation of subshocks, while the strength of
the shocks is reduced and the dynamics of the post-shock regions smoothed when
the Hall parameter is negative. 

An analytic estimate of the shock positions is necessary for the calculation
of the simple model described in subsection \ref{num:simp} and as an initial
guess for the iterative routine in subsection \ref{num:iter}; these are
derived in subsections \ref{s:xd} and \ref{s:xc} for the magnetic diffusion
and centrifugal shocks respectively. The shock positions depend upon the Hall
parameter, as the outwards diffusion of the field leads to the formation of
the magnetic diffusion shock, and the magnetic braking determines the position
of the centrifugal shock, and Hall diffusion is important to both of these
processes. For simplicity, the subshocks shall not be examined analytically at
this point, as they were adequately described in the preceding sections, and
their positions are usually well-estimated by those of the principal shock
they follow. The Hall diffusion parameter does not affect the jump conditions
applied at the shocks (described in subsection \ref{s:jump}), although its
influence on the shock position and the values of the density and magnetic
field at this point does change the intensity of the shock. 

\subsection{Magnetic diffusion shock position}\label{s:xd}

The first shock encountered by the collapsing flow is the decoupling front in
which flux freezing breaks down and the magnetic field behaviour comes to be
dominated by Hall and ambipolar diffusion. Similarly to the calculation of the
magnetic diffusion shock position in the ambipolar diffusion model in Section
\ref{ad}, $x_d$ is estimated by examining the vertical magnetic field
behaviour on both sides of the shock. In the outermost regions of the core the
material is collapsing inward at a near-free fall rate, ideal
magnetohydrodynamics holds true, and the vertical field component is given by
$b_{z,low}$ defined in Equation \ref{bzlow}. Inwards of the magnetic diffusion
shock, the magnetic diffusion terms are important to the field transport and
the drift of the magnetic field depends on the coupling between field lines
and charged species; in this region the vertical field component may be
approximated by $b_{z,high}$, which was given in Equation \ref{bzhigh}. 

When calculating the position of the magnetic diffusion shock for the
ambipolar diffusion model, it was noted that downstream of the shock the
thickness of the disc was controlled by magnetic squeezing. The scale height
of the disc may then be approximated by the relation 
\begin{equation}
    h \approx \frac{2\sigma}{b_{r,s}^2};
\label{xd-hbad}
\end{equation}
this is then substituted into Equation \ref{bzhigh} to give 
\begin{equation}
    b_{z,high} \approx \frac{m}{\sqrt{2}x}\left(\tilde{\eta}_H b_1
     \frac{b_{\phi,s}}{b_{r,s}} + \tilde{\eta}_A\right)^{-1},
\label{xd-bzhighbad}
\end{equation}
where $b_1$ is the coefficient of $b$ such that $b = b_1b_z$. If the field
components are approximately equal interior to the magnetic diffusion shock as
expected then $b_{r,s} \approx b_{z,high}$ and $b_{\phi,s} \approx -\delta
b_{z,high}$, and the magnitude of the total field strength (in terms of $b_z$)
is then
\begin{equation}
    b_1 \approx \sqrt{2 + \delta^2}.
\label{xd-b1bad}
\end{equation}
These relationships are then substituted into Equation \ref{xd-bzhighbad},
which becomes
\begin{equation}
    \frac{\psi}{x^2} \approx \frac{m}{\sqrt{2}x} \left( \tilde{\eta}_A -
     \delta\tilde{\eta}_H\sqrt{2 + \delta^2} \right)^{-1}.
\label{xd-bzhigh2bad}
\end{equation}
Assuming that the mass-to-flux ratio is still constant in this regime and
close to its initial value (this is not strictly true, as flux freezing does
break down upstream of the magnetic diffusion shock), it is possible to solve
this equation and obtain an estimate of the position of the magnetic diffusion
shock: 
\begin{equation}
    x_d \approx \frac{\sqrt{2}}{\mu_0} \left( \tilde{\eta}_A -
     \delta\tilde{\eta}_H\sqrt{2 + \delta^2} \right),
\label{xdbad}
\end{equation}
which reduces to Equation \ref{adxd} in the ambipolar diffusion limit. 

\begin{figure}[htp]
  \centering
  \vspace{-4mm}
  \includegraphics[width=4.1in]{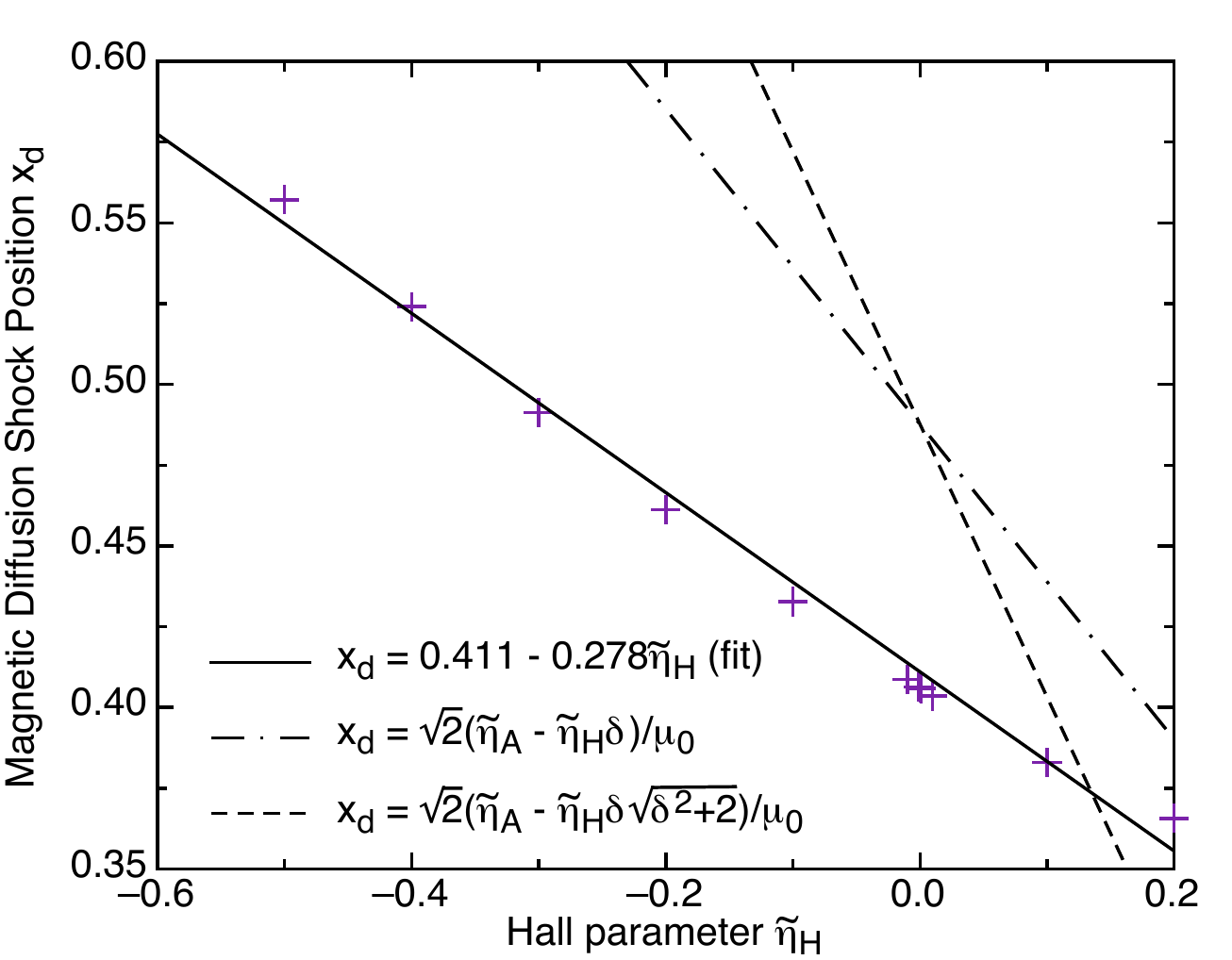}
  \vspace{-5mm}
  \caption[Linear fit to $x_d$ against $\tilde{\eta}_H$]{The position of the
magnetic diffusion shock $x_d$ depends in a seemingly near-linear manner on the
nondimensional Hall parameter $\tilde{\eta}_H$; crosses denote the values of
the shock position for the solutions tabulated in Table \ref{tab-shocks}, where 
$\tilde{\eta}_A = \delta = 1$. The solid line is the best linear fit to this
data; the dashed line is that given by Equation \ref{xdbad} for these
parameters; and the dot-dashed line is a simplified theoretical approximation
in which $b_1 = 1$.}  
\label{fig-xdbad}
  \centering
  \vspace{4mm}
  \includegraphics[width=4.1in]{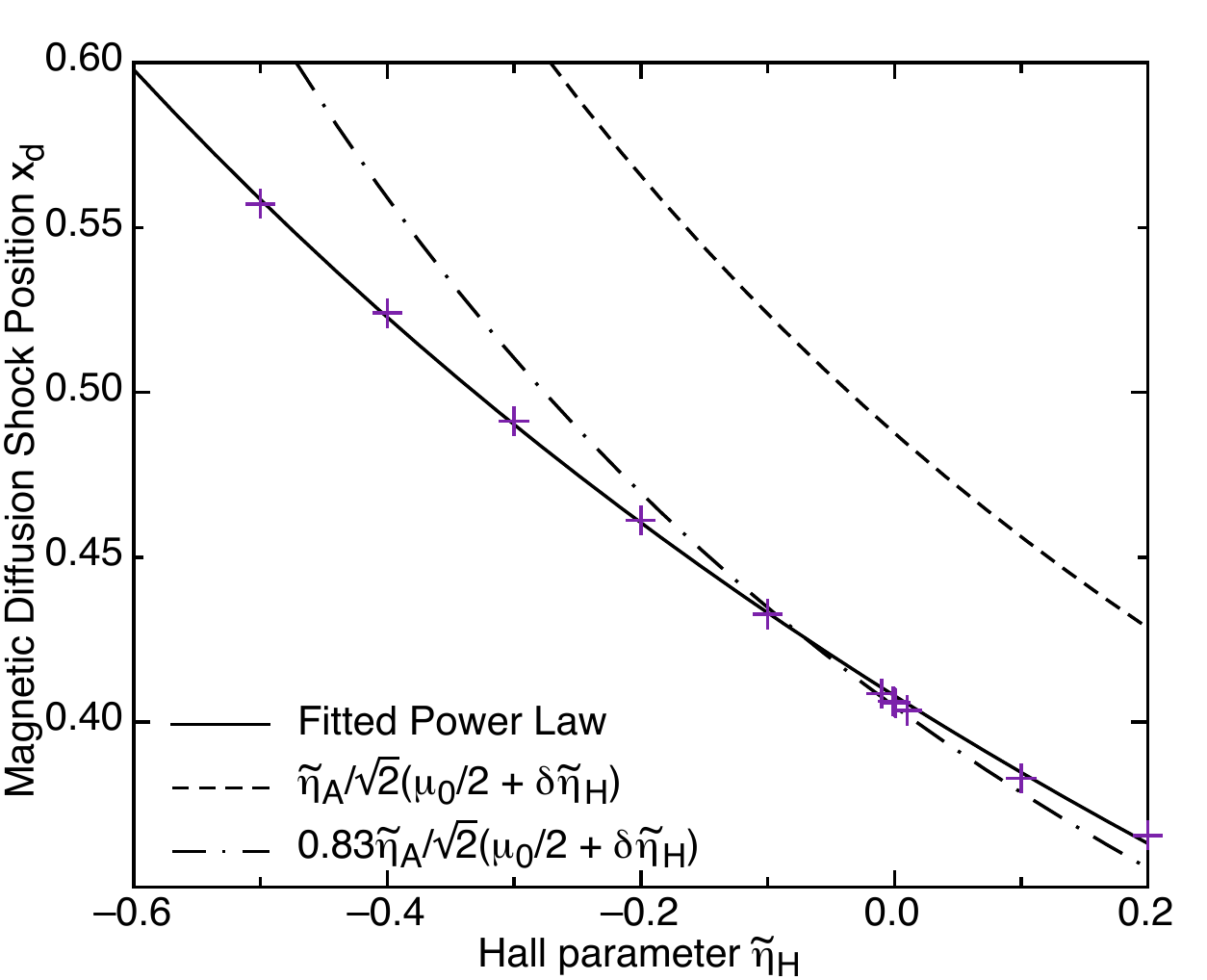}
  \vspace{-5mm}  
  \caption[Nonlinear fit to $x_d$ against $\tilde{\eta}_H$]{As in Figure
\ref{fig-xdbad} the position of the magnetic diffusion shock $x_d$ from the
similarity solutions is plotted as a series of crosses against the
nondimensional Hall parameter $\tilde{\eta}_H$ for solutions in which
$\tilde{\eta}_A = \delta = 1$. The dashed curve is the estimate to the shock
position given by Equation \ref{xdgood} for these parameters, while the
dot-dashed curve is Equation \ref{xdgood} multiplied by the constant 0.83. The
solid curve is the best fit of a curve defined by Equation \ref{xdfit} to the
data, using the parameters in Table \ref{tab-xdgood}.} 
\label{fig-xdgood}
  \vspace{-5mm}
\end{figure}

It was shown in subsection \ref{ad:shock} that this equation was a poor match
to the similarity solution in which $\tilde{\eta}_A = 1.0$, $\tilde{\eta}_H =
0$ and $v_0 = 0.73$ (Figure \ref{fig-admod}), although it did provide an
adequate approximation to the shock position when the initial azimuthal
velocity is lower. It is also a poor match for those similarity solutions with
both ambipolar and Hall diffusion calculated in this work: although the
magnetic diffusion shock positions plotted against the nondimensional Hall
diffusion parameter in Figure \ref{fig-xdbad} show that the relationship
between the two is near-linear, the dashed plot of Equation \ref{xdbad} is
clearly a very poor match to the data, as is the dot-dashed plot which shows
the result of deriving Equation \ref{xdbad} under the assumption that $b_1 =
1$ rather than the value given by Equation \ref{xd-b1bad}. The solid line is
the best linear fit to the data (using an equation of the form $x_d =
c\tilde{\eta}_H + d$); the parameters of this fit are given in Table
\ref{tab-xdbad}. 
\begin{table}[t]
\begin{center}
\begin{tabular}{ll}
\toprule
 estimated parameter & fitted parameter\\
\midrule
 $c = -0.845$ & $c = -0.278$ \\
 $d = 0.488$  & $d = 0.411$  \\
 \bottomrule
 \end{tabular}
\end{center}
\vspace{-5mm}
 \caption[Linear fit parameters to the magnetic diffusion shock position
$x_d$]{As shown in Figure \ref{fig-xdbad}, the position of the magnetic
diffusion shock, $x_d$, does not follow the linear relationship described in
Equation \ref{xdbad}. Fitting a straight line of the form $x_d =
c\tilde{\eta}_H + d$ to the data from the similarity solutions gives the
parameters to the fit outlined above (with $\chi^2 = 0.26 \times 10^{-3}$),
which are clearly very different to those estimated by Equation \ref{xdbad}.}
\label{tab-xdbad}
\end{table} 

Clearly the linear estimation to the magnetic diffusion shock position given
by Equation \ref{xdbad} is a poor match to the data from the similarity
solutions (with a typical error of $\Delta x_d \approx 20\%$); even the
simplified form of this equation with $b = b_z$ is unable to fit the data. A
more accurate estimation of the shock position is therefore required, and the
largest source of error in this calculation is the approximation to
$b_{\phi,s}$ used in $b_{z,high}$ (that $b_{\phi,s} = -\delta b_z$). 

On the upstream side of the magnetic diffusion shock $b_{\phi,s} \approx
-\delta b_{z,low}$ (as can be seen in the solutions as Section \ref{hall}),
however downstream of the shock the azimuthal field component is not equal to
$-\delta b_{z,high}$ as was assumed. The magnetic field lines are compressed
in the shock resulting in an increase in the vertical field and the azimuthal
field as the field lines are twisted up by the slowing of the compressed gas.
Conservation of flux ensures that the radial field component does not change
in the shock.

The magnetic diffusion shock represents a continuous increase in the magnetic
field strength, and though the matter is slowed in the post-shock region, $w$
is near-constant and $w > 1$ throughout the magnetic diffusion shock itself.
While the vertical field component increases from $b_{z,low}$ to $b_{z,high}$,
the azimuthal component downstream of the shock is given by 
\begin{equation}
    b_{\phi,sd} \approx wb_{\phi,su} \approx -w\delta b_{z,low},
\label{bphisd1}
\end{equation}
using the subscripts $u$ and $d$ to indicate the upstream and downstream sides
of the shock. Equation \ref{bzlow} is then substituted into this equation, and
this is solved for $b_{\phi,s}$ downstream of the shock:
\begin{equation}
    b_{\phi,s} \approx -x_d \delta b_{r,s}.
\label{bphisd2}
\end{equation}

As $x_d < 1$ in all of the similarity solutions, $b_{\phi,s} < b_{r,s}$ in
this region and there is no need to include the azimuthal field component in
the magnetic squeezing; downstream of the shock the disc scale height may
still be approximated by Equation \ref{xd-hbad} and the magnitude of the
magnetic field (in terms of $b_z$) is 
\begin{equation}
    b_1 \approx \sqrt{2}.
\label{b1good}
\end{equation}
These are then substituted into Equation \ref{xd-bzhighbad} for the downstream
vertical field component
\begin{equation}
    b_{z,high} \approx \frac{m}{x} \left(\sqrt{2}\tilde{\eta}_A - 
     2\delta x_d \tilde{\eta}_H\right)^{-1}. 
\label{bzhighgood}
\end{equation}
It is then possible (using the approximation that $b_{z,high} = b_{r,s}$) to
rearrange this into a new estimation of the magnetic diffusion shock position 
\begin{equation}
    x_d \approx \frac{\tilde{\eta}_A}{\sqrt{2}}
     \left(\frac{\mu_0}{2} + \delta\tilde{\eta}_H\right)^{-1},
\label{xdgood}
\end{equation}
which again reduces to Equation \ref{adxd} in the ambipolar diffusion limit.

The dashed curve in Figure \ref{fig-xdgood} illustrates Equation \ref{xdgood};
this is a much better fit to the shape of the data from the similarity
solutions although it is still an overestimation of the actual magnetic
diffusion shock positions which are represented by crosses. Multiplying
Equation \ref{xdgood} by a constant $0.83$ gives an approximation to the
magnetic diffusion shock position (shown as the dot-dashed curve in the
figure) that is good enough that it may be used in the simple model for
calculating the initial value of the variables at $x_m$; this constant may be
assumed to replace some missing physics in this approximation. Table
\ref{tab-xdgood} compares the derived equation to the parameters for the solid
curve of best fit in Figure \ref{fig-xdgood}, which takes the form
\begin{equation}
    x_d \approx c(d + \tilde{\eta}_H)^{-1}
\label{xdfit}
\end{equation}
where $c$ and $d$ are constants; these parameters are similar to their
equivalents from the derived estimation to the magnetic diffusion shock
position, with the largest different arising in the calculation of the
constant $c$ as expected.
\begin{table}[t]
\begin{center}
\begin{tabular}{ll}
\toprule
 estimated parameter & fitted parameter \\
\midrule
 $c = 0.707$	& $c = 0.742$ \\
 $d = 1.450$	& $d = 1.821$ \\
 \bottomrule
 \end{tabular}
\end{center}
\vspace{-5mm}
 \caption[Nonlinear fit parameters to the magnetic diffusion shock position
$x_d$]{As shown in Figure \ref{fig-xdgood}, a power law of the form described
by Equation \ref{xdfit} may be fit to the position of the magnetic diffusion
shock. The estimated parameters are those for Equation \ref{xdgood} for the
similarity solutions with $\tilde{\eta}_A = \delta = 1$. The value of $\chi^2
= 6.7\times10^{-5}$ suggests that this is a good fit to the data.}
\label{tab-xdgood} 
\end{table}

The curved approximation to $x_d$ (when multiplied by the constant 0.83) is
accurate to within $10\%$ of the true value for all of the data in Figure
\ref{fig-xdgood}, and to within $5\%$ for $-0.3 \le \tilde{\eta}_H \le 0.2$.
Even Equation \ref{xdgood} without the constant is accurate to within $20\%$
of the true values, which is much better than the linear analytic estimate
given by Equation \ref{xdbad}. It is the nonlinear estimate of $x_d$ (using
the factor of 0.83) that is used in the simple model, and in calculating the
initial estimate of the centrifugal shock position $x_c$. 

\subsection{Centrifugal shock position}\label{s:xc}

The presence of Hall diffusion in the model complicates the calculation of the
centrifugal shock position as its dependence on the prescription for
$b_{\phi,s}$ (Equation \ref{ha-b_phis}) makes it much more difficult to solve
the angular momentum equation (\ref{cam}) explicitly, to the point where the
method used to estimate $x_c$ in the ambipolar diffusion model is not
duplicated in this chapter. Instead, the approximation to the centrifugal
shock position from the ambipolar diffusion model is taken and modified to
include Hall diffusion in a simplistic manner. 

\begin{figure}[t]
  \centering
  \vspace{-3mm}
  \includegraphics[width=4.7in]{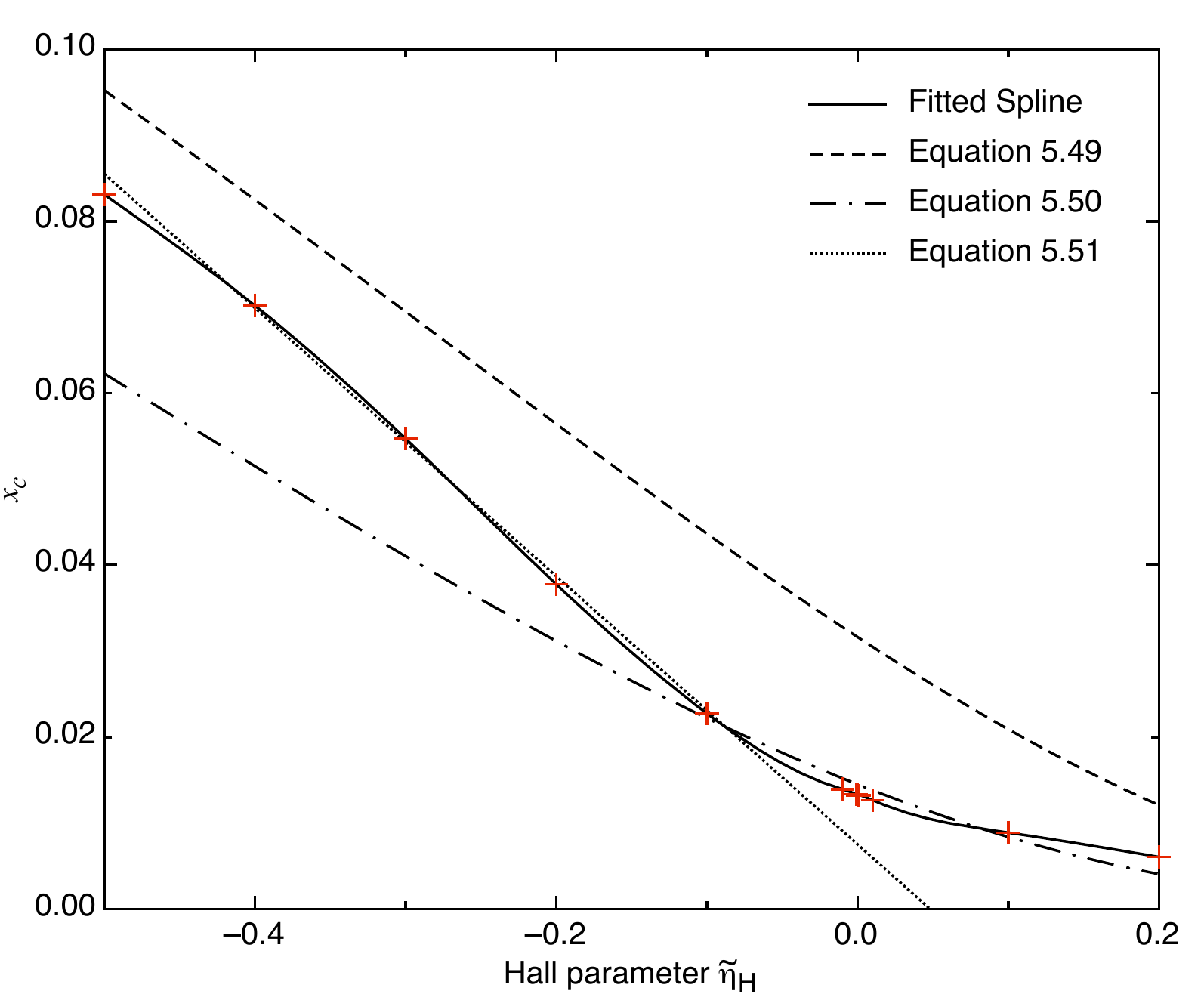}
  \vspace{-3mm}
  \caption[Estimations of $x_c$ against $\tilde{\eta}_H$]{The positions of the
centrifugal shock position $x_c$ are plotted against the nondimensional Hall
parameter $\tilde{\eta}_H$ for similarity solutions in which $\tilde{\eta}_A =
\delta = 1$. The dashed curve is the estimate to the shock position given by
Equation \ref{xcbad} for these parameters, while the dot-dashed curve is
Equation \ref{xcbetter} and the dotted line is the linear relation given in
Equation \ref{xcline}.} 
\label{fig-xc}
\end{figure}

In order to find the centrifugal shock position, the iterative routine
requires an estimate of $x_c$ that is accurate to around 20\%, as the routine
described in subsection \ref{num:iter} is very flexible and able to cope with
most poor initial guesses of this shock position. Starting then with Equation
\ref{adxc} for the centrifugal shock in the limit where $\tilde{\eta}_H = 0$,
\begin{equation}
    x_c \approx \frac{v_0^2}{A^2}m_c 
     \exp\left[-\sqrt{\frac{2^{3/2}m_c}{\mu_0\tilde{\eta}_A^3}}\right],
\label{xc-ad}
\end{equation}
the nondimensional ambipolar diffusion parameter is replaced by a term
$(\tilde{\eta}_A - \delta\tilde{\eta}_H)$ that includes both Hall and
ambipolar diffusion, as well as the magnetic braking parameter:
\begin{equation}
    x_c \approx \frac{v_0^2}{A^2}m_c 
     \exp\left[-\sqrt{\frac{2^{3/2}m_c}{\mu_0
	(\tilde{\eta}_A-\delta\tilde{\eta}_H)^3}}\right].
\label{xcbad}
\end{equation}
This equation, depicted as the dashed line in Figure \ref{fig-xc}, is a poor
match to the shock position data from the similarity solutions (listed in
Table \ref{tab-shocks}). 

Although Equation \ref{xcbad} gives a position that is typically close
enough to the true value that the routine will succeed in finding $x_c$, a
better match to the behaviour of $x_c$ is sought as finding the position of
$x_c$ is one of the slower parts of the model to compute. Equation \ref{xcbad}
can be somewhat improved by multiplying the magnetic diffusion term by the
constant 0.83 that was used to improve the determination of $x_d$; this then
gives the equation 
\begin{equation}
    x_c \approx \frac{v_0^2}{A^2}m_c 
     \exp\left[-\sqrt{\frac{2^{3/2}m_c}{\mu_0
	[0.83(\tilde{\eta}_A-\delta\tilde{\eta}_H)]^3}}\right],
\label{xcbetter}
\end{equation}
which is shown as the dot-dashed line in Figure \ref{fig-xc}. As can be seen
from the plot, this approximation to $x_c$ is only close enough to be used as
a first guess in the shock-finding routine when $\tilde{\eta}_H \gtrsim -0.1$.

The remainder of the Hall parameter space is mapped by the linear relation
plotted as the dotted line in Figure \ref{fig-xc}, which has the equation
\begin{equation}
    x_c \approx -0.156\tilde{\eta}_H + 0.01\tilde{\eta}_A - 0.0025;
\label{xcline}
\end{equation}
while this approach is simplistic, and required the calculation of a number of
similarity solutions before it could be adopted, it provides an acceptable
guess of the centrifugal shock position to the iterative routine described in
\ref{num:iter} for solutions with $\tilde{\eta}_H \lesssim -0.1$ (and
$\tilde{\eta}_A \approx 1$). This linear relationship is a more refined form
of the linear extrapolation (from the two closest solutions in
$\tilde{\eta}_H$-space) that was used to estimate the shock position before an 
analytic estimate was available. 

All of the shocks that require explicit calculation of the jump conditions
share the same set of jump conditions, which are outlined in the following
subsection. 

\subsection{Jump conditions}\label{s:jump}

As in the first ambipolar diffusion solution of Chapter \ref{ch:nohall}, the
magnetic diffusion shock in these similarity solutions is smooth and
continuous, and does not require any explicit calculation of shock conditions.
The shock itself is a transition between the two approximations to the
vertical field component, $b_{z,low}$ and $b_{z,high}$, and in the shock front
only $b_z$, $b_{\phi,s}$ and $h$ are changed, with $b_{\phi,s}$ downstream of
the shock given by $b_{\phi,s} \approx -w x_d \delta b_{z,low}$ (see Figure
\ref{fig-etah0.2-xd2} and subsection \ref{s:xd}). At the shock $h$ is
dramatically compressed, suggesting that a breakdown of the vertical
hydrostatic equilibrium occurs at this point. In reality, the enhanced
magnetic squeezing during the shock front is unable to reduce the disc
thickness so dramatically over the fluid transit time through the shock, and
the magnetic pressure far exceeds the gas pressure so that any breakdown of
isothermality would not greatly affect the collapse. 

On the other hand, the jump conditions at the centrifugal shock must be
explicitly calculated at the position found using the iterative routine
described in subsection \ref{num:iter}. The shock conditions at $x_c$ are
those derived in Section \ref{nonmag} for the nonmagnetic solutions; these
were found by integrating the conservation of mass and radial momentum
equations (\ref{cm} and \ref{crm}) at the shock position to give the relations
\begin{align}
    \sigma w &= \text{constant} \label{shock1} \\
    \text{and } \sigma \left(w^2 + 1\right) &= \text{constant.} \label{shock2}
\end{align}
These equations are then solved simultaneously to give the non-trivial
solution to the jump conditions
\begin{align}
    w_d &= \frac{1}{w_u} \label{jump1} \\
    \text{and } \sigma_d &= \sigma_uw_u^2, \label{jump2} 
\end{align}
where $u$ and $d$ denote the upstream and downstream sides of the shock. 

The magnetic field strength does not change at the centrifugal shock front
itself, as the magnetic pressure and tension terms are not large enough during
the shock transition to change the field behaviour, however it typically
increases in the regions immediately interior to the shock as the other
variables settle to their asymptotic values. The centrifugal shock is the
boundary of the rotationally-supported disc and as the position of this shock
depends on the Hall diffusion parameter so too does the strength of the shock,
with stronger shocks occurring with increasing Hall parameter. The increased
magnetic diffusion can also cause the formation of subshocks --- rings of
sharply-enhanced density in the post-centrifugal shock region --- as the
outward-moving flux causes the infalling gas to be slowed. 

The jump conditions defined in Equations \ref{jump1} and \ref{jump2} are also
used for any subshocks that may be encountered downstream of the principal
shocks. Although the magnetically-squeezed jump conditions defined in Section
\ref{imhd} could be applied at the magnetic diffusion subshock, they produce
numerically similar results to the jump conditions defined above and so no
benefit was seen to adding this complexity to the code. The second set of jump
conditions derived in Section \ref{imhd} only apply when IMHD holds true and
could not be used for any of the subshocks encountered in this model, as
magnetic diffusion is always active near to the shock fronts. 

Having summarised the behaviour, jump conditions and the ways in which both
the principal and subshocks affect the numerical routines that integrate the
ordinary differential equations that define the self-similar collapse problem,
it is now possible to calculate the similarity solutions.

\section{Hall Solutions}\label{hall}

Superficially, the similarity solutions within this section appear to be
similar to those of the ambipolar diffusion model in Section \ref{ad}. They
all have the same parameters and boundary conditions, outlined in Table
\ref{tab-bc}, which match those in Figure \ref{fig-admod} and figure 7 of
\citet{kk2002}. The nondimensional ambipolar diffusion parameter is held
constant at $\tilde{\eta}_A = 1.0$, as is the cap on the magnetic braking
$\delta = 1.0$ and the magnetic braking parameter $\alpha = 0.08$, so that the
only changes between the similarity solutions presented here and the ambipolar
diffusion-only solution in Figure \ref{fig-admod} are those wrought by the
addition of Hall diffusion. The values of the variables at the matching point
$x_m$ for a selection of similarity solutions are given in Table \ref{tab-xm};
interpolating between these it is possible to obtain a good initial guess of
the variables for the calculation of a similarity solution with any value of
the Hall parameter $\tilde{\eta}_H \in [-0.5,0.2]$. 
\begin{table}[t]
\begin{center}
\begin{tabular}{ccccc}
\toprule
 boundary condition & value & \hspace{10mm} & parameter & value \\
\midrule
 $\mu_0$ & 2.9  & & $\tilde{\eta}_A$ & 1.0 \\
 $v_0$	 & 0.73 & & $\delta$	     & 1.0 \\
 $u_0$	 & $-1$ & & $\alpha$	     & 0.08 \\
 $A$	 & 3 \\
 \bottomrule
 \end{tabular}
\end{center}
\vspace{-5mm}
 \caption[Boundary conditions and parameters for the Hall similarity
solutions]{Boundary conditions and parameters for all the Hall similarity
solutions presented in this chapter, and those in Appendix
\ref{ch:extrasols}.} 
\label{tab-bc}
\end{table}

\begin{figure}[htp]
  \centering
  \vspace{-2mm}
  \includegraphics[width=5.2in]{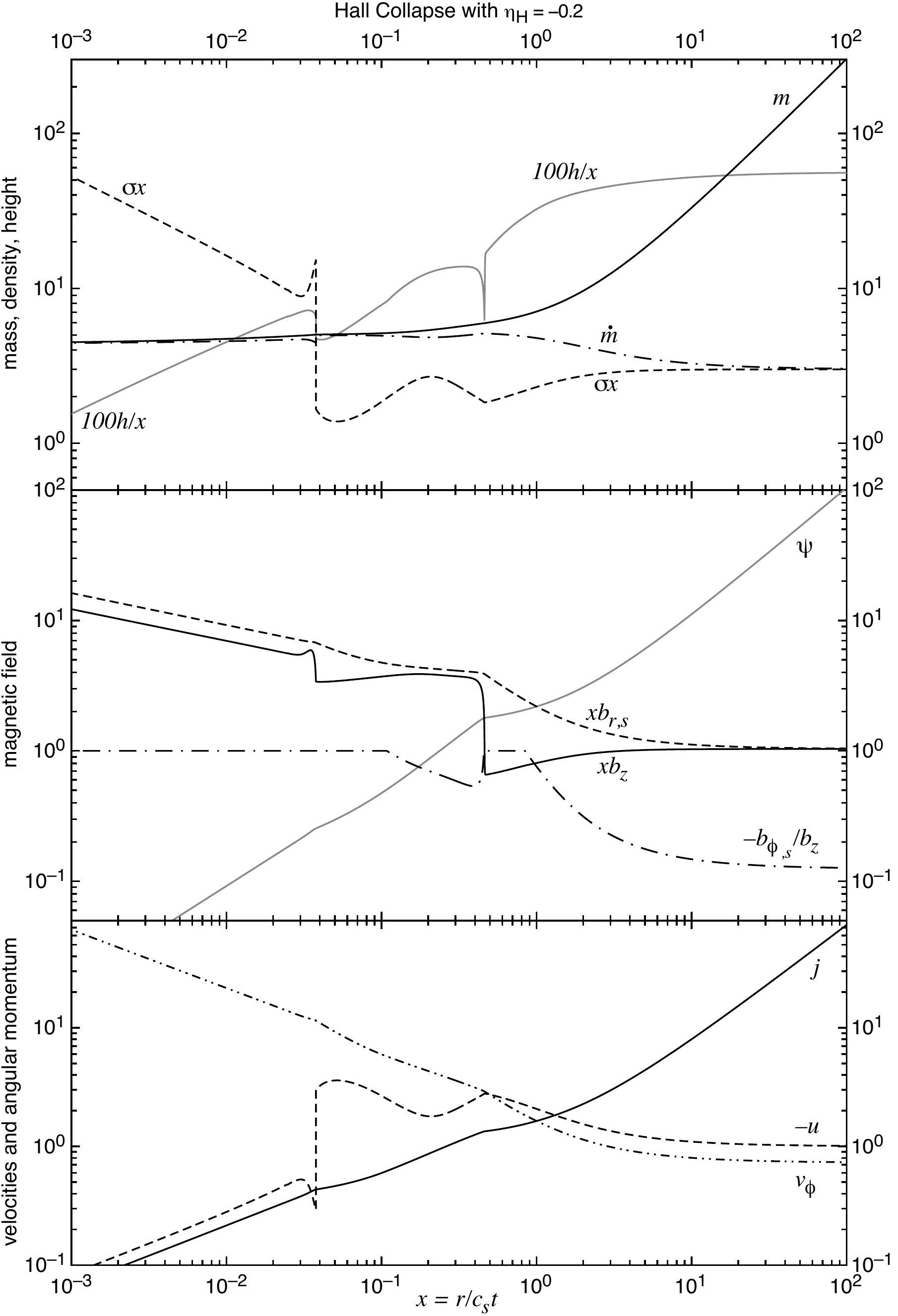}
  \vspace{-5mm}
  \caption[Hall and ambipolar diffusion collapse with $\tilde{\eta}_H = -0.2$
]{Similarity solution for Hall collapse with $\tilde{\eta}_H = -0.2$. The
displayed variables are the same as those in Figure \ref{fig-admod}; the
parameters and boundary conditions are given in Table \ref{tab-bc}. The
nondimensional central mass is $m_c = 4.23$; the magnetic diffusion and
centrifugal shocks are located at $x_d = 0.461$ and $x_c = 3.78 \times
10^{-2}$ respectively; both of these have increased from the non-Hall
positions, and the post-shock regions have been smoothed by the presence of
Hall diffusion.}   
\label{fig-hall-0.2}
\vspace{-3mm}
\end{figure}
The first of these similarity solutions is that presented in Figure
\ref{fig-hall-0.2} for the self-similar collapse of a molecular cloud with
$\tilde{\eta}_H = -0.2$. Those similarity solutions that possess a negative
Hall parameter have more radial diffusion of the magnetic field against the
neutral fluid and the charged grains (see the induction equation, \ref{in}),
so that the magnetic pressure builds up much earlier in the collapse process,
triggering the formation of the magnetic diffusion shock. The negative Hall
parameter also increases the initial rate of magnetic braking (by increasing
the first term in the brackets of Equation \ref{ha-b_phis}) so that
$b_{\phi,s}$ attains its capped value much earlier in the collapse, and the
magnetic braking is then determined by the strength of the vertical field
component (see the right term in the brackets of Equation \ref{ha-b_phis}). 

As in the previous solutions, at the outer edge of the collapse the matter is
falling in supersonically under IMHD, bringing the magnetic field with it. As
the surface density builds up the field does so too, causing the magnetic
pressure terms to become important, while the magnetic braking transports
angular momentum from the infalling gas to the external envelope by twisting
the magnetic field lines. The angular momentum and enclosed mass start to
plateau as the dominant force on the radial velocity changes between the
self-gravity of the disc and the gravity of the central mass, which in turn
causes the accretion rate to taper off. The formation of the magnetic
diffusion shock at $x_d = 0.461$ (\textit{cf.}\ $x_d = 0.406$ in the ambipolar
diffusion similarity solution with the same parameters) is caused by the
decoupling of the field from the neutral particles. 

The magnetic diffusion shock in this solution is weaker than that in Figure
\ref{fig-admod} as most of the neutral particles and the grains have already
been decoupled from the magnetic field, so that the vertical field component
increases by only 4.5 times (as opposed to 6.2 times in Figure
\ref{fig-admod}). The scale height is less compressed by the field, producing
a thinner shock, while $b_{\phi,s}$ does not grow rapidly under the build up
of flux and braking in the shock. Within the shock the field is further
decoupled from the neutral particles and charged grains, allowing Hall and
ambipolar diffusion to become more important downstream of the shock and
throughout the remainder of the solution. The field lines straighten as in
Figure \ref{fig-adsilverfish}; although the radial field component is still
dominant, the vertical component increases in the shock until it is just
smaller than $b_{r,s}$.  

Downstream of the magnetic diffusion shock the surface density gradually
increases as the infall velocity is slowed by the increased magnetic support.
This post-shock region is smoother than that without Hall diffusion,
presenting a gentler transition to the free fall collapse that occurs outside
of the rotationally-supported disc. The vertical field scales with $x^{-1}$
during the post-magnetic diffusion shock region as the increased radial
magnetic diffusion means that there are fewer field lines in total moving
against the flow of the neutral particles. In this region the disc scale
height is dominated by the magnetic squeezing; the gravity of the central mass
becomes the most important radial force on the gas at the end of the
post-shock region (the peak in the surface density) and the matter is
accelerated inwards until it is in near-free fall collapse. The magnetic
braking decreases the angular momentum rapidly until $b_{\phi,s}$ attains its
capped value and $j$ begins to plateau once more. 

The centrifugal force builds up as the matter falls inwards until it exceeds
the gravitational force, triggering the formation of the centrifugal shock at
$x_c = 3.78 \times 10^{-2}$ (\textit{cf.}\ $x_c = 1.32 \times 10^{-2}$ in
Figure \ref{fig-admod}). This shock is a discontinuity in the surface density
and the radial velocity, which is again less strong than that in the solution
without Hall diffusion, and inwards of this the vertical and azimuthal field
components increase steeply throughout the post-shock reaction of the field
(although $b_{\phi,s}$ remains capped at $-\delta b_z$). Downstream of the
centrifugal shock the variables tend (with overshoots in $b_z$) towards their
asymptotic values. 

The inner disc is in Keplerian rotation and satisfies Equations
\ref{in-m}--\ref{in-f} for the given parameters. The central mass is given by
$m_c = 4.42$, which is decreased from the non-Hall solution (\textit{cf.}\
$m_c = 4.67$) and corresponds to an accretion rate onto the central star of
$\dot{M}_c = 7.21 \times 10^{-6}$ M$_\odot$ yr$^{-1}$. The scalings with
respect to $x$ of the other variables are the same as in Figure
\ref{fig-admod} (see Section \ref{ha:inner}), however the surface density is
increased by the changed magnetic diffusion parameter $f = 1.72$ (compared to
$f(\tilde{\eta}_H = 0) = 1.\dot{3}$) and the increased radial magnetic
diffusion causes the strength of the magnetic field threading the disc to be
decreased from that in the ambipolar diffusion-only solution. This in turn
means that less matter can lose its angular momentum by magnetic braking and
fall onto the central mass, so that the gas is at a higher surface density in
this larger Keplerian disc. 

\begin{figure}[htp]
  \centering
  \vspace{-2mm}
  \includegraphics[width=5.2in]{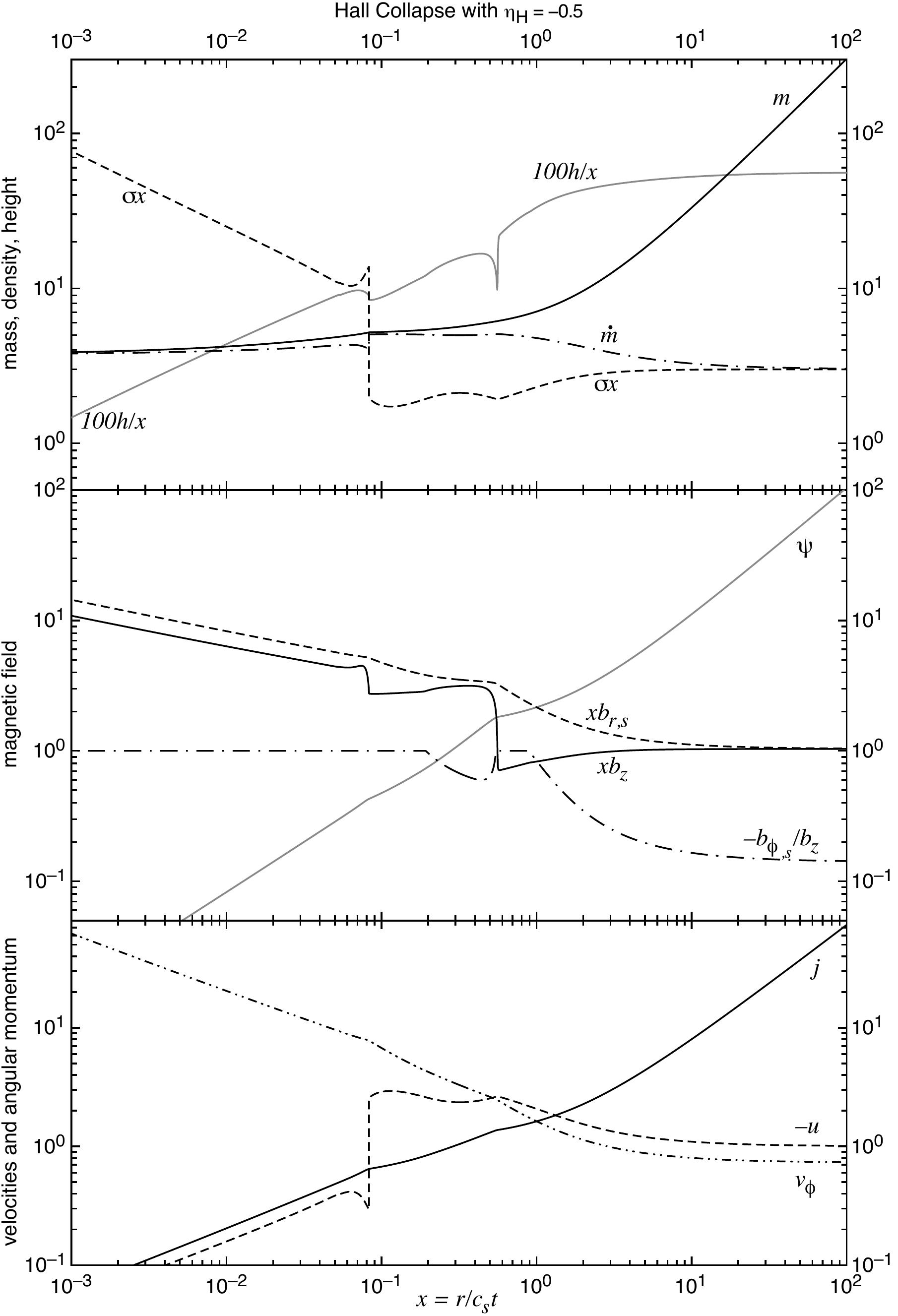}
  \vspace{-5mm}
  \caption[Hall and ambipolar diffusion collapse with $\tilde{\eta}_H = -0.5$
]{Similarity solution for Hall collapse with larger negative Hall parameter
$\tilde{\eta}_H = -0.5$. The  boundary conditions and parameters match those
in Figure \ref{fig-hall-0.2} (see Table \ref{tab-bc}). In this similarity
solution the nondimensional central mass is reduced to $m_c = 3.77$; the
magnetic diffusion and centrifugal shocks have moved outwards to $x_d = 0.557$
and $x_c = 8.31 \times 10^{-2}$ respectively; and the increased radial
magnetic diffusion has smoothed the post-shock regions and increased the size
of the inner Keplerian disc.} 
\label{fig-hall-0.5}
\vspace{-6mm}
\end{figure}

The next similarity solution, presented in Figure \ref{fig-hall-0.5}, shows
the results of a calculation with $\tilde{\eta}_H = -0.5$ on the same scale
and with the same parameters as Figure \ref{fig-hall-0.2}. The total radial
magnetic diffusion is further increased in this solution so that many of the
neutral particles have decoupled from the field before the diffusion shock;
this causes the intensity of the magnetic diffusion shock at $x_d = 0.557$ to
drop further so that the increase in the vertical field strength is only 4
times its original value. There is less of a magnetic wall at this point as
less of the field needs to be decoupled from the neutrals within the shock
itself. 

As in the previous solution with a negative Hall parameter the post-magnetic
diffusion shock region is smoothed, with even less change occurring to the
surface density and radial infall. The gas is still somewhat slowed by the
magnetic diffusion shock, but the gravity of the central mass quickly
overcomes this and pulls the fluid inwards. The radial velocity downstream of
the post-shock region is accelerated as the fluid nears the protostar, however
it remains below the free fall velocity at all times. The mass and angular
momentum both plateau in this region before the increasing centrifugal force
causes the formation of the centrifugal shock. 

The centrifugal shock occurs earlier in this similarity solution at $x_c =
8.31 \times 10^{-2}$, which is a large increase over that in the
$\tilde{\eta}_H = -0.2$ similarity solution. This change is brought about by
the decreased values of $b_z$ and $b_{\phi,s}$ in the free fall region, which
reduce the amount of magnetic braking that takes place and cause the
centrifugal force to become important earlier in the collapse. Inwards of this
shock is a much wider post-shock region of adjustment in which the variables
tend towards the inner disc solution. 

The Keplerian disc in this similarity solution is substantially larger than
that in the previous solution, containing $\sim 38\%$ of the mass of the
central protostar. The surface density of this disc has also increased as the
magnetic diffusion parameter $f = 2.31$, while the lowered central mass $m_c =
3.77$ corresponds to an accretion rate of $\dot{M}_c = 6.15 \times 10^{-6}$
M$_\odot$ yr$^{-1}$ onto the central star. Again, the larger disc corresponds
to a lower accretion rate, as the reduced magnetic braking makes it more
difficult for the fluid to lose rotational support and fall inwards. 

\begin{figure}[htp]
  \centering
  \vspace{-2mm}
  \includegraphics[width=5.2in]{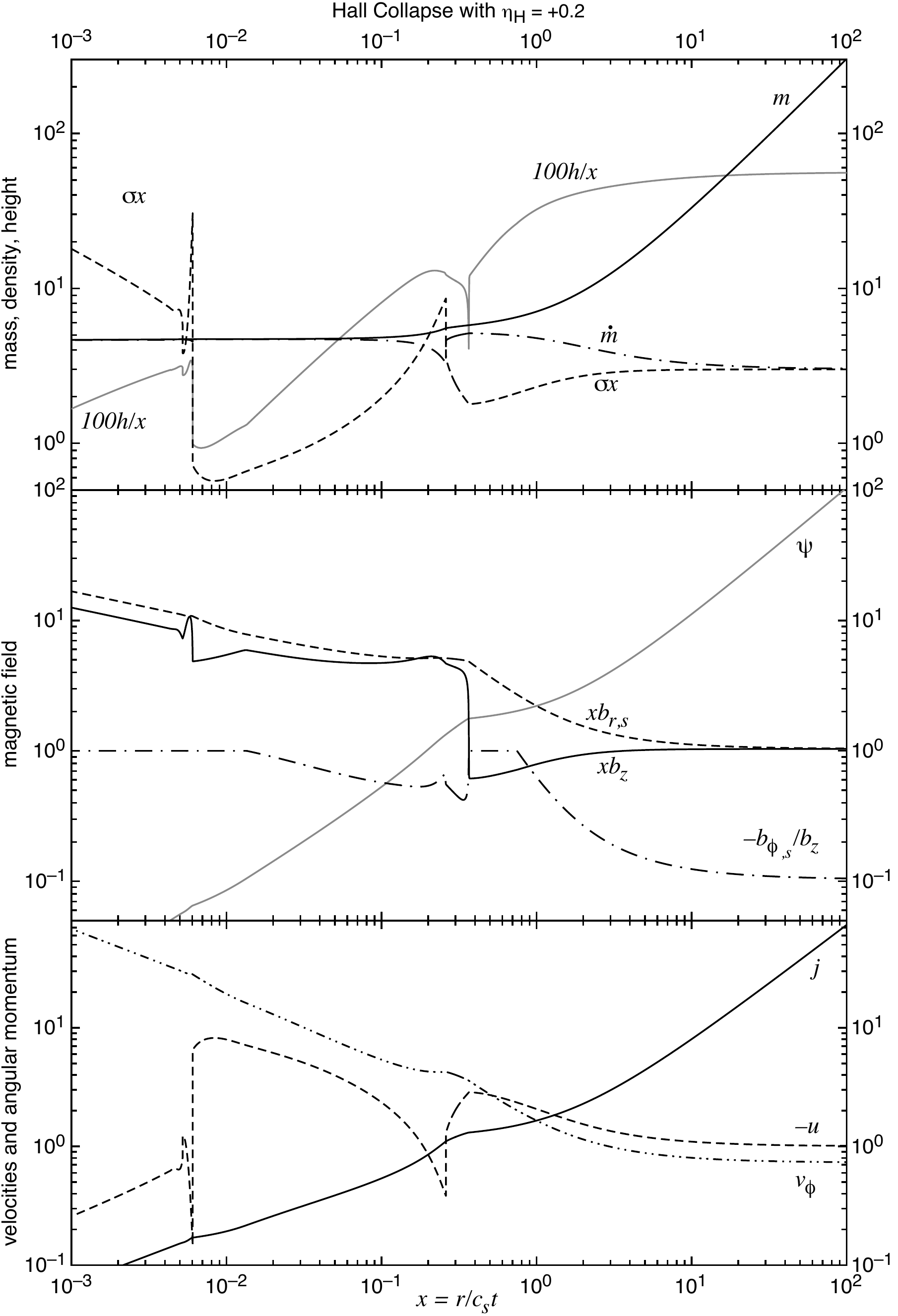}
  \vspace{-5mm}
  \caption[Hall and ambipolar diffusion collapse with $\tilde{\eta}_H = +0.2$ 
]{Similarity solution for gravitational collapse with positive Hall parameter
$\tilde{\eta}_H = +0.2$. The boundary conditions and parameters otherwise
match those in Figure \ref{fig-hall-0.2} and Table \ref{tab-bc}; the
nondimensional central mass is $m_c = 4.63$. The positive Hall term causes the
formation of subshocks downstream of the principal shocks: the magnetic
diffusion shocks are located at $x_d = 0.365$ and $x_{d2} = 0.260$; the
centrifugal shocks at $x_c = 6.05 \times 10^{-3}$ and $x_{c2} = 5.21 \times
10^{-3}$.} 
\vspace{-4mm}
\label{fig-hall0.2}
\end{figure}

The final similarity solution presented in this chapter is that with
$\tilde{\eta}_H = +0.2$ in Figure \ref{fig-hall0.2}, which is the most
dynamically different solution from the ambipolar diffusion solutions in
Chapter \ref{ch:nohall}. Although the initial conditions and parameters remain
the same as in the previous solutions, the change in the sign of the Hall
parameter, which corresponds to a reversal of the orientation of the initial
magnetic field with respect to the direction of rotation, introduces many
changes to the collapse dynamics. 

These begin at the magnetic diffusion shock, which is located inwards of the
position for the $\tilde{\eta}_H = 0$ solution in Figure \ref{fig-admod} at
$x_d = 0.366$. This shock is of increased intensity due to the reduced radial
magnetic diffusion prior to the passage of the shock, which causes a larger
increase in $b_z$ in the shock front. The magnetic braking downstream of the
shock is increased by the presence of a stronger field and the sign of the
Hall term in the expression for $b_{\phi,s}$ (\ref{ha-b_phis}). The disc is
more sharply compressed as the field lines straighten at the shock front, and
the fluid is so slowed by the increase in the magnetic pressure that a second
shock front forms at $x_{d2} = 0.260$. 

In the magnetic diffusion subshock, which was shown at higher resolution in
Figure \ref{fig-etah0.2-xd2} and described in subsection \ref{num:sub}, the
fluid is abruptly slowed until the radial velocity is low and subsonic. The
surface density increases under the jump conditions given in Section
\ref{s:jump}; this ring of matter contains approximately 18\% of the mass in
the central protostar. The azimuthal field component and the disc scale height
also increase in the subshock, while the derivative $db_z/dx$ decreases
steeply. The infall region downstream of the subshock is much wider in
logarithmic similarity space than in the previous solutions, with the
increased magnetic braking reducing the angular momentum more rapidly as the
radial velocity increases under the gravitational pull of the central mass.
The surface density drops as the fluid rapidly falls in and magnetic squeezing
dominates the vertical compression until the gravity of the central mass takes
over near to the centrifugal shock. 

The centrifugal shock occurs at $x_c = 6.05 \times 10^{-3}$, half that of the
$\tilde{\eta}_H = 0$ solution, dramatically decreasing the size of the
rotationally-supported disc. This change is brought about by the increased
magnetic braking caused by the positive Hall term, which reduces the
angular momentum so that the centrifugal force cannot become dynamically
important until the gas is very close to the protostar. The centrifugal shock
is followed by another subshock at $x_{c2} = 5.21 \times 10^{-3}$, which was
shown in more detail in Figure \ref{fig-etah0.2-xc2} and also described in
subsection \ref{num:sub}. Upstream of this shock the fluid is rapidly
accelerated inward as the magnetic field increases in the very thin region 
between the two shocks, forcing further magnetic braking and a rapid drop in
the centrifugal force. The surface density drops as the radial velocity
becomes supersonic once more, and as the fluid nears the protostar the
centrifugal force becomes important once more --- triggering the formation of
the subshock. At the subshock discontinuity the matter is slowed until the 
infall rate is low and subsonic once more. 

Downstream of the subshock, the variables settle with overshoots to the
asymptotic disc behaviour. The central mass is given by $m_c = 4.63$, which
corresponds to an accretion rate onto the central star of $\dot{M}_c = 7.53
\times 10^{-6}$ M$_\odot$ yr$^{-1}$, and the small Keplerian disc contains
only $1.6\%$ the mass of the central disc. The magnetic diffusion parameter in
this disc is $f = 0.945$, decreasing the surface density in the disc as the
strength of the magnetic field is increased. 

The magnetic diffusion and centrifugal subshocks both occur only in those
solutions in which $\tilde{\eta}_H$ is positive, where the increased magnetic
pressures and braking caused by the increase in the magnetic field falling
inwards force the gas to rapidly change in radial velocity and density. The
number of subshocks downstream of the principal centrifugal shock increases
with increasing $\tilde{\eta}_H$ --- three subshocks have been observed in one
similarity solution that was not properly converged at the time of
publication. The restriction that $f > 0$ (see Chapter \ref{ch:asymptotic})
also limits the range of positive $\tilde{\eta}_H$ that can be explored; as
$\tilde{\eta}_H$ increases, the size and surface density of the
rotationally-supported disc decrease, and the rings of gas formed by the
subshocks are likely to be gravitationally unstable.
\raggedbottom

The inclusion of Hall diffusion into the calculations causes many changes to
the dynamics of gravitational collapse, with the sign of $\tilde{\eta}_H$
introducing or suppressing the formation of subshocks downstream of the
principal shocks. The intensity and position of these shocks change with the
Hall parameter, as does the mass of the central protostar. Additional
similarity solutions with $\tilde{\eta}_H \in [-0.5,0.2]$ that demonstrate
these trends in more detail are presented in Appendix \ref{ch:extrasols}. 

\section{Summary}\label{ha:summary}

\begin{figure}[htp]
  \centering
  \includegraphics{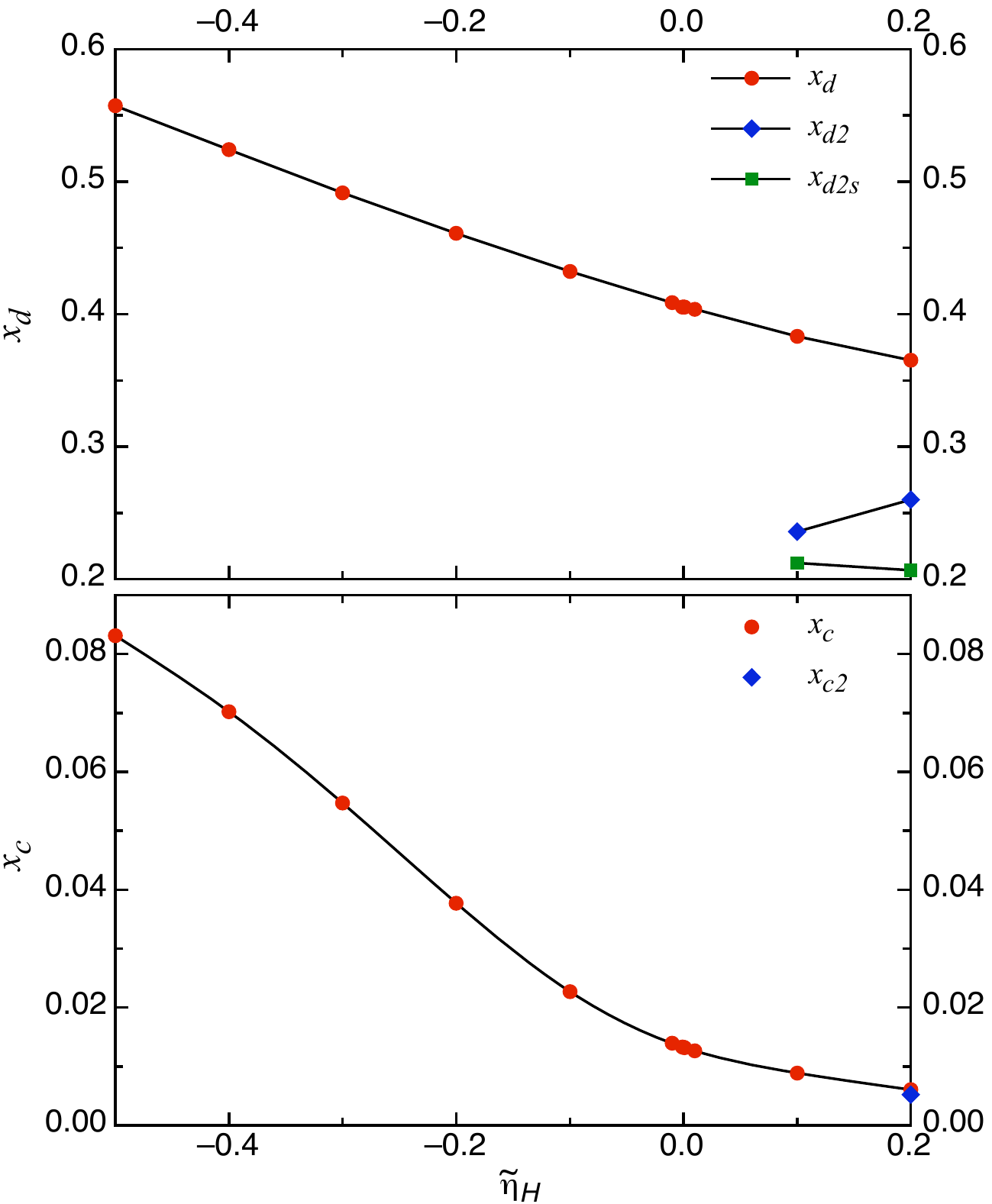}
\vspace{-4mm}
  \caption[Variance of shock positions with $\tilde{\eta}_H$]{The relationship
between the shock positions and the nondimensional Hall parameter
$\tilde{\eta}_H$. The top panel shows the position of the magnetic diffusion
shock $x_d$, which decreases as $\tilde{\eta}_H$ becomes more positive, as
well as the position of the subshock $x_{d2}$ and its associated sonic point
$x_{d2s}$. The lower panel shows the position of the centrifugal shock $x_c$,
which also decreases with increasing $\tilde{\eta}_H$, and the position of the
subshock at $x_{c2}$ for the similarity solution with $\tilde{\eta}_H = +0.2$,
which is the only solution that possesses a subshock to the centrifugal shock.
The sonic point between the centrifugal shock and subshock is not plotted due
to the resolution of this graph. All of the similarity solutions have initial
conditions matching those in Table \ref{tab-bc}, values of the variables at
$x_m$ and shock positions as described in Appendix \ref{ch:params}, and are
plotted individually in Appendix \ref{ch:extrasols}. } 
\vspace{-4mm}
\label{fig-shocks}
\end{figure}

Building on the work of the previous chapter, this chapter oversaw the
calculation of similarity solutions to the MHD equations for gravitational
collapse including Hall and ambipolar diffusion. Hall diffusion introduces new
subshocks and sonic points downstream of the shocks (where the radial speed 
changes between being sub- and supersonic), as seen in Figure
\ref{fig-shocks}, requiring a more complicated approach to finding the 
similarity solutions. A simplified model for calculating the initial guess of
the values of the variables at the matching point and the model used for
performing the innermost integration were presented, as were the analytic
expressions for the initial guesses of the magnetic diffusion and centrifugal
shock positions. Finally a number of similarity solutions with initial
conditions identical to those of the first ambipolar diffusion-only similarity
solution were discussed, showing the importance of Hall diffusion to the
collapse dynamics. 

The presented similarity solutions showed that Hall diffusion has a profound
effect on the complexity of the solution, depending on the orientation of the
magnetic field with respect to the rotational axis. When the Hall parameter
$\tilde{\eta}_H$ was negative, Hall diffusion smoothed the post-shock regions,
and the solutions matched onto the inner disc solution with fewer overshoots.
The magnetic diffusion in the radial direction was increased, so that the
magnetic diffusion shock formed much earlier in the collapse; this shock was
less strong than in the solution without Hall diffusion with smaller changes
to $b_z$ and $h$, and the post-shock slowing of the radial velocity was also
weakened. Similarly, the reduced magnetic braking caused $b_{\phi,s}$ to
attain its capped value much earlier in the collapse, causing the centrifugal
force to become dynamically important sooner and triggering the formation of a
larger rotationally-supported disc, as shown in Figure \ref{fig-shocks} which
demonstrates the relationship between the shock positions and
$\tilde{\eta}_H$. As accretion through the Keplerian disc is slow, larger
discs tend to correlate with lower accretion rates onto the protostar.
\flushbottom

On the other hand, when the Hall parameter was positive the similarity
solutions were complicated by the introduction of subshocks driven by the
addition of Hall diffusion. The increased magnetic pressure caused by the
reduced radial magnetic diffusion enhanced the strength of the principal
shocks and caused them to occur much later than in the non-Hall similarity
solutions. The azimuthal magnetic field component $b_{\phi,s}$ reached its
capped value much later in the collapse process, changing the behaviour of the
angular momentum so that the size of the innermost Keplerian disc was much
reduced from that in the similarity solution with no Hall diffusion. The
smaller rotationally-supported disc surrounded a larger protostar, as the
increased magnetic braking allowed more gas to lose its angular momentum and
fall onto the protostar. 

The accretion rate and the size of the rotationally-supported disc both depend
upon the Hall parameter; this relationship shall be discussed further in the
following chapter. The importance of the sign of the Hall term with respect to
the axis of rotation shall be analysed in the context of the magnetic braking
catastrophe and potential future observations of young stellar objects.

\cleardoublepage

\chapter{Discussion and Conclusions}\label{ch:discuss}

The similarity solutions have shown clearly that Hall diffusion changes the
structure and dynamics of the collapse of molecular cloud cores into
protostars and protostellar discs. The size of the rotationally-supported disc
and the accretion rate onto the protostar are determined by the ratio of the
Hall and ambipolar diffusivities, which influences the amount of magnetic
braking affecting the rotation of the collapsing core. It is also clear that
Hall diffusion can inhibit disc formation by enhancing the magnetic braking,
or by counteracting ambipolar diffusion to the point that the field starts to
infall faster than the fluid, increasing the magnetic pressure. These
behaviours depend both upon the size of the Hall parameter and its sign (which
represents the orientation of the field with respect to the direction of
rotation and that of the Hall current). 

This final chapter explores the trends that exist between the similarity
solutions in Chapters \ref{ch:nohall} and \ref{ch:hall}, and those in Appendix
\ref{ch:extrasols}, and contrasts these results with simulations from the
literature in Section \ref{discuss}. The results show that there is a
preferred handedness to the combination of the magnetic field alignment, the
sign of the Hall diffusivity and the axis of rotation in disc formation
calculations, which could be observed using new instruments such as ALMA. The
role of Hall diffusion in resolving the magnetic braking catastrophe is
expounded in Section \ref{catastrophe}, and the Hall effect is shown to induce
rotation in initially nonrotating collapsing cores. 

Options for future work will be outlined in Section \ref{future} --- these
include improving upon the limitations and assumptions of the self-similar
model; adopting alternative scalings of the Hall diffusivity $\eta_H$ as
functions of the similarity variable $x$; and further explorations of the 
influence of Hall diffusion on the rotation of the collapsing gas. Large
regions of parameter space remain unexplored, as the similarity solutions in
this work differed only in the magnitude of the Hall parameter, while keeping
the initial angular momentum, density and ambipolar diffusion parameters
constant. Several directions for further explorations of parameter space are
considered, as well as alternate methods for modelling the vertical angular
momentum transport, which was here limited by the cap on the azimuthal field.
Finally, a summary of the results and their astrophysical implications are
discussed in Section \ref{conclusions}. 

\section{Star Formation and the Hall Effect}\label{discuss}

\begin{figure}[htp]
  \centering
  \includegraphics[width=5.4in]{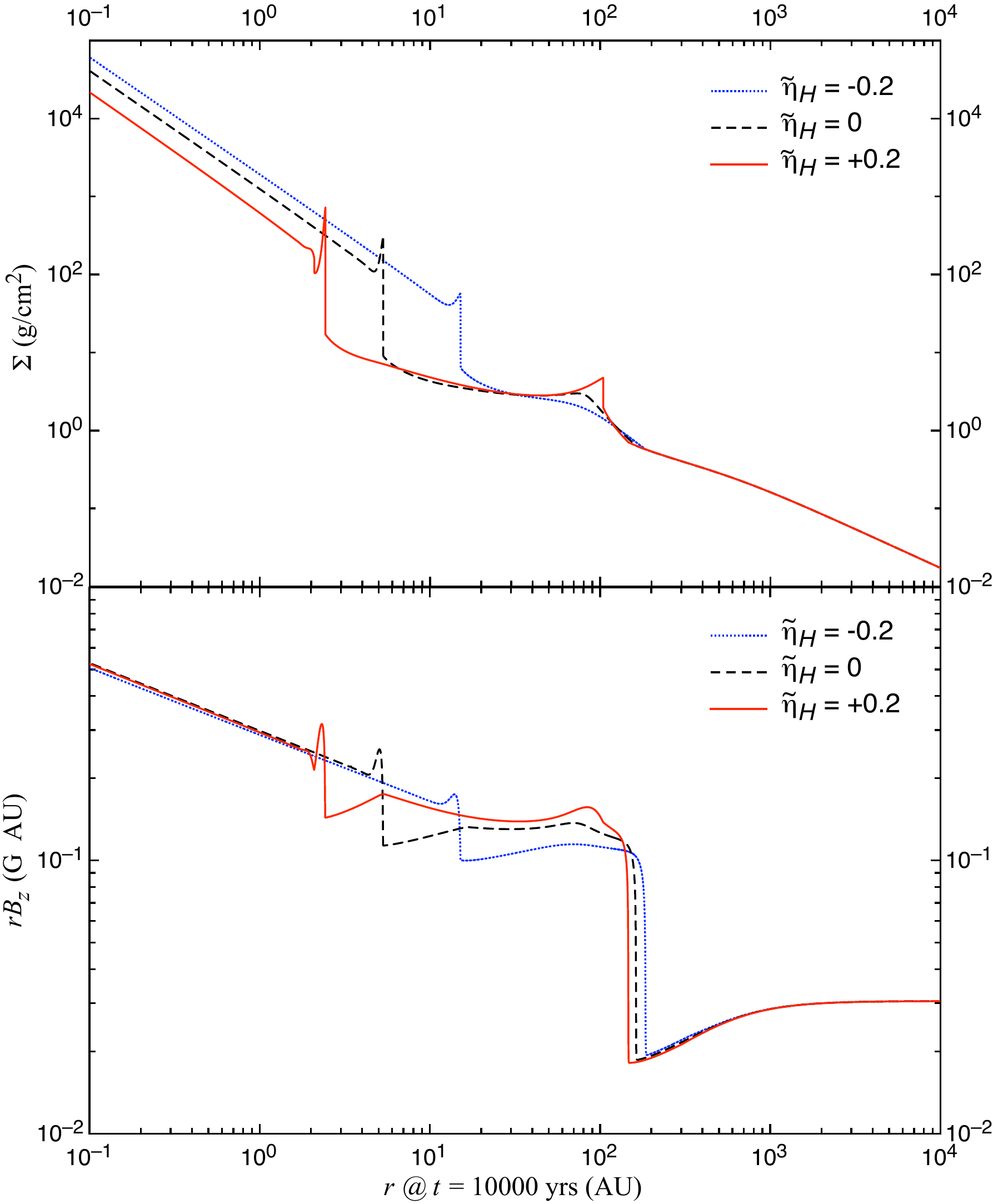}
  \vspace{-2mm}
  \caption[Comparison of $\Sigma$ \& $B_z$ with $r$ at $t = 10^4$ years]{The
surface density $\Sigma$ and the vertical magnetic field component $B_z$ (more
specifically $rB_z$) plotted against radius at $t = 10^4$ yr for the
similarity solutions with $\tilde{\eta}_A = 1.0$ and $\tilde{\eta}_H = -0.2$
(dotted blue line), $0$ (black dashed line) and $+0.2$ (solid red line). These
solutions were plotted individually in nondimensional form against $x$ in
Figures \ref{fig-hall-0.2}, \ref{fig-admod} and \ref{fig-hall0.2}
respectively.}
\label{fig-comp}
\end{figure}

\begin{table}[t]
\begin{center}
\begin{tabular}{rrrllr}
\toprule
 $\tilde{\eta}_H$ & $\Sigma$ (g cm$^{-2}$) & $B_z$ (G) &
	$M_c$ (M$_\odot$) & $M_{disc}$ (M$_\odot$) & $R_c$ (AU)\\
 \midrule
 $-0.2$	    & $1,920\quad$ & $0.289\;\;$ & $7.21 \times 10^{-2}$ & $9.99 \times
	10^{-3}$ & $15.10\quad$ \\ 
 $ 0\;\;\;$ & $1,250\quad$ & $0.299\;\;$ & $7.62 \times 10^{-2}$ & $3.75 \times
	10^{-3}$ & $5.31\quad$ \\ 
 $ 0.2$     & $620\quad$   & $0.304\;\;$ & $7.54 \times 10^{-2}$ & $1.24 \times
	10^{-3}$ & $2.43\quad$ \\ 
 \bottomrule
 \end{tabular}
\end{center}
\vspace{-3mm}
 \caption[$\Sigma$ and $B_z$ at $r = 1$ AU, and $M_c$, $M_{disc}$ and $R_c$ 
when $t = 10^4$ years for similarity solutions with $\tilde{\eta}_H = 0, \pm
0.2$]{Values of the surface density and vertical field component in the
Keplerian disc at $r = 1$ AU and $t = 10^4$ years, as well as the mass of the
central object and the size and mass of the rotationally-supported disc at the
same time, for the similarity solutions depicted in Figure \ref{fig-comp}. The
masses and the centrifugal shock radius all increase linearly with time, while
the surface density $\Sigma \propto t^{1/2}$ and the vertical magnetic field
is $B_z \propto t^{1/4}$.} 
\label{tab-Sigma}
\end{table}
The dependence of the similarity solutions upon the orientation of the
magnetic field and the sign of the Hall diffusion parameter $\tilde{\eta}_H$
(more specifically upon the sign of $\tilde{\eta}_H (\mathbf{B}\cdot\Omega)$,
although in this work only the sign of the Hall diffusion parameter is
altered) gives rise to two different patterns of collapse behaviours. Three
similarity solutions, with $\tilde{\eta}_H = 0$, $\pm 0.2$ (originally plotted
in nondimensional form in Figures \ref{fig-admod}, \ref{fig-hall0.2} and
\ref{fig-hall-0.2} respectively) are converted to dimensional form and plotted
against the radius $r$ (at a time $t = 10^4$ years) in Figure \ref{fig-comp},
with the surface density $\Sigma$ plotted in the upper panel and the vertical
magnetic field strength plotted as $rB_z$ in the lower. The three curves in
each panel have the same outer boundary conditions and parameters (listed in
Table \ref{tab-bc}), and the surface density and vertical field strength (at
$r = 1$ AU), as well as the central mass, and the mass and size of the inner
disc (all at $t = 10^4$ years) of the solutions are listed in Table
\ref{tab-Sigma}. The outer regions of the collapse are near-identical, as the
collapse is dominated by IMHD (so that diffusion is not important) and it is
only near the magnetic diffusion shock at $r \approx 100$ AU that the changes
brought on by Hall diffusion become apparent. 

The dotted blue lines in Figure \ref{fig-comp}, corresponding to the negative
Hall solution in Figure \ref{fig-hall-0.2}, show the formation of a large
rotationally-supported disc that has a radius $R_c \approx 12$ AU at $t =
10^4$ years, and the highest inner Keplerian disc surface density of all the
solutions. The dashed black lines are the $\tilde{\eta}_H = 0$ ambipolar
diffusion-only solution from Figure \ref{fig-admod}, which possesses a disc
half the size of the negative Hall solution. The surface density of the
Keplerian disc in this solution is decreased from that in the negative Hall
solution by a constant factor as the accretion rate increases (this change in
$\dot{M}_c$ is discussed below). Finally, the solid red curves characterise
the similarity solution with a positive Hall parameter $\tilde{\eta}_H = 0.2$
(illustrated in Figure \ref{fig-hall0.2}) which has a Keplerian disc that is
almost an order of magnitude smaller than that in the equivalent negative Hall
case. This disc is bounded by a thin ring of enhanced density, which rapidly
drops off as the magnetic field peaks; the material is then shocked again and
comes to match onto the inner solution. The density in this smaller disc is
much lower than in the previous solutions, and the disc grows at a much slower
rate. 

The similarity solutions span many orders of magnitude in both radius and
density, and the inclusion of a Hall parameter that is only 20\% that of the
ambipolar diffusion parameter has a large effect on the behaviour of the
magnetic field. In the radial direction, the drift of the field lines with
respect to the fluid takes place with velocity 
\begin{equation}
    V_{Br} = \frac{1}{H}\left[ \frac{\eta_H}{B}\,B_{\phi,s} 
	+ \frac{\eta_P}{B^2} \left(B_{r,s} - H \frac{\partial B_z}{\partial r}
	\right)\! B_z \right]. \label{V_Br}
\end{equation}
In the intermediate region between the magnetic diffusion shock (at $R_d
\approx 100$ AU in Figure \ref{fig-comp}) and the centrifugal shock ($R_c
\approx 1-10$ AU), when $\tilde{\eta}_H$ is negative, the azimuthal field
tension causes the Hall drift to enhance the radial diffusion of the field
lines. This torque creates a radial force on the infalling neutral particles
that slows the infall speed. The magnetic diffusion shock occurs earlier in
the collapse and is less dynamic, that is, there is less of an increase in
$B_z$ than in the other solutions, as much of the field has already been
decoupled from the fluid. 

However, when $\tilde{\eta}_H$ is positive then Hall diffusion acts to reduce
the net radial diffusion, resulting in magnetic walls and subshocks that
disrupt the flow. The magnetic field carried in towards the protostar
increases, and the magnetic pressure and tension terms remain important 
throughout the collapse. Any twist in the field lines causes an increase in
the magnetic pressure gradient, so that the net amount of radial diffusion
drops off as the magnetic braking slows the rotation by twisting the field
above the pseudodisc. 

There is a similar duality to the azimuthal field drift, which occurs with a
drift velocity as defined in Equation \ref{vb3}:
\begin{equation}
    V_{B\phi} = -\frac{1}{H}\left[ \frac{\eta_H}{B} \left(B_{r,s} 
	- H \frac{\partial B_z}{\partial r}\right) - \frac{\eta_P}{B^2}\, 
	B_zB_{\phi,s} \right]. \label{V_Bphi}
\end{equation}
Again looking at the intermediate regions between the two shocks, when
$\tilde{\eta}_H$ is negative the bracketed terms, which are proportional to
the vertically integrated component of the current density, cause the Hall
drift in the azimuthal direction, which twists up the field lines in the
pseudodisc and creates a leading torque on the neutral rotation. The reduced 
value of $B_z$ in this region causes the azimuthal field component to reach
its capped value $B_{\phi,s} = -\delta B_z$ sooner in the collapse, and the
magnetic braking, which depends upon $B_zB_{\phi,s}$, is also reduced. Because
of this, less angular momentum is removed from the pseudodisc, causing the
centrifugal force to become dynamically important earlier and a larger
rotationally-supported disc to form.

In the other orientation, that is, when $\tilde{\eta}_H$ is positive, Hall and
ambipolar diffusion act together to untwist the field lines in the pseudodisc.
In these similarity solutions $B_z$ is larger, and so while it takes longer
for $B_{\phi,s}$ to achieve its capped value there is more total magnetic
braking and the angular momentum is further reduced. A smaller Keplerian disc
forms due to the reduced centrifugal force, and both shocks have subshocks
where the magnetic forces alter the radial velocity of the fluid. Downstream
of the magnetic diffusion shock the radial magnetic pressure gradient causes
the fluid to be slowed in the radial and azimuthal directions, while
downstream of the centrifugal shock the gas is accelerated inward as the
increase in $B_z$ causes a burst of magnetic braking that disrupts the disc
and causes the formation of a subshock. 

While the angular momentum behaviour between the magnetic diffusion and
centrifugal shocks is changed by the inclusion of Hall diffusion, the cap on
$B_{\phi,s}$ acts to ensure that the angular momentum in the inner disc is
that expected for a Keplerian disc. The cap, while physically motivated,
replaces unspecified disc physics such as reconnection, a disc wind, or
turbulence, which would act to prevent the azimuthal field component from
greatly exceeding the vertical component; the magnitude at which it ought to
act to limit $B_{\phi,s}$ is uncertain. It is also unclear if such limiting
of the azimuthal field component happens in real collapsing cores, and the
numerical simulations illustrating disc formation that are discussed in the
next section do demonstrate tightly-wound magnetic fields
\citep[e.g.][]{mim2008}. The azimuthal field cap limits the similarity 
solution set explored in the previous chapter to those in which discs form,
however despite this Hall diffusion has been shown to restrict disc formation
if the Hall diffusion is too strong in comparison to ambipolar diffusion and
$\tilde{\eta}_H$ has the ``wrong'' sign. 

It is in the region between the magnetic diffusion and centrifugal shocks that
the influence of Hall diffusion on the field behaviour becomes more obvious in
this comparison. In the negative Hall solution this region is  characterised
by a steady increase in $B_z$ with decreasing $r$, while the surface density
gradually varies, decreasing near both shocks where the radial velocity is
high. In the ambipolar diffusion-only solution the peak in $\Sigma$ interior
to the magnetic diffusion shock is sharpened, and causes a subtle change in
the slope of $B_z$. The positive Hall solution sees the peak in the surface
density become so strong that a subshock forms interior to the magnetic
diffusion shock. This subshock is a sharp increase in the surface density and
a marked change in the gradient of the magnetic field. The region between the
two shocks is extended as the gas is slowed by the subshock, and the rate of
magnetic braking in this solution has been increased by the Hall diffusion,
which acts to increase $B_z$. 

\begin{figure}[tp]
  \centering
  \vspace{-5mm}
  \includegraphics[width=5in]{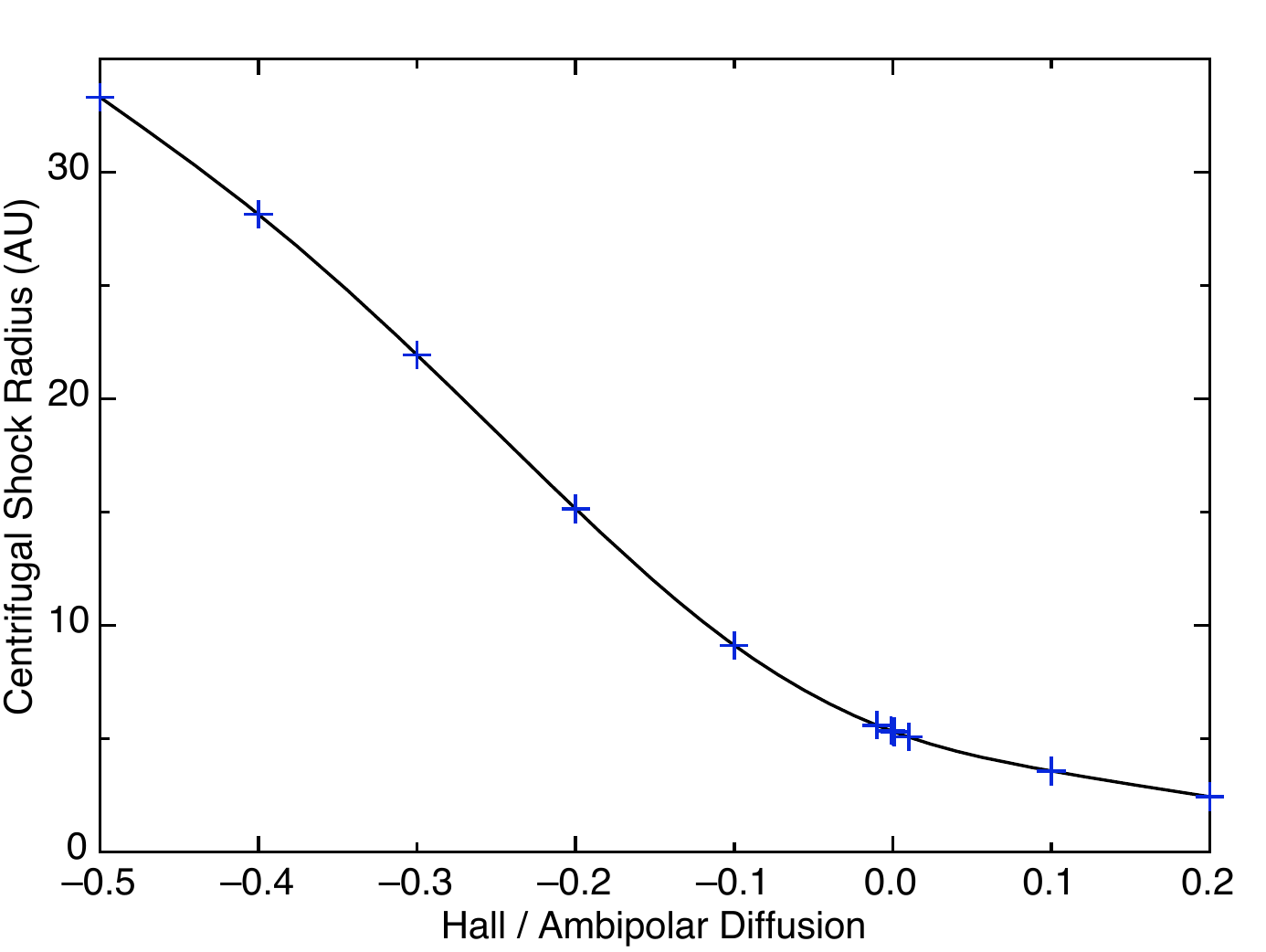}
  \vspace{-4mm}
  \caption[Centrifugal shock radius against
$\tilde{\eta}_H/\tilde{\eta}_A$]{The dependence of the centrifugal shock
radius (in AU at $10^4$ years) on the ratio of the nondimensional Hall to
ambipolar diffusion parameters for the calculated similarity solutions. The
shock positions are tabulated in Appendix \ref{ch:params}.} 
\label{fig-rc}
  \vspace{2mm}
\end{figure}

As shown in Figure \ref{fig-comp}, there exists a clear relationship between
the Hall diffusion parameter and the radius of the centrifugal shock. This is
further demonstrated in Figure \ref{fig-rc}, which plots the shock position
for similarity solutions with initial parameters equal to those in Table
\ref{tab-bc} (at a time $t = 10^4$ years) against the ratio of the Hall to
ambipolar diffusion parameters. The centrifugal shock marks the edge of the
Keplerian disc, and it clearly decreases in radius as the ratio of diffusion
parameters increases; this behaviour was noted in the analytic estimation of
the shock position derived in Section \ref{shocks}. While both directions of
Hall drift contribute to the change in the size of the disc, the radial drift
of $B_z$, which increases when the Hall parameter is negative (see Equation
\ref{V_Br}), has a greater effect on the radius of the centrifugal shock than
the torque caused by the azimuthal Hall drift. 

The constant velocity of the shocks in Figure \ref{fig-comp} is given by $V_s
= x_sc_s$, where the nondimensional shock positions $x_s$ are those listed in
Table \ref{tab-shocks} for these values of $\tilde{\eta}_H$, and $c_s = 0.19$
km s$^{-1}$ is the isothermal sound speed. All of the solutions form a
protostar of around $0.7$ solar masses and a protostellar disc of radius $R_c
\sim$10--$150$ AU and mass $M_d \sim 10^{-2}$--$10^{-1}$ M$_\odot$ in $t =
10^5$ years; these are the same order of magnitude expected from observations
of Class I YSOs \citep[e.g.][]{jetal2007}. 

The solutions compared in Figure \ref{fig-comp}, and those in Appendix
\ref{ch:extrasols}, have surface densities $\Sigma \propto r^{-3/2}$ and
$\Sigma \propto t^{1/2}$ in the inner Keplerian disc, with a value $\Sigma(r =
1$ AU, $t = 10^4$ years$) \sim 10^3$ g cm$^{-2}$; the exact value of $\Sigma$
at this point for each of these solutions are given in Table \ref{tab-Sigma}.
These values of the surface density are consistent with what is thought to
have occurred in the solar nebula \citep[e.g.\ the minimum mass solar nebula
has $\Sigma = 1700$ g cm$^{-2}$ at $r = 1$ AU;][]{w1977}. Although the scaling
of the surface density with radius appears to be imposed upon the disc by the
simple model used to integrate to the inner boundary (see Section
\ref{numerics}), the surface density tends asymptotically towards this
behaviour before the simple model is activated. The parameter space explored
was chosen to correspond to such disc-forming similarity solutions; variations
in the model parameters (particularly the cap on the azimuthal field $\delta$)
could cause the disc surface density to behave differently, however no other
disc-forming asymptotic solutions to the collapse equations were found in
Chapter \ref{ch:asymptotic}.

With increasing $\tilde{\eta}_H$ the increasing number of subshocks force the
calculations to become so unstable that they cannot converge on the true
similarity solution, which makes it difficult to speak of trends within the
solution space. However, for the other polarity (where $\tilde{\eta}_H$ is
negative) the solutions tend towards the behaviour exhibited in the slow IMHD
collapse solutions discussed in Section \ref{imhd}; this trend can be seen
across the similarity solutions in Appendix \ref{ch:extrasols}. The region
between the magnetic diffusion and centrifugal shocks becomes smaller as the
magnetic pressure gradient becomes less important to the infall rate. As
$\tilde{\eta}_H$ becomes increasingly large and negative, Hall diffusion
continues to be important to the azimuthal field diffusion and the magnetic
braking, controlling the amount of angular momentum in the inner regions of
the collapse and determining the size of the Keplerian disc. 

The value of $\Sigma$ in the Keplerian disc depends quite sensitively on the Hall
parameter $\tilde{\eta}_H$, even though the vertical field component $B_z$
increases only very slightly with increasing $\tilde{\eta}_H$. As the magnetic
field is decoupled from the fluid there is a corresponding increase in the
magnetic force, which slows down the inflow and reduces the inflow rate (see
Equation \ref{kepeqnsid}). The inflow rate decreases with decreasing
$\tilde{\eta}_H$ and $\tilde{\eta}_A$, and the decrease in radial velocity
corresponds to an increase in the disc surface density. The amount of radial
field diffusion determines the accretion rate $\dot{M}_c$, as well as the disc
surface density; this behaviour was noted by \citet{cck1998} for their
ambipolar diffusion solutions without rotation. 

The vertical magnetic field behaves the same (and has very similar magnitudes)
in all three similarity solutions in Figure \ref{fig-comp} on the inner
Keplerian disc; this is merely an artefact of the choice of illustrated
solutions, as the additional Hall similarity solutions in Appendix
\ref{ch:extrasols} show that the vertical field in the disc drops off with
increasingly large negative Hall parameter. The accretion onto the central
protostar depends upon the magnetic tension so that $\dot{M}$ and $B_z$ are
tied with $\dot{M} \sim \delta B_z^2/J$, and as Table \ref{tab-Sigma} shows,
there is little change in either of $B_z$ or $M_c$ across these three
solutions. Similarly, in the outer regions of the disc where there is very
little magnetic field diffusion and IMHD generally holds true, the vertical
field magnitude is the same across all three solutions. 

\begin{figure}[t]
  \centering
  \vspace{-5mm}
  \includegraphics[width=5in]{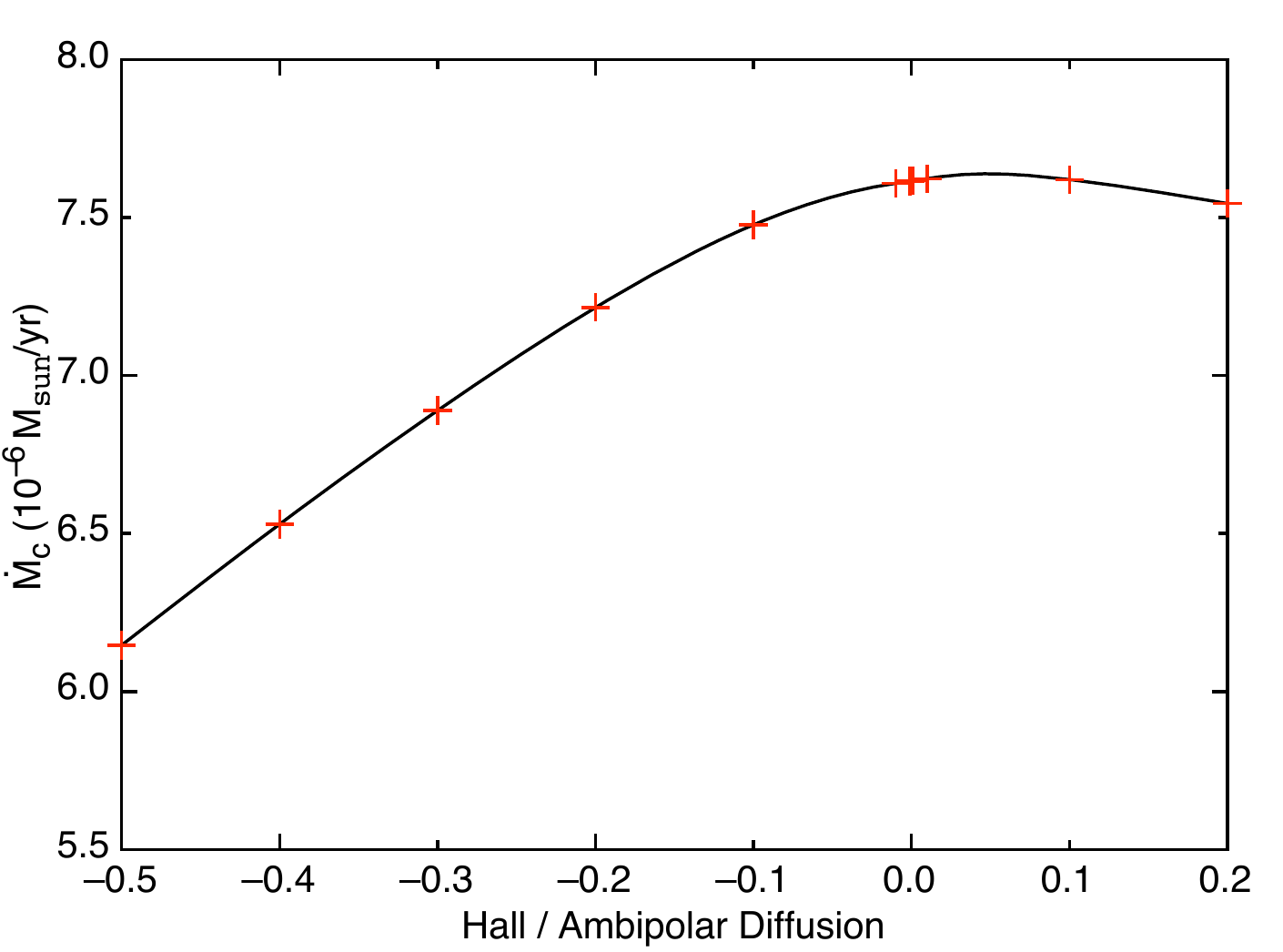}
  \vspace{-4mm}
  \caption[Central mass accretion rate against $\tilde{\eta}_H /
\tilde{\eta}_A$]{The dependence of the accretion rate onto the central star
(in $10^{-6}$ M$_\odot$/yr) on the ratio of the nondimensional Hall to
ambipolar diffusion parameters. The values of $\dot{M}_c$ are calculated using
the converged self-similar accretion rates $m_c$ tabulated in Appendix
\ref{ch:params}.} 
\label{fig-mdot}
\end{figure}

There also exists a correlation between the accretion rate onto the central
protostar and the ratio of the Hall to ambipolar diffusion parameters, as can
be seen in Figure \ref{fig-mdot}. The behaviour of the accretion rate depends
upon the disc radius, with larger discs corresponding to lower accretion rates
and vice versa; this behaviour was also noted by \citet{asl2003b}. The infall
rate through the Keplerian disc depends upon the radial diffusion of field
lines (see Section \ref{kepdisc}), with larger negative Hall parameters
causing an increased drag on the neutrals and a reduction of the radial
velocity of the neutral fluid. The disc radius (and by extension the accretion
rate) depends also upon the initial rotational velocity of the molecular cloud
core, as the centrifugal force becomes important sooner and a larger disc
forms when there is more initial angular momentum; this relationship was
demonstrated in Chapter \ref{ch:nohall} for the similarity solutions without
Hall diffusion. 

The accretion rate onto the central protostar appears to turn over at around 
$\dot{M}_c = 7.6 \times 10^{-6}$ M$_\odot$ yr$^{-1}$ once the ratio of the
nondimensional Hall to ambipolar diffusion parameters becomes positive and
greater than 0.05. This is likely due to the introduction of the subshocks in
the solutions with $\tilde{\eta}_H$ positive. In the subshocks the density is
enhanced, as is $dm/dx = \sigma x$, and so the accretion rate drops and less
matter can be accreted onto the central mass. As the positive Hall parameter
is increased, further subshocks are introduced into the solutions, which cause
there to be an even lower inflow rate; and as $\tilde{\eta}_H$ tends towards
the value that would see the diffusion parameter 
\begin{equation}
    f = \frac{4}{3}\tilde{\eta}_A 
	- \delta\tilde{\eta}_H\sqrt{\frac{25}{9} + \delta^2} 
\label{6f}
\end{equation}
become $0$, $\Sigma$ in the disc also tends towards 0 (see Section
\ref{kepdisc}). The surface density and the accretion rate through the disc
depend upon $f$ because the accretion is regulated by the magnetic diffusion
and associated torques, and this parameter places constraints on the possible
values of $\Sigma$ that exist in rotationally supported discs.

Clearly, the diffusion parameter of the disc, $f$, cannot be negative or equal
to zero, as this implies that the field moves inward faster than the neutral
particles; this causes an increase in the magnetic forces that would inhibit
further collapse and disc formation. Disc formation is also inhibited when
$\delta$ is too large and $\tilde{\eta}_H$ is positive, as again $f$ would
tend towards 0, constraining the azimuthal field twisting in a
rotationally-supported disc. Similar constraints on the launching of disc
winds were found by \citet{skw2011}, who showed that self-consistent disc wind
solutions only exist for particular combinations of the field polarity and the
ratio of the Hall to ambipolar diffusion parameters. 

It has recently been argued that there exists a handedness to observations of
transverse gradients in the Faraday rotation measure across the base of jets
associated with active galactic nuclei \citep[AGN;][]{cckg2009}. The majority
of sources in which it was possible to determine the transverse gradients were
found to have clockwise gradients, which imply that the outflow has a helical
magnetic field with a preferred positive magnetic polarity (i.e.\ that $B_z$
is parallel to the rotation vector). One explanation of this behaviour is that
the Hall effect is important in the inner accretion disc, acting to cause the
formation of a jet when the field has a positive magnetic polarity, and to
suppress jet formation when the polarity is negative \citep{ka2010}. This
explanation fits the limited available data well, although it can only be
confirmed by future observations at higher resolutions and sensitivities,
which will illuminate the properties of the dense molecular gas in the
accretion discs of AGNs and show whether the Hall current is important to the
jet formation. 

Similarly, it may be possible to show observationally the importance of the
sign of the Hall parameter in determining the properties of protostars and
their discs. Measuring the polarisation of the magnetic field with respect to
the axis of rotation using Zeeman observations of newly-forming stars and
their discs would provide a means of testing whether Hall diffusion plays an
important role in star formation --- should larger discs and lower accretion
rates be correlated in observations with a particular field orientation then
the Hall effect will have been shown to be important to the collapse process.
ALMA (among other next-generation instruments) shall be capable of imaging
nearby dense prestellar cores and their envelopes in both dust and molecular
line emission, and also use polarised dust emission to map the magnetic field
configuration in such cores \citep{alma}. Such observations could be sensitive
enough to observe if there is any difference in the field alignment between
protostellar discs and their envelopes, and whether there is any correlation
between disc size and the direction of rotation in the disc. 

None of the solutions calculated in this work explored the effects of very
weak magnetic diffusion and strong magnetic braking on the core, where the
angular momentum of the collapsing flow is expected to be so reduced that near
to the protostar the fluid is nonrotating (or rotating very slowly). This
behaviour, and its predictions for star and protostellar disc formation are
discussed in the following section. 

\section{The Magnetic Braking Catastrophe}\label{catastrophe}

In gravitational collapse models with magnetic fields two problems arise. The
first occurs in nonrotating collapse simulations where the mass-to-flux ratio
is constant and there is no mechanism for preventing the magnetic field from
falling in, so that as the mass of the protostar builds up so too does its
field. This results in the formation of a star with a very large (by many
orders of magnitude in comparison to observations) magnetic field, and is
referred to as the classical ``magnetic flux problem'' of star formation,
which was described by \citet{cf1953} and outlined in Section \ref{lr:mfp}. 
This is resolved by the inclusion of magnetic diffusion, which allows the
field to drift against the flow of the infalling fluid. 

However, there also exists a secondary problem tying the magnetic flux problem
and the angular momentum problem (described in Section \ref{lr:amp}) together,
in which strong magnetic braking removes all of the angular momentum from the
collapse, preventing the formation of a rotationally-supported disc. This
behaviour, dubbed the ``magnetic braking catastrophe'', affects many recent
numerical collapse simulations \citep[see e.g.][]{asl2003b, ml2008, ml2009,
hc2009}, whereas observations of young stars show that protostellar discs are
common \citep[e.g.][]{jetal2007}. 

The cap placed upon the magnitude of $B_{\phi,s}$ restricted the amount of
magnetic braking in the solutions presented in the previous chapters and is
perhaps the greatest limitation of the semianalytic model constructed in this
work. While it is reasonable to assume that physical processes other than
those included in the self-similar equations will restrict the growth of the
azimuthal field, the fact that $B_{\phi,s}$ is capped for almost all of the
solutions interior to the magnetic diffusion shock suggests that the results
are sensitive to this prescription. Were the cap to be lifted, stronger
magnetic braking could remove all of the angular momentum from the infalling
gas, preventing disc formation. 
\begin{figure}[t]
  \centering
  \vspace{-2mm}
  \includegraphics[width=4in]{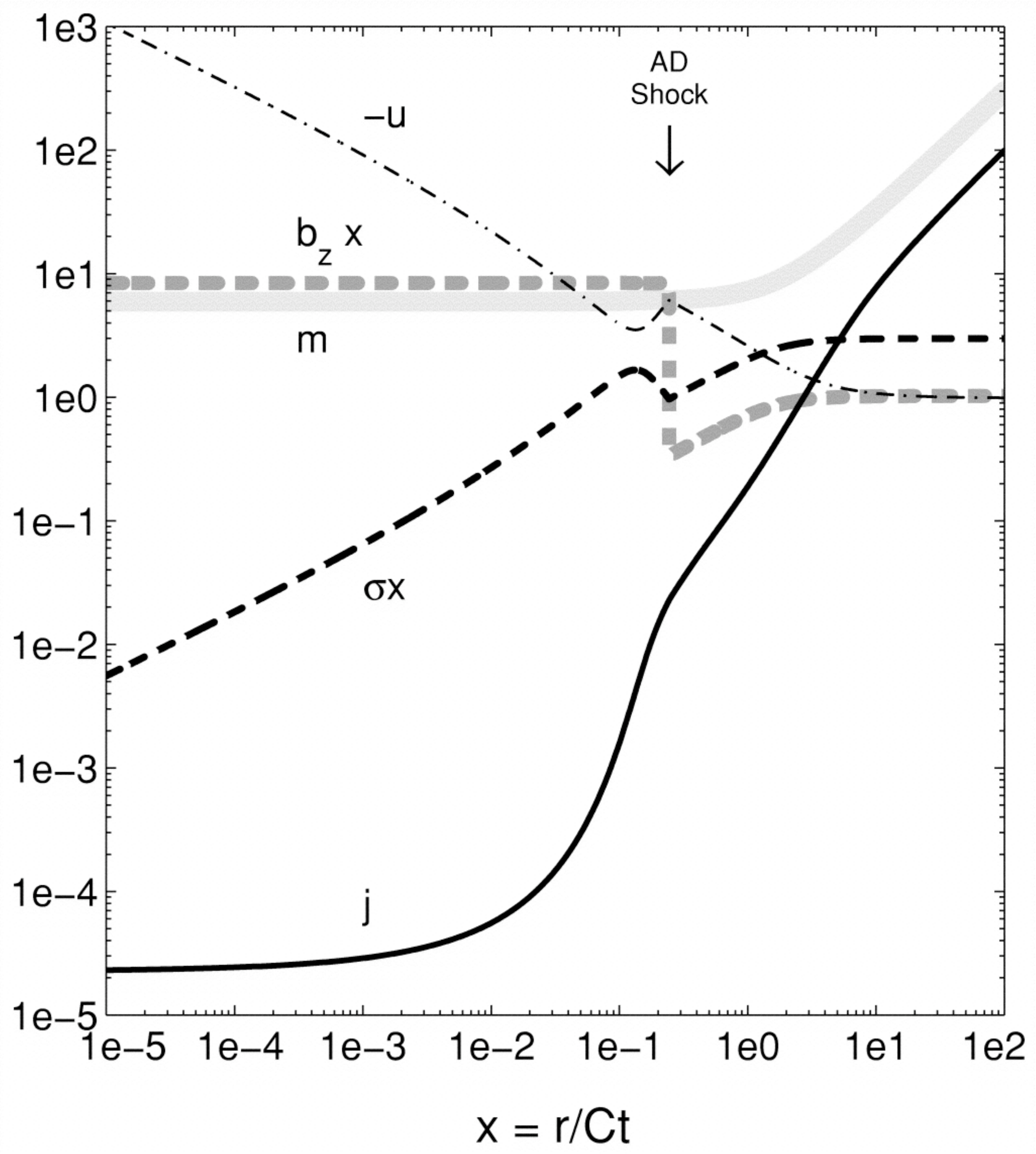}
  \vspace{-5mm}
  \caption[Strong braking solution from \citet{kk2002}]{Figure 10 from
\citet{kk2002}, showing a similarity solution with very strong magnetic
braking. The model parameters are $\tilde{\eta}_A = 0.5$, $v_0 = 1.0$, $\alpha
= 10$ and $\delta = 10$, with the remainder as in Table \ref{tab-bc}. The
magnetic diffusion shock (labelled ``AD Shock'') is located at $x_d = 0.24$,
and the central mass is $m_c = 5.9$.} 
\label{fig-kk10}
  \vspace{-2mm}
\end{figure}

In their semianalytic solutions to gravitational collapse with ambipolar
diffusion \citet{kk2002} found solutions in which disc formation was inhibited
by the magnetic braking. These ``strong braking'' solutions were found by
raising the azimuthal field cap, $\delta$, and the magnetic braking parameter,
$\alpha$ (where $\alpha$ is the ratio of the sound speed $c_s$ to the external
Alfv\'en speed $V_{A,ext}$, defined in Equation \ref{alpha}), so that the
magnetic braking reduces the centrifugal force until it is incapable of
supporting the gas against collapse. The high value of the azimuthal field cap
($\delta = 10$) allows the magnetic field to become tightly wrapped around the
rotational axis \citep[this has also been seen in numerical simulations such
as][]{mim2008}, however this wrapping may not be possible in real cores. 

This behaviour was illustrated in figure {10} of \citet{kk2002}, reproduced
here in Figure \ref{fig-kk10}, which is characterised by $\delta = \alpha =
10$, $\tilde{\eta}_A = 0.5$ and $v_0 = 1.0$, with the rest of the parameters
as listed in Table \ref{tab-bc}. In the outer region of the core the collapse
follows that for the similarity solutions with capped braking in Figure
\ref{fig-admod}, as IMHD is dominant. However, as the field is decoupled from
the neutrals the magnetic braking rapidly starts to reduce the angular
momentum. The magnetic diffusion shock at $x_d = 0.24$ sees an increase in the
vertical magnetic field component that accelerates the magnetic braking
further, and the angular momentum of the infalling material drops to a small
nonzero value $j_{pl} \approx 2 \times 10^{-5}$ which depends upon the
ambipolar diffusion parameter and the initial rotation rate. In this ambipolar
diffusion solution the neutral component must be rotating relative to the
charged component of the fluid in order to feel the magnetic braking torque;
this causes the formation of a plateau from which the angular momentum cannot
be reduced further. As there is no rotational support the fluid falls in with
a radial velocity $u$ close to the free fall velocity, and no disc forms. 

Such ``strong braking'' solutions were not duplicated in this work, however it
is clear that if the magnetic braking parameter and the azimuthal field cap
are raised then the similarity solutions with Hall diffusion will display
similar behaviour to that in Figure \ref{fig-kk10}. The inner asymptotic
behaviour of the solutions would follow that outlined by Equations
\ref{ff-ssimm}--\ref{ff-ssim} in Section \ref{ff} for the supersonic
magnetically-diluted free fall onto the central protostar. No
rotationally-supported disc would form, clearly demonstrating the magnetic
braking catastrophe.

The contrast in behaviour between the ambipolar diffusion solutions of
\citet{kk2002} demonstrating disc formation and those of \citet{ml2009}
illustrating the magnetic braking catastrophe is discussed in the following
subsection.

\subsection{Case study: ambipolar diffusion collapse}
\begin{figure}[p]
  \centering
  \includegraphics[width=4.9in]{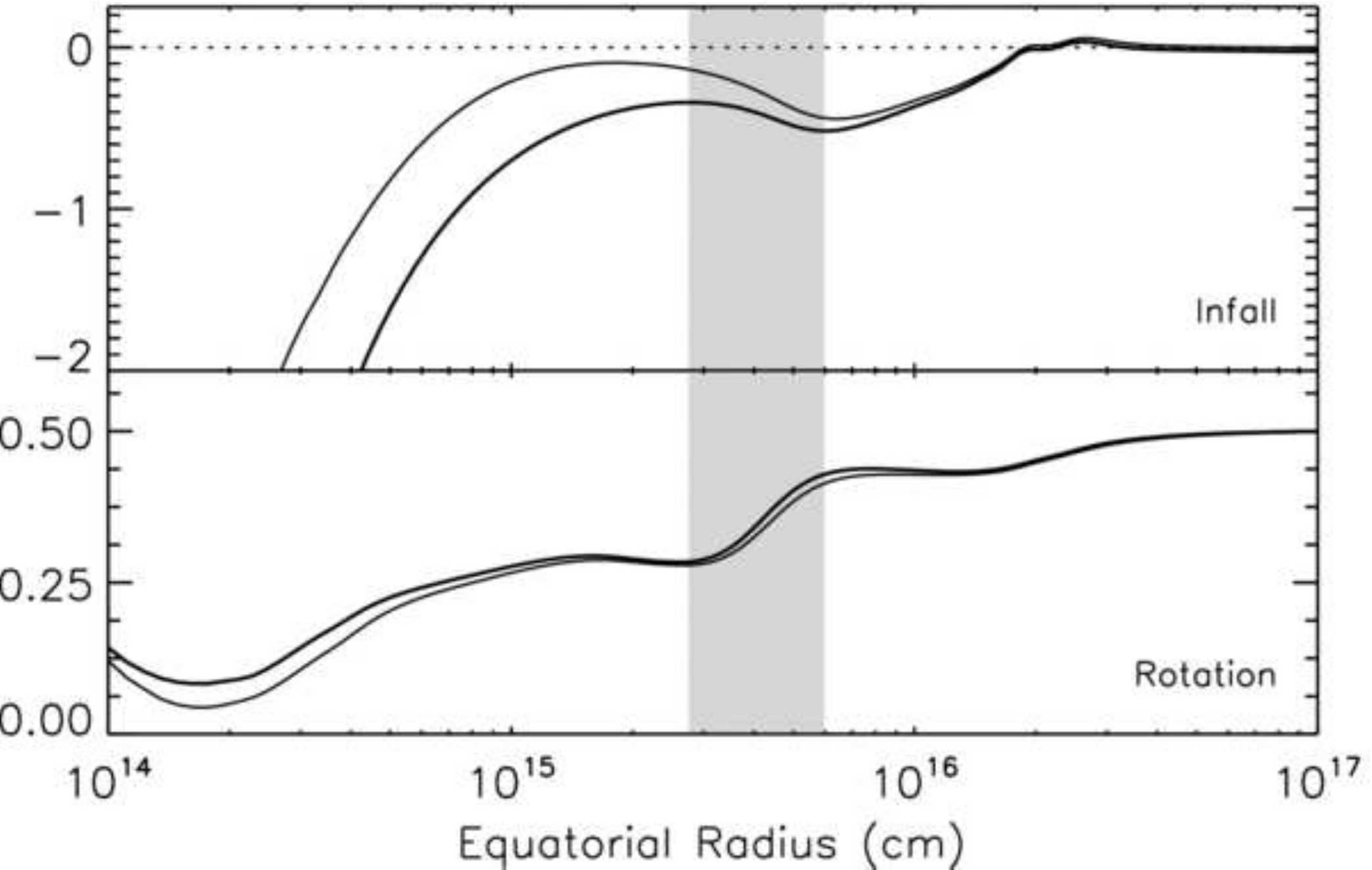}
  \vspace{-5mm}
  \caption[Velocity profiles from \citet{ml2009}]{Radial and rotational
velocities from the standard model of \citet[figure 3]{ml2009} in units of the
sound speed at a time $t = 5.85 \times 10^{11}$ s. The rotational velocity
decreases with decreasing radius, indicating that there is strong magnetic
braking, particularly in the deceleration region (which has been shaded). The
thick lines are the velocities of the neutral particles, and the thin lines
those of the ions.} 
\label{figml}
  \includegraphics[width=4.7in]{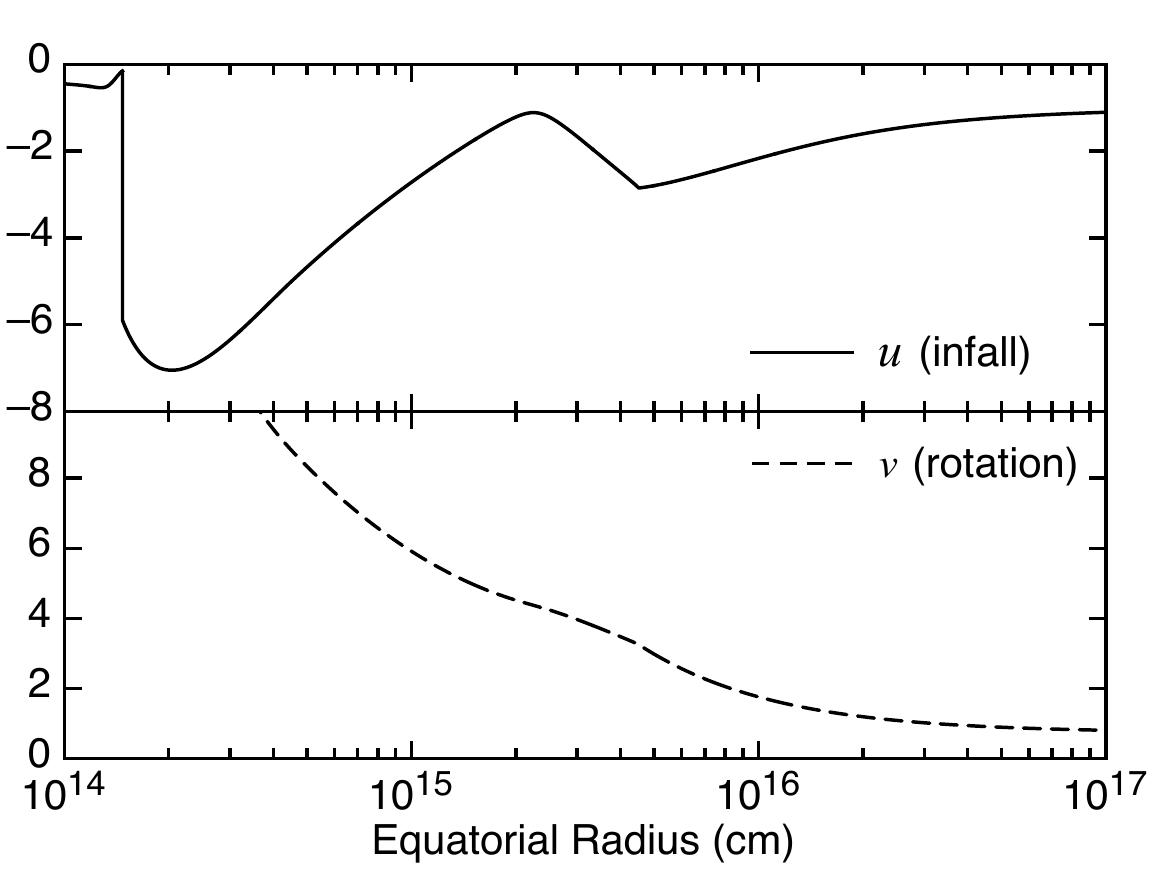}
  \vspace{-3mm}
  \caption[Velocity profiles from \citet{kk2002}]{Rotational and radial
velocity profiles in units of the sound speed at a time $t = 5.85 \times
10^{11}$ s from the fiducial solution of \citet{kk2002} (which was presented
in Figure \ref{fig-admod} of Chapter \ref{ch:nohall}). In this solution the
rotational velocity increases as the increased ambipolar diffusion reduces the
magnetic braking so that a centrifugal disc forms at $r \approx 10^{14}$ cm.}
\label{figmlkk}
  \vspace{-5mm}
\end{figure}

The similarity solution calculated in Chapter \ref{ch:nohall} for the
ambipolar diffusion-only collapse \citep[reproducing the fiducial result
of][]{kk2002} used a cap on the azimuthal field component to limit the
magnetic braking and ensure disc formation. This is in direct contrast with
the numerical simulations of \citet{ml2009}, who were unable to observe disc
formation in their simulations of collapsing cores with ambipolar diffusion
unless the cosmic ray ionisation rate was unreasonably low or the magnetic
field was particularly weak. To explore the discrepancy in disc forming
behaviour between the two models, the velocity profiles of both the standard
solution of \citet[figure 3]{ml2009} and the fiducial solution of
\citet{kk2002} at a time $t = 5.85 \times 10^{11}$ s ($1.85 \times 10^4$
years) are shown in Figures \ref{figml} and \ref{figmlkk} respectively. 

The sharp decrease in the infall velocity in Figure \ref{figmlkk} at $r \sim
10^{14}$ cm marks the boundary of the Keplerian disc in the similarity
solution, interior to which the rotational velocity increases further. No such
deceleration of the infall velocity occurs in Figure \ref{figml}, where the
uncapped magnetic braking prevents disc formation on this scale by removing 
all rotational support from the collapsing core. Both plots demonstrate
similar infall velocity behaviour in the region of the magnetic diffusion
shock at $r \sim 4\times 10^{15}$ cm, while the rotation velocity is always
quite different between the two solutions. Near the inner boundary of Figure
\ref{figml} the rotation velocity does start to increase, however this is
sufficiently far from the Keplerian value that any disc would have to form
inside the inner boundary of the calculations. While it is most likely the
cap on the magnetic braking in the similarity solution of Figure \ref{figmlkk}
that is responsible for the difference in disc-forming behaviour, there are
other discrepancies between the two models that explain why one is able to
form discs and the other cannot. 

The simulations of \citet{ml2009} were an extension of their previous work on
the collapse of molecular cloud cores under IMHD \citep{ml2008}, which was
modified to include ambipolar diffusion of the magnetic field. A spherical
coordinate system under the assumption of axisymmetry was adopted, and the
computational grid extended from $10^{14}$ to $10^{17}$ cm in $r$, and from
$0$ to $\pi$ in $\phi$. The resolution of the solution was characterised by
120 grid points in the radial direction and 60 angular points; these grid
points were logarithmically spaced in radius. The collapse solutions were not
fully isothermal; a broken power law equation of state was used to describe
the behaviour of the fluid so that the gas was isothermal below $\rho =
10^{-13}$ g cm$^{-3}$ and adiabatic with $\gamma = 7/5$ above.

The outer boundary of the collapse simulations of \citet{ml2009} was taken to
be the standard ``outflow'' condition that was implemented as a feature of the
ZeusMP MHD code in \citet{hetal2006}. The inner boundary was a modified
outflow condition, with the mass accreted across the boundary added to the
central point mass and a torque-free condition imposed upon the magnetic field
(that $B_\phi = 0$). The initial conditions of the core were idealised as a
rotating self-similar isothermal toroid supported against gravity partially by
thermal pressure and partially by the magnetic field at the moment of point
mass formation. These initial conditions are compared in Table \ref{tab-ml} to
those used for the fiducial solution of \citet{kk2002} (reproduced in Section
\ref{ad} as Figure \ref{fig-admod}). 
\begin{table}[t]
\begin{center}
\begin{tabular}{ccc}
\toprule
 Parameter & \citet{ml2009} & \citet{kk2002}\\
       & Figure \ref{figml} & Figure \ref{figmlkk}\\
 \midrule
	 $v_0$    	  & 0.5  	& 0.73\\
	 $u_0$    	  & 0    	& $-1.0$\\
	 $\mu_0$  	  & 4    	& 2.9\\
	 $A$      	  & 2.8		& 3\\
         $m_c$		  & 4.5		& 4.7\\ 
	 $\tilde{\eta}_A$ & 0.087 	& 1\\
         \\
      $c_s$ (km s$^{-1}$) & 0.3		& 0.19\\
       $\xi$ (s$^{-1}$) & 10$^{-17}$  & 10$^{-17}$\\
 \bottomrule
 \end{tabular}
\end{center}
\vspace{-5mm}
 \caption[Comparison of initial conditions and parameters between AD
models]{Comparison of initial conditions and collapse parameters between the
models illustrated in Figures \ref{figml} and \ref{figmlkk}. All of these are
given in the nondimensional form used in this work (see Section \ref{eqouter}
for details).} 
\label{tab-ml}
\end{table}

There are two major discrepancies between the initial conditions and collapse
parameters between the two models: the first is the infall rate at the outer
boundary $V_r = c_su_0$, and the second is the nondimensional ambipolar
diffusion parameter $\tilde{\eta}_A$, which changes by more than an order of
magnitude between the models. The initial velocity at the outer boundary was
found to have no qualitative effect on the simulations of collapse with IMHD
by \citet{ml2008}; this is used to justify their choice of no initial infall
rate as the outer boundary condition in this model. However, the ambipolar
diffusion parameter is key to the disc formation behaviour of the two models
and this large variation between the two models, in addition to the capped
magnetic braking, is more than enough to explain the differences between them.

Although both models cite the same cosmic ray ionisation rate (see Table
\ref{tab-ml}), there is an order of magnitude difference between the ambipolar
diffusion parameters adopted by each model. This occurs because \citet{kk2002}
chose an anomalously large value of $\tilde{\eta}_A = 1$ to use in their
fiducial solution. In exploring the applicability of the assumption that
$\tilde{\eta}_A$ is a constant in their model, \citet{kk2002} stated that
$\tilde{\eta}_A \approx 0.2 \xi^{-1/2}_{-17}$ in the outer regions of the
collapse where the density is typically between $\sim 10^4$ and $\lesssim
10^7$ cm$^{-3}$, and $\tilde{\eta}_A \approx 0.07$ in the inner regions where
the density lies between $10^7$ and $10^{12}$ cm$^{-3}$. These values are
clearly much closer to the value used by \citet{ml2009} in their standard
model, and indeed when \citet{ml2009} dropped their ionisation rate to $\xi =
10^{-18}$ s$^{-1}$ (corresponding to $\tilde{\eta}_A = 0.87$) their collapse
model was able to form a centrifugal disc around the protostar. 

In Section \ref{ad}, it was shown that the position of the centrifugal shock
depends quite clearly upon the ambipolar diffusion parameter. Substituting the
fiducial model parameters of \citet{ml2009} into Equation \ref{adxc} gives an
estimate of the shock position at the time $t = 5.85\times10^{11}$ s that is
many orders of magnitude smaller than the inner boundary of the collapse (due
to the exponential in the equation). Although it is a poor approximation to
the shock position in this region of parameter space, Equation \ref{adxc}
suggests that no rotationally-supported disc larger than the inner boundary of
Figure \ref{figml} ($R = 10^{14}$ cm) will form unless $\tilde{\eta}_A >
0.68$. Earlier in this section it was demonstrated that a large decrease in
the ambipolar diffusion parameter such as that between the two solutions,
coupled with the corresponding increase in the rate of magnetic braking,
prevented disc formation in the similarity solutions of \citet{kk2002}, as the
magnetic braking catastrophe also inhibited disc formation in their similarity
solution with $\tilde{\eta}_A = 0.5$ (shown in Figure \ref{fig-kk10}). 

The fiducial similarity solution of \citet{kk2002} in Figure \ref{figmlkk}
also possessed a higher initial rotation velocity than the simulation of
\citet{ml2009} in Figure \ref{figml}, and it was shown in Chapter
\ref{ch:nohall} that the value of $v_0$ directly affects the size of the
rotationally-supported disc, with larger discs corresponding to larger initial
rotational velocities. Contrasting the two simulations, as well as the strong
braking solutions of \citet{kk2002} in Figure \ref{fig-kk10}, shows that the
cap on $B_{\phi,s}$ is most likely responsible for the differences in the
magnetic braking behaviour. 

Further exploration of parameter space, particularly in matching the
parameters used in numerical and self-similar models, is needed to explore
whether adopting realistic values and scalings of the ambipolar and Hall
diffusion parameters will resolve the magnetic braking problem. In the
following subsection several mechanisms for reducing the magnetic braking in
the collapsing flow and their expected results shall be discussed. 

\subsection{Proposed Solutions}\label{mbc:sols}

Keplerian discs form in those simulations in which there is enough angular
momentum that the centrifugal force is able to support the fluid against
collapse. The magnetic braking catastrophe occurs because the twisting of the
field lines caused by the rotation of the fluid in the disc transports angular
momentum from the core to the envelope, slowing the rotation and inhibiting
disc formation. This can be seen in the results of Sections \ref{nonmag} and
\ref{imhd}, where the introduction of ideal MHD to the previously nonmagnetic
collapse saw magnetic braking reduce the size of the rotationally-supported
disc, and prevents disc formation entirely in both the IMHD strong braking
similarity solution in figure 6 of \citet{kk2002} and many numerical
simulations \citep[see e.g.][]{asl2003b, glsa2006, ml2008}

Various approaches to solving the magnetic braking catastrophe have been
adopted with different degrees of success. It is possible to reduce the
magnetic braking by changing the alignment of the magnetic field with respect
to the rotational axis, limiting the twisting of the field within the core or
the amount of angular momentum that can be accepted by the envelope and
employing magnetic diffusion to change the coupling between the magnetic field
and the fluid. Magnetic diffusion reduces the amount of braking by allowing
the field lines at the midplane to slip against the neutrals rather than
rotating with them around the core. 

The amount of magnetic braking affecting the core depends upon the product
$B_zB_{\phi,s}$, and so any change in the alignment of the field with respect
to the axis of rotation shall influence the magnetic braking behaviour.
\citet{mp1980} showed that the magnetic braking affecting a disc of fixed
height was more efficient when the magnetic field and the axis of rotation
were perpendicular. The rotational waves in their discussion propagated
perpendicular to the axis of rotation, affecting matter in the external medium
with increasing moments of inertia as the moment of inertia of the core itself
decreased, and this lead to a higher magnetic braking efficiency.

\citet{hc2009} found that the opposite was true. In their three-dimensional
simulations including IMHD the height of the pseudodisc and inner Keplerian
disc increases with the angle of alignment between the field and the
rotational axis as the magnetic field lines are twisted about an axis parallel
to the plane of the pseudodisc. The rotation of the core itself also impedes
contraction along the magnetic field lines, further increasing the scale
height of the core. This increase in scale height causes a reduction in the
magnetic braking as the braking time is shorter, so that larger Keplerian
discs form when the angle between the rotation and magnetic field axes is
90$^\circ$. 

This conclusion was supported by the results of \citet{ch2010}, which showed
that the angle between the rotation and magnetic field axes also influences
the formation of outflows and jets, and can suppress these when the axes are
perpendicular. Any misalignment causes the jets to precess, which further
reduces the efficiency of the magnetic torque in removing angular momentum
from the pseudodisc and allows a larger Keplerian disc to form when the angle
is perpendicular. 

Other mechanisms for limiting the magnetic braking in the collapsing core and
enabling disc formation include capping the azimuthal field and insisting that
the core possess a finite boundary, as disc formation is clearly enabled by
limiting the degree of magnetic braking in the collapsing core. In
\citet{kk2002} and this thesis it was assumed that the external envelope of
the core was effectively infinite and that the magnetic field lines were
anchored into this external medium. This meant that the magnetic braking could
transfer as much angular momentum from the core as the cap on $B_{\phi,s}$
allowed, which (as shown in Figure \ref{fig-kk10}) could inhibit disc formation
when the braking was sufficiently strong. However, this limitation is
unrealistic, as the forming protostar is expected to accrete or dissipate the
envelope, and the cap on $B_{\phi,s}$ is itself merely a way of representing
unknown disc physics that needs to be better understood, which should be
replaced in future work on the subject (discussed further in Section
\ref{future}). 

Simulations of the collapse of a core in a finite host cloud under IMHD
\citep{glsa2006} and resistive MHD \citep{mim2007, mim2010} were able to form
Keplerian discs, as the magnetic braking became less effective as the core
envelope was depleted. In these simulations the early phase of collapse was 
characterised by slow circumstellar disc growth regulated by magnetic braking;
in the later stages the envelope is depleted by accretion onto the pseudodisc,
so that magnetic braking is no longer effective and a large disc is able to
form. Limiting the amount of material that can accept the angular momentum
from the core clearly allows a rotationally-supported disc to form, and so the
boundary conditions of numerical simulations must be carefully considered in
light of the magnetic braking catastrophe. 

Magnetic field diffusion is also important to resolving the magnetic braking
catastrophe, as less magnetic braking occurs in a pseudodisc that has less
magnetic flux. In calculations where the Ohmic resistivity is taken to be
spatially-constant, such as those by \citet{sglc2006} and \citet{kls2010}, the
Ohmic diffusivity required to allow disc formation must be anomalously large.
\citet{kls2010} in particular required an Ohmic resistivity three orders of
magnitude larger than that expected in molecular cloud cores and many
numerical reconnection events in order to form a rotationally-supported disc
that was $\sim100$ AU during the Class 0 stage of star formation. In reality
the importance of Ohmic dissipation is expected to vary with density (see
Figure \ref{eta-regimes}) in the inner regions of the core; by taking it to be
spatially-constant across the collapsing core a higher value is required in
order to avoid the magnetic braking catastrophe.

When the Ohmic diffusivity is not spatially-constant, and instead varies with
the density of the neutrals, it is able to prevent the removal of all the
angular momentum from the core. \citet{mim2010} found that their simulations
with Ohmic dissipation formed rotationally-supported discs with radii $\sim
100$ AU at $t_c = 10^5$ yr (where $t_c$ is the time after disc formation) as
the decoupling of the field from the charged particles in the innermost
regions of the collapse halted the magnetic braking in these regions.
\citet{db2010}, following on from \citet{mim2007}, demonstrated that a
canonical amount of Ohmic dissipation was able to reduce the amount of braking
in the first (adiabatic) core so that a centrifugal disc formed in their
axisymmetric numerical calculations. The Ohmic resistivity adopted in their
calculations \citep[and in those of][]{mim2010} depended upon the density in
the core and never greatly exceeded that expected from ionisation calculations
such as those by \citet{wn1999}. 

Reducing the inner field strength by ambipolar diffusion allows larger discs
to form than in IMHD simulations with equivalent initial conditions and
parameters (demonstrated in Section \ref{ad}), however, this required the use
of a cap on $B_{\phi,s}$. Without this cap, ambipolar diffusion is unable to
completely resolve the magnetic braking catastrophe \citep[e.g. Figure
\ref{fig-kk10};][]{kk2002, ml2009}. The simulations of \citet{ml2009}
demonstrating the magnetic braking catastrophe were discussed in detail in the
previous subsection, however these showed a possible resolution of the
catastrophe by reducing the cosmic ray ionisation rate (increasing the
ambipolar diffusion parameter) in the core. (\citet{km2010} were able to form
disc-like structures in their simulations in which the field was decoupled
from all species by ambipolar diffusion before Ohmic dissipation became
important, however their solutions were nonrotating and so no Keplerian disc
could form.) 

The results of the semianalytic model presented in this work showed that the
Hall effect also plays a critical role in determining the amount of magnetic
braking that affects the collapsing flow. It is difficult to include Hall
diffusion in numerical simulations, due to the small time steps required to
trace it accurately, however some work is being done in this area
\citep[see][]{kls2010-2}. The size of the rotationally-supported disc that
formed in the solutions presented in the previous chapter depends upon the
direction of the magnetic field, and varied by almost an order of magnitude
between the solutions with $\tilde{\eta}_H = -0.2$ and $\tilde{\eta}_H =
+0.2$. As will be discussed further in subsection \ref{nonrot}, Hall diffusion
can increase the magnetic braking affecting the core or instead act in the
opposite direction to increase the amount of rotation in the disc, despite the
cap that has been placed upon $B_{\phi,s}$. 

Although much of the focus of this thesis has been on the magnetic braking
caused by Hall and ambipolar diffusion, it is also the case that in weakly
ionised cores with no initial rotation the Hall effect can induce rotation via
\textit{magnetic acceleration}. Some preliminary work has been performed to
calculate such similarity solutions; in subsection \ref{nonrot} it shall be
shown that Hall diffusion could induce disc formation in an initially
nonrotating core.

\subsection{Hall-driven spin-up of collapsing cores}\label{nonrot}

In \citet{wn1999} it was suggested that the Hall current in a collapsing
molecular cloud core would cause $B_\phi$ to grow and force the fluid to pick
up a toroidal component of momentum, which could contribute to the
gravitational support of the core. \citet{kls2010-2} have recently published
the results of a three-dimensional simulation of the collapse of an initially
nonrotating core under the influence of Hall diffusion, showing that the Hall
effect causes the collapsing material to spin-up. They demonstrated that
azimuthal torques powered by the Hall effect twist the ions and by extension
the neutral particles (as the magnetic field is anchored in the external
medium) as expected; conservation of momentum implies then that the gas and
envelope will both spin-up in opposite directions, which was observed in their
simulation. 

The full model code from Chapter \ref{ch:hall} has been employed here to
similarly demonstrate such Hall spin-up of an initially nonrotating molecular
cloud core. This model has yet to converge on the true similarity solution,
as the shock-finding routine outlined in Section \ref{numerics} does not
converge properly in these calculations without modification. The variables in
the outer regions and those downstream of the magnetic diffusion shock before
the gravity of the central mass dominates the radial velocity of the fluid are
quite trustworthy, while the inner regions of the calculations are unreliable
representations of the core behaviour.

\begin{figure}[htp]
  \centering
  \vspace{-1mm}
  \includegraphics[width=5.2in]{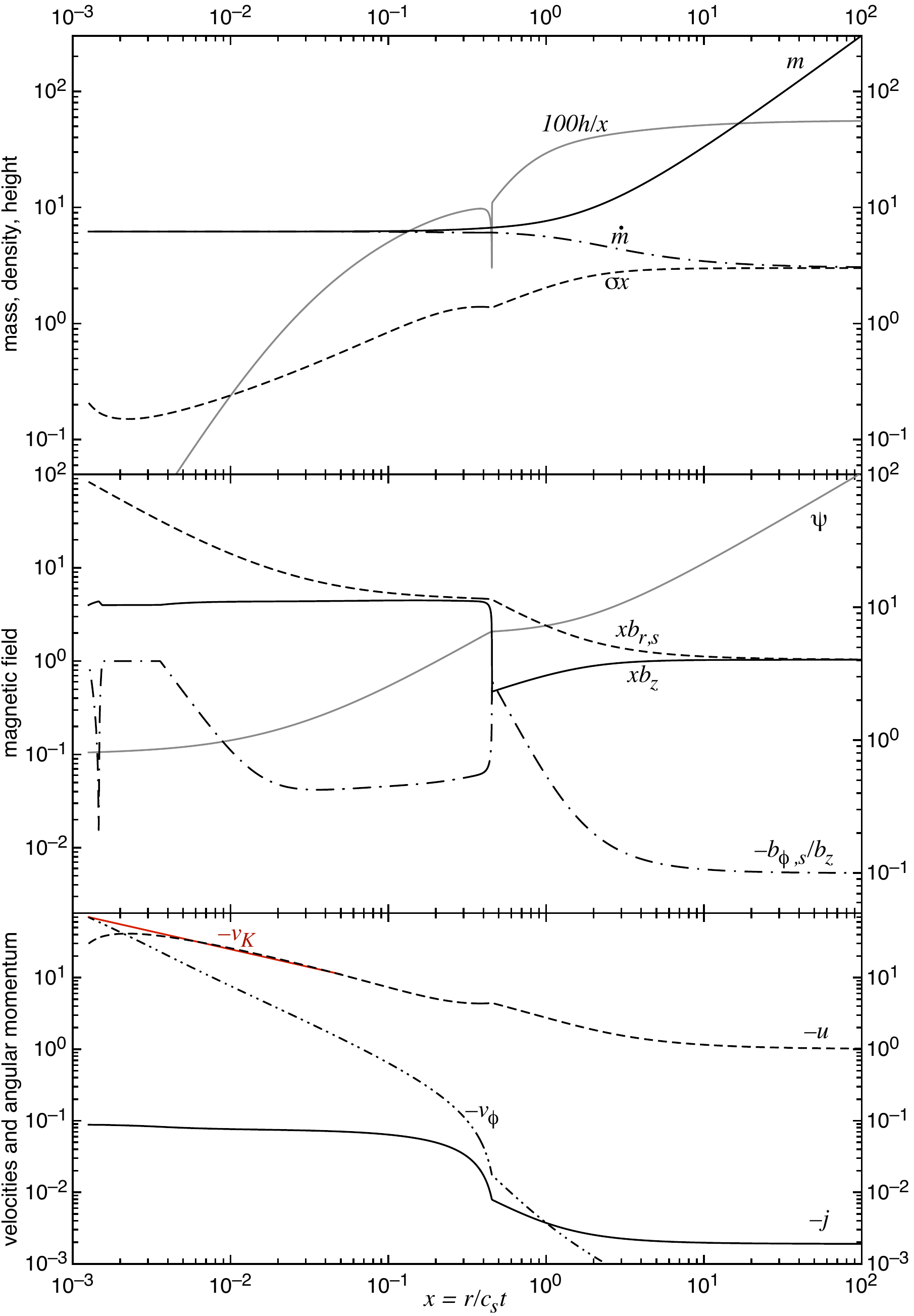}
  \vspace{-4mm}
  \caption[Non-converged initially nonrotating similarity
solution]{Non-converged similarity solution with $\tilde{\eta}_H = -0.1$ and
initial rotational velocity $v_0 = 0$. The other initial conditions and
parameters are those in Table \ref{tab-bc}. The model code used to calculate
the similarity solutions in Chapter \ref{ch:hall} was unable to locate the
inner centrifugal shock position and converge on the true solution; however,
the angular momentum has increased downstream of the magnetic diffusion shock
until it is comparable to that for the similarity solutions with nonzero
$v_0$. The solid red line in the lower panel is the nondimensional Keplerian
velocity $v_K = \sqrt{m_c/x}$.}  
\vspace{-5mm}
\label{fig-nr}
\end{figure}

Figure \ref{fig-nr} shows this non-converged solution to the outer boundary 
conditions, which (as in Figure \ref{fig-adsimp}) fails on the inward
integration interior to the magnetic diffusion shock. This calculation has
nondimensional Hall parameter $\tilde{\eta}_H = -0.1$ and ambipolar diffusion
parameter $\tilde{\eta}_A = 1.0$ (matching those in Figure
\ref{fig-hall-0.1}), which were chosen to demonstrate the spin-up behaviour 
caused by the Hall effect, even when $\tilde{\eta}_H$ is small, is enough that
a rotationally-supported disc could form in the fully converged solution. The
solution satisfies the outer boundary conditions for the parameters listed in
Table \ref{tab-bc}, save for the initial rotational velocity which is taken to
be $v_0 = 0$ in the initially nonrotating core. The central mass is estimated
by the plateau value $m_c \approx m_{pl} = 6$ defined in Equation \ref{nmmpl}. 

In the outermost regions of Figure \ref{fig-nr} the material undergoes
near-ideal MHD collapse; the small divergences from flux-freezing cause the
angular momentum of the gas to rapidly increase from zero at the outer
boundary $x_{out} = 10^4$ (which is outside the domain of the figure) to a
plateau value determined by the Hall diffusion parameter. The value of this
plateau is estimated from the conservation of angular momentum equation
(\ref{ssam}), in the limit where $j$ is constant and given by the plateau
values $j = j_{pl}$: 
\begin{equation}
    \frac{dj}{dx} = \frac{j_{pl}}{w} - \frac{x^2b_zb_{\phi,s}}{m} = 0.
\label{conam0}
\end{equation}
In the large $x$ limit the infall velocity is small compared to $x$ so that $w
= x$, and the other variables are given by the outer asymptotic power law
relations defined in Section \ref{eqouter}. These relations are substituted
into Equation \ref{conam0}, which is rearranged to give the expression 
\begin{equation}
    j_{pl} = \frac{xb_{\phi,s}}{\mu_0}\,.
\label{jplprelim}
\end{equation}

The azimuthal field component $b_{\phi,s}$ in this region is determined by
both the Hall diffusion and the vertical compression of the disc, which in
turn depend upon the self-gravity of the pseudo disc and the magnetic
squeezing caused by $b_{r,s}$. The vertical hydrostatic equilibrium equation
(\ref{ssvhe}) is 
\begin{equation}
    \left(\frac{\sigma m_c}{x^3} - b_{r,s}\frac{db_z}{dx}\right)h^2 
    + \left(b_{r,s}^2 + b_{\phi,s}^2 + \sigma^2\right)h -2\sigma = 0;
\label{jplhold}
\end{equation}
in the limit of large $x$, both $db_z/dx$ and $b_{r,s}$ are inversely
proportional to $x$ and as such their product may be dropped from the
equation. Similarly the azimuthal field component is small in comparison to
the poloidal components, so that Equation \ref{jplhold} may be simplified into
\begin{equation}
    \left(b_{r,s}^2 + \sigma^2\right)h = 2\sigma,
\label{jplhsimp}
\end{equation}
which becomes
\begin{equation}
    h = \frac{2}{\sigma(1 + \mu_0^{-2})}
\label{jplh}
\end{equation}
upon substitution of the outer boundary flux-freezing condition for the radial
field component. Physically, in the outermost regions of the core the scale
height is dominated by tidal compression and magnetic squeezing associated
with the radial field component $b_{r,s}$.

As the azimuthal field component and angular momentum are small when $x$ is
large, Equation \ref{ssb_phis} may then be written as
\begin{equation}
    b_{\phi,s} = \frac{2\alpha\tilde{\eta}_H \psi bb_{r,s}h^{1/2}}{x^2\sigma^{3/2}}
     \left[1 + \frac{2\alpha\tilde{\eta}_A\psi b_zh^{1/2}}{x^2\sigma^{3/2}}\right]^{-1}.
\label{jplb_phisimp}
\end{equation}
The total field magnitude may be approximated by $b = \sqrt{2}b_z$ as the
azimuthal component is small. The outer asymptotic relations (Equations
\ref{outersig}--\ref{outeru}) are then substituted into Equation
\ref{jplb_phisimp}, so that $b_{\phi,s}$ may then be approximated by 
\begin{equation}
    b_{\phi,s} = \frac{4\alpha A\tilde{\eta}_H}{x\mu_0} 
     \left[\mu_0^2\sqrt{1+\mu_0^{-2}} 
     + 2\sqrt{2}\alpha\tilde{\eta}_A\right]^{-1}.
\label{jplb_phis}
\end{equation}
This is then substituted into Equation \ref{jplprelim} to give an estimate of
the angular momentum plateau,
\begin{equation}
    j_{pl} = \frac{4\alpha A\tilde{\eta}_H}{\mu_0^2}
     \left[\mu_0^2\sqrt{1+\mu_0^{-2}} + 2\sqrt{2}\alpha\tilde{\eta}_A\right]^{-1}.
\label{jpl}
\end{equation}
\begin{table}[t]
 \begin{center}
  \begin{tabular}{cccccc}
   \toprule
    $\tilde{\eta}_A$ & $\tilde{\eta}_H$ & $j_{pl,est}$ & $j_{pl,sol}$ 
             & $(-b_{\phi,s}/b_z)_{est}$ & $(-b_{\phi,s}/b_z)_{sol}$\\
   \midrule
    1.0 & $-0.6$ & $-7.508\times10^{-3}$ & $-7.506\times10^{-3}$ & 0.02105
        & 0.02146 \\
    1.0 & $-0.1$ & $-1.251\times10^{-3}$ & $-1.251\times10^{-3}$ & $3.508\times10^{-3}$ 
        & $3.577\times10^{-3}$ \\
   \bottomrule
  \end{tabular}
 \end{center}
 \vspace{-5mm}
 \caption[Estimated vs actual values of $j_{pl}$ in the Hall spin-up
calculations]{Calculated and actual values of the angular momentum plateau and
the relation $-b_{\phi,s}/b_z$ in the non-converged (though matching on the
outer boundary) solution for an initially nonrotating core. The values from
the solutions are taken at $x=100$; and the parameters used in the estimations
are those for the similarity solutions presented in Chapters \ref{ch:nohall}
and \ref{ch:hall} ($\alpha = 0.08$, $A = 3$ and $\mu_0 = 2.9$).} 
 \label{tab-jpl}
\end{table}

The value of the angular momentum plateau and the near-constant ratio of the
azimuthal to vertical field components in the plateau region for the
non-converged similarity solution in Figure \ref{fig-nr} and a second
calculation in which $\tilde{\eta}_H = -0.6$ (which is also matched to the
outer boundary only) are compared to the theoretical estimations from
Equations \ref{jplb_phis}--\ref{jpl} in Table \ref{tab-jpl}. The estimated
values match the calculated ones very well, with $j_{pl}$ matching to $\sim
0.1\%$ and $b_{\phi,s}/b_z$ to $\sim 2\%$, demonstrating that the physics of
the region is well accounted for by the approximations. All of the angular
momentum in the core, and the twisting of the field lines is caused by the
Hall effect, as the radial component of the magnetic pressure gradient causes
azimuthal drift of the field lines, which in turn creates the torque that
spins up the ions and neutrals. The value of the specific angular momentum
plateaus once the induced rotation is rapid enough that the magnetic torque is
balanced by the rotational torque. As the density of the gas is quite limited
in the outer regions, the Hall spin-up of the disc is limited until the
variables start to diverge from their asymptotic behaviour and IMHD breaks
down.

The behaviour of the fluid downstream of the magnetic diffusion shock at $x_d
= 0.453$ (which is close to the value $x_d = 0.435$ estimated by multiplying
Equation \ref{xdgood} by the factor 0.83; see Section \ref{shocks}) is less
clear, as the solution has not yet converged and the position of the
centrifugal shock is only approximately known. As in the initially rotating
solutions in Chapter \ref{ch:hall}, the fluid is slowed at the shock, until
the gravity of the central mass starts to dominate the radial velocity and the
fluid is accelerated inwards once more. Hall diffusion quickly spins up the
gas in the post-shock region, until again it reaches a plateau which is
similar in value to that experienced in the rotating solution in Figure
\ref{fig-hall-0.1}. Near the region where the centrifugal force becomes
important, the azimuthal component of the magnetic field $b_{\phi,s}$ reaches
its capped value, and the angular velocity is quite close to that expected
from Keplerian rotation (depicted as the solid red line in the lower panel of
Figure \ref{fig-nr}). 

Much remains to be done on this model to ensure that the similarity solution
converges to the true solution and matches the expected asymptotic behaviour
on both boundaries of the collapse. The spin-up behaviour can clearly only
exist when there is Hall diffusion of the magnetic field, as when
$\tilde{\eta}_H = 0$ there is no mechanism for inducing rotation in an
initially nonrotating core and neither the angular momentum or the azimuthal
field component can acquire a nonzero value. Although disc formation has yet
to be observed, the solution in Figure \ref{fig-nr} shows that the rotational
velocity of the collapsing material reaches the rotationally-supported value
of $v = v_K$ near the point where the code breaks down, suggesting that disc
formation may be enabled by the Hall effect in the collapsing gas, supporting
the numerical results of \citet{kls2010-2}.

The direction of rotation within the collapsing flow is determined by the sign
of $\tilde{\eta}_H$. The spin in the outer regions of the core, given by
Equation \ref{jpl}, shows this; and as the field increases the magnetic
acceleration causes the gas to spin-up further in this direction. While this
solution is clearly a special case, such spin-up behaviour caused by the Hall
effect could in fact resolve the magnetic braking catastrophe. Hall diffusion
causes a rotational torque on the gas that can enhance or reduce the magnetic
braking that removes angular momentum from the disc and it does not stop
acting once all of the angular momentum reaches zero --- indeed, it could
act at this point to start spinning the fluid back up in the opposite
direction to the initial rotation of the core. This may produce enough angular
momentum to form a Keplerian disc, in complete defiance of the expected
behaviour from non-Hall solutions in which there is no way to increase the
rotational energy of the flow.

The magnetic braking catastrophe prevents the formation of
rotationally-supported discs around protostars by the removal of angular
momentum from the collapsing flow, in contrast with observations that show
protostellar discs are common. This causes the collapse downstream of the
magnetic diffusion shock to occur in near-free fall, limited only by the
magnetic pressure, which is insufficient to prevent collapse, particularly
when the field is allowed to diffuse outward against the fluid. Various
approaches have been adopted to solve this problem, including limiting the
magnetic braking by placing a cap upon $B_{\phi,s}$ and introducing Ohmic
diffusion, which decouples the field from the fluid and limits the
effectiveness of magnetic braking in the inner regions. 

Hall diffusion is capable of inducing spin in an initially nonrotating core,
and could resolve the magnetic braking catastrophe by further inducing spin in
the opposite direction once magnetic braking has removed the initial angular
momentum from the core. Further work must be done to study the role of the
Hall effect on the magnetic field diffusion in star formation, particularly
using the semianalytic model constructed in this thesis, and possible
directions for further study of Hall diffusion in collapse are discussed in
the following section.

\section{Future Work}\label{future}

Much still needs to be done both with the self-similar model of gravitational
collapse as presented in this thesis and with the model modified by improved
assumptions about such physics as the vertical angular momentum transport and
the way in which the Hall diffusion parameter scales with the density in the
collapsing flow. This section shall explore some of the work that is already
underway and the directions yet to be examined in future studies of magnetized
gravitational collapse. 

\subsection{Limitations and assumptions}\label{limit}

The self-similar model is quite limited in that it is effectively only two
dimensional with much of the physics simplified so as to produce a
self-similar semianalytic solution of infinite resolution to the equations of
gravitational collapse, with azimuthal and vertical effects in the disc
sacrificed for ease of calculation. Numerical solutions can trace the 
behaviour of the collapse until the density of the inner regions becomes so
high compared to the resolution in space, and the time step required to trace
the fluid so short as to make further integration possible (although the
application of a sink particle at the centre of the collapse can allow some
further calculation). It is also the case that some numerical models do not
possess the resolution necessary to see the formation of
rotationally-supported discs of equivalent size to those that form within the
self-similar model at a given system age. However, numerical simulations are
able to include nonaxisymmetric and vertical effects, which allow them to
better explore the magnetic braking catastrophe and outflows. 

The assumptions of isothermality, axisymmetry and that the magnetic field and
axis of rotation are aligned are all required to ensure the self-similarity of
collapse, and cannot be lifted. Isothermality keeps the sound speed in the
fluid constant, and although it breaks down in the late stages of collapse
when $n = 10^{10}$ cm$^{-3}$ \citep{g1963} this only occurs in the innermost
regions of the rotationally-supported disc in the similarity solutions. The
only other place in the solutions where isothermality is likely to break down
is in the magnetic diffusion shock, however the magnetic pressure is more than
an order of magnitude larger than the gas pressure in the shock front and this
is not likely to greatly affect the nature of the collapse. Axisymmetry, which
reduces the dimensionality of the problem, causes the suppression of disc
physics such as turbulence and gravitational instabilities; these could
potentially be included in the model using a scaled prescription, but it is
not clear that any such prescription would add meaningfully to the physics in
the core. \citet{hc2009} showed that varying the angle between the magnetic
field and the axis of rotation affected the efficiency of disc formation,
however such behaviour is fully three-dimensional and could never be included
in self-similar collapse models. 

The assumptions that $b_{r,s}$ and $g_r$ can be approximated by monopoles is
another that could be improved upon with more careful calculation. Adding an
additional cycle of computation to the convergence routine that would find
these terms more precisely \citep[as in][]{cck1998} would slow the calculation
of similarity solutions down further; however, this could be offset by some
parallelization of the full code which is made possible by recent increases in
the computing power and number of processors on individual CPUs. Such
calculations would improve the accuracy of the determination of the radial
gravitational and magnetic field components, although it is not clear that
these approximations need to be replaced.

The thin disc assumption made it possible to adopt linear scalings of the
radial and azimuthal field components with height in the disc and average the
other parameters with respect to the disc scale height to produce a set of
equations that depended only upon $r$ and $t$. This averaging prevents the
examination of any behaviours that occur within the thin disc, such as the MRI
and other turbulent effects, that may be important to the angular momentum
transport. The $\alpha$-viscosity of \citet{ss1973} provides an axisymmetric
average of turbulence within the disc, which can be scaled with the
self-similar variables and could in future be included in the model. The thin
disc assumption also allows terms of order $H/r$ to be dropped from the
equation set, on the grounds that they are usually small. Reintroducing these
terms would allow the structure of the collapsing flow to be refined without
adding very much to the computation costs. It is not clear what changes (if
any) these small terms would bring to the self-similar model of collapse. 

The transportation of angular momentum in the vertical direction is perhaps
the most limiting of the simplifications adopted in the model. Several
mechanisms for better calculating the behaviour of the azimuthal field and the
angular momentum transport are outlined in the following subsection. 

\subsection{Vertical angular momentum transport}\label{angmom2}

The azimuthal field in this model was limited by an artificial cap that was
specifically designed to replace physics missing from the rest of the
model. Based on the assumption that $B_z$ is the dominant component of the
field, it is unclear what the magnitude and applicability of this cap ought to
be. For the purpose of properly solving the magnetic braking catastrophe,
different ways of replacing or calculating this cap need to be determined. 

The assumption of axisymmetry precludes the addition of magnetohydrodynamic
instabilities such as internal kinks and turbulence to the self-similar model;
such behaviour would prevent $B_{\phi,s}$ from greatly exceeding the poloidal
field components. These effects could potentially be parameterised according
to the way in which they scale with the field strength, the density and other
variables in numerical solutions, and then included into the self-similar
model as a mechanism for constraining $B_{\phi,s}$.

In the inner Keplerian disc $B_{r,s} / B_z > 1/\sqrt{3}$, which is the
launching condition for a cold, centrifugally-driven wind, and the radial
scaling of the magnetic field components is identical to that of the
radially-self-similar wind solution of \citet{bp1982}. Such a disc wind, which
was described in appendix C of \citet{kk2002}, would be the dominant mechanism
for the transfer of angular momentum from the disc to the envelope. A disc
wind must be included in future self-similar collapse simulations in order to
explore the influence a wind may have on the angular momentum transport, the
magnetic braking catastrophe and furthermore to simply improve the accuracy of
the models, as disc winds and jets are known to occur in numerical simulations
\citep[e.g][]{t2002, mim2007, ml2009, ch2010} of collapsing protostellar
discs. 

\subsection[Scaling the magnetic diffusivities]{Scaling the magnetic
diffusivities}\label{eta_H} 

For an idealised fluid with no grains, the Hall and ambipolar diffusivities
are given in dimensional form by
\begin{align}
    \eta_H &= \frac{cB}{4\pi e n_e} \label{6eta_H}\\
    \text{and }\eta_A &= \frac{D^2B^2}{4\pi\rho_i\nu_{in}}\,, \label{6eta_A}
\end{align}
where $D \equiv \rho_n/\rho$ is the neutral density fraction \citep[which
tends towards $1$ in weakly ionised fluids such as are found in molecular
clouds cores;][]{c1957, w1999, pw2008}. In their numerical solutions
\citet{kls2010-2} used a spatially constant coefficient $Q = \eta_H/B$ for the
Hall diffusivity, so that the behaviour of the Hall effect was not directly
dependent upon the density of the disc. The Hall term spins up the inner part
of the collapsing flow to Keplerian speeds, with the direction of the field
determining the direction of the Hall-induced magnetic torque, and
rotationally-supported disc formation was found to be possible in their
calculations only when $Q \gtrsim 3 \times 10^{20}$ cm$^2$ s$^{-1}$ G$^{-1}$.

The self-similarised Hall diffusivity $\eta_H' = (c_s^2t)^{-1}\eta_H$ from the
full set of fluid equations solved in this work was taken to scale with the
surface density and scale height of the disc as
\begin{equation}
    \frac{\eta_H'}{b} = \tilde{\eta}_Hb\left(\frac{h}{\sigma}\right)^{3/2},
\label{etahscale1}
\end{equation}
where $\tilde{\eta}_H$ was a constant. The decision to employ this scaling was
a pragmatic one, as the Hall diffusion then becomes important with the
ambipolar diffusion (which also scaled as 
\begin{equation}
    \frac{\eta_A'}{b^2} = \tilde{\eta}_A\left(\frac{h}{\sigma}\right)^{3/2}
\label{etaascale1}
\end{equation}
in self-similar space), and the drift of both grains and ions are important to
the collapse. The ratio of the diffusion parameters was then critical to all
discussions of the similarity solutions, and by keeping the ambipolar diffusion
coefficient constant it was possible to explore the individual influence of
the Hall term. In general though, both diffusion parameters could scale as any 
function of the similarity variable $x$ and the fluid variables without
breaking self-similarity. 

For example, a second choice for scaling the Hall parameter that could be
adopted in future is 
\begin{equation}
    \frac{\eta_H'}{b} = \tilde{\eta}_H\sqrt{\frac{h}{\sigma}}\,,
\label{etahscale2}
\end{equation}
which would be the behaviour expected if 
\begin{equation}
    \eta_H \sim \frac{B}{n_i},
\label{etahscale3}
\end{equation}
that is, if the only important particles in the collapsing core are neutrals
and ions, without grains. This change would see the Hall diffusion become
important earlier in the collapse, as the density increases and would also
change the degree of magnetic braking. Such a formulation would also change
the behaviour in the inner Keplerian disc, and a new set of asymptotic inner
boundary conditions describing this disc behaviour would need to be derived.

There is considerable scope for examining similarity solutions where the
relationship between the two diffusivities changes with $x$, so that each may
become important at different points in the collapse process. This is the case
in reality; Figure \ref{eta-regimes} showed that Hall diffusion is expected to
become important when the density and the magnetic field are both of
intermediate strength, while ambipolar diffusion is dominant where the density
is low and the field strength high, and these could be better reflected in the
collapse solutions. The diffusivities may be made to scale with any function
of the self-similar variables and $x$, in order to mimic the behaviour
expected from ionisation equilibrium calculations, and these scalings could
include changing the sign as well as the magnitude of the Hall diffusivity
with increasing density.   

\subsection{Exploring parameter space}\label{params}

The Hall similarity solutions presented in Chapter \ref{ch:hall} explored only
a very narrow region of parameter space where the outer boundary conditions
and the magnetic parameters were held constant as the Hall diffusion parameter
was varied in both directions. The similarity solutions without Hall diffusion
from Chapter \ref{ch:nohall} and Section \ref{catastrophe} showed that
reducing the initial rotational velocity of the core altered the size of the
Keplerian disc; that raising the azimuthal field cap and the magnetic braking
parameter caused all of the angular momentum to be removed from the core so
that no disc formed; and that reducing the ambipolar diffusion parameter
caused the size of the rotationally-supported disc and the accretion rate onto
the protostar to be reduced further. Future work on the Hall similarity
solutions must reproduce this diversity of parameter space, in order to better
understand the importance of Hall diffusion on the collapse process. 

Perhaps the most interesting region of parameter space to explore is that in
which Hall diffusion is the dominant form of flux transport. Without changing
the magnetic braking parameter and the cap on $b_{\phi,s}$, it is only
possible to find similarity solutions in which rotationally-supported discs
form when the Hall diffusion parameter is negative and large (see Section
\ref{discuss} and Equation \ref{6f}). If Hall diffusion were dominant, it
would further change the overall structure of the collapse, as it appears in
the equations for determining both the rotational velocity of the collapsing
gas as well as the magnetic support in the flow. In reality Hall diffusion is
not necessarily expected to exceed ambipolar diffusion except in the innermost
regions of the collapse \citep{w2004, w2007}, but this presents an interesting
goal for theoretical calculations as such similarity solutions would further
demonstrate the importance of the Hall effect on the collapse. 

Further to this, it would be of interest to calculate solutions without
ambipolar diffusion at all, in which all of the magnetic field diffusion is
performed by the Hall diffusion term, as in \citet{kls2010-2}. Hall diffusion
would change the nature of the collapse, as all of the field diffusion would
then depend upon the $\mathbf{J}\times\mathbf{B}$ terms, so the radial field
diffusion depends on the azimuthal field component (which may be capped) as
opposed to the whole of the field, and the azimuthal diffusion depends upon
the radial component. Such similarity solutions would again be not strictly
realistic, however in the regions of the collapse interior to the magnetic
diffusion shock (which would likely occur later in the collapse due to the
absence of ambipolar diffusion) the density and field strength are both of
intermediate value and Hall diffusion could already be more important than
ambipolar diffusion (see Section \ref{lr:mhd}). 

As mentioned in the previous section, more needs to be done to explore the
importance of the magnetic braking parameter $\alpha$ and the cap on the
azimuthal field component $\delta$. Should these be increased it seems likely
that the increased magnetic braking will remove all angular momentum from the
collapse and the solutions will tend towards the asymptotic free fall collapse
described in Section \ref{ff}. In that inner asymptotic solution Hall
diffusion changes the field strength in the disc, while the surface density
and radial velocity depend only upon the central mass. Such similarity
solutions would provide further insight into the magnetic braking catastrophe
and should be determined in future work on this topic.  

In Chapter \ref{ch:nohall} it was shown that the initial rotational velocity
plays an important role in determining the size of the Keplerian disc in those
solutions without Hall diffusion. Further work should be done to confirm that
larger discs form if the core is initially rapidly-rotating in the similarity
solutions with Hall diffusion. Although the Hall effect does change the
behaviour of the magnetic braking in the disc, it is not expected to make any
dramatic changes to the collapse behaviour otherwise as the initial angular
momentum is varied. The orientation of the field, and in consequence the
strength of the magnetic braking, could influence the effect of the initial
rotational velocity of the molecular cloud core on the centrifugal shock
position; obviously work must be done to properly explore this concept. 

\section{Conclusions}\label{conclusions} 

This thesis described a semianalytic self-similar model of the gravitational
collapse of rotating magnetic molecular cloud cores with both Hall and
ambipolar diffusion, presenting similarity solutions that showed that the Hall
effect has a profound influence on the dynamics of collapse. The solutions
satisfied the vertically-averaged self-similar equations for MHD collapse
under the assumptions of axisymmetry and isothermality, matching onto the
self-similar power law relations describing an isothermal core at the moment
of point mass formation on the outer boundary and a Keplerian disc on the
inner boundary. 

Two power law similarity solutions were derived that satisfy the collapse
equations on the inner boundary. The first of these represents a Keplerian
disc in which accretion through the disc depends upon the magnetic diffusion;
with an appropriate value of the nondimensional Hall diffusion parameter
$\tilde{\eta}_H$ a stable rotationally-supported disc forms in which the
surface density $\Sigma$ scales as $r^{-3/2}$ and vertical field strength $B_z
\propto r^{-5/4}$. These are the scalings expected from other simulations of
protostellar discs to which the solutions calculated in this work compare
favourably. No disc may form when the Hall parameter is large (in comparison
to the ambipolar diffusion parameter) and has the wrong sign (which indicates
the orientation of the magnetic field with respect to the axis of rotation),
as the diffusion in these solutions is too strong and causes disruptive
torques that form subshocks in the full similarity solutions. This
seemingly-odd behaviour occurs because the response of the fluid to Hall
diffusion is not invariant under a global reversal of the magnetic field.  

The second power law similarity solution describes the behaviour of the infall
when the magnetic braking is efficient at removing angular momentum from the
flow and no rotationally-supported disc forms. The matter falls rapidly onto
the central protostar and what little angular momentum remains is that induced
by the Hall effect; the direction of rotation depends upon the sign of the
Hall parameter. This behaviour is indicative of the ``magnetic braking
catastrophe'' that occurs in many magnetic collapse simulations. 

The size of the rotationally-supported disc in the full similarity solutions
was shown to vary with the amount of Hall and ambipolar diffusion affecting
the pseudodisc through their effect on the magnetic braking in the fluid. By
creating an additional torque on the disc, Hall diffusion can either increase
or decrease the angular momentum and rotational support in the infalling
fluid, leading to an order of magnitude change in the Keplerian disc radius
between the similarity solutions at the extremes of $-0.5 \le \tilde{\eta}_H /
\tilde{\eta}_A \le 0.2$ (where the ambipolar diffusion parameter,
$\tilde{\eta}_A = 1$). A small amount of Hall diffusion was shown have a large
effect on the solution because the dynamic range of collapse is itself many
orders of magnitude in space and time. Hall diffusion causes there to be a
preferred handedness to the field alignment and the direction of rotation in
forming a large Keplerian disc that could be observed using next-generation
instruments such as ALMA. 

The accretion rate onto the central point mass is similarly influenced by the
inclusion of Hall diffusion. This is a smaller effect than that on the disc
radius, as between $\tilde{\eta}_H = \pm 0.1 \tilde{\eta}_A$ (again with
$\tilde{\eta}_A = 1$) the accretion rate onto the protostar only changes by
6\%, or $0.2 \times 10^{-6}$ M$_\odot$ yr$^{-1}$. There exists a clear trend
in which the accretion rate drops off with increasingly negative Hall
parameters, as the reduced magnetic braking in these solutions causes a larger
Keplerian disc to form through which accretion onto the protostar is slow. 

It has also been shown that Hall diffusion can induce rotation in an initially
nonrotating molecular cloud core, as the azimuthal torques twist the field
lines, spinning up the core and envelope in opposite directions. The direction
of spin in the core is determined by the sign of the Hall parameter; changing
the sign of $\tilde{\eta}_H$ reverses the direction of rotation and shall also
affect the size of the disc as in the solutions where the initial rotation is
nonzero. As the density and flux rise the field decouples from the neutrals,
and the azimuthal Hall diffusion causes the angular momentum to increase to
the point where the centrifugal force roughly matches the gravitational force,
suggesting that it is possible to form a centrifugal disc around the
protostar. 

The magnetic braking catastrophe could be resolved by the inclusion of Hall
diffusion in numerical solutions, as with one sign of $\tilde{\eta}_H$ the Hall
effect acts to reduce the total amount of braking affecting the core,
preventing it from removing too much angular momentum from the collapse.
However, with the other sign of $\tilde{\eta}_H$ the magnetic braking is
increased so that more angular momentum is transported to the envelope. As
magnetic braking due to Hall diffusion does not stop acting once no angular
momentum remains (as ambipolar diffusion does) it could also then spin-up the
collapsing fluid back up in the opposite direction to the initial rotation.
This acceleration is only possible with Hall diffusion, and it has the
potential to completely resolve the magnetic braking catastrophe. 

Because of its tendency to pull the gas in unusual directions Hall diffusion
is usually overlooked in simulations of gravitational collapse and star
formation. It has been shown that the Hall effect is important to the dynamics
of the collapse, particularly the magnetic braking behaviour which determines
the existence and size of the rotationally-supported protostellar disc. The
handedness of the response of the collapse to the inclusion of the Hall effect
has obvious dynamical and potentially observable consequences for the
gravitational collapse of molecular cloud cores, which must be studied more
closely if the dynamics of the star formation process and the variations
observed across YSOs and their discs are to be properly understood. 

\cleardoublepage


\nocite{l2007,ppv-cbsbha}
\nocite{mmc1994, gpl1977, ls1997, pbd2008, skbt1994, tin1988a,
tsgtl2009, r1996}
\renewcommand\bibname{References}
\addcontentsline{toc}{chapter}{References}

\bibliography{thesis}{}
\bibliographystyle{abbrvnatme}

\cleardoublepage
\appendix
\chapter{Deriving the Inner Solutions}\label{ch:pq}

In this appendix the derivation of the inner asymptotic solutions discussed in
Chapter \ref{ch:asymptotic} is presented as part of the exploration of
$pq$-space that was proposed in Section \ref{indev}. Only the derivation of
physical solutions is presented here; while solutions in which the scale
height tends to zero or the surface density is negative may satisfy the
collapse equations they are unphysical and not pursued in this work. To
briefly recap the early stages of the derivation from \ref{ch:asymptotic}: the
inner asymptotic similarity solutions are assumed to take the form of power
laws in $x$, specifically 
\begin{align} 
    \sigma &= \sigma_1\,x^{-p}, \label{a-sigma} \\
    b_z &= b_{z1}\,x^{-q}, \label{a-bz} \\
    \text{and }j &= j_1\,x^{-r}, \label{a-j}
\end{align}
where $p$, $q$ and $r$ are real numbers, and $\sigma_1$, $b_{z1}$ and $j_1$
are constants. By substituting these power laws into the fluid equations and
taking the limit as $x \to 0$ it is possible to solve for all of the fluid
variables. The enclosed mass and flux may be written as 
\begin{align}
    m &= m_c + \frac{\sigma_1}{2-p}x^{2-p} \label{a-m} \\
    \text{and }\psi &= \frac{b_{z1}}{2-q}\,x^{2-q}, \label{a-psi}
\end{align}
and the radial field component is 
\begin{equation}
    b_{r,s} = \frac{b_{z1}}{(2-q)}\,x^{-q},
    \label{a-b_rs}
\end{equation}
which clearly scales with the vertical field component. 

Due to the cap on $b_{\phi,s}$, its precise value is not easily determined.
However, as $|b_{\phi,s}| \le \delta{b_z}$, the azimuthal field component is
taken to be its largest possible value, scaling with $b_z \sim x^{-q}$, until
it is possible to make refine this calculation. 

\begin{figure}[ht]
  \centering
  \vspace{-3mm}
  \includegraphics[width=5.in]{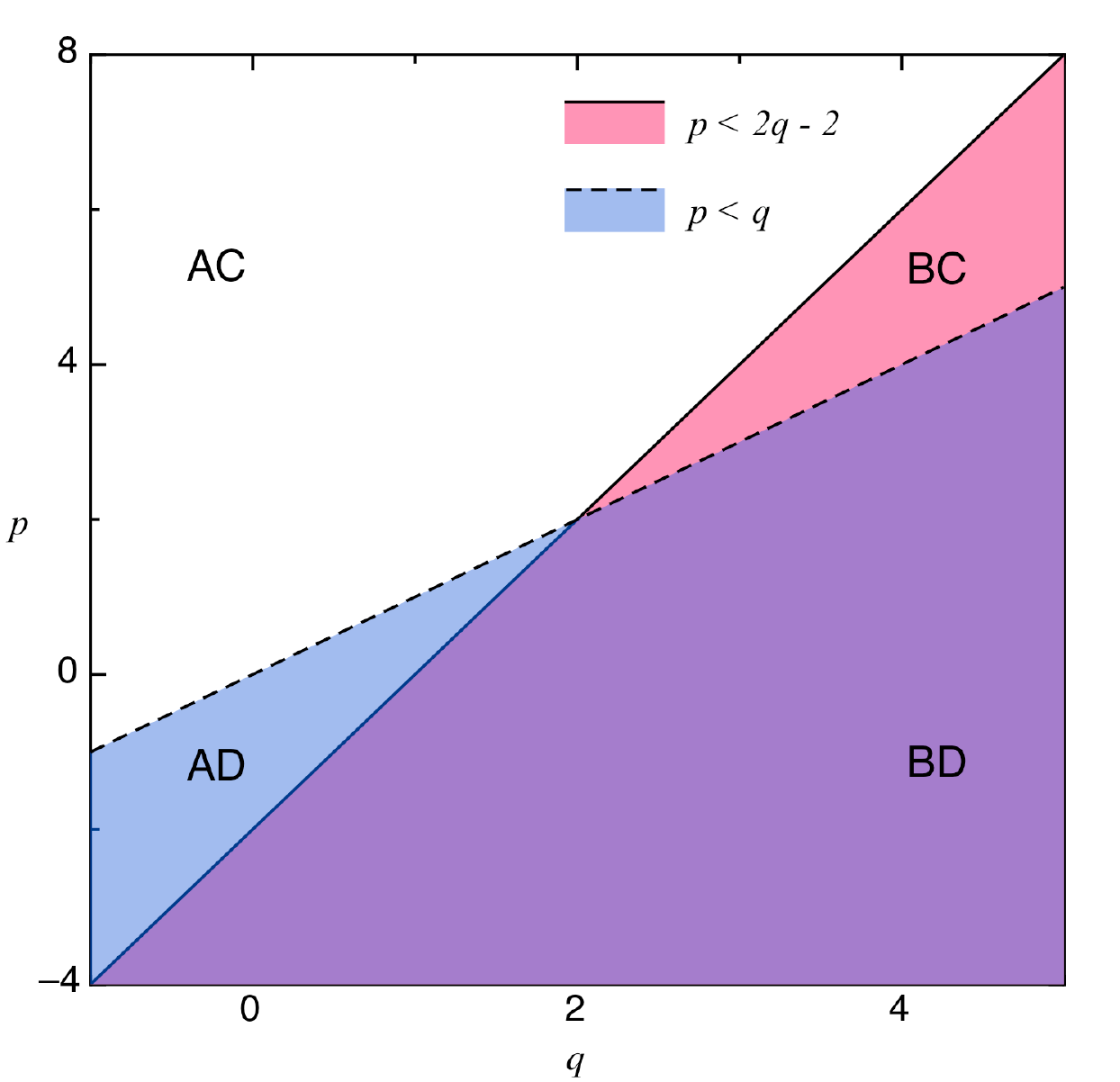}
  \vspace{-5mm}
  \caption[The $pq$-plane]{The $pq$-plane used to describe how the components
of $h$ behave with respect to the exponents of $\sigma$ and $b_z$. The
different regions have been colour coded, for example, the white area in the
upper left section of the plane represents the region AC where the
inequalities $p>2q-2$ and $p>q$ are satisfied.} 
\label{pq-plane2}
\end{figure}

The scale height is written as the solution to the quadratic equation
(\ref{ssvhe}): 
\begin{equation}
    h = \frac{\hat{\sigma}x^3}{2\hat{m}_c}\left[-1
          + \sqrt{1+\frac{8\hat{m}_c}{\hat{\sigma}^2x^3}}\,\right],
\label{ah-quad}
\end{equation}
where
\begin{equation}
    \hat{m}_c = m_c - x^3b_{r,s}\frac{db_z}{dx}
\label{m_chat}
\end{equation}
and
\begin{equation}
    \hat{\sigma} = \sigma + \frac{b_{r,s}^2 + b_{\phi,s}^2}{\sigma};
\label{sigmahat}
\end{equation}
for any combination of $p$ and $q$ the behaviour of $h$ can be determined.
Figure \ref{pq-plane}, reproduced here as \ref{pq-plane2}, shows the different
regions of $pq$-space in which $\hat{\sigma}$ and $\hat{m}_c$ take on the
following forms: 
\begin{align}
    &\text{A.} &\hat{m}_c &= m_c &(p &> 2q -2)\label{aa}\\
    &\text{B.} &\hat{m}_c &= -\frac{x^3b_{r,s}}{\sigma}\left(\frac{db_z}{dx}\right)
      &(p &< 2q-2)\label{ab}\\
    &\text{C.} &\hat{\sigma} &= \sigma &(p &> q)\label{ac}\\
    &\text{D.} &\hat{\sigma} &= \frac{(b_{r,s}^2 + b_{\phi,s}^2)}{\sigma}
      &(p &< q)\label{aad}
\end{align}

As explained in Section \ref{indev}, only those solutions where 
\begin{equation}
    \frac{8\hat{m}_c}{\hat{\sigma}^2x^3} \ge 1 \label{a-h1}
\end{equation}
are sought, so that
\begin{equation}
    h \sim \sqrt{\frac{2x^3}{\hat{m}_c}}. \label{a-h2}
\end{equation}
Any similarity solutions that do not satisfy this criteria, while
mathematically valid, are unphysical and so those regions of parameter space
are not explored in this work. Each of the four regions of the $pq$-plane
must be examined in order to find those physically possible similarity
solutions. 

\section{Region AC}\label{sub:AC}

This section of the $pq$-plane is characterised by the following inequalities:
\begin{align}
    p &> 2q-2, \label{pq-ac1}\\
    \text{and }p &> q; \label{pq-ac2}
\end{align}
and is represented by the white area in the upper left of Figure
\ref{pq-plane2}. In this region $8\hat{m}_c/\hat{\sigma}^2x^3$ scales as 
\begin{equation}
    \frac{8\hat{m}_c}{\hat{\sigma}^2x^3} \sim \frac{m_c}{\sigma^2x^3}
      \sim x^{2p-3},
\label{caseb-ac}
\end{equation}
so that $h$ is given by
\begin{equation}
    h = \frac{\sigma_1}{2m_c}x^{3-p}\left[-1 + 
      \sqrt{1 + \frac{8m_c}{\sigma_1^2}x^{2p-3}}\,\right].
    \label{ac-h}
\end{equation}
The desired behaviour of $h$ (denoted ``case b'' in Section \ref{indev})
occurs when
\begin{equation}
    2p - 3 < 0 \quad\Rightarrow\quad p,q < 3/2.
\label{caseb-ineq-ac}
\end{equation}
As these limits are well within the boundaries of region AC, the boundary case
where $2p - 3 = 0$ ($p = 3/2$) should also be examined. All of these
restrictions on the exponents imply that $2 - p \ge 1/2$, which means that as
$x \to 0$ 
\begin{equation}
    m = m_c,
\label{ac-m}
\end{equation}
while $h \sim x^{3/2}$ and $b_{\phi,s} \lesssim x^{-q}$. 

Substituting these power law expressions into the angular momentum equation
(\ref{ssam}) shows that its terms scale as  
\begin{align}
    \frac{dj}{dx} &\sim x^{-r-1},\nonumber\\
    \frac{j}{w} &\sim x^{1-r-p},\nonumber\\
    \text{and }\frac{x^2b_{z}b_{\phi,s}}{m} &\lesssim x^{2-2q}.
    \label{acj1}
\end{align}
It is not possible to directly compare the exponents of these terms at this
point, however it is clear that there are two possible solutions to this
equation, the first of which is 
\begin{equation}
    \frac{dj}{dx} = \frac{j}{w}.
\label{ac-jtype1}
\end{equation}
For this equation to be satisfied the exponents of the two terms must be
equal, which implies that
\begin{equation}
    -r-1 = 1-r-p \quad \Rightarrow \quad p = 2;
\label{ac-jbad}
\end{equation}
but as $p\le 3/2$ this is a contradiction and there can be no solution to
Equation \ref{ac-jtype1} satisfying the requirement that $h$ has a positive
and physically possible value. 

The second possible solution to the angular momentum equation must then be
adopted, which takes the form 
\begin{equation} 
    \frac{dj}{dx} = - \frac{x^2b_zb_{\phi,s}}{m};
\label{ac-jtype2}
\end{equation}
this equation cannot be solved fully until the scaling of $b_{\phi,s}$ is
properly known, however examination of the exponents allows for a limit on $r$
to be derived, as
\begin{align}
    -r-1 &\le 2-2q \\
    \therefore r &\ge 2q-3.
\label{ac-rineq}
\end{align}
This also implies that $1-r-p > 2-2q$, which can be used with the inequalities
in \ref{pq-ac1}--\ref{pq-ac2} to show that $r < 1$. 

Examining the scaling of the pair of terms $b_{r,s} - h(db_z/dx)$ that occur
in many of the equations shows that they scale with $x$ as
\begin{align}
    b_{r,s} &\sim x^{-q}\nonumber\\
    \text{and }h\frac{db_z}{dz} &\sim x^{1/2-q};
\label{ac-brsh}
\end{align}
the second exponent is clearly larger than the first, implying that this term
is the smaller as the limit $x \to 0$ is taken, and can be discarded wherever
it appears in the equations. 

The other terms of the radial momentum equation (\ref{sscrm}) then scale with
$x$ as
\begin{align}
    \frac{1}{\sigma}\frac{d\sigma}{dx} &\sim x^{-1},\nonumber\\
    \frac{w^2}{\sigma}\frac{d\sigma}{dx} &\sim x^{2p-3},\nonumber\\
    \frac{m}{x^2} &\sim x^{-2},\nonumber\\
    \frac{b_zb_{r,s}}{\sigma} &\sim x^{p-2q},\nonumber\\
    \frac{j^2}{x^3} &\sim x^{-2r -3},\nonumber\\
    \text{and }\frac{w^2}{x} &\sim x^{2p-3}.
\label{ac-rmterms}
\end{align}
It is clear from comparing the exponents that the first term is smaller than
the third and can be disregarded as small. The inequality \ref{pq-ac1} gives
$p - 2q > -2$, which means that the fourth term may also be disregarded,
leaving a simplified radial momentum equation, 
\begin{equation}
    -w^2\left(\frac{1}{x} + \frac{1}{\sigma}\frac{d\sigma}{dx}\right) = 
    -\frac{m}{x^2} + \frac{j^2}{x^3},
\label{ac-rmsimp}
\end{equation}
that cannot be easily solved. 

Without knowing the values of $p$ and $r$, there are multiple simplified
solutions to this equation:
\begin{align}
    \frac{m}{x^2} &= \frac{j^2}{x^3}
    &w^2\left(\frac{1}{x} + \frac{1}{\sigma}\frac{d\sigma}{dx}\right) 
        &= \frac{m}{x^2}
    &-w^2\left(\frac{1}{x} + \frac{1}{\sigma}\frac{d\sigma}{dx}\right) 
	&= \frac{j^2}{x^3}.
\label{ac-rmoptions}
\end{align}
The third of these requires that $2p -3 < -2$, which is satisfied whenever $p
< 1/2$. However, for $j$ to be a real number the left hand side of the
equation must be positive, which is only possible when $p > 1$. This
contradiction means that there can be no physical solution to this equation.

The second of the equations in \ref{ac-rmoptions} has the solution $p = 1/2$
and is satisfied when the inequality $p < -r$ holds true. This can be 
substituted into \ref{ac-rineq} to show that $p < 2q - 3$, which is a
contradiction of the inequality \ref{pq-ac1} that defines this region of
$pq$-space. Therefore this also cannot be a valid solution to the radial 
momentum equation. 

The only solution to the radial momentum equation is then the first equation
in \ref{ac-rmoptions}, which is solved to give
\begin{equation}
    j = \sqrt{m_cx} 
\label{ac-j}
\end{equation}
so that the exponent
\begin{equation}
    r = -\frac{1}{2}.
\label{ac-r}
\end{equation}
This exponent is then substituted into the inequality \ref{ac-rineq} to show
that $q \le 5/4$. 

The scaling of $b_{\phi,s}$ can now be determined by examining the scaling
with $x$ of its component terms. The non-constant term in the denominator of
the fraction on the left hand side of Equation \ref{ssb_phis} scales as 
\begin{equation}
    \frac{2\alpha\tilde{\eta}_Ah^{1/2}\psi{b_z}}{x^2\sigma^{3/2}}
    \sim x^{3/4+3p/2 -2q};
\label{ac-b_psdscale}
\end{equation}
as $p>q$, it is clear that $3p/2 - 2q > -q/2$, and that 
\begin{equation}
    3/4 + 3p/2 -2q > 3/4 - q/2. 
\label{ac-bpsineq1}
\end{equation}
Finally, as $q \le 3/2$, the inequality in \ref{ac-bpsineq1} becomes
\begin{equation}
    3/4+3p/2-2q > 0;
\label{ac-bpsineq2}
\end{equation}
this means that the non-constant term in the denominator of this fraction
becomes smaller than the constant term as $x \to 0$ and may be dropped from
the equation. 

The $h(db_z/dx)$ term has already been dropped from the equation for
$b_{\phi,s}$ (\ref{ssb_phis}), and the remaining terms in the numerator scale
as 
\begin{align}
    \frac{j}{x} &\sim x^{-1/2}\nonumber\\
    \text{and } \frac{\tilde{\eta}_Hbb_{r,s}h^{1/2} }{\sigma^{3/2}} 
	&\sim x^{3/4 + 3p/2 - 2q}.
\label{ac-b_psnscale}
\end{align} 
The inequality in \ref{ac-bpsineq2} shows that $j/x$ is therefore the dominant
term in the numerator of this fraction, so that the Hall term is not important
and the left hand side of the equation for $b_{\phi,s}$ scales with $x$ as
\begin{equation}
    \frac{2\alpha\psi{j}}{x^3} \sim x^{5/2-q-3} = x^{-q-1/2}.
\label{ac-b_pslhs}
\end{equation}
This is a larger term than that on the right hand side of the equation for
$b_{\phi,s}$, and so the right hand side is the minimum of these two sides,
giving 
\begin{equation}
    b_{\phi,s} = -\delta{b_z}.
\label{ac-b_ps}
\end{equation}
This solution illuminates the flaws in the way that $b_{\phi,s}$ was defined
and the method adopted for describing the vertical angular momentum transport.
It is not realistic for the azimuthal field to dominate the vertical field in
the disc, and limiting the value of $b_{\phi,s}$ so that it scales as $b_z$ is
a reasonable simplification to keep it from becoming too large in this inner
region. However, a better prescription for $b_{\phi,s}$ is needed in order to
properly understand and model the actual behaviour of the field as $x \to 0$,
as discussed in Section \ref{future}.

To find the value of $b_z$, Equations \ref{ac-m}, \ref{ac-j} and \ref{ac-b_ps}
are substituted into the simplified angular momentum equation
(\ref{ac-jtype2}) to obtain the equation 
\begin{equation}
    \frac{1}{2}\sqrt{\frac{m_c}{x}} = \frac{x^2\delta{b_z}^2}{m_c},
\label{ac-am-bz}
\end{equation}
which is rearranged and solved for $b_z$,
\begin{equation}
    b_z = \frac{m_c^{3/4}}{\sqrt{2\delta}}\,x^{-5/4} ,
\label{ac-bz}
\end{equation}
and the exponent
\begin{equation}
    q = \frac{5}{4}.
\label{ac-q}
\end{equation}

Curiously, $b_z$ does not depend upon the diffusion coefficients, although
they do influence the value of $m_c$ in the solutions presented in Chapters
\ref{ch:nohall} and \ref{ch:hall}, as discussed in Section \ref{discuss}. The
flux and the other field components are then described by the power laws:
\begin{align}
    \psi &= \frac{4}{3}\frac{m_c^{3/4}}{\sqrt{2\delta}}\,x^{3/4},\label{ac-psi}\\
    b_{r,s} &= \frac{4}{3}b_z 
	= \frac{4}{3}\frac{m_c^{3/4}}{\sqrt{2\delta}}\,x^{-5/4},\\
    b_{\phi,s} &= -\delta{b_z} 
	= -\sqrt\frac{\delta}{2}\,m_c^{3/4}x^{-5/4}\\
    \text{and } b &= b_z\sqrt{\frac{25}{9} + \delta^2}.
\label{ac-flux}
\end{align}

Finally the induction equation (\ref{ssin}) is examined, and as with the other
equations the $h(db_z/dx)$ term is small and may be disregarded as $x \to 0$.
The remaining terms in the equation scale with $x$ as 
\begin{align}
    \psi &\sim x^{3/4},\nonumber\\
    xwb_z &\sim x^{p-5/4},\nonumber\\
    \tilde{\eta}_Hxb_{\phi,s}b_zbh^{1/2}\sigma^{-3/2} &\sim x^{3p/2 - 2},\nonumber\\
    \text{and }\tilde{\eta}_Axb_{r,s}b_z^2h^{1/2}\sigma^{-3/2} &\sim x^{3p/2 - 2};
\label{ac-inscale}
\end{align}
as $3p/2 - 2 \le 1/4$, then $\psi$ is small compared to the other terms and is
dropped from the induction equation. The induction equation is then simplified
to 
\begin{equation}
    -xwb_z + \tilde{\eta}_Hxb_{\phi,s}b_zbh^{1/2}\sigma^{-3/2} 
	+ \tilde{\eta}_Axb_{r,s}b_z^2h^{1/2}\sigma^{-3/2} = 0;
\label{ac-insimp}
\end{equation}
which upon substitution of the scalings in \ref{ac-m}, \ref{ac-j}, \ref{ac-bz}
and \ref{ac-psi}--\ref{ac-flux} becomes 
\begin{equation}
    -\frac{m_c}{x\sigma} + \left(-\tilde{\eta}_H\delta\sqrt{\frac{25}{9} + \delta^2}
	+ \frac{4}{3}\tilde{\eta}_A\right)b_z^2h^{1/2}\sigma^{-3/2} = 0,
\label{ac-insub}
\end{equation}
which is tidied to give
\begin{equation}
    \frac{m_c\sigma^{1/2}}{x} = \left(\frac{4}{3}\tilde{\eta}_A 
	- \tilde{\eta}_H\delta\sqrt{\frac{25}{9} + \delta^2}\right)
	b_z^2h^{1/2}.
\label{ac-insimper}
\end{equation}
The equation is then rearranged so that $h$ is the subject:
\begin{equation}
    h^{1/2} = \frac{2\delta}{f}\sqrt{\frac{\sigma_1}{m_c}}\,x^{3/2-p/2};
\label{ac-hroot}
\end{equation}
the magnetic diffusion and azimuthal field cap parameters have been combined
into a single parameter, $f$, defined as
\begin{equation}
    f = \frac{4}{3}\tilde{\eta}_A 
    - \tilde{\eta}_H\delta\sqrt{\frac{25}{9} + \delta^2}.
\label{ac-f}
\end{equation}

As it is known that $h \sim x^{3/2}$, equating the exponents of $x$ in
Equation \ref{ac-hroot} gives
\begin{equation}
    p = \frac{3}{2}.
\label{ac-p}
\end{equation}
This straddles the boundaries between cases a and b for the behaviour of $h$
defined in Section \ref{indev}, and so all of the terms in the equation for
$h$ (\ref{ac-h}) must be kept. 

Squaring equation \ref{ac-hroot} and equating it with Equation \ref{ac-h}
gives 
\begin{equation}
    \left(\frac{2\delta}{f}\right)^2\frac{\sigma_1}{m_c}
	= \frac{\sigma_1}{2m_c}
	\left(-1+\sqrt{1+\frac{8m_c}{\sigma_1^2}}\,\right),
\label{ac-hderivs1}
\end{equation}
where $\sigma_1$ is the coefficient of $\sigma$. This is rearranged into
\begin{equation}
    2\left(\frac{2\delta}{f}\right)^2 + 1 = \sqrt{1+\frac{8m_c}{\sigma_1^2}},
\label{ac-hderivs2}
\end{equation}
and both sides are squared so that this equation becomes
\begin{equation}
    4\left(\frac{2\delta}{f}\right)^4 + 4\left(\frac{2\delta}{f}\right)^2 + 1
	= 1 + \frac{8m_c}{\sigma_1^2}.
\label{ac-hderivs3}
\end{equation}
Equation \ref{ac-hderivs3} may then be solved for the coefficient of $\sigma$:
\begin{equation}
    \sigma_1^2 = \frac{2m_c(f/2\delta)^2}{(2\delta/f)^2+1};
\label{ac-sigmasimp}
\end{equation}
so that the surface density is given by the equation
\begin{equation}
    \sigma = \frac{\sqrt{2m_c}f}{2\delta\sqrt{(2\delta/f)^2+1}}\,x^{-3/2}
\label{ac-sigma}
\end{equation}
and the scale height of the disc is
\begin{equation}
    h = \left(\frac{2}{m_c[1+(f/2\delta)^2]}\right)^{\!1/2}\,x^{3/2}.
\label{ac-hreal}
\end{equation}
The magnetic diffusion terms play an important role in these equations as
$\sigma$ cannot be negative; this requires that $f$ must always be greater
than zero. This in turn places limits on the size of the Hall diffusion
parameter with respect to the ambipolar diffusion parameter in order to ensure
that disc formation may take place. 

This solution to the disc equations represents the slowly-accreting Keplerian
disc that was discussed in detail in Section \ref{kepdisc}. 

\section{Region AD}\label{sub:AD}
Region AD is the blue area in the lower left of Figure \ref{pq-plane2}, where
$p$ and $q$ satisfy the inequalities
\begin{align}
    p &> 2q-2,\label{pq-ad1}\\
    \text{and }p &< p, \label{pq-ad2}
\end{align}
which together imply that
\begin{equation}
    p, q < 2.
\label{pq-ad3}
\end{equation}
In this region, $\hat{m}_c = m_c$ and $\hat{\sigma} = (b_{r,s}^2  + b_{\phi,s}^2)
/\sigma$, so that the scaling of $8\hat{m}_c/\hat{\sigma}^2x^3$ goes as 
\begin{equation}
    \frac{8\hat{m}_c}{\hat{\sigma}^2x^3} \sim 
      \frac{m_c}{x^3(x^{p-2q})^2} \sim x^{4q-2p-3}.
    \label{caseb-ad}
\end{equation}
In this region of the $pq$-plane, the desired case (b) for the behaviour of
the scaling of $h$ is defined by the inequality
\begin{equation}
    0 > 4q - 2p -3 
\label{caseb-ineq-ad}
\end{equation}
which can be rearranged into 
\begin{equation}
    p - 2q > -3/2.
\label{ad-caseb1}
\end{equation}
The boundary case when $p-2q = -3/2$ must also be examined, as this inequality
is satisfied whenever both $p$ and $q < 3/2$ in this region.

As in region AC this means that $h \sim x^{3/2}$, and as in the previous
solution $b_{r,s} \sim x^{-q}$ and $h(db_z/dx) \sim x^{1/2-q}$, so that once
more the latter term is the smaller of the two and can be disregarded when
compared with $b_{r,s}$ in the induction equation, the radial momentum
equation and the equation for the azimuthal component of the magnetic field. 

The remaining terms in the induction equation (\ref{ssin}) scale with $x$ as
\begin{align}
    \psi &\sim x^{2-q},\nonumber\\
    xwb_z &\sim x^{p-q},\nonumber\\
    \tilde{\eta}_Hxb_{\phi,s}b_zbh^{1/2}\sigma^{-3/2} &\lesssim x^{7/4+3p/2-3q},\nonumber\\
    \text{and }\tilde{\eta}_Axb_{r,s}b_z^2h^{1/2}\sigma^{-3/2} &\sim x^{7/4+3p/2-3q};
\label{ad-in-scale}
\end{align}
and as $p<2$, then $x^{2-q} < x^{p-q}$ in the small $x$ limit. Therefore
$\psi$ can be dropped from the induction equation as in region AC, and the
equation becomes 
\begin{equation}
    -w +\tilde{\eta}_Hb_{\phi,s}bh^{1/2}\sigma^{-3/2} 
        + \tilde{\eta}_Ab_{r,s}b_zh^{1/2}\sigma^{-3/2} = 0.
\label{ad-insimp}
\end{equation}
Equating the exponents of these remaining terms gives the exponent
\begin{equation}
    q = \frac{p}{4} + \frac{7}{8},
\label{ad-q-p}
\end{equation}
which can be substituted into the inequalities in \ref{pq-ad1}--\ref{pq-ad2}
and \ref{caseb-ineq-ad} to place further limits on $p$: 
\begin{equation}
    1/2 \le p \le 3/2.
\label{ad-p-limit}
\end{equation}

The rest of this solution depends critically upon the value of $b_{\phi,s}$,
so it must be the next focus of the derivation. The scaling with $x$ of the
variable term in the denominator of the fraction on the left hand side of the
equation for $b_{\phi,s}$, \ref{ssb_phis}, is 
\begin{equation}
    \frac{2\alpha\tilde{\eta}_A\psi{b_z}h^{1/2}}{\sigma^{3/2}x^2} 
	\sim x^{3/4+3p/2-2q} = x^{p-1},
\label{ad-bpsdenom}
\end{equation}
which cannot easily be compared with the constant as in the AC case. The
situations in which each of the terms in the denominator is dominant must be
examined separately, in order to properly survey $pq$-space. 

\subsection[$p<1$]{$\mathbf{p<1}$}\label{sub:ad-plt1}

The first case is the situation where $p < 1$ and the equation for
$b_{\phi,s}$ becomes 
\begin{equation}
    b_{\phi,s} = -\text{min}\left[\frac{\sigma^{3/2}}{\tilde{\eta}_Ab_zh^{1/2}}
	\left(\frac{j}{x} - \frac{\tilde{\eta}_Hbb_{r,s}h^{1/2}}{\sigma^{3/2}}\right);
	\delta b_z\right];
\label{ad-b_ps1}
\end{equation}
and the terms on the left hand side scale as
\begin{align}
    \frac{\sigma^{3/2}}{\tilde{\eta}_Ab_zh^{1/2}} &\sim x^{-5p/4+1/8},\nonumber\\
    \frac{j}{x} &\sim x^{-r-1}\nonumber\\
    \text{and }\frac{\tilde{\eta}_Hbb_{r,s}h^{1/2}}{\sigma^{3/2}} &\sim x^{p-1}.
\label{ad-b_pslhsplt1}
\end{align}
If $-r < p$ then the left hand side of Equation \ref{ad-b_ps1} scales as 
$x^{-5p/4 -7/8 -r}$ and is larger than the term on the right hand side (which
scales as $x^{-q = -p/4-7/8}$), and so the cap is applicable and $b_{\phi,s} =
-\delta b_z$. Alternatively, if $-r > p$ then both sides of Equation
\ref{ad-b_ps1} scale as $b_z$ and the coefficient of $b_{\phi,s}$ cannot be
determined until the coefficients of the other variables are known. In both
situations then $b_{\phi,s} \sim x^{-q}$.

Adopting this scaling of $b_{\phi,s}$ with $x$ and looking at the angular
momentum equation (\ref{ssam}) shows that its terms scale with $x$ as
\begin{align}
    \frac{dj}{dx} &\sim x^{-r-1},\nonumber\\
    \frac{j}{w} &\sim x^{1-r-p}\nonumber\\
    \text{and }\frac{xb_zb_{\phi,s}}{w\sigma} &\sim x^{1/4-p/2}.
\label{ad-amscale1}
\end{align}
As $p<1$, it is clear that $1-r-p > -r-1$ so that the angular momentum
equation may be simplified into
\begin{equation}
    \frac{dj}{dx} = \frac{xb_zb_{\phi,s}}{w\sigma}
\label{ad-amj1}
\end{equation}
and solved for the exponent of $j$,   
\begin{equation}
    r = \frac{p}{2} - \frac{5}{4}.
\label{ad-amr1}
\end{equation}
Again, the constant coefficient of $j$ cannot be determined until the
coefficient of $b_{\phi,s}$ is known. 

Turning finally to the radial momentum equation (\ref{sscrm}), it can be shown
that the terms scale as
\begin{align}
    \frac{1}{\sigma}\frac{d\sigma}{dx} &\sim x^{-1},\nonumber\\
    -\frac{w^2}{\sigma}\frac{d\sigma}{dx} &\sim x^{2p-3},\nonumber\\
    -\frac{m}{x^2}&\sim x^{-2},\nonumber\\
    \frac{b_zb_{r,s}}{\sigma} &\sim x^{p-2q} = x^{p/2 - 7/4},\nonumber\\
    \frac{j^2}{x^3} &\sim x^{-2r-3} = x^{-p-1/2},\nonumber\\
    \text{and }\frac{w^2}{x} &\sim x^{2p-3};
\label{ad-rmscale}
\end{align}
clearly $x^{-1} < x^{-2}$ in the small $x$ limit, and as $p<1$ it is trivial
to show that $-p-1/2>-3/2$. The lower limit on $p$ (see \ref{ad-p-limit}) is
then manipulated to show that $-3/2 < p/2 - 7/4$, so that the angular momentum
contribution to the radial support may also be dropped. The radial momentum
equation is then simplified into the form 
\begin{equation}
    \frac{m}{x^2} = w^2\left(\frac{1}{x} 
	+ \frac{1}{\sigma}\frac{d\sigma}{dx}\right).
\label{ad-rm1}
\end{equation}
From this equation the exponents of the power law relations can then be
equated to give 
\begin{equation}
    p = \frac{1}{2},
\label{ad-p1}
\end{equation} 
which can be substituted into Equations \ref{ad-q-p} and \ref{ad-amr1} to
solve for the other exponents
\begin{align}
    r &= -1\\ \text{and } q &= 1.
\label{ad-coeff1}
\end{align}
This is the only solution to the power law expansion of $\sigma$, $j$ and
$b_z$ in this region of the $pq$-plane that also satisfies $p<1$. The
coefficients of the variables and the conditions under which they satisfy the
fluid equations are derived below in subsection \ref{sub:ad-sol}. 

\subsection[$p>1$]{$\mathbf{p > 1}$}\label{sub:ad-pgt1}

The other behaviour of the denominator of the left hand side of $b_{\phi,s}$
occurs when $p > 1$, and so $b_{\phi,s}$ is given by
\begin{equation}
    b_{\phi,s} = -\text{min}\left[\frac{2\alpha\psi}{x^2}
	\left(\frac{j}{x} - \frac{\tilde{\eta}_Hbb_{r,s}h^{1/2}}{\sigma^{3/2}}\right);
	\delta b_z\right].
\label{ad-bps2}
\end{equation}
The bracketed terms on the left hand side scale as
\begin{align}
    \frac{j}{x} &\sim x^{-r-1}\nonumber\\
    \text{and }\frac{\tilde{\eta}_Hh^{1/2}b}{\sigma^{3/2}}\,b_{r,s} 
	&\sim x^{3/4+3p/2-2q} \sim x^{p-1},
\label{ad-bphisLHS}
\end{align}
and as the relationship between $p$ and $r$ has yet to be determined the
dominant term cannot be decided. As before, both possibilities must be
considered. 

When $-r<p$, the left hand side of Equation \ref{ad-bps2} scales as
$x^{-r-p/4-15/8}$, and if $-r < 1$ then this side is the larger of the two and
$b_{\phi,s}$ takes on the value of the right hand side. The rest of the
derivation then follows that outlined in subsection \ref{sub:ad-plt1}, giving
the solution $p = 1/2$, which is a contradiction of the requirement that $p >
1$. 

However, if $1 < -r < p$, then $b_{\phi,s}$ will then scale with $x$ as
$x^{-r-p/4-15/8}$, and the terms of the angular momentum equation will scale
as 
\begin{align}
    \frac{dj}{dx} &\sim x^{-r-1},\nonumber\\
    \frac{j}{w} &\sim x^{1-r-p},\nonumber\\
    \text{and }\frac{xb_{\phi,s}b_z}{w\sigma} &\sim x^{-r-p/2-3/4}.
\label{ad-amscale2}
\end{align}
Rearranging the limits on both $p$ and $r$ then allows for the derivation of
limits on the exponents of the scaling terms:
\begin{align}
    0&<-r-1,\\
    \frac{1}{2} &< 1-r-p < 1\\
    \text{and }-r&-\frac{p}{2}-\frac{3}{4} < 0;
\label{ad-amrpineq}
\end{align}
only one of these terms is less than 0, and as there are not two large terms
that can be equated there is no way to solve this equation. Therefore there is
no solution to the fluid equations when $-r<p$ in the region AD of the
$pq$-plane.

Finally, there is the case where $1<p<-r$. In this circumstance the left hand
side of Equation \ref{ad-bps2} scales as $x^{3p/4-15/8}$. As $p > 1$,  
\begin{align}
	\frac{3p}{4} - \frac{15}{8} &> -\frac{9}{8} \\
        \text{and } -\frac{p}{4}-\frac{7}{8} &< -\frac{9}{8}
\label{ad-bpsstuff}
\end{align}
so that $b_{\phi,s} \sim x^{3p/4-15/8}$. The scalings of the terms in the
angular momentum equation remain as in \ref{ad-amscale2}, save for the last 
term which is now
\begin{equation}
    \frac{xb_{\phi,s}b_z}{w\sigma} \sim x^{p/2-3/4}.
\label{ad-amscale3}
\end{equation}
As above, the first two terms both have exponents that are greater than zero,
while the final term has the exponent $p/2-3/4 <0$; again there is no solution
to this equation in the small $x$ limit when $p > 1$. The only valid solution
to the power law behaviour of the variables in the disc equations in this
region of the $pq$-plane is then that outlined in Equations
\ref{ad-p1}--\ref{ad-coeff1}. 

\subsection{Coefficients of the solution}\label{sub:ad-sol}

Having explored the entirety of Region AD, the only solution to the fluid
equations with a nontrivial value of the scale height of the pseudodisc is
that described by the series of power laws with exponents $p=1/2$, $q=1$ and 
$r=-1$. Substituting these into the radial momentum equation gives 
\begin{equation}
    w^2\left(\frac{1}{x} + \frac{1}{\sigma}\frac{d\sigma}{dx}\right) 
	= \frac{m}{x^2},
\label{ad-rmsolve1}
\end{equation}
which is rearranged into the form
\begin{equation}
    \left(\frac{1}{x} - \frac{1}{2x}\right) = \frac{\sigma^2}{m}
\label{ad-rmsolve}
\end{equation}
and then solved for the surface density, 
\begin{equation}
    \sigma = \sqrt{\frac{m_c}{2x}}.
\label{ad-sigma}
\end{equation}

The flux is given by Equation \ref{asym-psi} to be
\begin{equation}
    \psi = b_zx^2
\label{ad-psi}
\end{equation}
and the radial field component from Equation \ref{asym-b_rs} is
\begin{equation}
    b_{r,s} = b_z;
\label{ad-b_z}
\end{equation}
the azimuthal field component is critical to the determination of the
coefficients of the other variables, and Equation \ref{ad-b_ps1} is simplified
to give 
\begin{equation}
    b_{\phi,s} = -\text{min}\left[-\frac{\tilde{\eta}_H
		bb_{r,s}}{\tilde{\eta}_Ab_z}\,; \delta b_z\right].
\label{ad-bps1}
\end{equation}
As mentioned previously, both sides of $b_{\phi,s}$ scale with $x$ in the same
manner, and the precise value chosen depends entirely upon the constants that
describe the magnetic diffusion and the cap placed upon $b_{\phi,s}$. It is
also clear from Equation \ref{ad-bps1} that changing the sign of
$\tilde{\eta}_H$ changes the sign of $b_{\phi,s}$, however, changing the sign
of $b_{\phi,s}$ leads to a change in the signs of both $j$ and $b_z$, so that
the similarity solution is effectively upside down but otherwise unchanged.
Taking the absolute value of $\tilde{\eta}_H$ allows for the sign of
$b_{\phi,s}$ to be kept negative, so that the final value of the azimuthal
field component is given by 
\begin{equation}
    b_{\phi,s} = -\text{min}\left[\frac{|\tilde{\eta}_H|}{\tilde{\eta}_A}\,b\,;
	\delta b_z\right]. 
\label{ad-bps}
\end{equation}

Before exploring the two different values that $b_{\phi,s}$ can take (with
coefficients $b_{\phi 1}$ and $b_{\phi 2}$) in great detail, more generalised
solutions to the magnetic field components and scale height may then be
written down:
\begin{align}
    b_{\phi,s} &= -b_{\phi 1,2}\,x^{-1},\\
    b_z &= b_{z1,2}\,x^{-1},\\
    b_{r,s} &= b_{z1,2}\,x^{-1},\\
    b &= \sqrt{2b_z^2 + b_{\phi,s}^2} 
	= \sqrt{2b_{z1,2}^2 + b_{\phi 1,2}^2}\,x^{-1},
\label{ad-bs1}
\end{align}
and 
\begin{equation}
    h = h_{1,2} \,x^{3/2}
	= \frac{(b_{z1,2}^2 + b_{\phi 1,2}^2)}{\sqrt{2m_c^3}}
	\left[-1 
	+ \sqrt{1+\frac{4m_c^2}{(b_{z1,2}^2+b_{\phi 1,2}^2)^2}}\,\right]x^{3/2};
\label{ad-bs}
\end{equation}
the induction equation (\ref{ad-insimp}) then becomes
\begin{equation}
    \left(\tilde{\eta}_Ab_{z1,2}^2 
	+ \tilde{\eta}_Hb_{\phi 1,2}\sqrt{2b_{z1,2}^2 + b_{\phi 1,2}^2}\,
    \right)h_{1,2}^{1/2} = 2^{-1/4}m_c^{5/4}.
\label{ad-inreallysimp}
\end{equation}

The first of the two solutions is defined by $b_{\phi,s} = -b_{\phi1}\,x^{-1} = 
-|\tilde{\eta}_H|b/\tilde{\eta}_A$ which is solved to give
\begin{equation}
    b_{\phi,s} = - |\tilde{\eta}_H|b_{z}\sqrt{\frac{2}{\tilde{\eta}_A^2 -
	\tilde{\eta}_H^2}}\,.
\label{ad-bphis1}
\end{equation}
In this case
\begin{equation}
    b = b_{z1}\,x^{-1}\sqrt{\frac{2\tilde{\eta}_A^2}{\tilde{\eta}_A^2 -
	\tilde{\eta}_H^2}}\,;
\label{ad-b1}
\end{equation}
the coefficient of $h$ is given by 
\begin{equation}
     h_1 = \frac{f_1b_{z1}^2}{\sqrt{2m_c^3}}\left[-1+
	\sqrt{1+\frac{4m_c^2}{f_1^2b_{z1}^4}}\,\right],
\label{ad-h1}
\end{equation}
where the diffusion parameter $f_1$ is defined as
\begin{equation}
    f_1 = \frac{\tilde{\eta}_A^2 + \tilde{\eta}_H^2}
	{\tilde{\eta}_A^2 - \tilde{\eta}_H^2};
\label{ad-f1}
\end{equation}
the angular momentum coefficient, $j_1$, is determined by
\begin{equation}
    j_1 = \frac{b_{z1}}{m_c}\,\sqrt{\frac{2\tilde{\eta}_H^2}{\tilde{\eta}_A^2 -
	\tilde{\eta}_H^2}};
\label{ad-j1}
\end{equation}
and $b_{z1}$ is obtained by substituting these into Equation
\ref{ad-inreallysimp} and finding the single positive real root of the
polynomial
\begin{equation}
    b_{z1}^8 - \frac{m_c^2}{2\tilde{\eta}_A^2f_1}\,b_{z1}^6 -
	\frac{m_c^6}{4\tilde{\eta}_A^4f_1^4} = 0.
\label{ad-bz1}
\end{equation}
This similarity solution applies when the inequality 
\begin{equation}
    \sqrt{\frac{2\tilde{\eta}_H^2}{\tilde{\eta}_A^2 - \tilde{\eta}_H^2}} < \delta
\label{ad-bz1ineq}
\end{equation}
is satisfied, and corresponds to the asymptotic inner solution for the strong
braking case presented by \citet{kk2002} in the limit of pure ambipolar
diffusion. In their solution $\tilde{\eta}_H = b_{\phi,s} = 0$ and there is
effectively no angular momentum, although as can be seen in Figure
\ref{fig-kk10}, the angular momentum $j$ actually settles to a small
asymptotic plateau value. This particular solution applies in their
calculations when the magnetic braking parameter $\alpha$ is large, although
the parameter itself does not appear in any of the coefficients describing the
asymptotic power law solution.

The second solution exists when the other value of $b_{\phi,s}$ is chosen,
that is, when 
\begin{equation}
    b_{\phi,s} = -b_{\phi2}\,x^{-1} = -\delta \,b_z.
\label{ad-bphis2}
\end{equation}
In this case the coefficient of $h$ is determined by the equation
\begin{equation}
    h_2 = \frac{(1+\delta^2)b^2_{z2}}{\sqrt{2m_c^3}}
	\left[-1 + \sqrt{1 + \frac{4m_c^2}{(1+\delta^2)^2b_{z2}^4}}\,\right];
\label{ad-h2}
\end{equation}
the coefficient $j_2$ is simply
\begin{equation}
    j_2 = \frac{\delta b_{z2}^2}{m_c};
\label{ad-j2}
\end{equation}
and $b_{z2}$ is the single positive real root of the equation
\begin{equation}
    b_{z2}^8 - \frac{m_c^2(1+\delta^2)}{2f_2^2}\,b_{z2}^6 - \frac{m_c^6}{4f_2^4}
	= 0 
\label{ad-bz2}
\end{equation}
where $f_2$ is given by
\begin{equation}
    f_2 = \tilde{\eta}_A - \tilde{\eta}_H\delta\sqrt{2+\delta^2}.
\label{ad-f2}
\end{equation}
This particular similarity solution has no corresponding solution in the
results of \citet{kk2002} and is unique to this work. Both similarity
solutions represent a flow of matter onto the protostar at near-free fall
rates with little rotational momentum; these solutions are representative of
the magnetic braking catastrophe in star formation, and are discussed in more
detail in Section \ref{ff}.

\section{Region BC}\label{sub:BC}

This section of the $pq$-plane is painted pink in the upper right of Figure
\ref{pq-plane2} and defined mathematically by the inequalities
\begin{align}
    p &< 2q-2,\label{pq-bc1}\\
    p &> q,\label{pq-bc2}\\
    \text{and }p,q &> 2.\label{pq-bc3}
\end{align}
In this region $\hat{m}_c = -x^3b_{r,s}(db_z/dx)/\sigma$ and $\hat{\sigma} =
\sigma$, so that the term that determines the scaling of $h$ is given by
\begin{equation}
    \frac{8\hat{m}_c}{\hat{\sigma}^2x^3} \sim
      \frac{x^{2+p-2q}}{x^{3-2p}} \sim x^{3p-2q-1}.
\label{caseb-bc}
\end{equation}
In this region of the plane, case b (restricting $h$ to only physical
solutions) is defined by the inequality 
\begin{equation}
    3p-2q-1 < 0,
\label{caseb-ineq-bc}
\end{equation}
however, using the inequalities in \ref{pq-bc1}--\ref{pq-bc3} it can be seen
that 
\begin{equation}
	p - q > 0,
\label{bc-1}
\end{equation}
so that
\begin{equation}
	3p - 2q > q,
\label{bc-2}
\end{equation}
and
\begin{equation}
	3p - 2q - 1 > q - 1 > 1,
\label{bc-3}
\end{equation}
which is a clear contradiction of \ref{caseb-ineq-bc}. In this region then the
only similarity solutions that may exist are those in which $h$ very small and
may be unphysical.
The boundary case where $3p - 2q - 1 = 0$ is also unsatisfied in this region,
and so there can be no physical similarity solutions to the disc equations in
region BC of the $pq$-plane.

\section{Region BD}\label{sub:BD}

The large purple region at the lower right of Figure \ref{pq-plane2} is region
BD, which is characterised by the inequalities
\begin{align}
    p &< 2q-2\label{pq-bd1}\\
   \text{and } p &< q,\label{pq-bd2}
\end{align}
and has $\hat{m}_c = -x^3b_{r,s}(db_z/dx)/\sigma$ and $\hat{\sigma} =
(b_{r,s}^2 + b_{\phi,s}^2)/\sigma$. The term that determines the scaling of
$h$ with $x$ goes as
\begin{equation}
    \frac{8\hat{m}_c}{\hat{\sigma}^2x^3} \sim
      \frac{x^{2+p-2q}}{x^3(x^{p-2q})^2} \sim x^{2q -p -1},
\label{caseb-bd}
\end{equation}
so that in this region the desired case b for the scaling of $h$ applies when
the inequality 
\begin{equation}
    2q -p -1 \le 0
\label{caseb-ineqbd}
\end{equation}
is satisfied. However, as in region BC, the inequalities that define the
region (Equations \ref{pq-bd1}--\ref{pq-bd2}) can be rearranged to show
\begin{equation}
    0 < 2q -p -2.
\label{bd-1}
\end{equation}
This implies that
\begin{equation}
    1 < 2q -p -1,
\label{bd-2}
\end{equation}
which is a contradiction of \ref{caseb-ineqbd}, and so case b, and the
boundary case between cases a and b, cannot exist in this region. Therefore
there is no physical similarity solution to the fluid equations in region BD
of the $pq$-plane. 

\cleardoublepage
\chapter{Parameters and Shock Positions}\label{ch:params}

In order to assist any future researchers in this area who wish to duplicate
the results of this work, the shock positions and values of the variables at
the matching point are listed in this appendix for all of the similarity
solutions illustrated in Chapters \ref{ch:nohall} and \ref{ch:hall} and
Appendix \ref{ch:extrasols}. The centrifugal shock position and the
nondimensional central mass for the nonmagnetic and ideal MHD similarity
solutions that were calculated in Sections \ref{nonmag} and \ref{imhd} (which
were simply integrated inwards from the outer boundary) are given in Table
\ref{tab-simpshocks}. 

The positions of the centrifugal and magnetic diffusion shocks, as well as
the location of any subshocks that occur downstream of these, for the Hall
similarity solutions with nondimensional Hall parameter $\tilde{\eta}_H \in
[-0.2, 0.5]$ are presented in Table \ref{tab-shocks}. The converged (or
near-converged) values of the variables at the matching point $x_m$, and the
nondimensional central mass $m_c$, of the same solutions are listed in Table
\ref{tab-xm}. 

Unless otherwise indicated, all of the similarity solutions with both Hall and
ambipolar diffusion listed in Table \ref{tab-shocks} have boundary conditions
and parameters matching those in Table \ref{tab-bc}. The parameters for the
nonmagnetic and ideal MHD solutions are listed in the captions of their plots,
which are referenced in Table \ref{tab-simpshocks}. 

\begin{table}[ht]
\vspace{5mm}
\begin{center}
\begin{tabular}{ccccll}
\toprule
 Figure & $v_0$ & $\mu_0$ & $\alpha$ & $x_c$ & $m_c$\\
 \midrule
 \ref{fig-nmslow}  	 & 0.1 & N/A & N/A  & 7.65603527 $\times 10^{-3}$ & 0 \\
 \ref{fig-nmfast}  	 & 1.0 & N/A & N/A  & 6.31438364 $\times 10^{-1}$ & 0 \\
 \ref{fig-imhdslow}	 & 0.1 & 2.9 & 0.1  & 4.93181873 $\times 10^{-3}$ & 6.150 \\
 \ref{fig-imhdfast}	 & 1.5 & 2.9 & 0.1  & 1.40017396    & 0.625 \\
 \ref{fig-imhdfastalpha} & 1.5 & 2.9 & 0.01 & 1.59239598    & 0.050 \\
 \bottomrule
 \end{tabular}
\end{center}
 \caption[Shock position in the nonmagnetic and IMHD similarity solutions]
{Positions of the centrifugal shock and the central mass in the nonmagnetic
and IMHD similarity solutions presented in Sections \ref{nonmag} and
\ref{imhd}.} 
\label{tab-simpshocks}
\end{table}

\begin{table}[ht]
\vspace{5mm}
\begin{center}
\begin{tabular}{lllll}
\toprule
 $\tilde{\eta}_H$ & $x_d$ & $x_{d2}$ & $x_c$ & $x_{c2}$\\
 \midrule
 $-0.5   $& 0.5571562272 & -- & 0.08313609521 & -- \\
 $-0.4   $& 0.5240689226 & -- & 0.07022194425 & -- \\
 $-0.3   $& 0.4914751233 & -- & 0.05470475621 & -- \\
 $-0.2   $& 0.4609084538 & -- & 0.03772972380 & -- \\
 $-0.1   $& 0.4322428394 & -- & 0.02270585875 & -- \\
 $-0.01  $& 0.4086267555 & -- & 0.01391928200 & -- \\
 $-0.001 $& 0.4053600464 & -- & 0.01331861020 & -- \\
 $ 0     $& 0.4053600464 & -- & 0.01325501322 & -- \\
 $ 0.001 $& 0.4053600464 & -- & 0.01319153733 & -- \\
 $ 0.01  $& 0.4037364982 & -- & 0.01264383282 & -- \\
 $ 0.1   $& 0.3832127348 & 0.2359118494 & 0.008883741072 & -- \\
 $ 0.2   $& 0.3651949666 & 0.2600572227 & 0.006052121453 & 0.005212123123 \\
 $-0.105$*& 0.4499425849 & -- & 0.03155084518 & -- \\
 \bottomrule
 \end{tabular}
\end{center}
\vspace{-5mm}
 \caption[Shock positions in the Hall diffusion similarity solutions]
{Positions of the magnetic diffusion and centrifugal shocks, and any subshocks
that may occur downstream of these in the set of converged similarity
solutions with parameters and initial conditions equal to those given in Table
\ref{tab-bc}. Those solutions that were not discussed in Chapter \ref{ch:hall}
are illustrated without comment in Appendix \ref{ch:extrasols}. 

*The similarity solution with $\tilde{\eta}_H = -0.105$ has the
nondimensional ambipolar diffusion parameter $\tilde{\eta}_A = 1.05$, to
explore the dependence of $x_c$ and $x_d$ upon the ratio of the magnetic
diffusion parameters as discussed in Section \ref{future}. The other
parameters remain unchanged.} 
\label{tab-shocks}
\end{table}

\begin{sidewaystable}[ht]
\vspace{5mm}
\begin{center}
\begin{tabular}{llllllll}
\toprule
 $\tilde{\eta}_H$ & $x_m$ & $m(x_m)$ & $\sigma(x_m)$ & $j(x_m)$ & $\psi(x_m)$
& $b_z(x_m)$ & $m_c$\\
 \midrule
 $-0.5$*  & 0.32 & 5.662680151 & 6.618598505 & 1.050574060 & 1.124110348 &
9.826278713 & 3.769000000 \\
 $-0.4  $ & 0.3   & 5.621615058 & 7.353687835 & 1.035288751 & 1.116077330 &
11.11475226 & 4.003813245 \\ 
 $-0.3  $ & 0.3   & 5.623968680 & 7.669114095 & 1.055246418 & 1.175236751 &
11.78779585 & 4.224377980 \\ 
 $-0.2  $ & 0.3   & 5.627639776 & 7.969005835 & 1.079020941 & 1.239078304 &
12.45420198 & 4.424377980 \\
 $-0.1  $ & 0.3   & 5.632885457 & 8.204312082 & 1.106920520 & 1.308610399 &
13.09647510 & 4.584727813 \\ 
 $-0.01 $ & 0.3   & 5.638815298 & 8.301889633 & 1.135502323 & 1.375848755 &
13.63201678 & 4.665173497 \\ 
 $-0.001$ & 0.3   & 5.639438769 & 8.303864989 & 1.138496344 & 1.382679558 &
13.68253865 & 4.669442792 \\
 $  0   $ & 0.3   & 5.639508465 & 8.303988767 & 1.138830374 & 1.383440138 &
13.68812340 & 4.669966724 \\
 $ 0.001$ & 0.3   & 5.639576935 & 8.304102170 & 1.139163663 & 1.384195474 &
13.69365766 & 4.670171632 \\
 $ 0.01 $ & 0.3   & 5.640201745 & 8.304254061 & 1.142177131 & 1.391024425 &
13.74343547 & 4.674097803 \\
 $ 0.1  $ & 0.3   & 5.645909658 & 8.235961519 & 1.172119218 & 1.456191209 &
14.19615046 & 4.672541054 \\
 $ 0.2  $ & 0.3   & 5.650117515 & 8.073367472 & 1.202126470 & 1.516319914 &
14.56296008 & 4.625991391 \\
 $-0.105$** & 0.3 & 5.630104509 & 8.102443350 & 1.104424953 & 1.271312260 &
12.51898858 & 4.480831780 \\
 \bottomrule
 \end{tabular}
\end{center}
\vspace{-5mm}
 \caption[Variables at $x_m$ for the Hall diffusion similarity solutions]
{Values of the variables at the matching point $x_m$, and the central mass
$m_c$ to which the code converged in the calculations for similarity solutions
with $\tilde{\eta}_H \in [-0.5,0.2]$ and $\tilde{\eta}_A = 1.0$.

*The similarity solution with $\tilde{\eta}_H = -0.5$ has a changed matching
point $x_m = 0.32$, as $x = 0.3$ was too close to the local maximum in
$\sigma$ to allow convergence on the boundary conditions. 

**The similarity solution with $\tilde{\eta}_H = -0.105$ has $\tilde{\eta}_A =
1.05$. All the other parameters are unchanged.}
\label{tab-xm}
\end{sidewaystable}

\cleardoublepage
\chapter{Additional Similarity Solutions}\label{ch:extrasols}

More similarity solutions were calculated in this project than could be
presented in the thesis proper. This appendix contains plots of those
similarity solutions listed in Tables \ref{tab-shocks} and \ref{tab-xm}, for
the purpose of highlighting trends such as the increasing disc size with
decreasing Hall parameter and the appearance of subshocks. 

As in Appendix \ref{ch:params}, all of the similarity solutions match the
outer boundary conditions and collapse parameters that were listed in Table 
\ref{tab-bc} unless otherwise noted.

\begin{figure}[htp]
  \centering
  \includegraphics[width=5.2in]{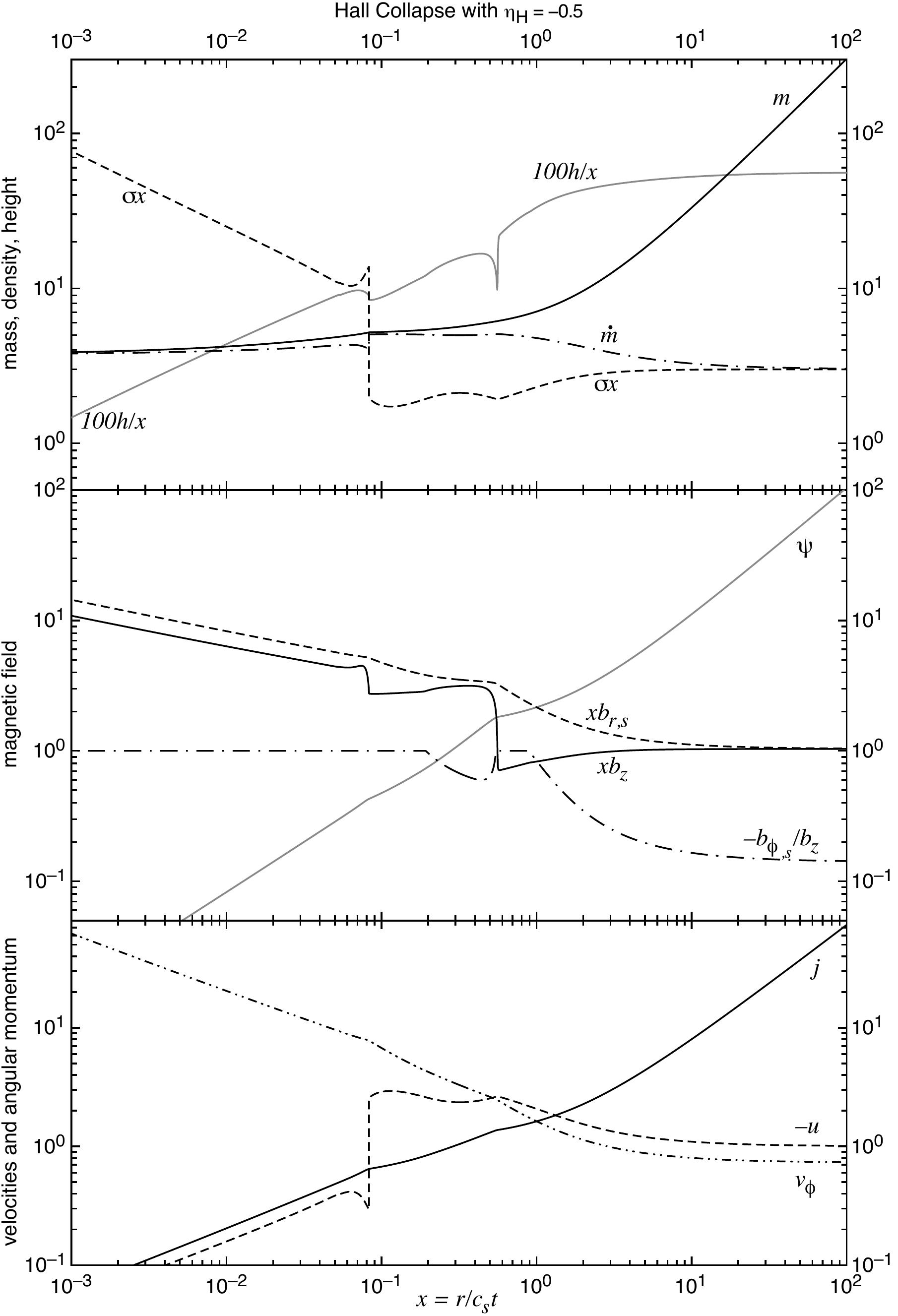}
  \caption[Hall and ambipolar diffusion collapse with $\tilde{\eta}_H = -0.5$
]{Similarity solution for Hall collapse with $\tilde{\eta}_H = -0.5$. The
shock positions and central mass are as listed in Appendix \ref{ch:params}.}   
\label{fig-hall-0.5b}
\end{figure}
\begin{figure}[htp]
  \centering
  \includegraphics[width=5.2in]{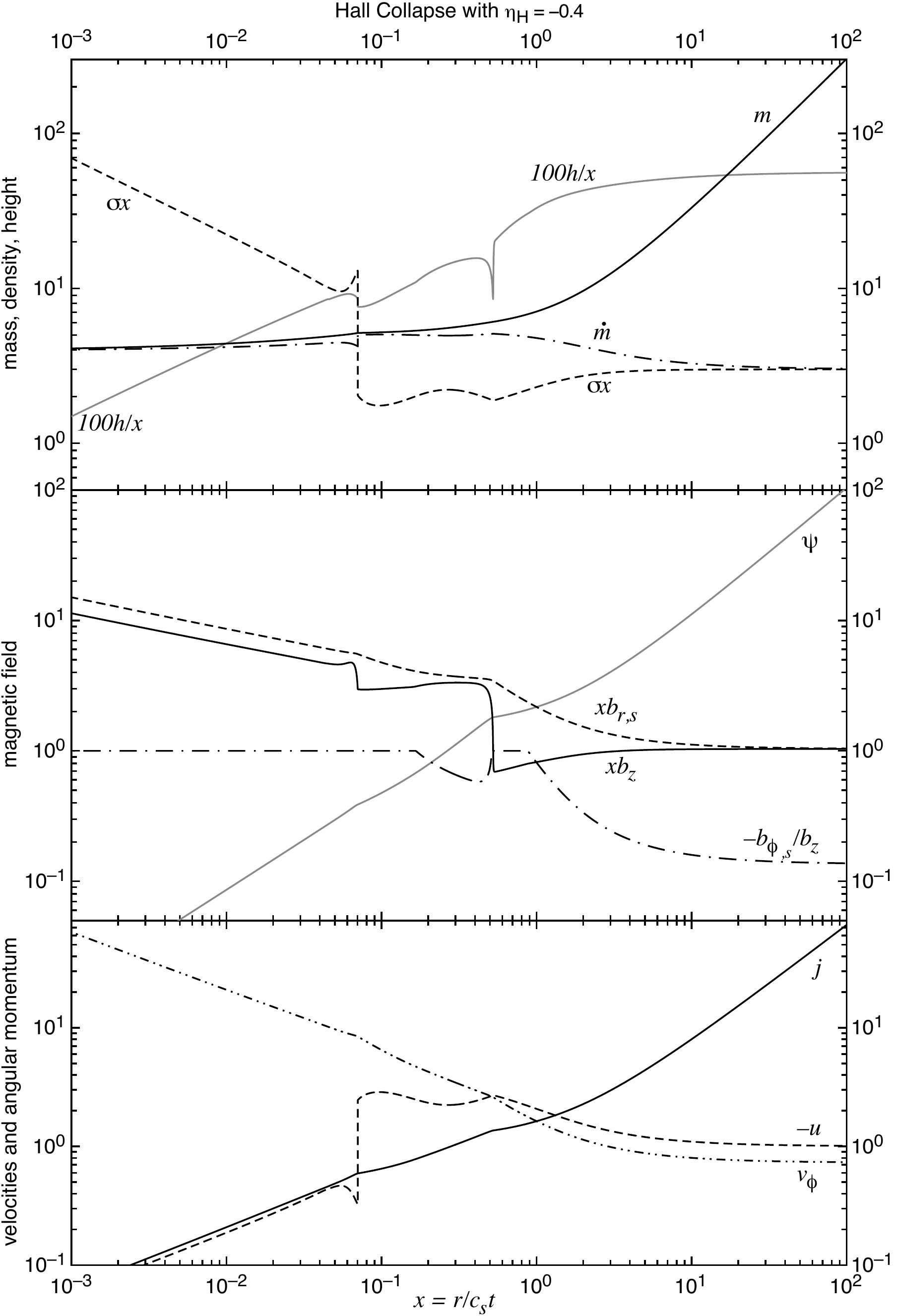}
  \caption[Hall and ambipolar diffusion collapse with $\tilde{\eta}_H = -0.4$
]{Similarity solution for Hall collapse with $\tilde{\eta}_H = -0.4$. The
shock positions and central mass are as listed in Appendix \ref{ch:params}.}   
\label{fig-hall-0.4}
\end{figure}
\begin{figure}[htp]
  \centering
  \includegraphics[width=5.2in]{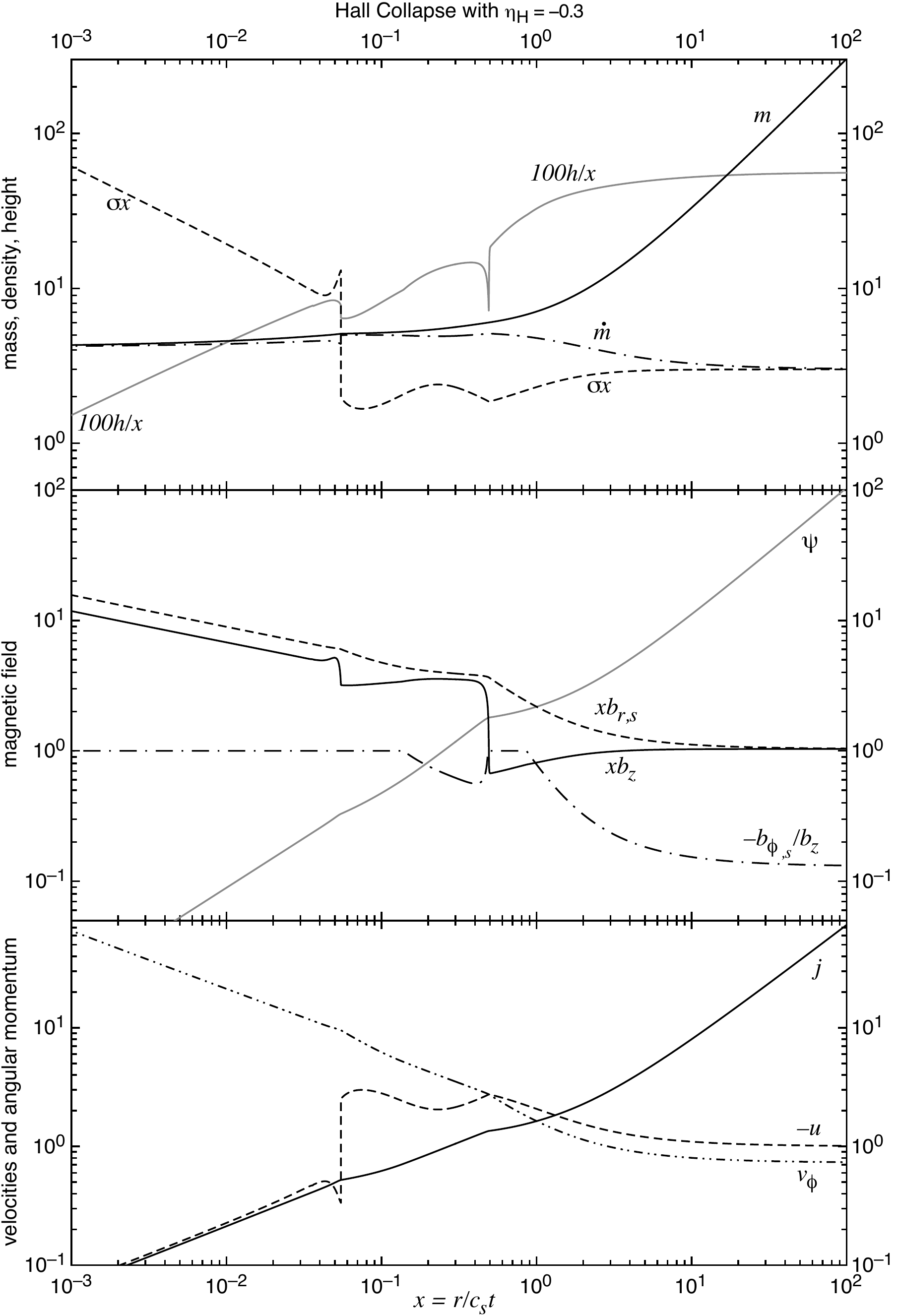}
  \caption[Hall and ambipolar diffusion collapse with $\tilde{\eta}_H = -0.3$
]{Similarity solution for Hall collapse with $\tilde{\eta}_H = -0.3$. The
shock positions and central mass are as listed in Appendix \ref{ch:params}.}   
\label{fig-hall-0.3}
\end{figure}
\begin{figure}[htp]
  \centering
  \includegraphics[width=5.2in]{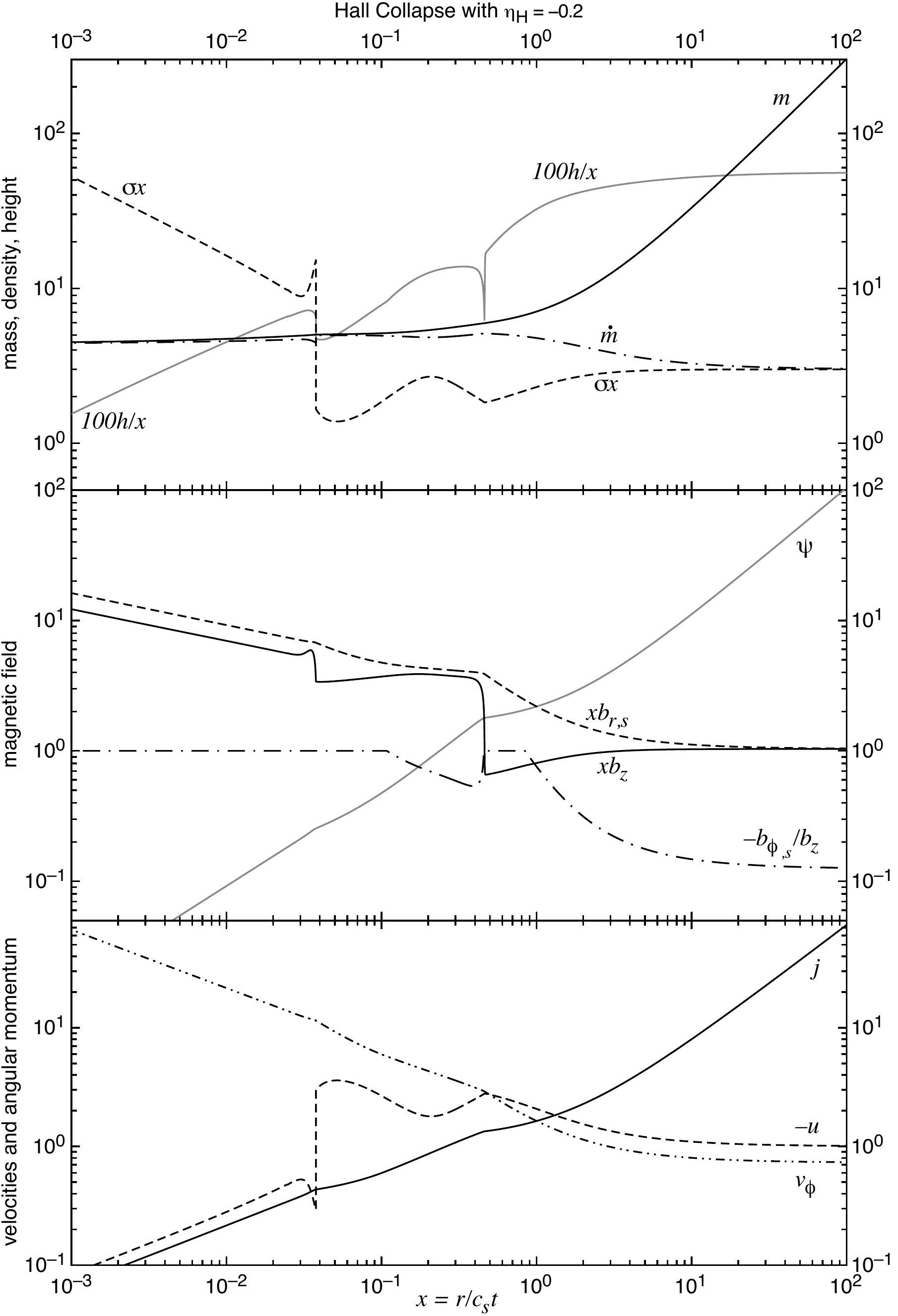}
  \caption[Hall and ambipolar diffusion collapse with $\tilde{\eta}_H = -0.2$
]{Similarity solution for Hall collapse with $\tilde{\eta}_H = -0.2$. The
shock positions and central mass are as listed in Appendix \ref{ch:params}.}   
\label{fig-hall-0.2b}
\end{figure}
\begin{figure}[htp]
  \centering
  \includegraphics[width=5.2in]{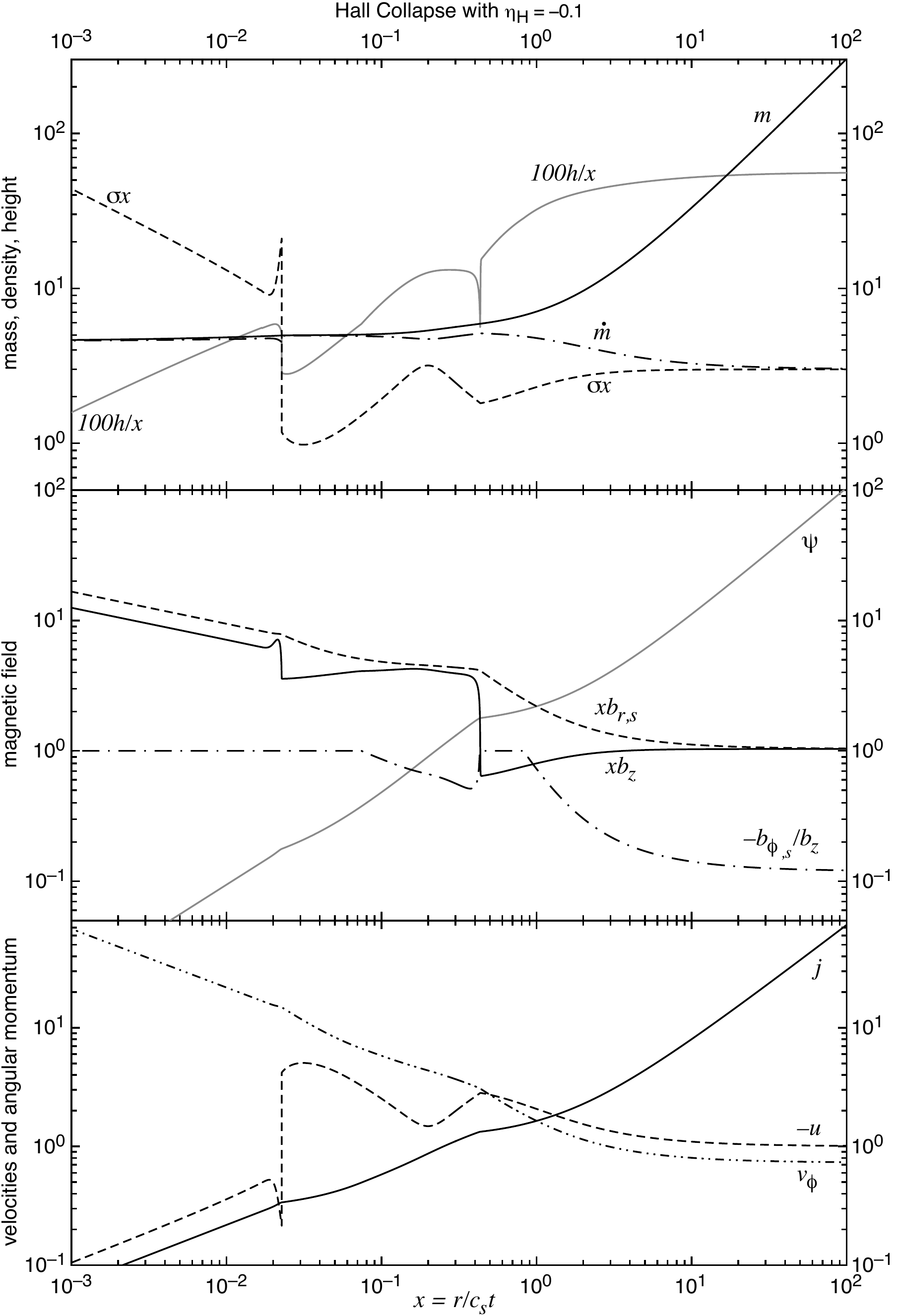}
  \caption[Hall and ambipolar diffusion collapse with $\tilde{\eta}_H = -0.1$
]{Similarity solution for Hall collapse with $\tilde{\eta}_H = -0.1$. The
shock positions and central mass are as listed in Appendix \ref{ch:params}.}   
\label{fig-hall-0.1}
\end{figure}
\begin{figure}[htp]
  \centering
  \includegraphics[width=5.2in]{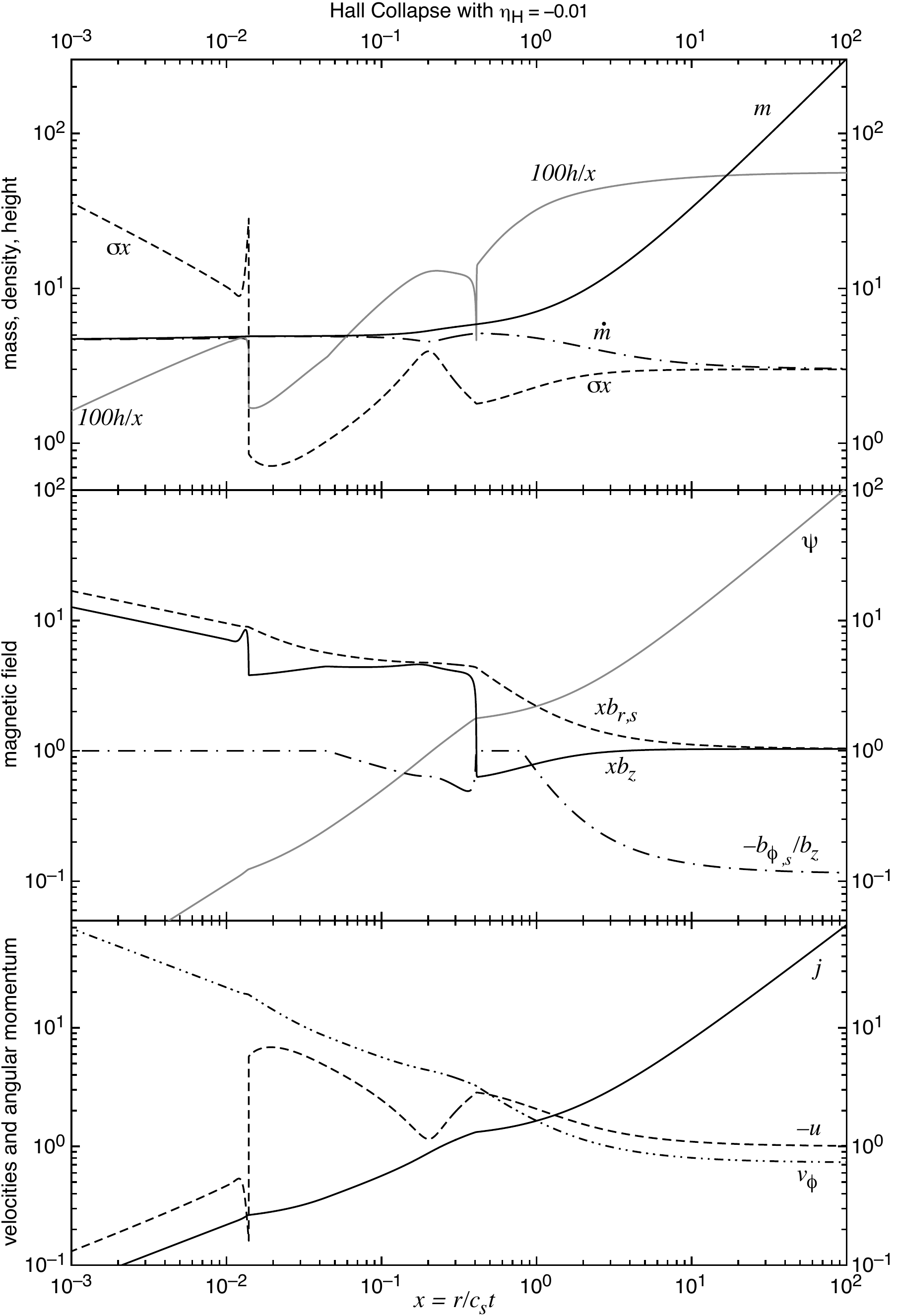}
  \caption[Hall and ambipolar diffusion collapse with $\tilde{\eta}_H = -0.01$
]{Similarity solution for Hall collapse with $\tilde{\eta}_H = -0.01$. The
shock positions and central mass are as listed in Appendix \ref{ch:params}.}   
\label{fig-hall-0.01}
\end{figure}
\begin{figure}[htp]
  \centering
  \includegraphics[width=5.2in]{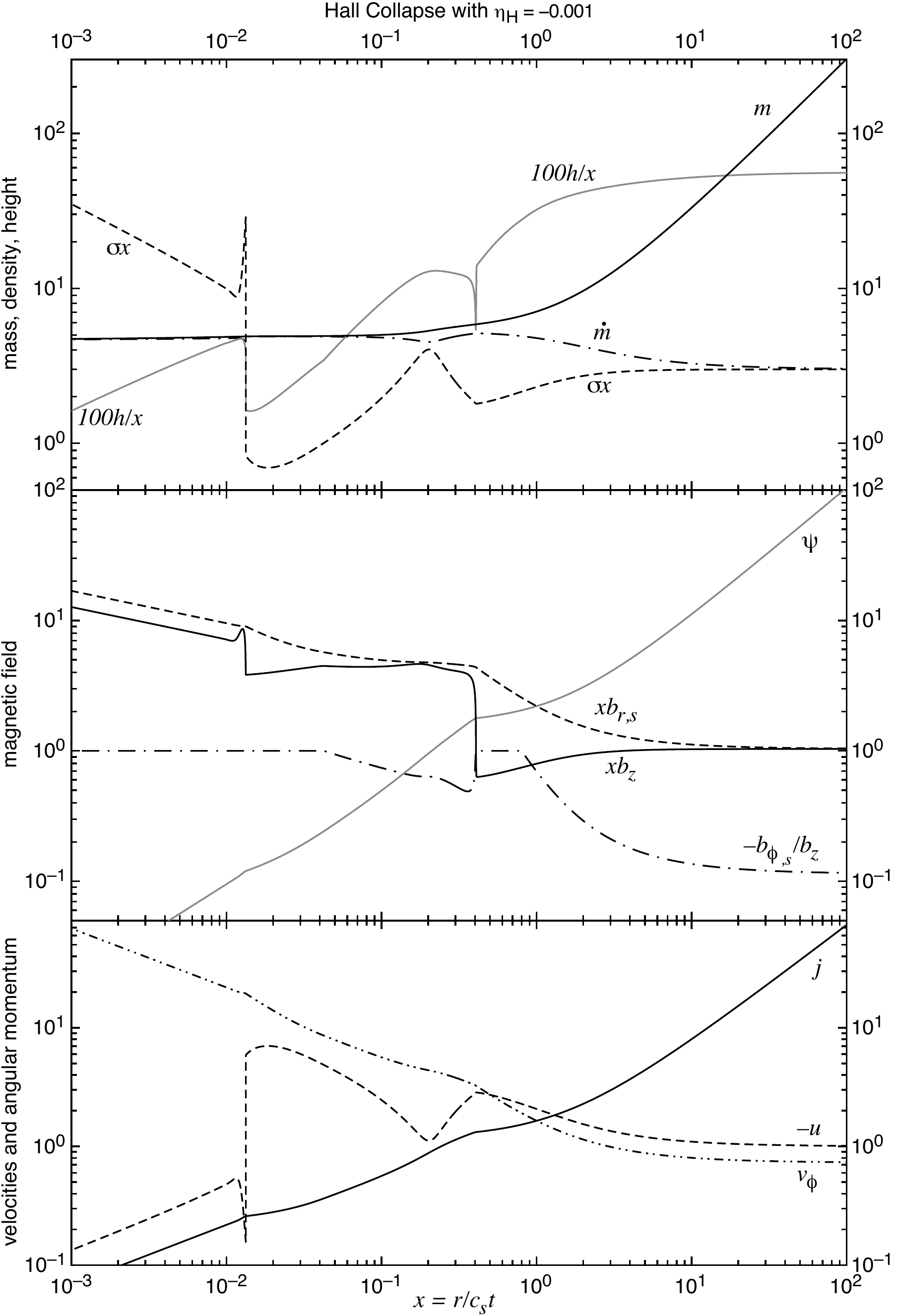}
  \caption[Hall and ambipolar diffusion collapse with $\tilde{\eta}_H = -0.001$
]{Similarity solution for Hall collapse with $\tilde{\eta}_H = -0.001$. The
shock positions and central mass are as listed in Appendix \ref{ch:params}.}   
\label{fig-hall-0.001}
\end{figure}
\begin{figure}[htp]
  \centering
  \includegraphics[width=5.2in]{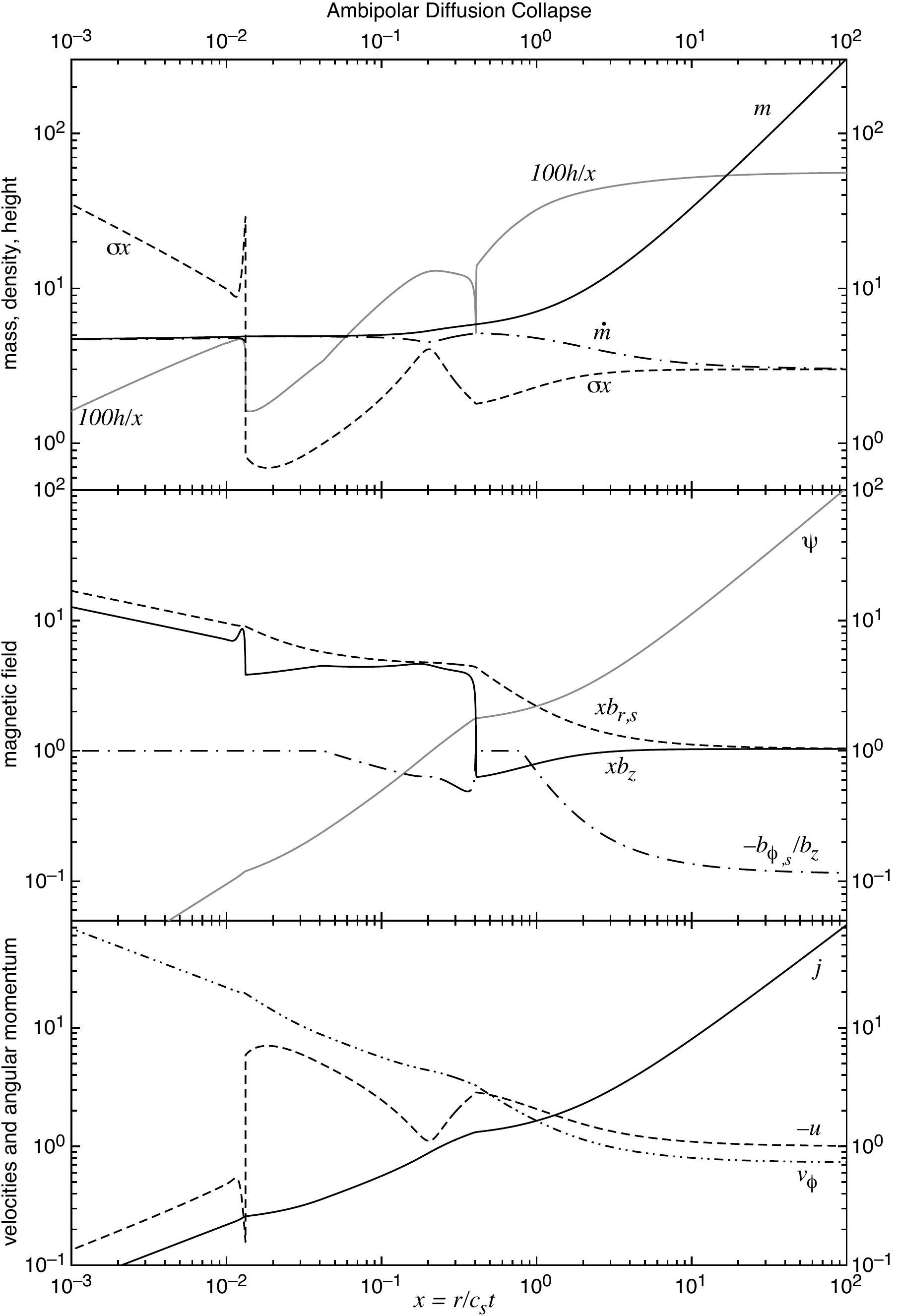}
  \caption[Ambipolar diffusion collapse with $\tilde{\eta}_H = 0$
]{Similarity solution for ambipolar diffusion-only collapse with
$\tilde{\eta}_H = 0$. The shock positions and central mass are as listed in
Appendix \ref{ch:params}.} 
\label{fig-hall0}
\end{figure}
\begin{figure}[htp]
  \centering
  \includegraphics[width=5.2in]{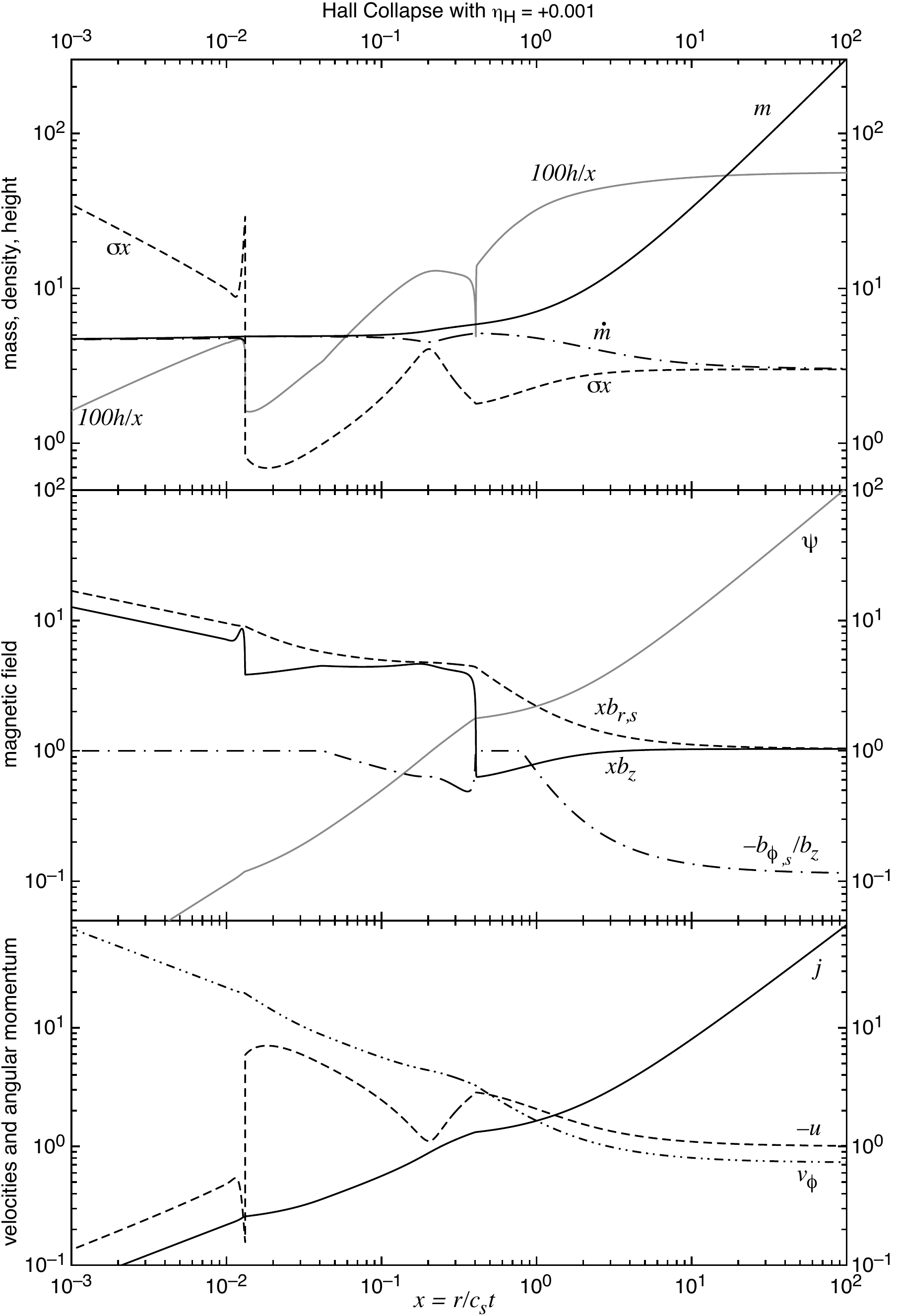}
  \caption[Hall and ambipolar diffusion collapse with $\tilde{\eta}_H = 0.001$
]{Similarity solution for Hall collapse with $\tilde{\eta}_H = 0.001$. The
shock positions and central mass are as listed in Appendix \ref{ch:params}.} 
\label{fig-hall0.001}
\end{figure}
\begin{figure}[htp]
  \centering
  \includegraphics[width=5.2in]{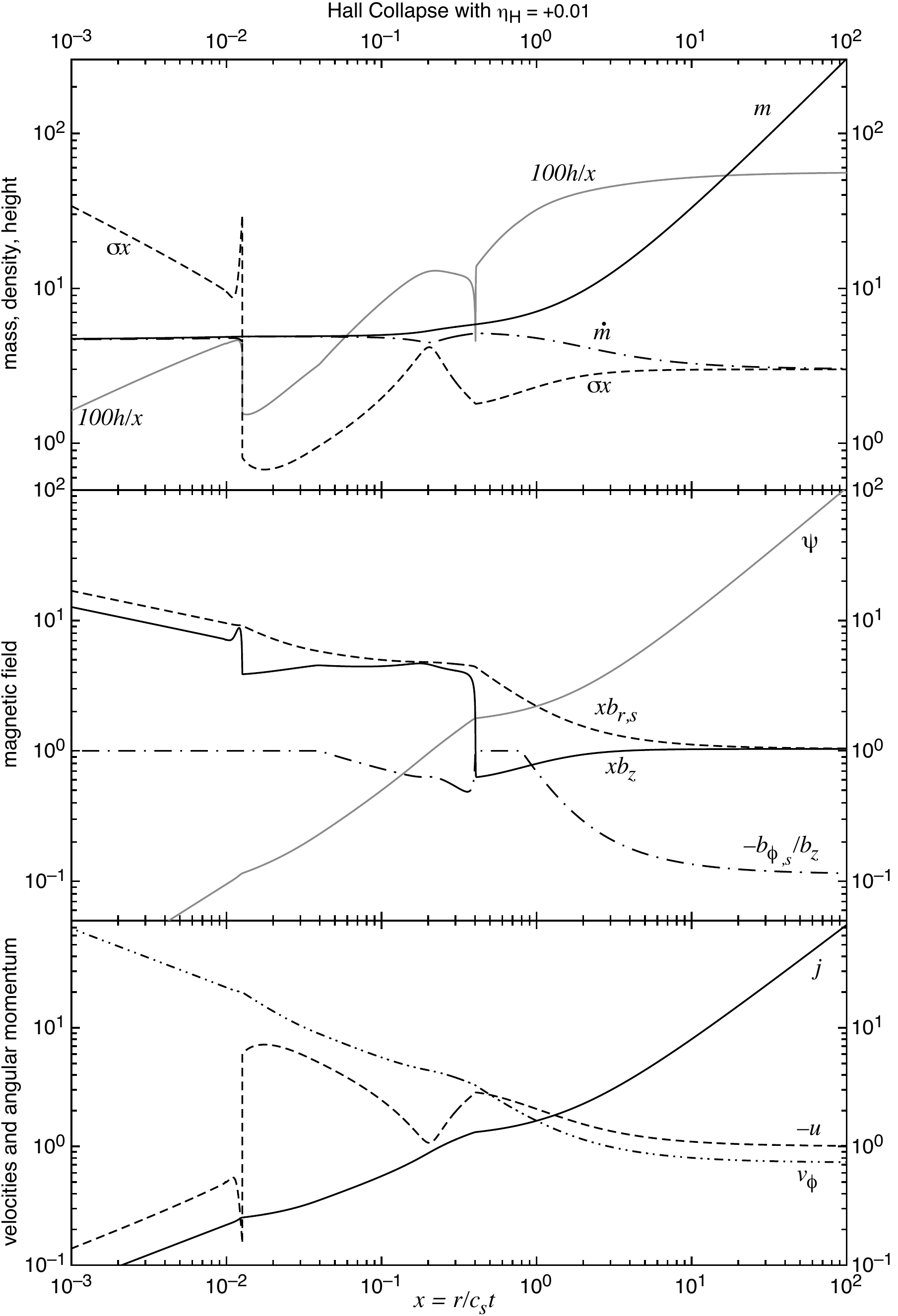}
  \caption[Hall and ambipolar diffusion collapse with $\tilde{\eta}_H = 0.01$
]{Similarity solution for Hall collapse with $\tilde{\eta}_H = 0.01$. The
shock positions and central mass are as listed in Appendix \ref{ch:params}.} 
\label{fig-hall0.01}
\end{figure}
\begin{figure}[htp]
  \centering
  \includegraphics[width=5.2in]{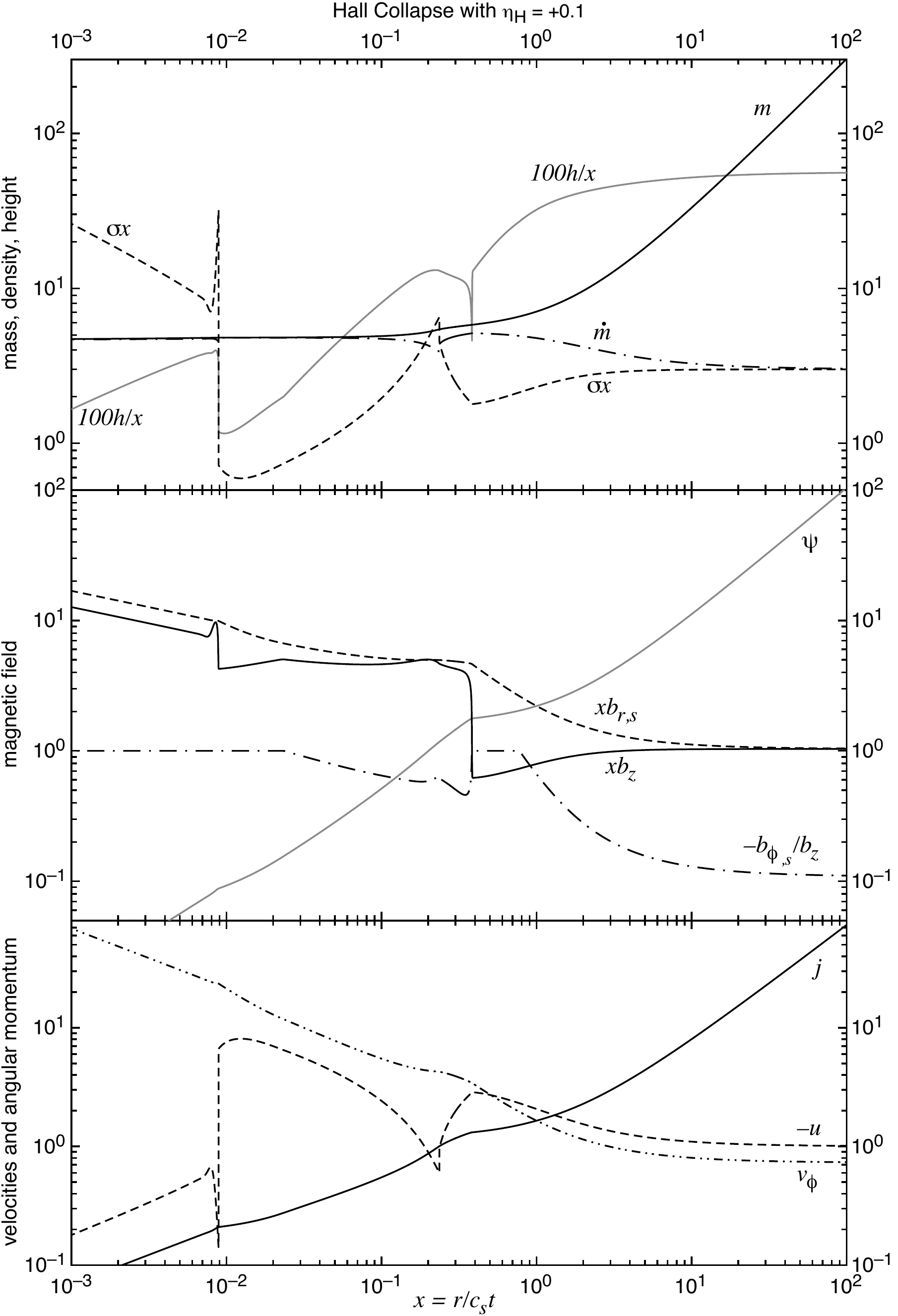}
  \caption[Hall and ambipolar diffusion collapse with $\tilde{\eta}_H = 0.1$
]{Similarity solution for Hall collapse with $\tilde{\eta}_H = 0.1$. The
shock positions and central mass are as listed in Appendix \ref{ch:params}.} 
\label{fig-hall0.1}
\end{figure}
\begin{figure}[htp]
  \centering
  \includegraphics[width=5.2in]{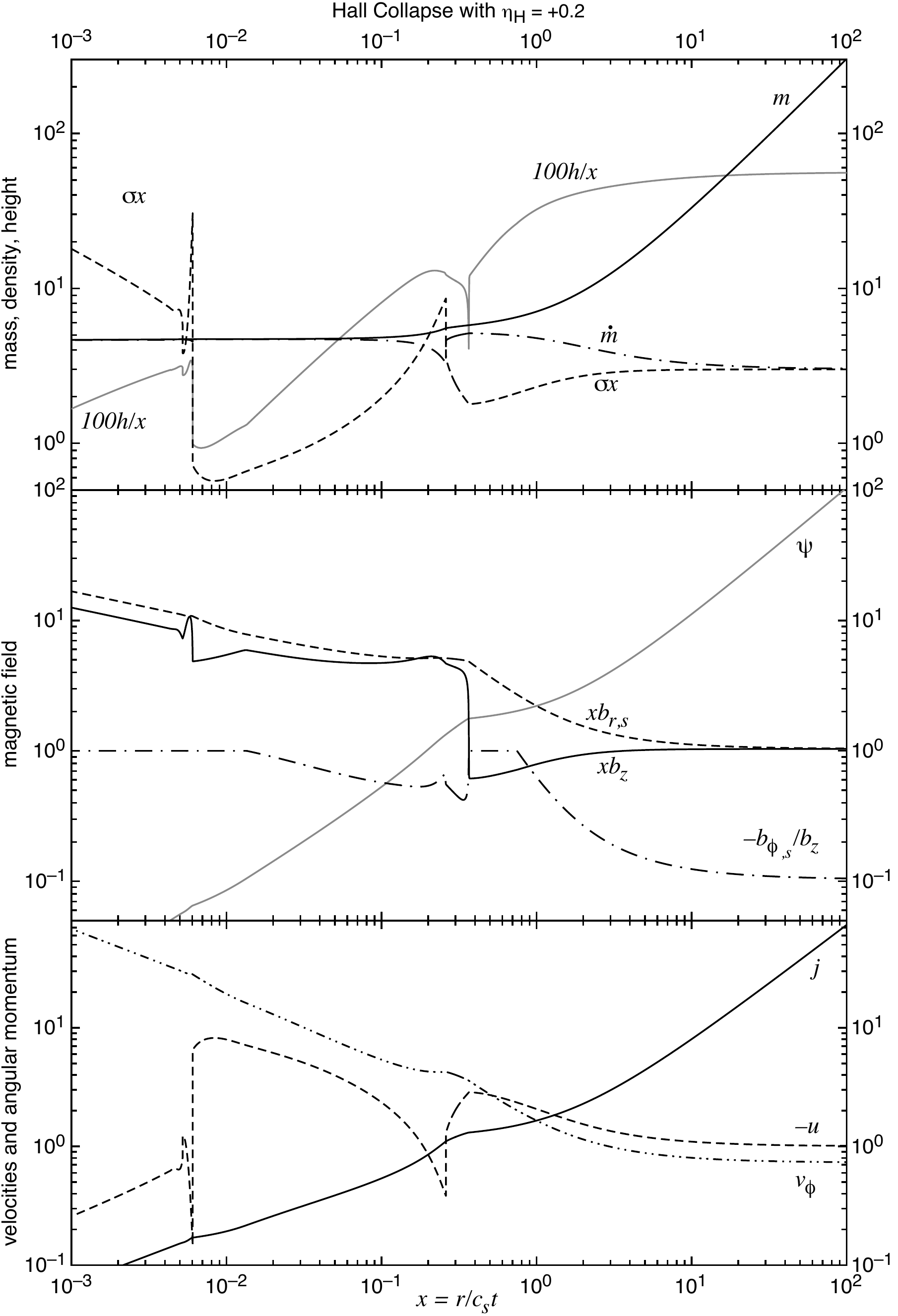}
  \caption[Hall and ambipolar diffusion collapse with $\tilde{\eta}_H = 0.2$
]{Similarity solution for Hall collapse with $\tilde{\eta}_H = 0.2$. The
shock positions and central mass are as listed in Appendix \ref{ch:params}.} 
\label{fig-hall0.2b}
\end{figure}


%
%


\end{document}